\documentclass[longauth]{aa}

\usepackage[varg]{txfonts}
\usepackage{placeins}
\usepackage{natbib}
\usepackage{ae,aecompl}
\usepackage{graphicx}   \usepackage{amsmath}    \usepackage{amssymb}    \usepackage{siunitx}    \DeclareSIUnit{\degree}{deg} \DeclareSIUnit{\arcmin}{arcmin} \DeclareSIUnit{\arcsec}{arcsec} \usepackage{color,verbatim,url}
\usepackage{tabularx}
\usepackage[x11names]{xcolor}
\usepackage{booktabs}
\usepackage{hyperref}

\usepackage{breqn}

\usepackage{fontawesome}
\usepackage{bm}

\definecolor{pink}{rgb}{0.858, 0.188, 0.478}
\definecolor{purple}{RGB}{76, 0,153}
\definecolor{forest}{RGB}{13, 55,13}

\newcommand{\red}[1]{{\color{red}{#1}}}
\newcommand{\orange}[1]{{\color{orange}{#1}}}
\newcommand{\green}[1]{{\color{green}{#1}}}
\newcommand{\blue}[1]{{\color{blue}{#1}}}
\newcommand{\pink}[1]{{\color{pink}{#1}}}
\newcommand{\purple}[1]{{\color{purple}{#1}}}

\newcommand{\HHblindtext}[1]{{\purple{\blindtext}}}
\newcommand{\KKblindtext}[1]{{\blue{\blindtext}}}
\newcommand{\MBblindtext}[1]{{\green{\blindtext}}}
\newcommand{\FGblindtext}[1]{{\orange{\blindtext}}}
\newcommand{\MRblindtext}[1]{{\red{\blindtext}}}
\newcommand{\AWblindtext}[1]{{\pink{\blindtext}}}

\newcommand{\beps}{\bm{\varepsilon}}
\newcommand{\eps}{\varepsilon}

\newcommand{\uband}{{$u$-band}}
\newcommand{\gband}{{$g$-band}}
\newcommand{\rband}{{$r$-band}}
\newcommand{\iband}{{$i$-band}}
\newcommand{\ione}{{$i_1$}}
\newcommand{\itwo}{{$i_2$}}

\newcommand{\zband}{{$Z$-band}}
\newcommand{\yband}{{$Y$-band}}
\newcommand{\jband}{{$J$-band}}
\newcommand{\hband}{{$H$-band}}
\newcommand{\ksband}{{$K_{\rm s}$-band}}
\newcommand{\utor}{{ugr}}
\newcommand{\utoi}{{ugri_{1}i_{2}}}

\newcommand{\utok}{{ugri_{1}i_{2}ZY\!J\!H\!K_{\rm s}}}

\newcommand{\ztok}{{ZY\!J\!H\!K_{\rm s}}}

\newcommand{\kidz}{{KiDZ}}
\newcommand{\kids}{{KiDS}}

\newcommand{\kidslegacy}{{KiDS-Legacy}}

\newcommand{\vst}{{VST}}
\newcommand{\vista}{{VISTA}}
\newcommand{\vircam}{{VIRCAM}}
\newcommand{\ultravista}{{UltraVISTA}}
\newcommand{\video}{{VIDEO}}
\newcommand{\viking}{{VIKING}}
\newcommand{\euclid}{{\it Euclid}}
\newcommand{\gama}{{GAMA}}
\newcommand{\gaia}{{\textit{Gaia}}}
\newcommand{\sdss}{{SDSS}}
\newcommand{\twomass}{{2MASS}}
\newcommand{\tdflens}{{2dFLens}}
\newcommand{\wigglez}{{WiggleZ}}
\newcommand{\vipers}{VIPERS}
\newcommand{\vvds}{VVDS}
\newcommand{\ctrtwo}{C3R2}
\newcommand{\deeptwo}{DEEP2}
\newcommand{\zcosmos}{zCOSMOS}
\newcommand{\goods}{GOODS}
\newcommand{\gaap}{{\sc GAaP}}
\newcommand{\theliraw}{{\sc theli}}
\newcommand{\theli}{{\sc theli}-Lens}
\newcommand{\atlas}{{\sc theli}-Phot}
\newcommand{\astrowise}{{\sc AstroWISE}}
\newcommand{\kvpipe}{{\sc KVpipe}}
\newcommand{\photopipe}{{\sc PhotoPipe}}
\newcommand{\lensfit}{{{\it lens}fit}}
\newcommand{\sourceextractor}{{\sc Source Extractor}}
\newcommand{\lambdar}{{\sc lambdar}}
\newcommand{\swarp}{{\sc SWarp}}
\newcommand{\pulcinella}{{\sc Pulecenella}}

\newcommand{\nir}{{NIR}}
\newcommand{\casu}{{CASU}}
\newcommand{\bpz}{{BPZ}}
\newcommand{\fwhm}{{\rm FWHM}}
\newcommand{\mask}{{\rm MASK}}

\newcommand{\drfour}{{\sc dr4}}
\newcommand{\drfive}{{\sc dr5}}
\newcommand{\photoz}{{photo-$z$}}
\newcommand{\specz}{{spec-$z$}}
\newcommand{\zb}{{z_{\rm B}}}
\newcommand{\zspec}{{z_{\rm spec}}}

\newcommand{\kidskidz}{{\kids$+$\kidz}}
\newcommand{\kidsviking}{{\kids$+$\viking}}
\newcommand{\sqdeg}{{deg$^2$}}

\newcommand{\snr}{{\sc s/n}}
\newcommand{\fitclass}{{\tt fitclass}}
\newcommand{\slr}{{SLR}}
\newcommand{\zeropseven}{{$0\farcs7$}}
\newcommand{\onepzero}{{$1\farcs0$}}

\newcommand{\pawprint}{{paw-print}}
\newcommand{\pawprints}{{paw-prints}}
\newcommand{\blindtext}{\lipsum[1-1]}

\makeatletter
\newcommand{\github}[1]{\href{#1}{\faGithubSquare}}
\makeatother

\begin{document}

\nolinenumbers

\defcitealias{kuijken/etal:2019}{[DR4]}

\title{The fifth data release of the Kilo Degree Survey:\\ Multi-epoch optical/NIR imaging covering wide
and legacy-calibration fields}
\titlerunning{KiDS-DR5}

\author{ Angus~H.~Wright        \inst{1}\thanks{awright@astro.rub.de} \and 
Konrad~Kuijken  \inst{2} \and 
Hendrik~Hildebrandt \inst{1} \and 
Mario~Radovich  \inst{3} \and 
Maciej~Bilicki  \inst{4} \and 
Andrej~Dvornik  \inst{1} \and 
Fedor~Getman\inst{5} \and  
Catherine~Heymans       \inst{1,6} \and 
Henk~Hoekstra\inst{2} \and 
Shun-Sheng~Li   \inst{2} \and 
Lance~Miller    \inst{7} \and 
Nicola~R.~Napolitano    \inst{8} \and 
Qianli~Xia \inst{6} \and 
Marika~Asgari   \inst{9,10} \and 
Massimo~Brescia \inst{11} \and 
Hugo~Buddelmeijer       \inst{12} \and 
Pierre~Burger   \inst{13} \and 
Gianluca~Castignani     \inst{14,15} \and 
Stefano~Cavuoti \inst{5} \and 
Jelte~de~Jong   \inst{2,16} \and 
Alastair~Edge   \inst{17} \and 
Benjamin~Giblin \inst{6,18} \and 
Carlo~Giocoli   \inst{15,19} \and 
Joachim~Harnois-D\'eraps \inst{20} \and 
Priyanka~Jalan  \inst{4} \and 
Benjamin~Joachimi       \inst{21} \and 
Anjitha~John~William    \inst{4} \and 
Shahab~Joudaki  \inst{22,23} \and 
Arun~Kannawadi  \inst{24} \and 
Gursharanjit~Kaur       \inst{4} \and 
Francesco~La~Barbera    \inst{15} \and 
Laila~Linke     \inst{13,25} \and 
Constance~Mahony        \inst{1} \and 
Matteo~Maturi   \inst{26} \and 
Lauro~Moscardini        \inst{14,15,27} \and 
Szymon~J.~Nakoneczny    \inst{28,29} \and 
Maurizio~Paolillo       \inst{11} \and 
Lucas~Porth     \inst{13} \and 
Emanuella~Puddu \inst{5} \and 
Robert~Reischke \inst{1} \and 
Peter~Schneider \inst{13} \and 
Mauro~Sereno    \inst{15} \and 
HuanYuan~Shan   \inst{30,31} \and 
Crist\'obal~Sif\'on     \inst{32} \and 
Benjamin~St\"olzner     \inst{1} \and 
Tilman~Tr\"oster        \inst{33} \and 
Edwin~Valentijn \inst{16} \and 
Jan~Luca~van~den~Busch  \inst{1} \and 
Gijs~Verdoes~Kleijn     \inst{16} \and 
Anna~Wittje     \inst{1} \and 
Ziang~Yan       \inst{1} \and 
Ji~Yao  \inst{30} \and 
Mijin~Yoon      \inst{1} \and 
Yun-Hao~Zhang   \inst{2,6} 
}
\authorrunning{The KiDS Collaboration}
\institute{
        Ruhr University Bochum, Faculty of Physics and Astronomy, Astronomical Institute (AIRUB), German Centre for Cosmological Lensing, 44780 Bochum, Germany \and Leiden Observatory, Leiden University, Niels Bohrweg 2, 2333 CA Leiden, The Netherlands      \and Instituto Nazionale di Astrofisica (INAF) - Osservatorio Astronomico di Padova, via dell'Osservatorio 5, 35122 Padova, Italy   \and Center for Theoretical Physics, Polish Academy of Sciences, al. Lotników 32/46, 02-668 Warsaw, Poland      \and Instituto Nazionale di Astrofisica (INAF) - Osservatorio Astronomico di Capodimonte, Via Moiariello 16, I-80131 Napoli, Italy \and Institute for Astronomy, University of Edinburgh, Royal Observatory, Blackford Hill, Edinburgh, EH9 3HJ, UK. \and Dept of Physics, University of Oxford, Denys Wilkinson Building, Keble Road, Oxford OX1 3RH, U.K.    \and School of Physics and Astronomy, Sun Yat-sen University, Guangzhou 519082, Zhuhai Campus, P.R. China \and  E.A Milne Centre, University of Hull, Cottingham Road, Hull, HU6 7RX, United Kingdom \and Centre of Excellence for Data Science, AI, and Modelling (DAIM), University of Hull, Cottingham Road, Kingston-upon-Hull, HU6 7RX, United Kingdom \and Department of Physics E. Pacini, University of Napoli Federico II, Via Cintia 6, I-80126, Napoli, Italy \and Institut f\"ur Astrophysik, T\"urkenschanzstrasse 17, 1180 Wien, Austria \and Argelander-Institut f\"ur Astronomie, Auf dem H\"ugel 71, 53121 Bonn, Germany \and Dipartimento di Fisica e Astronomia ``Augusto Righi'', Alma Mater Studiorum Universit\'a di Bologna, Via Gobetti 93/2, I-40129 Bologna, Italy \and Istituto Nazionale di Astrofisica (INAF) - Osservatorio di Astrofisica e Scienza dello Spazio (OAS) di Bologna, via Piero Gobetti 93/3, I-40129, Bologna, Italy \and Kapteyn Astronomical Institute, University of Groningen, PO Box 800, 9700 AV Groningen, The Netherlands\and Centre for Extragalactic Astronomy, Department of Physics, Durham University, South Road, Durham DH1 3LE, UK \and Instituto de Ciencias del Cosmos (ICC), Universidad de Barcelona, Martí i Franqu\'es, 1, 08028 Barcelona, Spain\and Instituto Nazionale di Astrofisica (INAF) - Sezione di Bologna, Viale Berti Pichat 6/2, 40127 Bologna, Italy\and School of Mathematics, Statistics and Physics, Newcastle University, Herschel Building, NE1 7RU, Newcastle-upon-Tyne, UK \and Department of Physics and Astronomy, University College London, Gower Street, London WC1E 6BT, UK\and Waterloo Centre for Astrophysics, University of Waterloo, 200 University Ave W, Waterloo, ON N2L 3G1, Canada\and Department of Physics and Astronomy, University of Waterloo, 200 University Ave W, Waterloo, ON N2L 3G1, Canada      \and Department of Astrophysical Sciences, Princeton University, 4 Ivy Lane, Princeton, NJ 08544, USA\and Universit\"at Innsbruck, Institut f\"ur Astro- und Teilchenphysik, Technikerstr. 25/8, 6020 Innsbruck, Austria\and Institute of Theoretical Astrophysics, Heidelberg University, Philosophenweg 12, 69120 Heidelberg, Germany \and Istituto Nazionale di Fisica Nucleare (INFN) - Sezione di Bologna, viale Berti Pichat 6/2, I-40127 Bologna, Italy\and Division of Physics, Mathematics and Astronomy, California Institute of Technology, 1200 E California Blvd, Pasadena, CA 91125, USA\and Department of Astrophysics, National Centre for Nuclear Research, Pasteura 7, 02-093 Warsaw, Poland \and Shanghai Astronomical Observatory (SHAO), Nandan Road 80, Shanghai 200030, China \and University of Chinese Academy of Sciences, Beijing 100049, China\and Instituto de F\'isica, Pontificia Universidad Cat\'olica de Valpara\'iso, Casilla 4059, Valpara\'iso, Chile   \and Institute for Particle Physics and Astrophysics, ETH Z\"urich, Wolfgang-Pauli-Strasse 27, 8093 Z\"urich, Switzerland }

\date{Accepted 12/12/2121}

\graphicspath{{./figures/}}

%\doparttoc \faketableofcontents \noptcrule  

\abstract{ 
  We present the final data release of the Kilo-Degree Survey (\kids-\drfive), a public European Southern Observatory (ESO) wide-field imaging survey optimised for weak 
  gravitational lensing studies.  We combined matched-depth multi-wavelength observations from the VLT Survey Telescope and the VISTA Kilo-degree INfrared Galaxy (\viking) 
  survey to create a nine-band optical-to-near-infrared survey spanning $1347$ deg$^2$.  The median $r$-band $5\sigma$
  limiting magnitude is 24.8 with median seeing $0\farcs7$. 
  The main survey footprint includes $4$ deg$^2$ of overlap with existing deep spectroscopic surveys. We complemented these data in \drfive\ with a targeted campaign to secure an 
  additional $23$ deg$^2$ of KiDS- and VIKING-like imaging over a range of additional deep spectroscopic survey fields. From these fields, we extracted a catalogue of 
  $126\,085$ sources with both spectroscopic and photometric redshift information, which enables the robust calibration of photometric redshifts across the full survey footprint.  
  In comparison to previous releases, \drfive\ represents a $34\%$ areal extension and includes an $i$-band re-observation of the full footprint, thereby increasing the effective 
  $i$-band depth by $0.4$ magnitudes and enabling multi-epoch science.  Our processed nine-band imaging, single- and multi-band catalogues with masks, and homogenised photometry and 
  photometric redshifts can be accessed through the ESO Archive Science Portal. 
}

\keywords{cosmology: observations -- gravitational lensing: weak -- galaxies: photometry -- surveys -- catalogs}

\maketitle

\section{Introduction}
\label{sec:intro}

Modern astronomical sky surveys are intended to be useful for a wide array of science objectives, extending both beyond the remit
of their primary science goals and the lifetime of the survey's operation. Such extended science use is typically referred to as
the `legacy' value of a survey. Given the ever-increasing monetary and temporal costs required to undertake large
programmes at state-of-the-art observatories, the allocation of large programmes is increasingly contingent on the inclusion of 
planning for legacy science. 

In order to take advantage of the full legacy value of any modern astronomical dataset, however, said dataset ought to 
follow the `FAIR' principle: being Findable, Accessible, Interoperable (i.e. able to be easily combined with other data), 
and Reusable. Conformity with these principles is generally achieved
by the storage and release of the data in an easily interfaceable system, with relevant accompanying documentation, 
such as in the European Southern Observatory (ESO) archive. 

While all aspects of the FAIR principle are important, 
the `reusable' principle is particularly relevant here. This manuscript details the fifth data
release of the Kilo Degree Survey \citep[\kids;][]{dejong/etal:2013}, with a particular focus on ensuring that the 
legacy value of the \kids\ dataset is preserved. This includes the survey's  
on-sky overlap with the VISTA Kilo-degree Infrared Galaxy Survey \citep[VIKING;][]{edge/etal:2013} and 
coverage of wide and deep spectroscopic survey fields. 

\begin{figure*}
    \centering
    \includegraphics[width=0.8\textwidth]{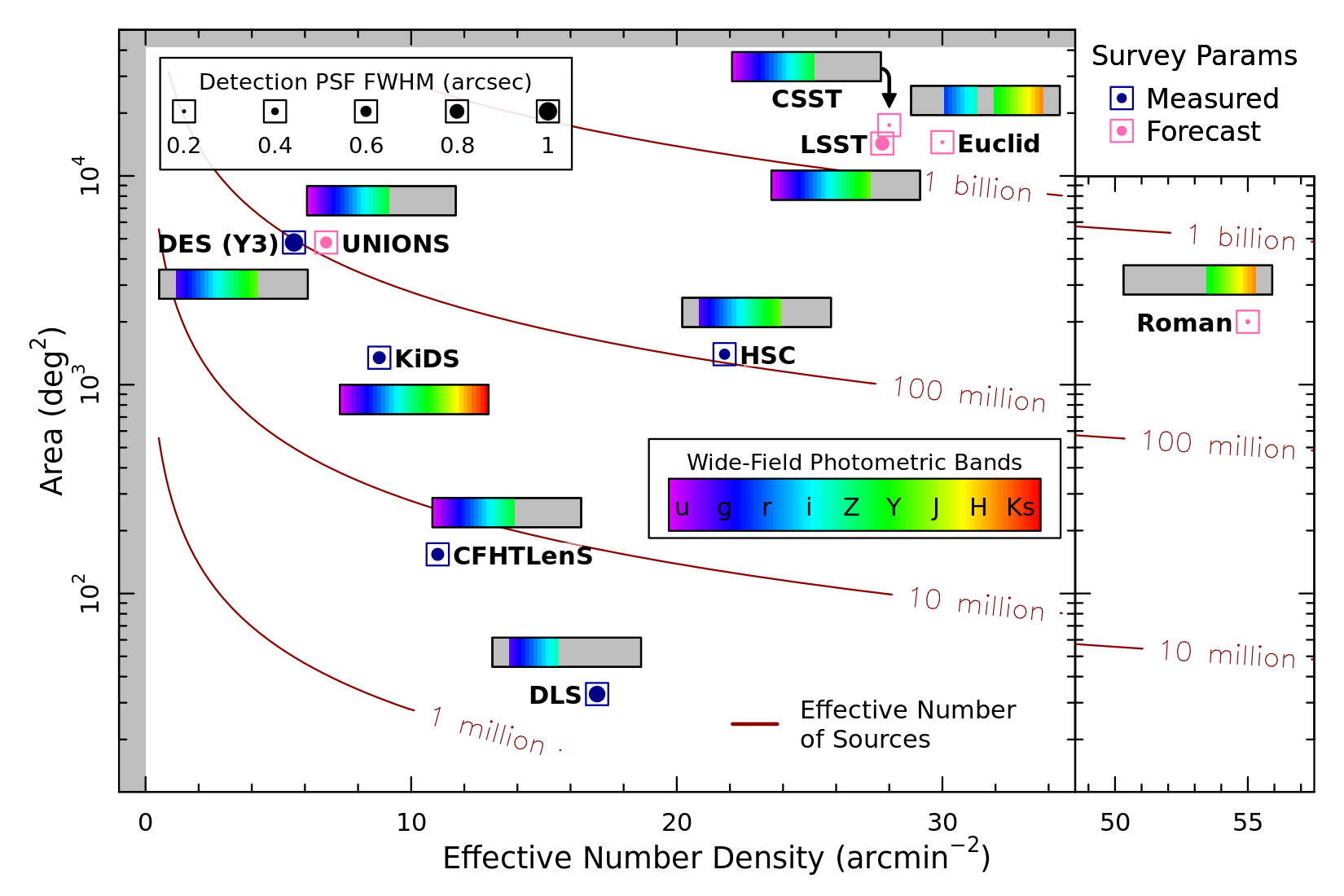}
    \caption{Summary statistics for stage-II \citep{heymans/etal:2012,jee/etal:2016}, stage-III 
    \citep[this work;][]{sevilla/etal:2021,hikage/etal:2018,guinot/etal:2022}, and stage-IV 
    \citep{scaramella/etal:2022,mandelbaum/etal:2018,gong/etal:2019,eifler/etal:2021} cosmological imaging surveys, comparing the survey area and 
    effective number density of sources for weak lensing analyses (square markers).  The red curves show the total effective number of sources, 
    which principally determines the cosmological constraining power. The colour bars indicate each survey's wavelength coverage, 
    which principally determines the photometric redshift quality of a survey. The size of the circular point within each survey's square marker shows the typical seeing in the 
    lensing band, which principally determines the accuracy of shape measurements.  
    The statistics presented for future surveys are forecasts, as indicated in pink.  The final DES-Y6 data release will 
    increase in depth, relative to the DES-Y3 statistics shown, raising the final effective number density of sources.}
    \label{fig:surveycomp}
\end{figure*}

\kids\ is most notable as a so-called stage-III imaging 
survey for cosmology \citep[using nomenclature defined in the Dark Energy Task Force white paper;][]{albrecht/etal:2006}. There are 
three principal stage-III imaging surveys for cosmology: \kids, the Dark Energy Survey \citep[DES;][]{sevilla/etal:2021}, and 
the Hyper-Suprime Camera (HSC) survey 
\citep[][]{aihara/etal:2018}. These surveys predominantly differ in their complementary combinations 
of imaging depth, survey area, and wavelength coverage. This is illustrated in Fig. \ref{fig:surveycomp}: the HSC survey 
is the deepest of the stage-III surveys, DES has the largest sky coverage, and \kids\ spans the greatest range in wavelength. 

The stage-III surveys are directly comparable to earlier stage-II surveys, such as the Canada-France-Hawaii Telescope Lensing 
Survey \citep[CFHTLenS;][]{heymans/etal:2012} and the Deep Lens Survey \citep[DLS;][]{jee/etal:2016}, which are notable due to their limited area but 
impressive photometric depth. Finally, the next generation of cosmological imaging surveys, stage-IV, stand out due to their 
combination of large areas and exceptional depths: \euclid\ \citep{laureijs/etal:2011,scaramella/etal:2022}, the \textit{Vera C. Rubin }Observatory Legacy Survey of Space and Time 
\citep[LSST;][]{ivezic/etal:2019,mandelbaum/etal:2018}, the Chinese Space Station Optical Survey \citep[CSS-OS;][]{gong/etal:2019}, and 
the \textit{Nancy G. Roman} Telescope High Latitude
Survey \citep[\textit{Roman};][]{eifler/etal:2021}. As with the stage-III surveys, each of these stage-IV surveys fills a particular niche combination of 
depth, resolution, wavelength coverage, and area on the sky. Figure \ref{fig:surveycomp} also includes the ground-based partner survey to \euclid, 
the Ultraviolet Near-Infrared Optical Northern Survey \citep[UNIONS;][]{guinot/etal:2022}, which has the capability to become a stage-III-like weak lensing 
survey in its own right. 

As the final data production manuscript for the \kids\ and \viking\ surveys, this paper is designed as a one-stop
reference for researchers and students who wish to use this dataset for their science. The manuscript therefore
includes a full description of the production processes for all data products included in the fifth data release (\drfive), 
including documentation of the input data, the analysis methods and settings, and the final data products. 

The manuscript is arranged approximately in the order that the production itself takes place, which we describe in
Sect. \ref{sec:flowchart}. Section \ref{sec:data} presents details of the optical reduction and photometric processing. 
Section \ref{sec:nirobservations} presents details of the near-infrared (NIR) data reduction. Section \ref{sec:specz} details the compilation of spectroscopic data in the \kids\ \drfive\ fields. Section \ref{sec:multiband} presents details of the NIR photometric
processing, the combination into the full multi-wavelength dataset, and the computation of photometric redshifts. 
Section \ref{sec:shapemeasurement} describes the shape measurements for weak gravitational lensing that are contained in the dataset, which will be used for cosmic
shear analyses with the final \kids\ \drfive\ weak-lensing sample, named the \kidslegacy\  sample. Finally, Sect. \ref{sec:datarelease} outlines the contents of the full data release, detailing catalogues and imaging,
important practical information for access and catalogue use, and summary statistics for the full dataset.

\section{Production workflow and updates}
\label{sec:flowchart}

To assist with the understanding of the production process (and with the navigation of this paper) we start by providing 
an overview of the production workflow, and direct readers to the relevant sections of the paper that describe the 
workflow stages. Details of surveys, technical concepts, algorithms, and other associated jargon are provided in the 
relevant sections, as are the appropriate references to the literature. 

The fifth data release of \kids\ (\drfive) includes a collection of new observations, previously released data that 
have been re-reduced, and previously released data that are unchanged. As such, this release 
supersedes the previous releases of \kids; the fourth data release (\drfour) is not simply a subset of \drfive. 
Of particular note in this release is the
inclusion of an additional $30\%$ of survey area, a complete second-pass of \iband\ observations, and $25$ square degrees of
imaging data over deep spectroscopic-survey fields (which are useful for photometric redshift calibration, and so are 
given the label `\kidz'). 
Each of these additions has particular importance for scientific analyses with \kids, as discussed in the following sections. 

\begin{figure}
  \centering
  \includegraphics[width=\columnwidth]{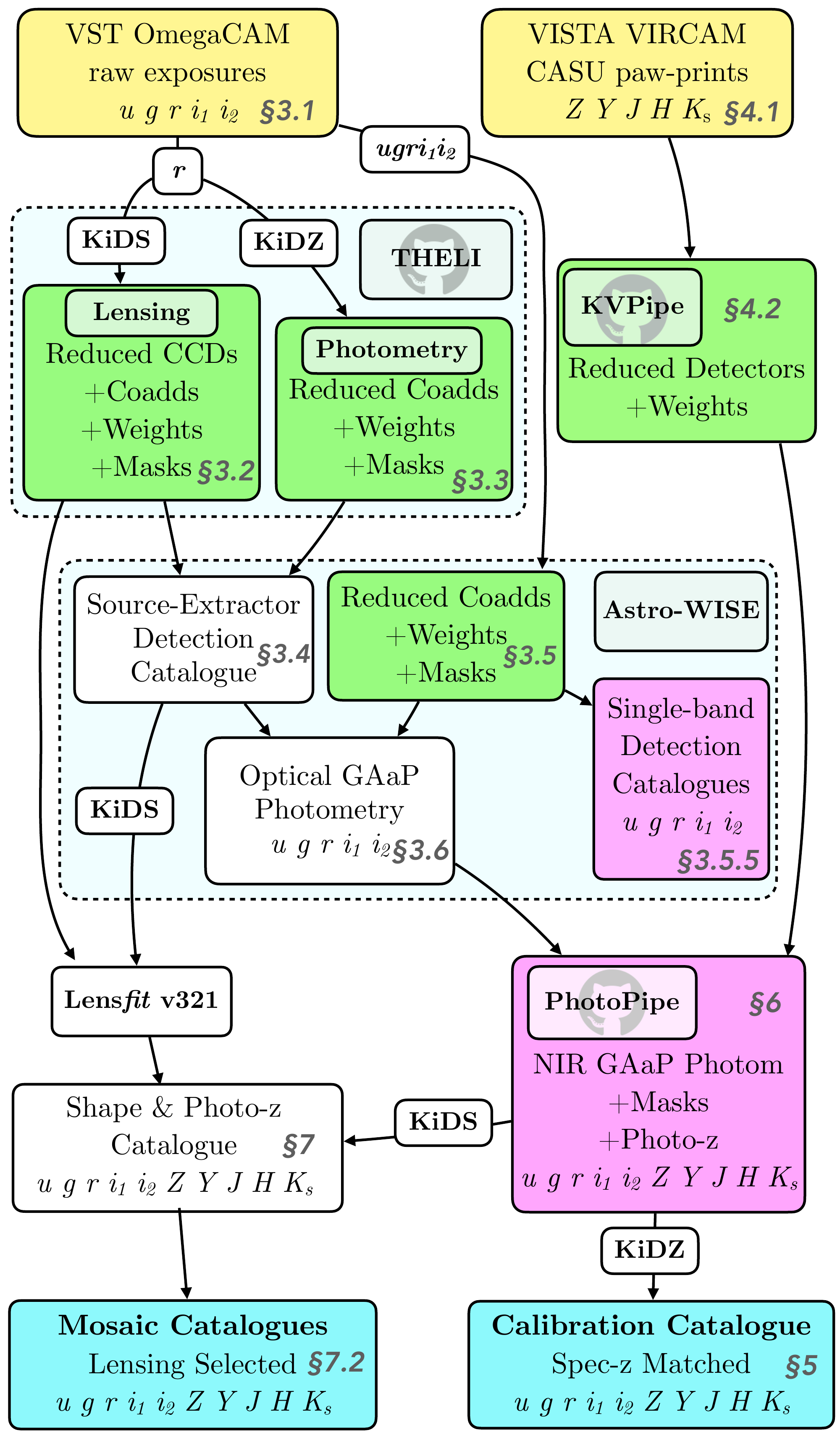}
  \caption{Chart showing the data processing path that is taken for the \kids\ and \kidz\ data in \drfive, from optical
  and NIR imaging (top) through to final mosaic catalogues (bottom). Each step outlined in this graph is discussed in
  the correspondingly annotated section. Yellow boxes show raw data products and data products input into our reduction pipelines. Green
  boxes contain imaging data products released as part of this data release. Pink boxes contain per-tile catalogue-level
  data products released as part of this data release. Blue boxes contain mosaic catalogue-level data products released
  as part of this data release.
  }\label{fig: flowchart}
\end{figure}

Figure \ref{fig: flowchart} shows the updated data-production flowchart for \drfive. The flowchart represents a slight
development of a similar chart for \drfour\ presented in Fig. 1 of \cite{kuijken/etal:2019}, with added items for the processing of 
the \kidz\ data, the \nir-reduction pipeline (\kvpipe\ \github{https://github.com/AngusWright/VIKINGProcessingPipeline}), 
the unified multi-band photometric processing pipeline (\photopipe\ \github{https://github.com/KiDS-WL/PhotoPipe}), and 
the final mosaic and compilation catalogue construction. 
Raw and input data products to the data-production pipeline are shown in yellow (accessible via e.g. the ESO archive), 
calibrated and released imaging data products are provided in green (accessible via the ESO archive and \kids\ collaboration web pages), 
released per-tile catalogue data products are shown in pink (accessible via the ESO archive),
and released mosaic data products are shown in blue (accessible via the \kids\ collaboration web pages). The flowchart is
annotated with the sections of the manuscript containing the description (and relevant changes with respect to previous
work) related to each stage. We also outline the sections here for ease of reference. 

\begin{table*}
    \caption{Summary of imaging data released in \kids-\drfive.}
    \label{tab:summary}
    \centering
    \begin{tabular}{clrrcccc}
        \hline\hline
        Telescope \& & Filter & $\lambda_{\rm cen}$ & Exp. time & Mag. Lim. (corr) & Mag. Lim. (uncor) & PSF FWHM & Area  \\
        Camera & & $(\AA)$ & (s) & ($5\sigma\,2^{\prime\prime}$ AB) & ($5\sigma\,2^{\prime\prime}$ AB) & ($^{\prime\prime}$) & (deg$^2$) \\
        \hline
                 & $u$         & $3\,550 $ & $1000$  & $24.17\pm0.10$ & $24.26\pm0.10$& $1.01 \pm 0.17$ & $1\,312.1$ \\
       VST       & $g$         & $4\,775 $ & $900 $  & $24.96\pm0.11$ & $25.15\pm0.12$& $0.88 \pm 0.15$ & $1\,270.7$ \\
       (OmegaCAM)& $r$         & $6\,230 $ & $1800$  & $24.79\pm0.13$ & $25.07\pm0.14$& $0.70 \pm 0.12$ & $1\,233.8$ \\
                 & $i_1$       & $7\,630 $ & $1200$  & $23.41\pm0.26$ & $23.66\pm0.25$& $0.81 \pm 0.18$ & $1\,246.2$ \\
                 & $i_2$       & $7\,630 $ & $1200$  & $23.49\pm0.28$ & $23.73\pm0.30$& $0.81 \pm 0.18$ & $1\,264.4$ \\
        \hline
                 & $Z$         & $8\,770 $ & $480 $  & $23.47\pm0.20$ & $23.79\pm0.20$ & $0.90\pm0.10$ & $1\,262.0$ \\
        VISTA    & $Y$         & $10\,200$ & $400 $  & $22.72\pm0.18$ & $23.02\pm0.19$ & $0.86\pm0.09$ & $1\,261.4$ \\
        (VIRCAM) & $J$         & $12\,520$ & $400 $  & $22.53\pm0.20$ & $22.72\pm0.20$ & $0.85\pm0.07$ & $1\,262.7$ \\
                 & $H$         & $16\,450$ & $300 $  & $21.94\pm0.24$ & $22.27\pm0.24$ & $0.88\pm0.09$ & $1\,258.3$ \\
                 & $K_{\rm s}$ & $21\,470$ & $480 $  & $21.75\pm0.19$ & $22.02\pm0.19$ & $0.87\pm0.08$ & $1\,258.1$ \\
        \hline
    \end{tabular}
    \tablefoot{ 
    Limiting magnitudes are constructed in two ways as described in Appendix \ref{sec:lambdar}. One method accounts for pixel-to-pixel correlations in the noise background (`corr'), but is likely contaminated by contamination by source flux and therefore underestimates the true depth of the imaging. The second method does not account for pixel-to-pixel correlations in the noise background (`uncor') but it is robust to contamination from sources, and therefore likely overestimates the true depth of the imaging. These estimates therefore bracket the true depth of the imaging released in \kids-\drfive. For quoted limiting magnitudes and PSF sizes, the uncertainties describe the variations in these metrics between the $1 \deg^2$ fields on-sky. 
    }
\end{table*}

Our production workflow starts, naturally, with raw observations made with the Very Large Telescope  (VLT) Survey Telescope (VST; Sect. \ref{sec:vst}). Observations
are then transferred to two separate reduction pipelines: the full set of photometric bands is sent to the \astrowise\
pipeline, and the \rband\ data are also sent to the \theli\ (for \kids\ observations) or \atlas\ (for \kidz) pipelines.
The \theli\ (Sect. \ref{sec:theli}) and \atlas\ (Sect. \ref{sec:atlas}) pipelines are responsible for producing the
imaging (and associated products) that are used for source detection in the \kids\ and \kidz\ fields, respectively, while
the \theli\ pipeline also produces calibrated individual detector images that are used by our fiducial shape measurement code
\lensfit\ (Sect. \ref{sec:shapemeasurement}). 

Within \astrowise\ (Sect. \ref{sec:astrowise}) we performed four main tasks: the primary \kids\ and \kidz\ source
detection (Sect. \ref{sec:sourcedetection}), production of co-added images, weight-maps, and masks in each of the optical filters (Sect.
\ref{sec:astrowise}), extraction of individual per-filter source catalogues (for stand-alone `single-band catalogues', Sect.
\ref{sec:singleband}), and measurements of forced photometry using the Gaussian Aperture and PSF (\gaap) code on the optical images (Sect. \ref{sec:gaap}).
For the primary source extraction \theli\ and \atlas\ \rband\  co-adds are ingested into \astrowise, and extraction is made
on these images using \sourceextractor\ (Sect. \ref{sec:sourcedetection}). These catalogues form the basis of essentially all subsequent catalogues used in
\kids\ and \kidz. For some use cases, however, it may be desirable to have individual source extractions in each of the
optical filters (for example, for the detection of \rband\ drop-out sources). As such, the single band catalogues in
each of the optical bands are also an important output of the reduction process.  After the definition of the source
lists, however, \astrowise\ falls back to the use of its own optical imaging for photometric measurements. 

After the measurement of optical forced aperture photometry, the optical \gaap\ catalogues are  passed to our
unified multi-band photometry and \photoz\ measurement pipeline \photopipe\ (Sect. \ref{sec:photopipe}). 
This pipeline combines the optical data with the \nir\ images from the \viking\ survey, starting with the 
`paw-prints' provided by the Cambridge Astronomical Survey Unit (CASU).  For all sources
in \kids\ and \kidz, \photopipe\ performs the forced photometry of the \nir\ bands (Sect. \ref{sec:nirgaap}), 
construction of \nir\ mosaics, combination of optical and \nir\ photometric catalogues, 
estimation of photometric redshifts (Sect. \ref{sec:photoz}), and construction of
combined multi-band masks (Sect. \ref{sec:strongselection}). 
These catalogues are the primary data product that is released to ESO (Sect.
\ref{sec:datarelease}). For weak-lensing applications, however, we required robust shape measurements: for tiles in the
main \kids\ footprint, these are performed with the \lensfit\ algorithm (Sect. \ref{sec:shapemeasurement}). 

Combined `mosaic catalogues' are then prepared for both the \kids\ and \kidz\ datasets. These are the primary
science catalogues that are used by the majority of the cosmology team in \kids. In the \kids\ (i.e. main survey) areas,
these catalogues are constructed by combining all per-pointing catalogues after source selection based on masks, shape
measurements, photometry, and more (Sect. \ref{sec:sampleselection}). In the \kidz\ fields, a final `calibration catalogue'
is constructed by combining all catalogues from the \kidz\ fields and matching them to a master
spectroscopic compilation (Sect. \ref{sec:specz}). These mosaic and calibration catalogues are released to the community
via the \kids\ database\footnote{\url{https://kids.strw.leidenuniv.nl/sciencedata.php}} after the conclusion
of the primary cosmological analyses. 

Finally, for convenience of reference, we also provide here a summary of the data release in terms of imaging quality and other relevant statistics (Table~\ref{tab:summary}). 

\section{Optical dataset and reduction}
\label{sec:data}
In this section we detail the optical portion of the \kids\ \drfive\ dataset, its reduction, and its differences  with the equivalent data released in \drfour.  Of particular note: new data have been observed with the \vst\
(Sect.~\ref{sec:vst}), \theli\ data have been fully re-reduced (Sect.~\ref{sec:theli}), new \kidz\ data have been reduced with
the \atlas\ pipeline (Sect.~\ref{sec:atlas}), source detection has been recomputed for the full dataset (Sect.~\ref{sec:sourcedetection}), 
\astrowise\ data have been partially re-reduced (Sect.~\ref{sec:astrowise}), and multi-band
\gaap\ photometry has been re-extracted and the post-processing updated (Sect.~\ref{sec:gaap}). As a result, despite
\drfive\ representing largely the union of existing (from \citealt{kuijken/etal:2019}) and new \astrowise\ imaging, it is
unlikely that source-lists, photometry, and/or higher-level data products will be identical (when comparing \drfour\ to
\drfive) for any individual source, population, or tile.

\subsection{VST observations}\label{sec:vst} 

\begin{table} \centering
    \caption{Run numbers of all \vst\ observations taken in the \kids\ and \kidz\ fields.}
    \label{tab:kidskidzobs}
    \begin{tabular}{c|ccc}
      Survey & Run Number & Start Date & Number of OBs \\
      \hline 
              &   60.A-9038  & 2011-08-09 &    1 \\
              &  177.A-3016  & 2011-08-26 & 6618 \\
      \kids   &  177.A-3018  & 2011-09-19 &   63 \\
              &  177.A-3017  & 2011-10-02 &   48 \\
              &  094.B-0512  & 2014-11-20 &    4 \\
              & 0103.A-0181  & 2019-05-03 &    1 \\
      \hline 
              &   177.A-3016 & 2012-11-25 & 105 \\
      \kidz   &   096.B-0501 & 2015-11-06 &  10 \\
              &   098.B-0298 & 2016-11-21 &   6 \\
              &  0100.B-0148 & 2017-11-24 &   4 \\
    \end{tabular}
    \tablefoot{
        Run numbers include observations taken as part of guaranteed time observations programmes.
      }
\end{table}

\begin{table*}\caption{\kidskidz\ observing strategy: observing condition constraints and exposure times.}
  \label{tab:ObservingConstraints}
  \centering
  \begin{tabular}{l c c c c c c c c}
    \hline\hline
    Filter & Max. lunar & Min. Moon & Max. seeing & Max. airmass & Sky transp. & Dithers & Total Exp. & Num \\
    ~ & illumination & distance [deg] & [arcsec] & ~ & ~ & ~ & time per OB [s] & OBs \\
    \hline
    \emph{u} & 0.4 & 90 & 1.1 & 1.2 & CLEAR & 4 & 1000 & 1 \\
    \emph{g} & 0.4 & 80 & 0.9 & 1.6 & CLEAR & 5 &  900 & 1 \\
    \emph{r} & 0.4 & 60 & 0.8 & 1.3 & CLEAR & 5 & 1800 & 1 \\
    \emph{i} & 1.0 & 60 & 1.1 & 2.0 & CLEAR & 5 & 1200 & 2 \\
    \hline
  \end{tabular}
\end{table*}

The VST \citep[][]{capaccioli/etal:2005} is a 2.6m modified Ritchey-Chr\'etien telescope on an alt-az
mount, located at the ESO’s Cerro Paranal observatory in Chile. The VST hosts a single
instrument at its Cassegrain focus: the 300-megapixel OmegaCAM CCD mosaic imager \citep{kuijken:2011}. The $32$ thinned
CCDs that make up the `science array' of the camera consist of $4102\times\,2048$-pixel e2v $44-82$ devices, which
uniformly sample the focal plane at a scale of $0\farcs214$ per $15\mu$m pixel and have very narrow chip-gaps
($25\arcsec$ and $85\arcsec$). The telescope and camera were co-designed for optimal image quality, and as a result they are capable of 
producing imaging with a point-spread function (PSF) that is quite round and shows little variation across the focal plane: seeing ellipticities 
(defined as $e=1-b/a$, where $a$ and $b$ are the major and minor axes of the PSF, respectively) across all \kids\ \vst\
observations are consistently less than $\langle e\rangle=0.08$ per exposure. 

Observations with the \vst\ are split into observing blocks (OBs), which each target a single tile with dithered observations
using a particular filter, and which were queued in order to optimise the use of the VST (as it simultaneously carried
out a suite of surveys). Observations for \kids\ and \kidz\ made with the \vst\ were performed under the programmes listed
in Table~\ref{tab:kidskidzobs}. Each tile was observed in four filters
($ugri$), with bandpasses very similar to the $ugri$ filters from the Sloan Digital Sky Survey
\citep[SDSS;][]{york/etal:2000}\footnote{The effective wavelengths of the \vst\ and SDSS filters agree to approximately $2\AA$, $6\AA$, $100\AA$, and $50\AA$ in the $ugri$-bands, respectively.}, and with various requirements regarding observational conditions.
Table~\ref{tab:ObservingConstraints} summarises the observing constraints, number of OBs, number of dithers per OB, and
total exposure time for each filter.  $r$-band observations were made under only the best atmospheric conditions
(seeing full-width at half-maximum of less than $0\farcs8$) during dark-time (maximum lunar illuminations of
$40\%$). The $g$- and $u$-band observations similarly required dark-time, but with less stringent seeing requirements:
allowing maximal seeing of $\fwhm\leq 0\farcs9$ and $\fwhm\leq 1\farcs1,$ respectively. The $i$-band observations
were required to be performed under the same seeing conditions as the $u$-band ($\fwhm\leq 1\farcs1$), but, unlike
the other bands, they could be obtained in grey or bright-time (i.e. with no restriction on maximum lunar illumination). 

The observing strategy for \kids\ and \kidz\ can best be described as `depth first, area second':
each OB immediately achieves the full depth in a given filter on a tile, and that tile is not revisited with that filter
unless the observations subsequently fail quality control. The one exception to this rule is the second \iband\
pass, which is revisited by design (see Sect.~\ref{sec:kidsacquisition}). There is no requirement that the observations
with a different filter on a given tile be taken within a certain time period, although a bias towards observing
partially completed tiles was built into the observing scheduler. 

In November 2020, towards the end of the survey data taking and after the \vst\ was recommissioned
following an extended shutdown during the COVID-19 pandemic, one of the 32 CCDs in the OmegaCAM mosaic was found to have failed. This 
resulted in a rectangular gap near the centre of the mosaic for all observations taken after this date. This defect
affects mostly the second \iband\ pass of the \kidz\ fields, which were almost exclusively observed after this date
(Sect.~\ref{sec:kidzacquisition}). 

\subsubsection{\kids\ data acquisition}\label{sec:kidsacquisition} 

\begin{figure*}\centering
  \includegraphics[width=\textwidth]{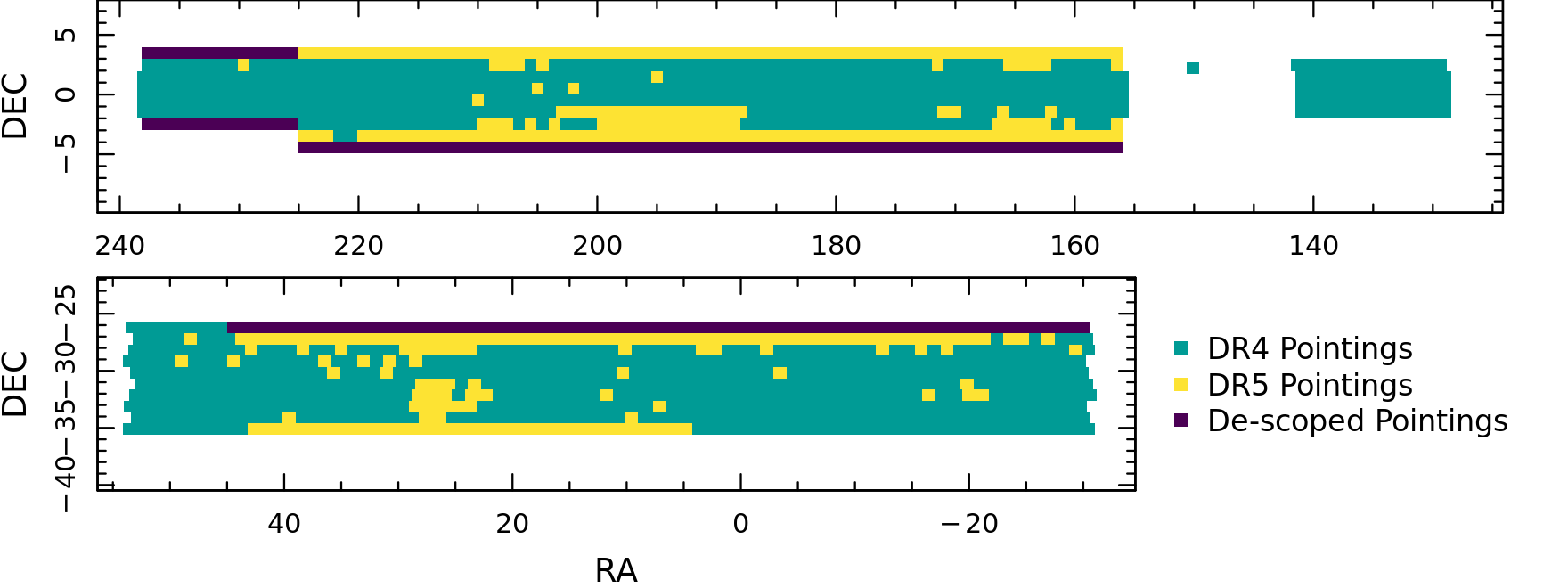}
  \caption{Distribution of \kids\ \drfive\ pointings on-sky. The figure shows the distribution of \drfour\ pointings
  (dark cyan) and new \drfive\ pointings (yellow). Pointings that were originally included in the $1500$ \sqdeg\ \kids\
  footprint, but which were subsequently de-scoped due to the limited area observed by \viking, are shown in dark purple. The
  combination of the green and yellow data therefore show the $1347$ \sqdeg\ of \kids\ \drfive\ observations.}
  \label{fig:tilesky}
\end{figure*}

The \kids\ survey footprint consists of $1347$ tiles, of $1\deg\times\,1\deg$ each, corresponding to the field of view of
the OmegaCAM CCD mosaic camera. It
covers two regions of the sky, roughly coincident with the main survey area of the 2dF Galaxy Redshift Survey
\citep{colless/etal:2001}: a $70\deg\times\,9\deg$ stripe (in RA$\times\,$Dec) across the South Galactic Pole, and a
$85\deg\times\,8\deg$ stripe across the celestial equator in the North Galactic Cap, with extensions to include the
Galaxy and Mass Assembly (GAMA) 
$9$hr field \citep[$12\deg\times\,5\deg$,][]{driver/etal:2011} and the Cosmic Evolutions Survey field \citep[COSMOS;
$1\deg\times\,1\deg$,][]{scoville/etal:2007}. The layout of the  \kids\ survey fields is shown in
Fig.~\ref{fig:tilesky}. 

The original plan for the \kids\ survey, as endorsed by ESO's Public Surveys Panel in 2005, was to cover $1500$
square degrees in $ugri$-bands with the \vst\ \citep{dejong/etal:2013}. This proposed area is shown in
Fig.~\ref{fig:tilesky} as the full span of coloured tiles. Subsequently, in 2006, ESO approved the \viking\ survey
\citep{edge/etal:2013}, which would add five \nir\  filters to the same proposed survey area. The observations for \kids\ and \viking\ started in 2011 and 2009, respectively, soon after their
corresponding telescopes were commissioned. 

The initial \viking\ time allocation resulted in about 90\% of the planned area of sky being observed, at which point
ESO decided to truncate that survey in favour of new projects (see Sect~\ref{sec:viking}). Given the slightly reduced
area that \viking\ would cover, it was decided also to truncate \kids\ to the same total area, which resulted in the
final footprint of $1347$ tiles. Observations that were removed after this area truncation are shown in purple in
Fig.~\ref{fig:tilesky}. Some of the observing time on the \vst\ originally allocated to the now-deprecated \kids\ and \viking\ tiles
was reallocated to observations of deep spectroscopic fields, forming the basis of the `\kidz' dataset
(Sect.~\ref{sec:kidzacquisition}). 

\begin{figure} \centering
  \includegraphics[width=\columnwidth]{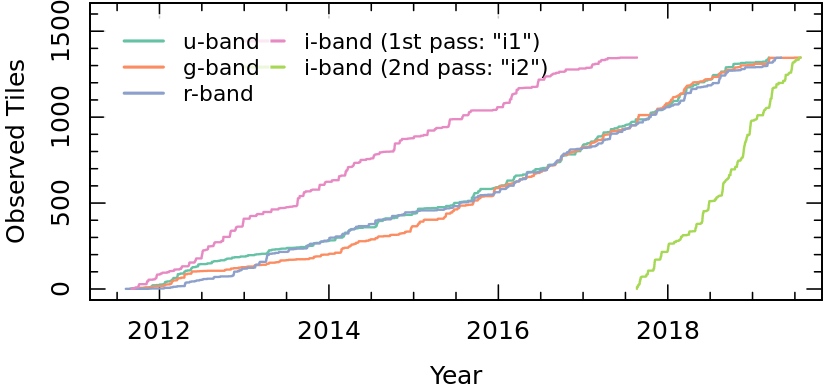}
  \caption{Progression of optical observations over the full \kids\ survey, for pointings within the final $1347$
  \sqdeg\ footprint. The total number of \drfive\ pointings is shown as the dashed grey line. The figure demonstrates the slow initial progress made by the survey, whereby only $\sim 5\%$ of
  planned \rband\ observations were completed in the initial $18$ months of data-taking, prompting changes in the
  queuing of (in particular) the dark-time observations. Conversely, at the end of the survey, the complete second-pass
  of \iband\ observations (designated \itwo) was completed in roughly the same time span.}\label{fig:obsprogress}
\end{figure} 

\begin{figure*} \centering
  \includegraphics[width=2\columnwidth]{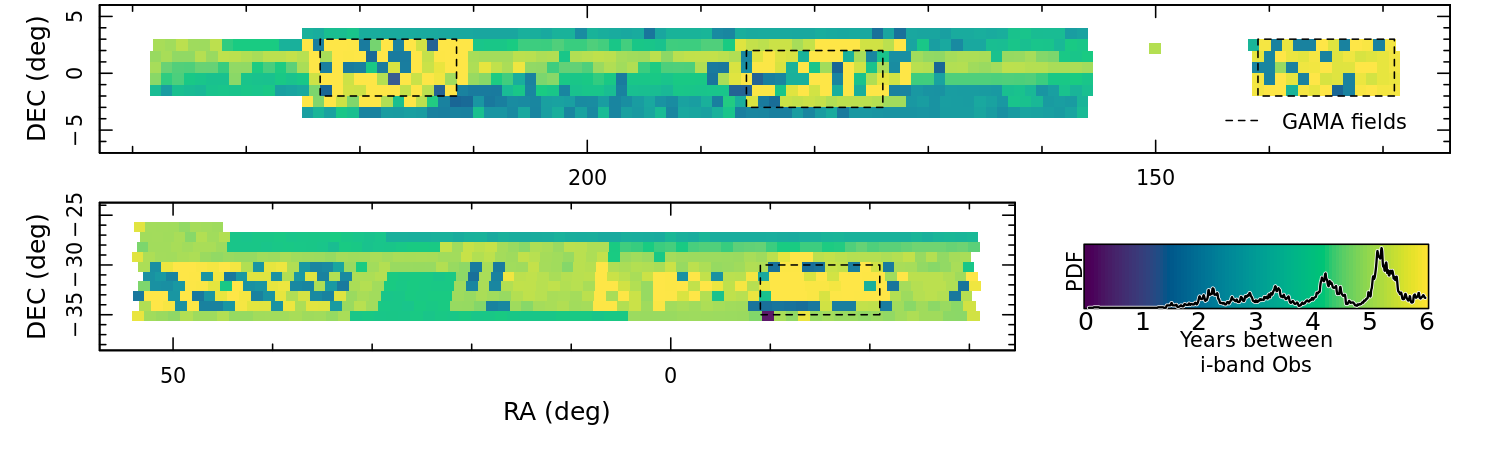}
  \caption{Temporal separation between the two \iband\ passes, as a function of position on-sky. The figure shows
  all \kids\ \drfive\ pointings, coloured by the number of years separating the two \iband\ passes. The overall
  distribution of these values is shown as a KDE within the colour bar, computed with a $0.1$-year rectangular
  bandwidth. The largest temporal separations occur typically in the \gama\ fields, where \kids\ observations were
  focussed during early survey operations. The shortest temporal separation between any two passes is at ${\rm RA}=350.3\,\deg$,
  ${\rm Dec}=-35.1\,\deg$, which was observed on 21 August 2017 and 7 October 2017 for the two \iband\ passes, respectively, for a
  separation of $50$ nights.
  }\label{fig:obsdatessky}
\end{figure*} 

A second component of the original \kids\ survey plan was to re-observe the entire footprint at the end of the survey in the
\gband, for proper motion and variability studies. As the survey progressed, however, it became clear that bright-time
was in considerably less demand than dark-time on the \vst, with the bright-time \iband\ observations advancing much
faster than those using the dark-time $u,g,r$ filters.  This can be seen in Fig.~\ref{fig:obsprogress}, which shows the
progression of observations in each of the optical filters over the lifetime of the \kids\ survey. Given the rapid
progression of the \iband\ data, it was decided to change the filter for the repeat pass to $i$, rather than $g$. The
second pass \iband\ observations are visible in the figure under the \itwo\ label. Similarly the \iband\ is unique in
Table~\ref{tab:ObservingConstraints} as being the only filter with more than one OB per tile. 

The choice to switch the filters used for the second pass comes with certain scientific
benefits. First, the additional depth in the \iband\ improves the quality of photometric redshifts
(particularly beyond $z=0.9$, where the $4000\,\AA$ break redshifts into the \iband).  Second, the repeat observations
somewhat counteract the variability in the seeing and sky brightness that is inherent to the \iband\ observations, due
to their less stringent constraints on observing conditions. Finally, all the second-pass observations were taken after
significant improvements to the telescope baffling in 2015, much reducing scattered light that particularly affected
bright-time observations. Due to the rapidity of the \iband\ progress during the first half of the survey, the scattered
light effect is particularly prevalent in the \iband.  Therefore, having the second pass be in this band is of particular
benefit. 

Given the primary focus of the repeated \iband\ observations on variability and transient science, it is worth noting
the somewhat correlated nature of the cadence between \iband\ observations on-sky. As early \kids\ data targeted the
\gama\ fields preferentially, the largest baselines in the \iband\ observation cadence are in those fields.
This can be seen in Fig.~\ref{fig:obsdatessky}, which shows the difference between \iband\ observation dates over the
survey, with the rectangular \gama\ patches (dashed outlines) clearly visible.  

A second noteworthy feature of Fig.~\ref{fig:obsprogress} is the particularly slow progress of observations taken in the
$r$-band during the first $18$ months of data-taking. This feature was traced to the queue scheduler,
which was overly conservative for dark-time observations with strict
seeing requirements. Changes to the scheduler in early 2013 addressed this problem and led to a speeding up of the dark-time \kids\  observations.

\subsubsection{\kidz\ acquisition}\label{sec:kidzacquisition} 

\begin{figure*}\centering
  \includegraphics[width=\textwidth]{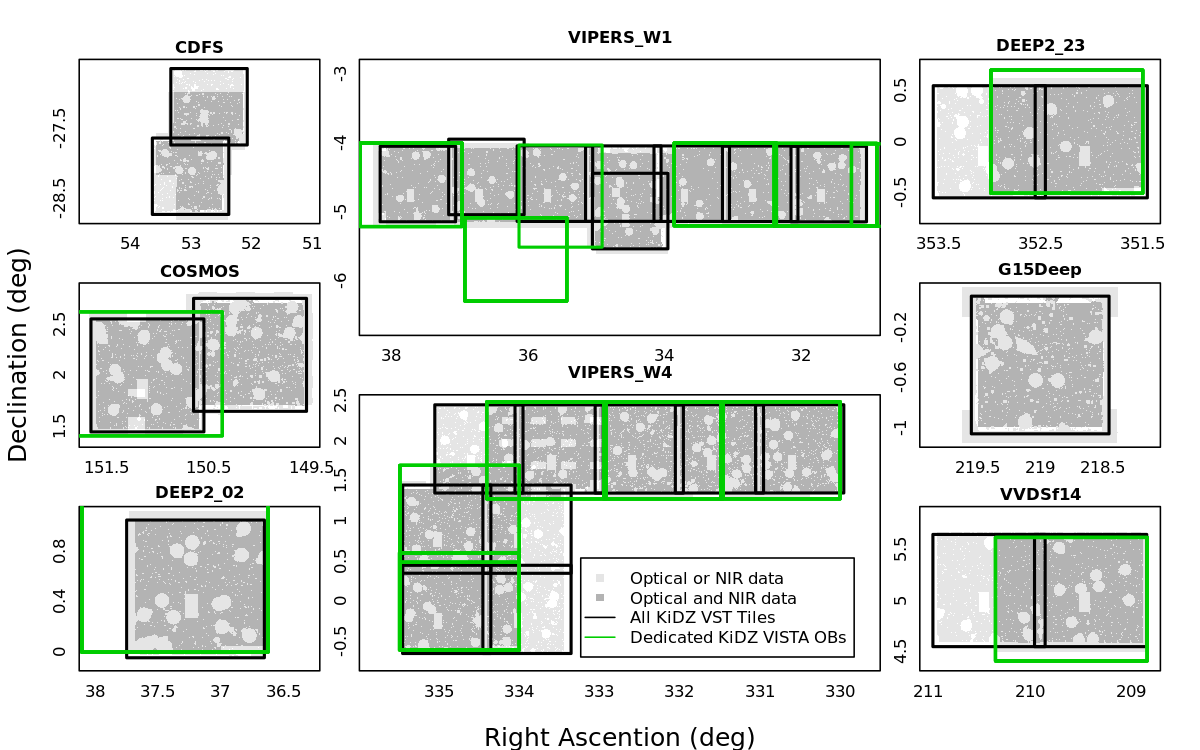}
  \caption{Footprints of the \kidz\ pointings. The figure shows the regions where
  there exist optical and/or \nir\ data in dark and light grey, respectively. The rectangles show the limits of 
  all observations made with the \vst\ (black) and dedicated observations made with \vista\ (green). Dark grey regions that are not covered by our dedicated \vista\ observations are either already
  contained within the \viking\ footprint (CDFS, G15-Deep) or have \viking-like observations constructed from existing
  deep observations (VIPERS, COSMOS; see Sect.~\ref{sec:kidznir}). 
}\label{fig:tilesky_KIDZ}
\end{figure*}

As mentioned in Sects.~\ref{sec:vst} and \ref{sec:kidsacquisition}, \drfive\ includes observations made by the \vst\ 
over deep spectroscopic survey fields that significantly enhance \kids\ science. Eight fields were identified for
these observations, as having both rich existing (or ongoing) spectroscopic campaigns, and being visible from Paranal:
\textit{Chandra} Deep Field South (CDFS), the COSMOS field, the DEEP2 Galaxy Redshift Survey (DEEP2) 
02hr and 23hr fields, the
GAMA G15-Deep field, the Visible Multi-object Spectrograph (VIMOS) Public Extragalactic
Redshift Survey (VIPERS) W1 and W4 fields, and the VIMOS VLT Deep Survey (VVDS) 14hr field.
Specific details of the spectroscopic samples available in these fields are provided in Sect. \ref{sec:specz}. 
Of these fields, two already resided completely within the \kids\ \drfive\ footprint (CDFS, G15-Deep) and 
so required no additional observations. \kidz\ observations with the \vst\ were taken under the same strategy as was
used in the main survey (Sect.~\ref{sec:vst} and Table~\ref{tab:ObservingConstraints}), including 
the second \iband\ pass, and  \viking-like observations on these fields were taken with \vista. The \vst\ \kidz\ observations were 
taken between November $2015$ and September $2021$, and can be identified in the ESO archive using the run numbers
listed in Table~\ref{tab:kidskidzobs}.  The full set of targeted fields are shown in Fig.~\ref{fig:tilesky_KIDZ}, with
\vst\ OBs visible by the black outlines.

\subsubsection{Optical data quality metrics}

\begin{figure*}\centering
  \includegraphics[width=\columnwidth]{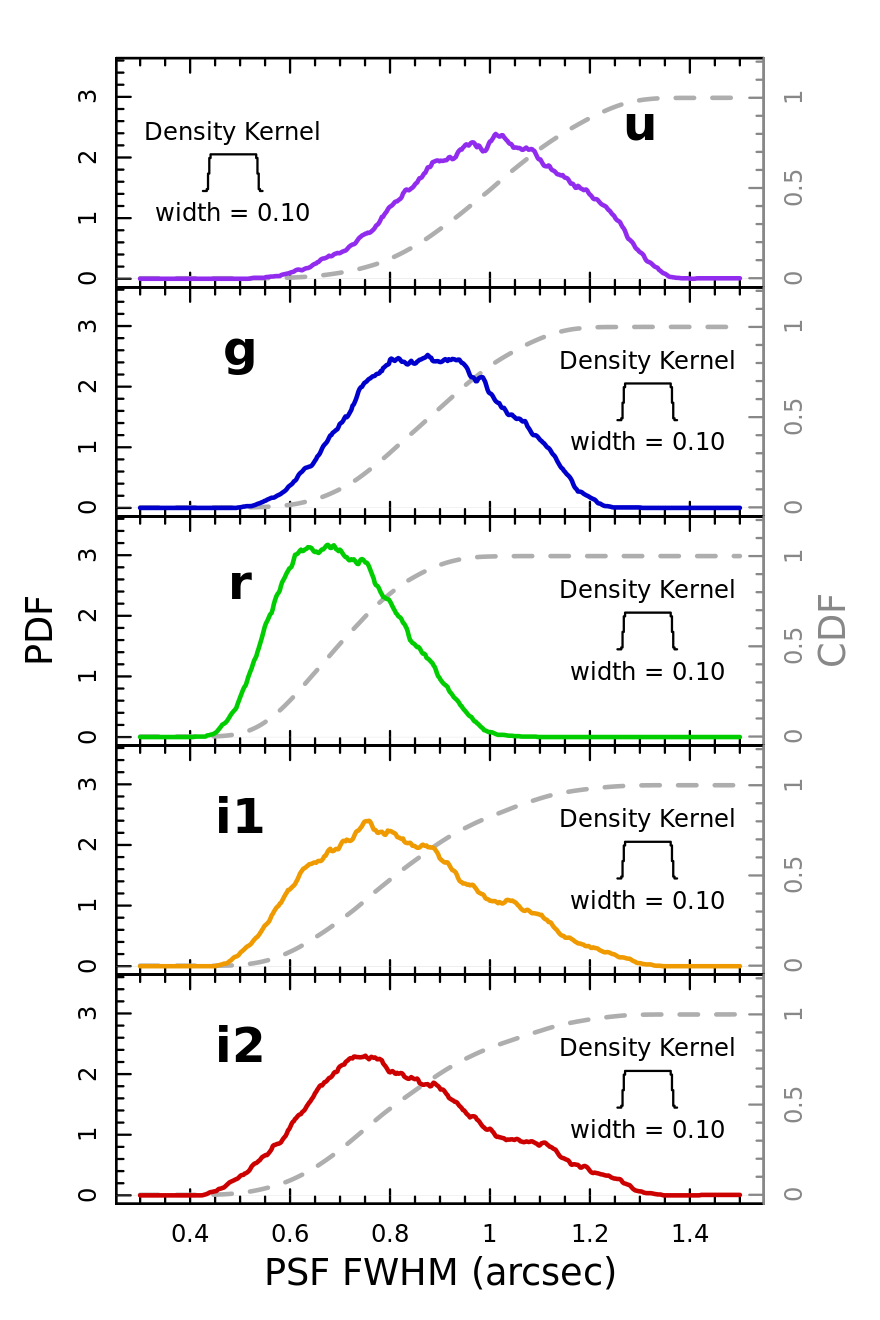}
  \includegraphics[width=\columnwidth]{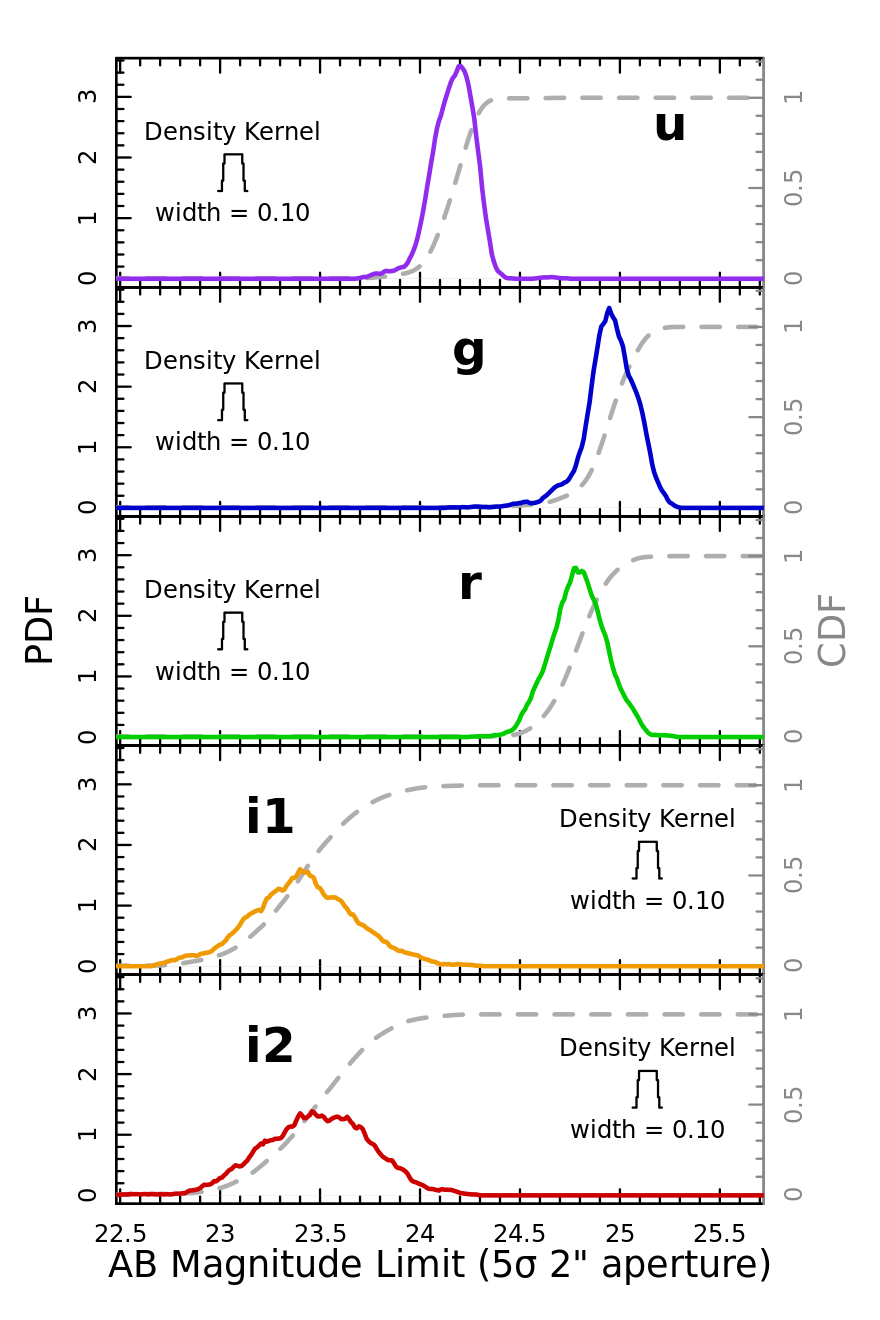}
  \caption{Primary observational properties of pointings in \kidskidz\ for observations in the four optical bands. 
  Observations in the \iband\ are split into the two epochs (labelled \ione\ and \itwo). 
  Each row shows the distribution of average PSF sizes (as reported by \astrowise; left) and limiting magnitude 
  ($5\sigma$ in a $2\arcsec$ diameter circular aperture; right) determined with a KDE using the annotated
  kernel. The corresponding cumulative distribution functions for each panel are shown as grey dashed lines.
  }\label{fig:psfsizes}
\end{figure*}

The two primary quality metrics that are relevant for our \kidskidz\ observations, made with the \vst, are the size of
the PSF and the brightness of the background. Combined, these two properties determine the effective depth of the
imaging, and the selection of source galaxies.  Figure~\ref{fig:psfsizes} shows the distribution of PSF size (full-width
at half-maximum) and limiting magnitude ($5\sigma$ in a $2\arcsec$ diameter circular aperture) for all images taken as
part of the \kidskidz\ observing campaigns. Limiting magnitudes in each of the bands were calculated from randomly
scattered apertures across the masked tiles using \lambdar\ \citep{wright/etal:2016}.  The details of the magnitude
limit calculation is given in Appendix~\ref{sec:lambdar}. 

The figures show both the distribution functions and cumulative distribution functions for each parameter.  The median
seeing in each optical filter is
$\{1\farcs01\pm0\farcs17,0\farcs88\pm0\farcs15,0\farcs70\pm0\farcs12,0\farcs81\pm0\farcs18,0\farcs81\pm0\farcs18\}$ for
the $\{u,g,r,i_1,i_2\}$-bands, respectively, and the corresponding median limiting magnitudes are
$\{24.17\pm0.10,24.96\pm0.11,24.77\pm0.13,23.41\pm0.26,23.49\pm0.28\}$\footnote{These magnitude limits are slightly
brighter than those presented in previous KiDS releases, due largely to their computation allowing for inclusion of
correlated noise. For a direct comparison with previous releases, see Appendix~\ref{sec:lambdar} and
Table~\ref{tab:summary}.} In both cases, the
uncertainties describe the tile-to-tile scatter in these metrics, computed with the normalised median absolute deviation
from the median (NMAD). The distribution of magnitude limits in the four bands can be seen to take a reasonably narrow
range of values; the scatter of the limiting magnitudes in the $ugr$-bands is consistently less than $0.15$ mag.  The
\iband, however, does show more variability in both seeing and limiting magnitude than is evident in the other bands,
largely due to these observations being taken in poorer conditions (Sect. \ref{sec:vst}). However, these metrics keep
the independent \iband\ passes separate, meaning that the effective limiting magnitude of the individual tiles in the
\iband\ would in-fact be approximately $0.4$ magnitudes deeper and have a $30\%$ reduction in scatter, bringing it
closer in line with the other bands (from $\sigma\approx 0.3$ mag to $\sigma\approx 0.2$ mag). We note that this
approximation is largely fair, as the individual distributions of PSF full-width at half-maximum (\fwhm) and limiting
magnitudes are extremely similar for the \ione\ and \itwo\ passes. 

Finally, systematic variation of observational conditions on-sky is an important nuisance in many scientific studies. As
such, we include visualisations of the on-sky distribution of both magnitude limit and PSF FWHM, for each of the optical filters, 
in Appendix~\ref{sec:vstonskymetrics}.

\subsection{\theli\ \rband\ reduction}\label{sec:theli} 

As with previous \kids\ analyses, the primary imaging used for lensing science (specifically shape measurement) in \drfive\ has
been produced with the \theli\ pipeline (version 1.3.0A). Since \drfour\ (which used \theli\ version 1.0.0A), \theli\ has undergone modifications to both the 
astrometric and photometric calibration routines. 
In this section we document, in particular, the specific changes to \theli\ (with respect to the \drfour\ version) that are relevant to 
\drfive\ science.

\subsubsection{Astrometric calibration with \gaia}\label{sec:astrometry} 

Until version 1.3.0A, \theli\ utilised two separate astrometric calibration samples for the northern and southern patches of
\kids: in the north \sdss\ \citep{york/etal:2000} was used as an absolute astrometric reference, whereas the southern fields were absolutely calibrated to
\twomass\ \citep{skrutskie/etal:2006}. In \drfive, \kids\ utilises a single astrometric calibration to \gaia\ DR2
\citep{gaia/dr2} stars, with positions translated to epoch J2000 by Vizier\footnote{Catalogue {\tt I/345/gaia2}, columns
{\tt RAJ2000 and DEJ2000}}.   
We note that the choice to calibrate \kids\ to a J2000 epoch can lead to some confusion when performing a direct match between 
\kids\ stars and \gaia\ catalogues that do not contain the {\tt RAJ2000/DEJ2000} columns. For an explanation of this issue, we 
direct the interested reader to Appendix~\ref{sec:astrometry_residuals}. 

\begin{figure*}
  \centering
  \includegraphics[width=0.8\textwidth]{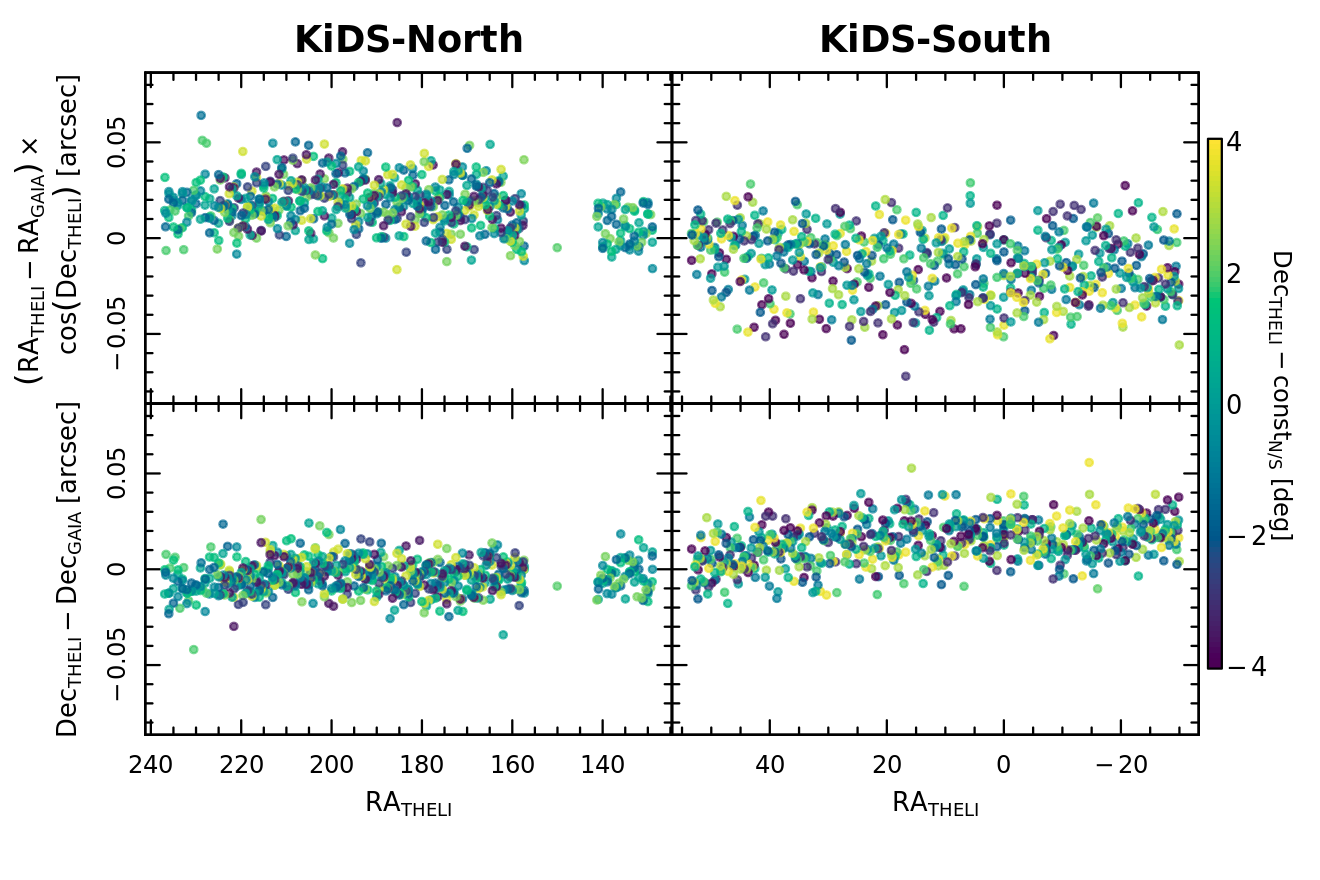}
  \caption{Astrometric calibration of \kids\ \drfive, with respect to \textit{Gaia}\ DR3 stars at epoch J2000. Values here show 
  the median astrometric residual per pointing for stars in the magnitude range $16.5\leq G\leq 19$. The colour bar shows the location of each field in the Dec direction, after subtracting a constant equal to the mean declination of all fields in the relevant hemisphere. Absolute residuals 
  are typically less than the $0\protect\farcs05$ level, and are therefore negligible. Residual offsets below this level are possibly  
  attributable to barycentric motion between the J2000 and J2015 epochs (see Appendix \protect \ref{sec:astrometry_residuals}).}\label{fig:astrometry}
\end{figure*}

\begin{figure*}
  \centering
  \includegraphics[width=2\columnwidth]{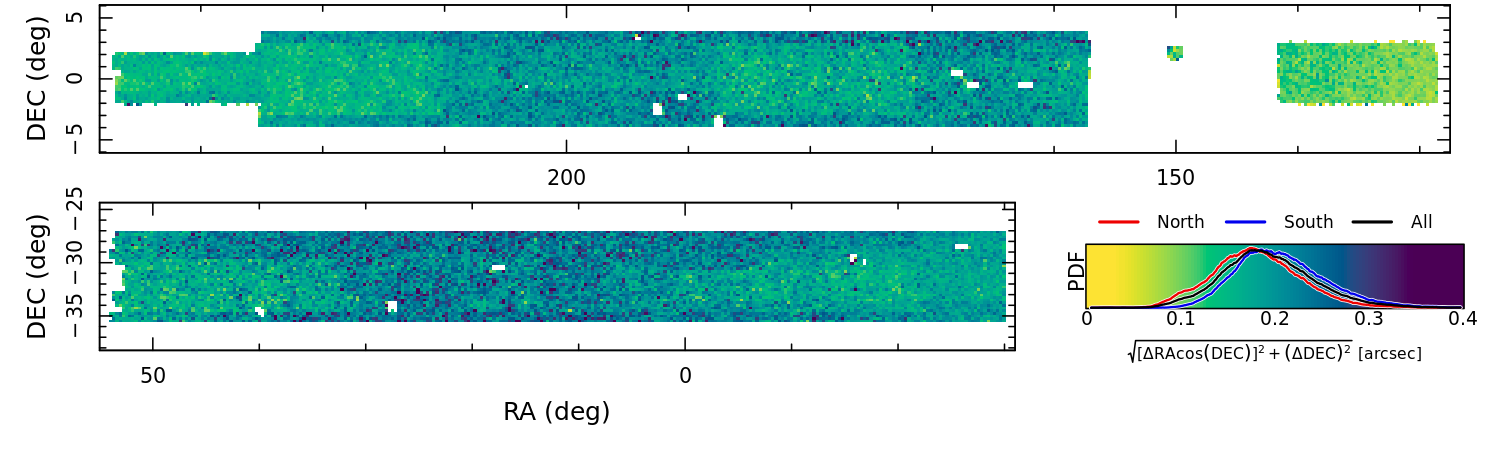}
  \caption{On-sky variation in the \kids\ \drfive\ astrometric calibration with respect to
  \gaia\ stars at the J2000 epoch. Residuals can be seen to fall around the earliest observations, due to 
  the reduced proper motion differences between our observations and the assumed epoch.}\label{fig:astrometryonsky}
\end{figure*}

\begin{figure}
  \centering
  \includegraphics[width=\columnwidth]{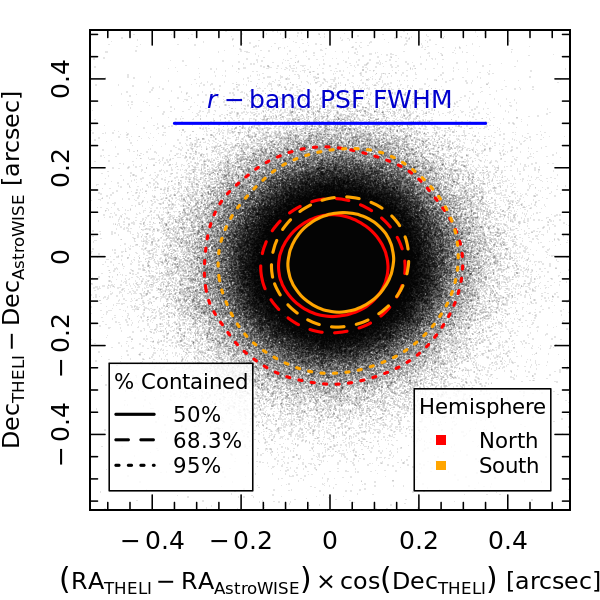}
  \caption{Astrometric agreement between sources extracted from \theli\ and \astrowise\ images. The sample of all
  sources has a median offset of \protect $0\protect\farcs014$ and \protect $-0\protect\farcs017$ in 
  the RA and Dec directions, respectively. The NMAD scatter in the RA and Dec directions is \protect $0\protect\farcs097$ and
  \protect $0\protect\farcs090,$ respectively.
  }\label{fig:astrometryastrowise}
\end{figure}

Figure~\ref{fig:astrometry} shows the accuracy of the \kids\ astrometric calibration to \gaia\ (here using \gaia\ DR3 stars, 
rather than the DR2 stars used to define the astrometric solutions), as a function of RA and Dec. Average absolute residuals are 
typically less than $0\farcs05$, demonstrating that the imaging has been successfully tied to \gaia. The small residuals 
that remain are systematically induced by the choice to calibrate \kids\ using \gaia\ stars at epoch J2000 positions: as 
\kids\ observations were made after the year 2000, stellar proper motions introduce noise to the astrometric calibration. 
This is seen most clearly in the on-sky distribution of astrometric residuals, shown in Fig. \ref{fig:astrometryonsky}: 
parts of \kids\ that were observed earliest (i.e. the \gama\ fields) clearly stand-out as having systematically smaller 
residuals than other (later) observations.  

Finally, we note that the change to the \gaia\ astrometric calibration introduces another possible source of bias: differing astrometric solutions 
between our \theli\ and \astrowise\ reductions. As \astrowise\ astrometry is computed with \twomass, it is possible that the locations of sources 
extracted from \theli\ images may have systematically different positions on-sky (as determined by our \gaia\ astrometry) to the same sources in \astrowise\ 
(as determined by the \twomass\ astrometry). To verify the consistency of the astrometry between our \theli\ and \astrowise\ reductions, we matched 
sources extracted from the reduced \astrowise\ \rband\ images (see Sect. \ref{sec:singleband}) to those in our master \theli\ source catalogue (Sect. \ref{sec:sourcedetection}). 
Figure \ref{fig:astrometryastrowise} shows the astrometric agreement between stars in these two catalogues, thereby validating the use of \theli\ defined positions 
for photometric extractions on \astrowise\ images. The median residual between stars in the two catalogues is $0\farcs014$ in the RA axis and $-0\farcs017$ in the 
Dec axis. The NMAD scatter in the RA and Dec directions is $0\farcs097$ and $0\farcs090,$ respectively.

\subsection{\atlas\ \rband\ reduction}\label{sec:atlas} 

As mentioned in Sect.~\ref{sec:kidzacquisition}, VST observations in the \rband\ were made in the \kidz\ fields between
November 2015 and September 2018. These observations were reduced with a slightly different version of the {\sc theli} pipeline, 
as the observations were not planned to be used for lensing (rather they would only be used for photometry in the context of 
redshift distribution calibration).   
This different version, which we designate `\atlas', is notable for its different 
astrometric calibration sample (\twomass) and the lack of individual exposures (which are used for shape measurement). 

\begin{figure}
  \centering
  \includegraphics[width=\columnwidth]{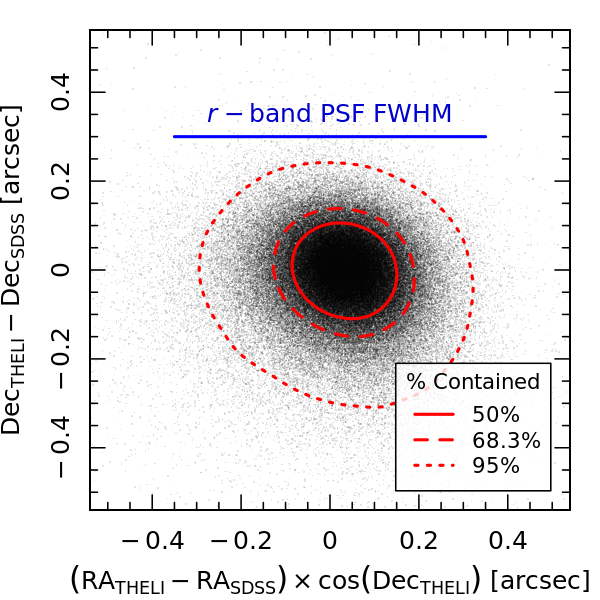}
  \caption{Overall astrometric calibration of \kids\ \drfive, calibrated to \gaia, with respect to \sdss\ in the KiDS-N field.
   The sample of all sources has a median offset of \protect $0\protect\farcs031$ and \protect $-0\protect\farcs006$ in 
  the RA and Dec directions, respectively. The NMAD scatter in the RA and Dec directions is \protect $0\protect\farcs094$ and
  \protect $0\protect\farcs084,$ respectively.
  }\label{fig:astrometrysdss}
\end{figure}

In practical terms, the \atlas\ pipeline is based on \theli\ version 1.2.0A, with changes made to the data products that are preserved 
at the end of each pointing's reduction. 
As mentioned previously, the primary distinction between these V1.2.0A and V1.3.0A is the use of \gaia\ as an astrometric reference (in V1.3.0A), as 
opposed to the use of \sdss\ (north) and \twomass\ (south) in V1.2.0A. This means that the astrometric solution for \kidz\ and \kids\ 
are nominally different, at the level of existing systematic differences between \gaia\ and \sdss\ or \twomass. 
Figure \ref{fig:astrometrysdss} shows the residual difference between the astrometric solutions of \kids\ stars (i.e. calibrated with \gaia) with respect to \sdss\ stars, 
demonstrating that there is no significant systematic bias in the two astrometric systems. As such, we conclude that there is unlikely to be any 
significant systematic effect imprinted on the \kidz\ data through the use of a different astrometric basis in the \atlas\ pipeline.

\subsection{Source detection}\label{sec:sourcedetection} 

Source detection in \kidskidz\ is performed with \sourceextractor\ \citep{bertin/arnouts:1996}, within the \astrowise\ environment. 
The sources are extracted from the \theli\ and \atlas\ \rband\ images, and these source lists are subsequently used for all 
primary science in \kids. Source extraction is performed in a relatively `hot' mode (i.e. significant fragmentation of sources), 
as the main targets of interest for \kids\ 
lensing science are small, faint sources at intermediate-to-high redshift. This means, however, that there is likely to be fragmentation 
of the largest sources within the footprint 
\citep[see e.g. the demonstration of the hot-mode shredding of galaxies in Figs. 5 and 6 of][]{andrews/etal:2017}. 
The parameters used to perform the \kids\ source extraction are provided in Table \ref{tab:sourceextraction}. 

\begin{table}
  \caption{Source Extraction parameters used in the creation of the \kids\ \drfive\ source lists within \astrowise.}
    \centering
    \begin{tabular}{c|c}
      Parameter  &  Value \\
      \hline
      {\tt ANALYSIS\_THRESH } & 1.5      \\
      {\tt DEBLEND\_MINCONT } & 0.001    \\
      {\tt DEBLEND\_NTHRESH } & 32       \\
      {\tt DETECT\_MINAREA  } & 4        \\
      {\tt DETECT\_THRESH   } & 1.5      \\
      {\tt FILTER           } & Y        \\
      {\tt FILTER\_NAME     } & default.conv \\
      {\tt THRESH\_TYPE     } & RELATIVE \\
      {\tt CLEAN            } & Y \\
      {\tt CLEAN\_PARAM     } & 3.0 \\
      {\tt BACK\_TYPE       } & MANUAL \\
      {\tt BACK\_VALUE      } & $[0.0,0.0]$ \\
    \end{tabular}
    \label{tab:sourceextraction}
    \tablefoot{
      This list shows only parameters most relevant to the reproduction of the source catalogues, and so is not exhaustive.
      }
\end{table}

\subsection{\astrowise\ reduction}\label{sec:astrowise} 

The \astrowise\footnote{\url{http://www.astro-wise.org}}\ environment is a distributed database and data processing
system, designed for calibrating and processing wide-field imaging data \citep{macfarland/etal:2013}. \kids\ has
utilised \astrowise\ for the production of reduced images, and forced optical photometry using \gaap\ \citep{kuijken:2008}, 
in the $ugri$-bands since the beginning of the survey \citep{dejong/etal:2015}.  

For \drfive\ there have been a few minor modifications to the \astrowise\
imaging reduction pipeline compared to that described in \cite{kuijken/etal:2019}, which are important to document here.

\subsubsection{Changes: Co-add production}\label{sec:co-add} 

There have been two changes to the co-add production for \kids\ \drfive: updates to the cross-talk coefficients (exclusively 
adding in new values), and a change to the polynomial order for defining the astrometric solution of the \uband\ co-adds. 

\begin{table}
  \caption{Applied cross-talk coefficients.}
  \label{tab:crosstalk}
  \centering
  \resizebox{\columnwidth}{!}{
    \begin{tabular}{c | c c | c c}
      \hline\hline
      Period & \multicolumn{2}{c|}{CCD \#95 to CCD \#96\tablefootmark{a}} & \multicolumn{2}{c}{CCD \#96 to CCD
      \#95\tablefootmark{a}} \\
      ~ & $a$ & $b$ ($\times10^{-3}$) & $a$ & $b$ ($\times10^{-3}$)\\
      \hline
      $2011.08.01$ --- $2011.09.17$ & $-210.12$ & $-2.5037$ & $ 59.44$ & $0.2739$ \\
      $2011.09.17$ --- $2011.12.23$ & $-413.05$ & $-6.8788$ & $234.81$ & $2.7275$ \\
      $2011.12.23$ --- $2012.01.05$ & $-268.00$ & $-5.1528$ & $154.26$ & $1.2251$ \\
      $2012.01.05$ --- $2012.07.14$ & $-499.87$ & $-7.8359$ & $248.92$ & $3.1095$ \\
      $2012.07.14$ --- $2012.11.24$ & $-450.86$ & $-6.9322$ & $220.72$ & $2.5348$ \\
      $2012.11.24$ --- $2013.01.09$ & $-493.10$ & $-7.2305$ & $230.28$ & $2.7224$ \\
      $2013.01.09$ --- $2013.01.31$ & $-554.17$ & $-7.5200$ & $211.86$ & $2.6099$ \\
      $2013.01.31$ --- $2013.05.10$ & $-483.69$ & $-7.0746$ & $224.66$ & $2.6283$ \\
      $2013.05.10$ --- $2013.06.24$ & $-479.09$ & $-6.9786$ & $221.06$ & $2.6380$ \\
      $2013.06.24$ --- $2013.07.14$ & $-570.00$ & $-7.7112$ & $228.88$ & $2.8397$ \\
      $2013.07.14$ --- $2014.01.01$ & $-535.62$ & $-7.4980$ & $218.87$ & $2.7008$ \\
      $2014.01.01$ --- $2014.03.08$ & $-502.23$ & $-7.1187$ & $211.62$ & $2.4288$ \\
      $2014.03.08$ --- $2014.04.12$ & $-565.77$ & $-7.5180$ & $215.12$ & $2.5782$ \\
      $2014.04.12$ --- $2014.08.12$ & $-485.06$ & $-6.8865$ & $201.67$ & $2.2365$ \\
      $2014.08.12$ --- $2014.01.09$ & $-557.89$ & $-7.5079$ & $204.17$ & $2.3036$ \\
      $2014.01.09$ --- $2015.05.01$ & $-542.55$ & $-7.5810$ & $219.87$ & $2.5347$ \\
      $2015.05.01$ --- $2015.07.25$ & $-439.28$ & $-6.9545$ & $221.48$ & $2.3951$ \\
      $2015.07.25$ --- $2015.08.25$ & $-505.57$ & $-7.5352$ & $229.71$ & $2.6048$ \\
      $2015.08.25$ --- $2015.11.10$ & $-475.21$ & $-7.3986$ & $218.05$ & $2.4449$ \\
      $2015.11.10$ --- $2016.06.17$ & $-457.77$ & $-6.8310$ & $201.57$ & $2.2117$ \\
      $2016.06.17$ --- $2016.06.25$ & $-351.80$ & $-4.9728$ & $165.33$ & $1.1678$ \\
      $2016.06.25$ --- $2016.09.08$ & $-476.33$ & $-6.9203$ & $200.42$ & $2.2018$ \\
      $2016.09.08$ --- $2017.08.01$ & $-465.32$ & $-6.5938$ & $184.68$ & $1.9801$ \\
      $2017.08.01$ --- $2018.11.01$ & $-492.33$ & $-6.4804$ & $169.90$ & $1.8016$ \\
      $2018.11.01$ --- $2021.10.01$ & $-506.90$ & $-6.4964$ & $172.11$ & $1.9174$ \\
      \hline
    \end{tabular}
  }
  \tablefoot{
    \footnotesize 
    \tablefoottext{a}{
      Correction factors $a$ and $b$ are applied to each pixel in the target CCD based on the pixel values in the source
      CCD:
      \begin{equation*}
        I^\prime_i =
        \begin{cases}
          I_i + a, &\text{if $I_j = I_{\rm{sat.}}$;}\\
          I_i + b  I_j, &\text{if $I_j < I_{\rm{sat.}}$,}\\
        \end{cases}
      \end{equation*}
      where $I_i$ and $I_j$ are the pixel values in CCDs $i$ and $j$, $I^\prime_i$ is the corrected pixel value in CCD $i$ due
      to cross-talk from CCD $j$, and $I_{\rm{sat.}}$ is the saturation pixel value.
    }
  }
\end{table}

Table~\ref{tab:crosstalk} presents the cross-talk coefficients for the \astrowise\ co-add production. These coefficients have 
been updated to include values spanning the final observation window for \kids\ and \kidz, including
the two lengthy shutdown periods of the \vst\ as a result of the COVID-19 pandemic. As a result, the final set of cross-talk 
coefficients span a considerably longer period than the previous sets. We have not explored whether this difference introduces 
any systematic effect in the efficiency of the cleaning of cross-talk in the post-shutdown images obtained by the \vst. However, 
as our source detection is performed on the \theliraw\ images (which were all observed prior to this period), we believe that this is 
unlikely to have a noticeable impact on our analyses. 

\begin{figure}
    \centering
    \includegraphics[width=\columnwidth]{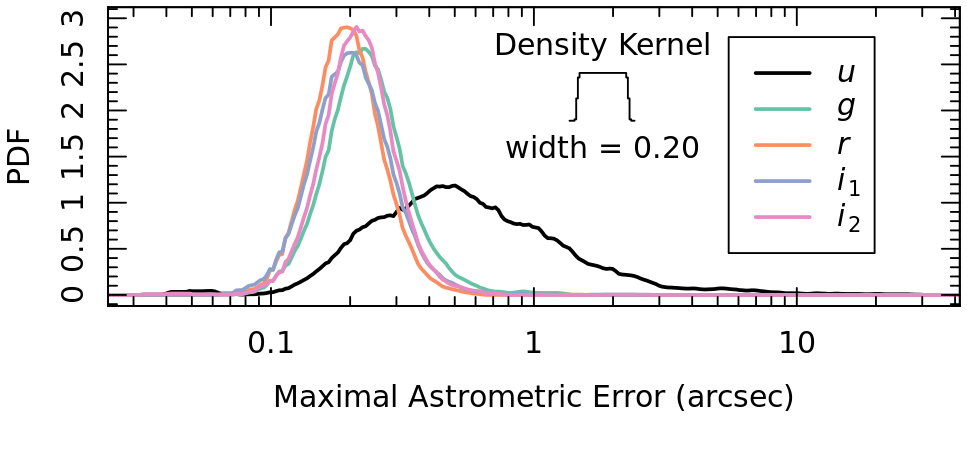}
    \caption{Maximal residual between chip corners within a single pointing, assuming all chips can be 
    shifted to a common RA and Dec centroid. Cases where the astrometric distortion parameters are poorly constrained (and so result in 
    unphysical chip distortions) manifest as large residuals. The distribution in the \uband\ when using third-order distortion polynomials is 
    clearly systematically larger than the other bands, with some chips having significant biases (greater than $10\arcsec$), due 
    to the lower number of stars creating instability in the polynomial fits. To 
    minimise the effect of this bias, we reprocessed all \uband\ tiles with maximal residuals greater than $1\arcsec$ using a 
    linear polynomial distortion order.  }
    \label{fig:distortions}
\end{figure}

The second change to the co-add production is related to the 
astrometry in the \uband. For all co-adds in \drfour, the polynomial order for the distortion was set to three. 
In \drfive, we found that the paucity of stars (per chip) detected in the \uband\ can occasionally lead to poorly constrained 
polynomial fits, leading to unphysical distortion solutions for some chips. Figure \ref{fig:distortions} shows the distribution of 
maximal astrometric residuals between the corners of all detectors in each pointing, assuming that all chips can be simply translated onto a 
common RA and Dec centroid. The figure shows that there is a clear systematic difference between the \uband\ and the other bands, whereby a significant number of pointings have chip corners that differ by more than $1\arcsec$; 
up to $24\arcsec$ in the worst case. 
Investigation of this effect led to the determination that failures in the \uband\ distortion parameters could be resolved by 
restricting the distortion polynomials to linear order. We opted to fit all \uband\ exposures with more than $1\arcsec$ chip-corner-residuals 
with a linear order polynomial fit instead. The $1\arcsec$ threshold was chosen because this is the median size of the PSF in the \uband, and (as the threshold is 
determined using the maximal distortion per pointing) 
this ensures any remaining systematic effects are below the seeing level for typical \uband\ images. 
The \uband\ co-adds that have been reprocessed in this way are listed in Appendix~\ref{sec:reprocess_list}. 

\subsubsection{Changes: Zero-point calibration}\label{sec:vstcalibration} 

One particular complexity that arises in \drfive\ is the treatment of the zero-point calibration in the presence of the
two \iband\ passes. In previous \kids\ data reductions, the calibration process has utilised a stellar locus regression
(SLR) between the four optical filters. We implemented the same fundamental procedure, outlined in Sect. 3.1.3 of \citet{kuijken/etal:2019} with further details in \citet{dejong/etal:2017}. Briefly, this process of SLR calibration 
within \astrowise\ proceeds as follows. First, so-called `principal colours' are constructed from the measured \gaap\ fluxes 
for bright, unsaturated stars in each tile \citep{ivezic/etal:2004}. Second, these colours are shifted so that straight sections  
of the colour-colour diagrams align with a set of fiducial templates. Finally, an overall zero-point correction is applied to all 
bands, in order to match the de-reddened $(r-G,g-i)$ diagram to a template constructed from the \sdss\ survey. 
For this purpose, $G$ band photometry is taken from \gaia\ DR2. 

\begin{figure*}
  \centering
  \includegraphics[width=2\columnwidth]{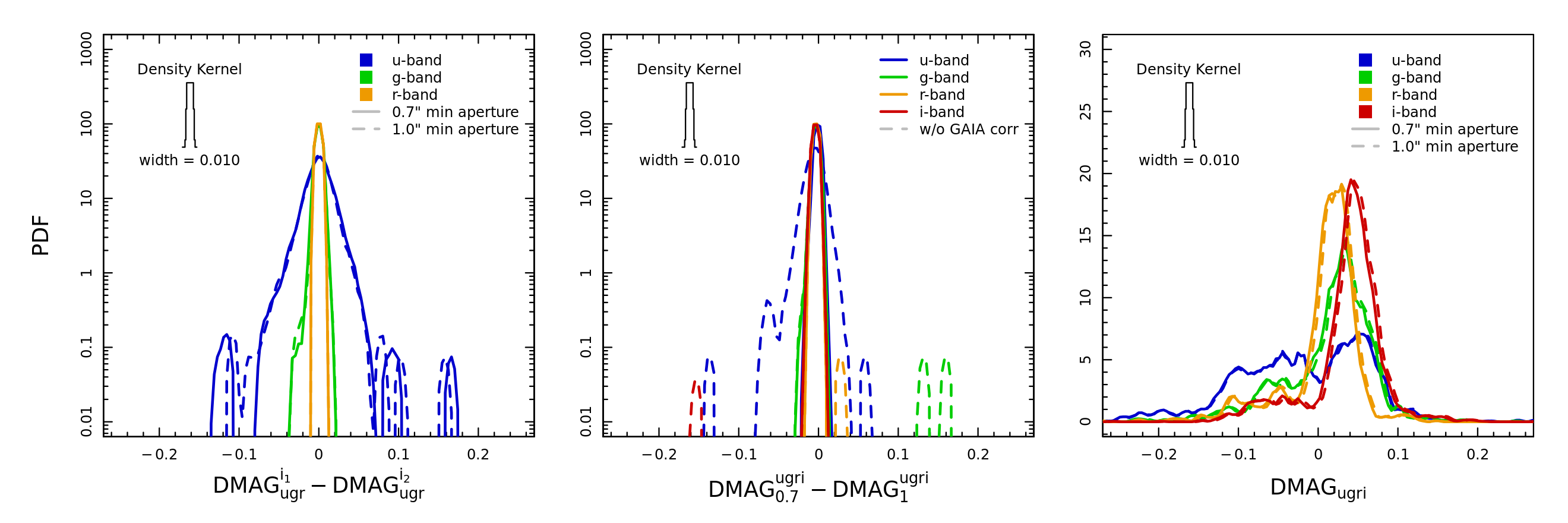}
  \caption{Zero-point corrections estimated in \kidskidz\ using SLR. {\em Left:} Difference
  between \slr\ offsets estimated using \ione\ photometry and \itwo\ photometry, prior to recalibration of the \uband\
  zero-points. {\em Centre:} Differences between \slr\ offsets estimated using \zeropseven\ minimum-radius aperture
  fluxes and \onepzero\ minimum-radius aperture fluxes (see Sect.~\ref{sec:gaap}). {\em Right:} Distribution of
  final \slr\ offsets used in \drfive\ (i.e. corrections to the nightly zero points derived from photometric standards), including \gaia\ recalibration. }\label{fig:dmags}
\end{figure*}

Given the additional \iband\ observation, we are required to run the SLR calibration process multiple times (i.e. running 
with the individual \iband\ passes separately). As a result, we have four SLR calibration estimates per tile: two for the 
\gaap\ minimum apertures of $0\farcs7$ and $1\farcs0$ (see Sect. \ref{sec:gaap}), each computed with either the \ione\ 
or \itwo\ imaging as the reference \iband. Performing the SLR in this way allows us to estimate
the systematic uncertainty introduced in the calibration of the $\utor$-bands by variations in the quality of the
\iband\ data, and to flag potentially bad calibrations. 

The difference in the zero-point calibration offsets as estimated with the \ione\ and \itwo\ fluxes, and with the two
\gaap\ minimum aperture sizes, are presented in Fig.~\ref{fig:dmags}. Looking firstly at the accuracy of the $g$- and
$r$-band calibrations, we can see that the zero-points are consistently reproduced to $\pm0.02$ magnitudes for all
but a few tiles in the $g$-band. Similarly, looking at the calibration measured between the different \gaap\ settings,
we see that, for all but a handful of outliers, the $g$- and $r$-band zero points are very consistently estimated.
However, we of course cannot overlook the outliers in the $gr$-band comparisons, nor the significantly larger dispersion
between zero-points estimated in the \uband. 

\begin{figure}
  \centering
  \includegraphics[width=\columnwidth]{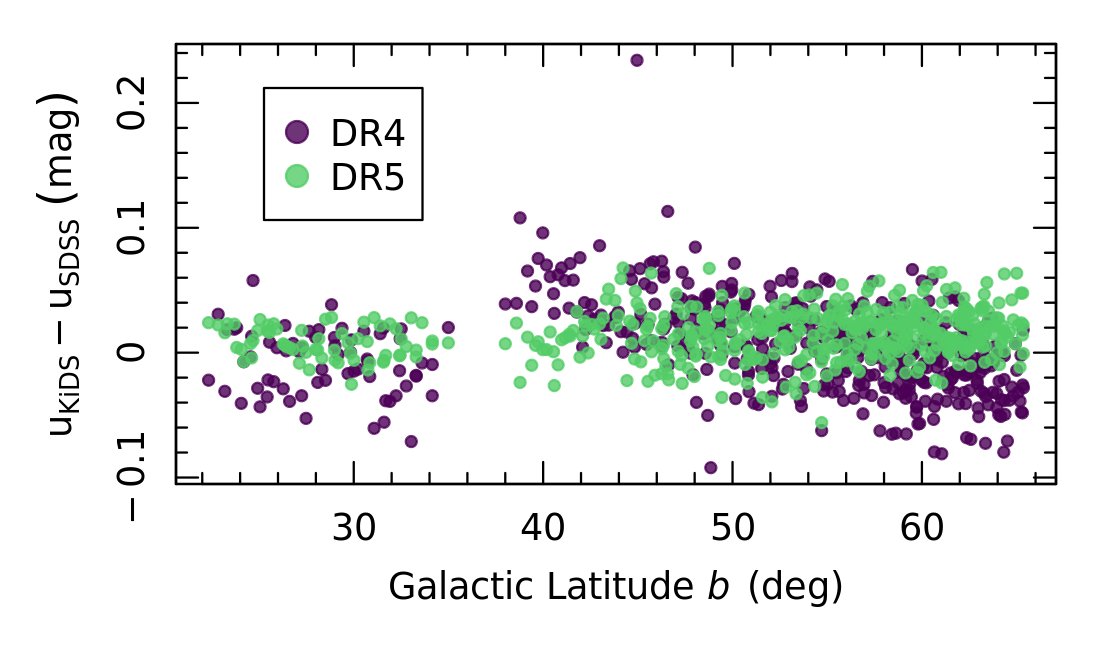}
  \caption{Comparison between the \uband\ zero-point calibration in \kids\ \drfour\ (green) and \kids\ \drfive\ (purple). 
  The updated \drfive\ calibration procedure produces zero-points that show considerably less systematic and random variation, 
  when compared to fluxes from SDSS.}\label{fig:ubandcompar}
\end{figure}

In the figure we can see that the \uband\ zero-point calibration is somewhat unreliable, as has been previously
documented \citep[see e.g.][]{kuijken/etal:2019}. In \drfour, this behaviour prompted a recalibration effort that
utilised \gaia\ and SDSS photometry to improve the \uband\ zero-point accuracies. For \drfive, we adopted a slightly
different approach to zero-point recalibration than was utilised in \drfour\ \citep[and presented
in][]{kuijken/etal:2019}, who based zero-point corrections solely on the \gaia\ DR2 `white-light' $G$-band. Our
zero-point corrections are instead now based on a combination of the \gaia\ white-light $G$-band, blue $G_{BP}$-band, and 
red $G_{RP}$-band magnitudes, available in \gaia\ Early Data Release 3 \citep[eDR3;][]{gaia/edr3}. Use of this information in our zero-point
calibration requires knowledge of the colour transformations between the \gaia\ and SDSS photometric systems, for which
we use relationships derived by \citet{riello/etal:2021}. However, the transformations in \citet{riello/etal:2021} were
only computed for the $gri$-bands; as such, we introduce a predictor of the SDSS $u$-band from available \gaia\
magnitudes, $\mathcal{U}$, in a manner similar to \citet{kuijken/etal:2019}: 
\begin{equation}\label{eqn:u}
\mathcal{U} = f(G, G_{BP}, G_{RP}) + g(E_{B-V}, b), 
\end{equation}
where the first term is the \gaia\ internal colour transformation, 
\begin{equation}\label{eqn:f}
f(G, G_{BP}, G_{RP}) = G + 2.8\ (G_{BP}-G_{RP}) - 0.6,
\end{equation}
and the second term is the \uband\ extinction, as a function of Galactic latitude $b$, 
\begin{equation}\label{eqn:g}
g(E_{B-V}, b) = -1.9\, E_{B-V} -  0.10\ |\sin(b)|\, + \, 0.09.
\end{equation}
The predictor $\mathcal{U}$ is calculated for sources that are bright ($G<19$) and occupy the heart of the stellar locus
$0.7 < G_{BP}-G_{RP} < 1.1$, in an effort to suppress non-linear terms in the preceding two equations. The dependence on
Galactic latitude $b$ was added as discussed in \cite{kuijken/etal:2019}, with the implicit assumption that the
Galactic latitude correction derived in \kids-North (i.e. where there is overlap with SDSS) can be extrapolated to
KiDS-S. The coefficients contained in Eqs. \ref{eqn:f} and \ref{eqn:g} are 
derived with two separate fits between \uband\ PSF magnitudes from the \sdss\ DR16 \citep{sdss_dr16:2020} and \gaia\ eDR3, selected
within $710$ \kids-North tiles that overlap with the \sdss. The \uband\ offset for each tile in both \kids-North and 
\kids-South were then computed as the median of the differences $\Delta u = \mathcal{U} - u_{\rm KiDS}$, measured for
\gaia\ stars. A comparison between the quality of the zero-point calibration in the \uband\ in \drfive\ and \drfour\ is 
presented in Fig. \ref{fig:ubandcompar}. The figure demonstrates that there is considerably less variability (systematic and 
random) in the \drfive\ \uband\ zero-points compared to \drfour. 
The  scatter in the \uband\ zero-point reduces from $\sim$ 0.035 in DR4 to $\sim$ 0.018 in DR5: 
we notice that a similar scatter ($0.016$) was  obtained by \citet{liang/etal:2023}, where
the calibration of \uband\ data in KiDS and other photometric surveys was improved  
using observed colours of blue Galactic halo stars.

The distribution of zero-point calibration offset difference between the two \iband\ passes (and the two minimum
aperture sizes) is presented in Fig.~\ref{fig:dmags}. Somewhat by construction, the various calibrations now all agree
with one-another, having been pegged to the external \gaia\ data. This is most apparent in the \uband, where we have
applied this recalibration to every pointing. However, equally relevant is the choice to apply this recalibration
process to the handful of tiles that show disagreement in the $gr$-bands. A similar recalibration approach can also be 
applied to the $gri$-bands, where we verified that robust results were obtained using the colour transformations
available in \cite{riello/etal:2021}. However, given the accuracy that we observe already in the $gr$-bands, we decided
to follow this option only for the few tiles where the SLR+\textit{Gaia} approach described above failed to produce reliable offsets.

After the definition of final zero point corrections, individual corrections are applied to the catalogues. As with \drfour, 
the corrections computed with the $0\farcs7$ minimum aperture are applied to fluxes computed with the same aperture; the same is 
done for the $1\farcs0$ minimum aperture corrections and sources. In \drfive, though, we must combine the corrections computed 
using the \ione\ and \itwo\ bands in the other (non-$i$) bands. In this case, we opted to take the straight arithmetic mean of the 
corrections estimated with the two \iband\ images for the $\utor$-bands. 

\begin{figure*}
  \centering
  \includegraphics[width=2\columnwidth]{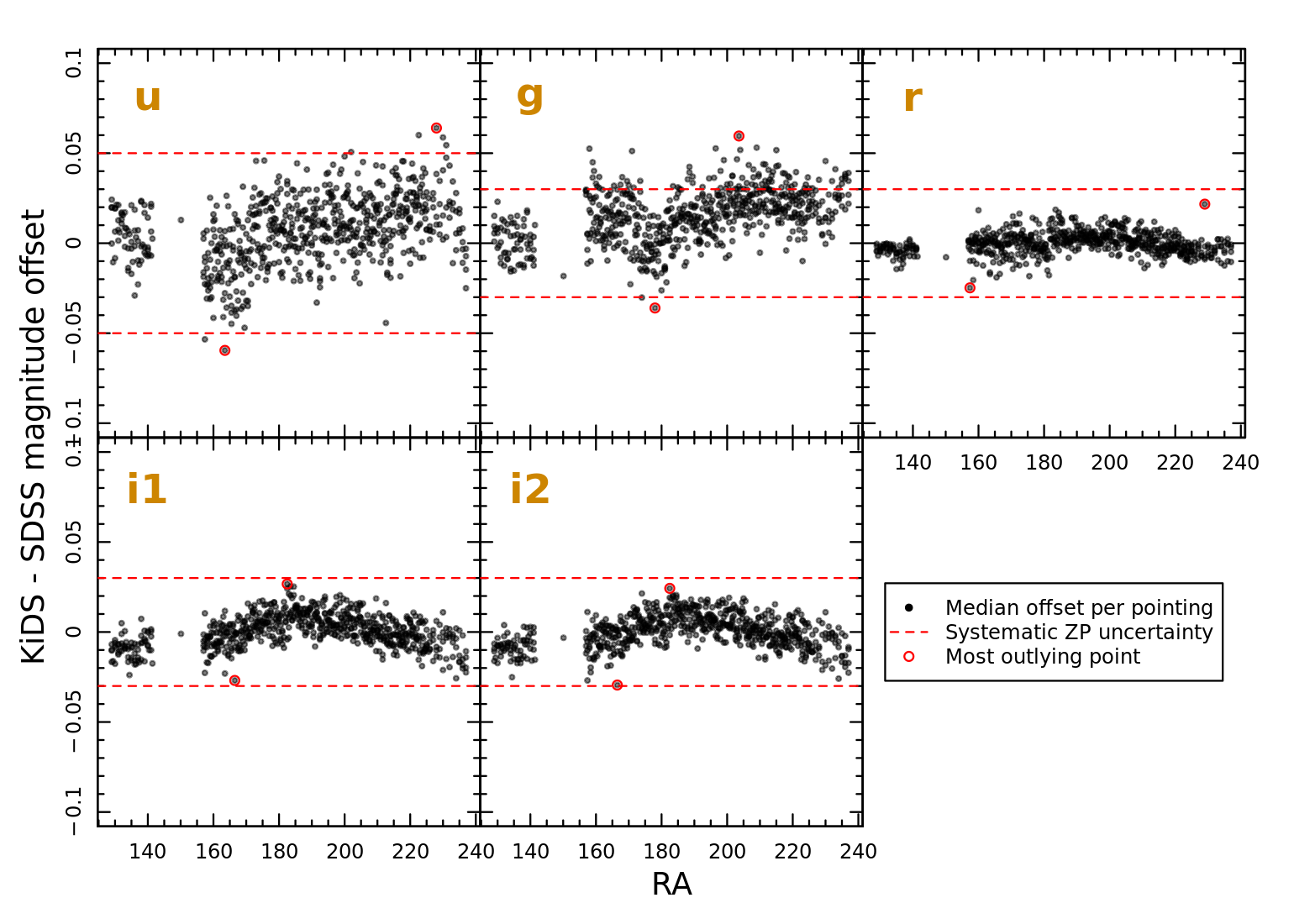}
  \caption{Distribution of the median offsets between stars in \kids\ \drfive\ tiles (in the magnitude range $16.5<r<19$) 
  and their counterparts from \sdss\ imaging, for tiles with $10\%$ or more unmasked data. 
  The offsets here were calculated after SLR corrections and \gaia\ recalibration, and therefore represent the final 
  quality of photometry in the survey. Horizontal dashed lines demonstrate the systematic zero-point uncertainty that is 
  included per-band in our scientific analyses (such as in the computation of \photoz), which are designed to 
  encapsulate any residual systematic variation in the photometry (such as can be seen as a function of RA, and whose origins are 
  unclear).}\label{fig:offsets}
\end{figure*}

Figure~\ref{fig:offsets} shows the residual median offsets between magnitudes for stars in \kids\ and SDSS, after the
full zero-point calibration and recalibration process.  Systematic residuals are below $0.05$ magnitudes in the \uband\ 
for the majority of the tiles ($<0.07$ mag for the two most outlying points). In the $gri$-bands, residuals are
 smaller: consistently less than $0.03$ magnitudes in the $r$- and $i$-bands, and typically less than $0.03$ 
in the $g$-band. For our subsequent analyses, we implemented systematic error floors of
$\sigma_u=0.05$, and $\sigma_X=0.03\,\forall\,X\in\{g,r,i_1,i_2\}$, which are designed to 
encapsulate any residual systematic variation in the photometry (such as can be seen as a function of RA). 

\subsubsection{Changes: \rband\ \pulcinella\ masks}\label{sec:pulcinella} 

The \astrowise\ pipeline includes the automatic masking of image artefacts around bright, saturated stars. This masking
is performed with the \pulcinella\ software, which separates the artefacts into different components
\citep{dejong/etal:2015}: saturated pixels in the cores of the stars, spikes caused by diffraction of the mirror
supports, spikes caused by the readout of saturated pixels, and up to three families of wide annular `ghost' reflection halos with spatially
dependent offsets around bright stars.

\begin{table} \caption{Updated parameters for the \rband\ \pulcinella\ masking in \kids\ \drfive.}\label{tab:awmasksettings}
  \begin{tabular}{c|cc}
    Parameter & DR4 Value(s) & DR5 Value(s) \\
    \hline                                
    {\tt halo1_area_min} &  850                   &   {\bf 1300 }                  \\
    {\tt halo2_area_min} &  950                   &   {\bf 1250 }                  \\
    {\tt halo3_area_min} &  950                   &   950                          \\
    {\tt spkd_profile [1]}  &  13.2025 &   13.2025  \\
    {\tt spkd_profile [2]}  &  34.9697 & 34.9697  \\
    {\tt spkd_profile [3]}  &  4.5 &  {\bf 4.0}   \\
    {\tt spkd_profile [1]}  &  102.386 &   102.386 \\
    {\tt spkd_profile [2]}  &  133.923 & 133.923  \\
    {\tt spkd_profile [3]}  &  4.5 &   {\bf 4.0}   \\
    {\tt spkd_profile [1]}  &  170.512 &   170.512 \\
    {\tt spkd_profile [2]}  &  192.542 & 192.542  \\
    {\tt spkd_profile [3]}  &  3.5 &  {\bf 3.0}   \\
    {\tt spkd_profile [1]}  &  195.332 &   195.332 \\
    {\tt spkd_profile [2]}  &  216.286 & 216.286  \\
    {\tt spkd_profile [3]}  &  4.0 &  {\bf 3.5}   \\
    {\tt spkd_profile [1]}  &  291.721   &   291.721  \\
    {\tt spkd_profile [2]}  &  309.5 & 309.5  \\
    {\tt spkd_profile [3]}  &  3.0   &  {\bf 2.5}   \\
    {\tt spkd_profile [1]}  &  310.561 &   310.561  \\
    {\tt spkd_profile [2]}  &  331.917 & 331.917  \\
    {\tt spkd_profile [3]}  &  3.5 &   {\bf 3.0}   \\
    {\tt spkd_profile [1]}  &  351.588 &   351.588 \\
    {\tt spkd_profile [2]}  &  369.336 &   369.336 \\
    {\tt spkd_profile [3]}  &  3.5 &   {\bf 3.0}   \\
    \hline
  \end{tabular}
  \tablefoot{
    These parameters define the minimum area of each halo ({\tt halo_area_min}) and describe the positioning of the
    diffraction spikes ({\tt spkd_profile}): starting angle ([1]), end angle ([2]), and spike length ([3]). Changed
    values are highlighted in bold face.
    }
\end{table}

\begin{figure} \centering
  \includegraphics[width=\columnwidth]{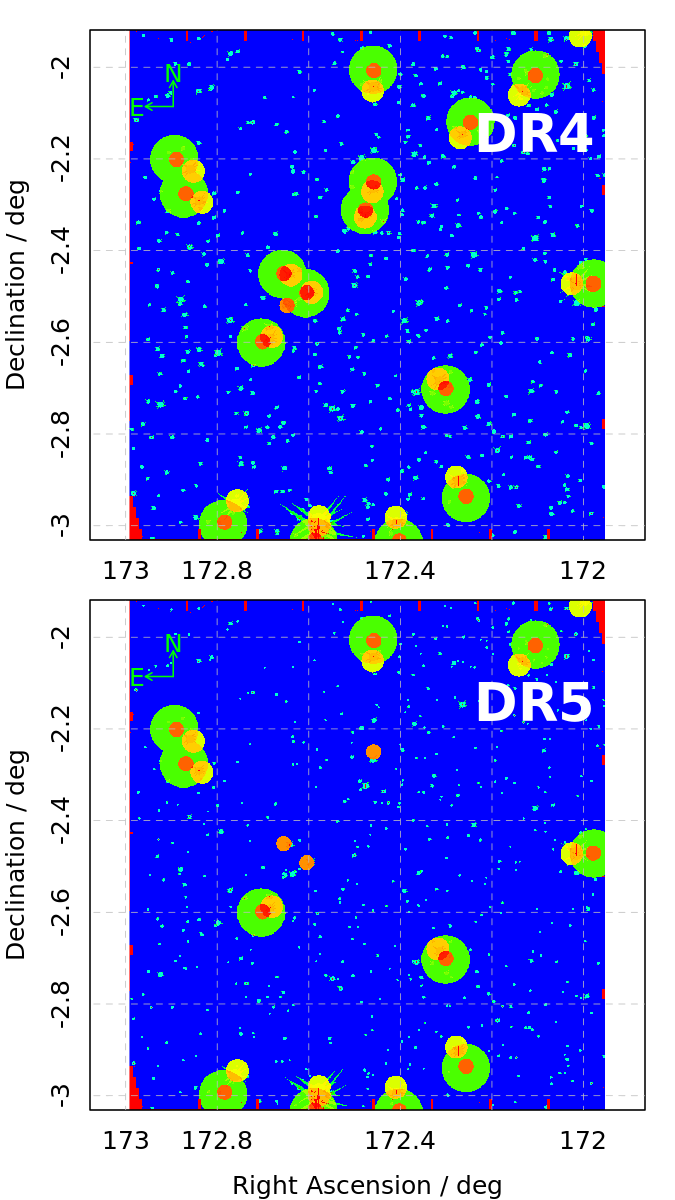}
  \caption{Comparison between the \astrowise\ \pulcinella\ masks for a selected \kids\ pointing. The images show the masks on a consistent linear colour scale, so pixels masked with the same bit(s) are shown with the same colour in each image. The \drfive\
  implementation of these masks was refined to use slightly different parameters (see Table~\ref{tab:awmasksettings}), 
  which more closely reproduce the masking behaviour of \theli\ in the \astrowise\ \rband. The primary effect is a reduction (in \drfive) of the size of stellar masks (cyan) and the frequency of masking of large reflection halos (green).}\label{fig:awmaskcompare}
\end{figure}

The properties of these components (the size of saturation cores, the size and offset of reflection halos, the
orientation of diffraction spikes) all depend on the brightness and position of the stars in the focal plane, 
in a way that is stable in time
for each photometric band. It is therefore possible to configure the modelling of each of these components by 
\pulcinella\ at the beginning of the survey \citep{dejong/etal:2015}. In \drfour, however, it was noted that some of 
these parameters were perhaps too conservative (particularly in the \rband), producing masks that were in better
agreement with the \theli\ `conservative' masks rather than the fiducial masks. As such, in \drfive\ we updated the
settings of the \pulcinella\ software, related particularly to the size of reflection halos and
orientation of diffraction spikes, to produce masks that are in better agreement with the fiducial masks generated by 
\theli. These changes are shown in Table~\ref{tab:awmasksettings}, and their effect on a sample \kids\ tile is displayed
in Figure~\ref{fig:awmaskcompare}. 

\subsubsection{Changes: $\utoi$-band manual masking} 

As a final step of the optical reduction process, compressed $1000\times\,1000$ pixel images of all $1347\times\,5$
\astrowise\ co-adds were visually inspected for artefacts, and manually masked by means of a polygon region file. The
main contaminating features discovered during this manual masking process were scattered light from bright sources
outside the field of view (primarily in the pre-mid-2015 data, when the telescope was still poorly baffled). These
artefacts can appear as fairly sharp-edged bright arcs across the focal plane (due to stars or planets), or as diffuse
patches (due to the Moon). Other artefacts discovered during this masking included satellite flares, airplanes, and
higher-order reflections from very bright stars. 

\begin{figure*}
  \centering
  \includegraphics[width=2\columnwidth]{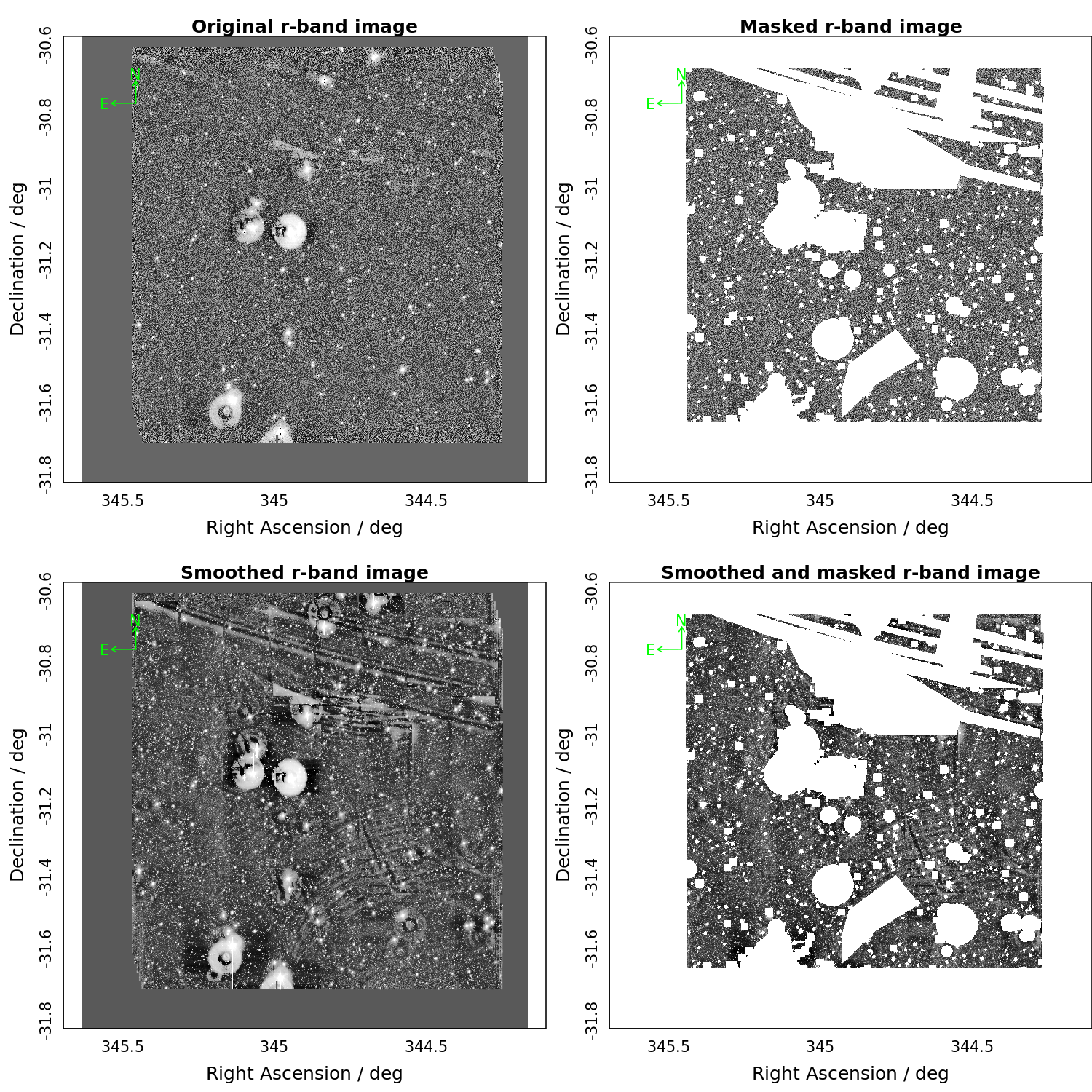}
  \caption{Examples of the new \astrowise\ manual masking implemented for \kids\ \drfive. The figure shows a heavily
  contaminated field, caused by scattered light from Formalhaut ($\alpha$PsA), a visible-magnitude star 
  ($V=1.16$) that is $\sim1.5\deg$ from the centre of the focal plane. The left column shows the tile before masking, 
  while the right column shows the tile after application of the manual and \pulcinella\ masks. The upper row shows the 
  tile at native resolution, while the bottom row shows the tile after smoothing with a $5\arcsec$ Gaussian filter.
  }\label{fig:manualmasks}
\end{figure*}

Figure~\ref{fig:manualmasks} shows an example of a heavily masked field in \kids\ \drfive. The field contains bright
reflections from the out-of-field star Fomalhaut ($\alpha$PsA, $V=1.16$), which resides $\sim1.5\deg$ NNW of
the centre of the focal plane. The field has been manually masked to remove the majority of bright contaminants. Faint
residual fluctuations are only visible after heavily smoothing (with a $5\arcsec$ Gaussian filter) and scaling the 
image to emphasise the background variations.  

\subsubsection{Single-band catalogues}\label{sec:singleband} 

As with previous \kids\ data releases, we provide source catalogues extracted from individual \astrowise\ co-adds in each band. 
These individual extractions are of particular interest for transient and variability studies, where one expects source lists 
to vary intrinsically between bands and/or observations. Of particular additional interest in \drfive\ is the distribution of single-band 
extractions in the two \iband\ passes, as sources with significant flux differences in the two \iband\ passes may indicate the presence of stationary transient features, such as extra-galactic supernova. 

These source extractions were performed using the same source detection parameters as used in the main \rband\ extraction on the 
\theli\ images, except they were applied to the \astrowise\ reduced co-adds. As a result, there are differences in the numbers 
of detected sources (and their observed properties), even when comparing the single-band \rband\ catalogues to the primary one. 
Also, it is important to note that cross-matching between single-band catalogues is unlikely to produce multi-band photometry 
that is accurate (at the same level as the dedicated multi-band photometry, Sect. \ref{sec:gaap}), because there is no guaranteed 
consistency of apertures and/or deblend solutions for the independent extractions. As such, multi-band photometry obtained from position-matching 
single-band catalogues should be treated with caution.  

Finally, as with previous releases, single-band catalogues are provided to the public largely `as-is'; they have not gone through the same rigorous quality control 
and testing as the lensing catalogues. More details of the contents of these catalogues are provided in Sect. \ref{sec:datarelease}.

\subsection{Optical multi-band \gaap\ photometry}\label{sec:gaap} 

One of the most important products released by the \kids\ consortium in each release is the multi-band photometry measured using the 
\gaap\ code \citep{kuijken:2011}. For a range of images (spanning various photometric bandpasses and PSFs), \gaap\ computes a (typically non-total) 
intrinsically consistent flux (i.e. probing the same pre-convolution extent in each image). This is performed as follows. 
Given an intrinsic source flux distribution, $f(x,y)$, an 
estimate of the flux can be made with a weighted aperture, using a Gaussian kernel. In \gaap, the Gaussian kernel is defined using an estimate of 
the source's major and minor axis lengths ($A$ and $B$, respectively), as measured in the \rband, and the 
source's orientation angle $\theta$. The flux estimated by \gaap\ is therefore 
\begin{equation}
    F_{\hbox{\gaap}}=\iint \mathrm{d} x \, \mathrm{d} y \, f(x,y) \, \exp\left\{\frac{-\left[\left(x^\prime/A\right)^2+\left(y^\prime/B\right)^2\right]}{2}\right\}, 
\end{equation}
where $x^\prime$ and $y^\prime$ are the $x$ and $y$ coordinates shifted to the centre of the source and rotated into the galaxy frame using the position angle $\theta$. 
In the presence of an arbitrary Gaussian PSF with standard deviation $\sigma_{\rm PSF}$, this intrinsic flux  
can be shown to be related directly to a Gaussian-weighted flux measured on the observed image, using a modified Gaussian kernel with 
\begin{align}\label{eqn:apers}
    A^\prime&\mapsto\sqrt{A^2-\sigma_{\rm PSF}^2}; \\
    B^\prime&\mapsto\sqrt{B^2-\sigma_{\rm PSF}^2}. 
\end{align}
This therefore provides a simple way to calculate fluxes across multiple images that probe the same intrinsic scales of a galaxy, provided 
that the image PSF is Gaussian (and assuming that sources are unblended). 

To ensure that image PSFs are Gaussian, we performed a step of Gaussianisation within \astrowise. This process involves accurately measuring the 
native PSF of each image, and using a shapelets expansion \citep{refregier:2003,kuijken:2006} to compute the (spatially varying) kernel required to convert the native PSF to 
a Gaussian (with minimal information loss). The native images are then convolved with the Gaussianisation kernel, producing images with a 
Gaussian PSF (at the cost of an increased correlation of the noise profile). The image PSF size is essentially unchanged in this process (as quantified 
via the FWHM), and the correlated noise caused by the convolution is propagated to the estimate of the flux uncertainty. 

As is clear from Eq. \ref{eqn:apers}, there is a physical limitation to the \gaap\ formalism when $\sigma_{\rm PSF}\geq [A,B]$. In this limit, the 
effective aperture sizes ($A^\prime,B^\prime$) become imaginary in one or both axes, and flux measurement is not possible. It is therefore sensible to define a minimum 
intrinsic aperture size for every source, that is larger than the expected PSF size in all bands. In \kids\ \drfive, as
in \drfour, we measured all 
sources with two intrinsic aperture sizes, which are a combination of the source intrinsic size (as the RMS of the flux distribution along the major and minor axes, in arcseconds, estimated from the \theli\ \rband\ imaging by \sourceextractor), a minimum aperture size $r_{\rm min}$, and a maximum aperture 
size of $2\arcsec$ (to limit blending effects). As such, the intrinsic aperture sizes for all sources in \drfive\ are defined as
\begin{align}
    A_{\mathrm{\gaap}}&=\min\left(\sqrt{A^2+r_{\rm min}^2},2\farcs0\right);\\
    B_{\mathrm{\gaap}}&=\min\left(\sqrt{B^2+r_{\rm min}^2},2\farcs0\right).
\end{align}
The two distinct sets of apertures come from the use of two distinct minimum aperture radii: 
$r_{\rm min}\in \left[0\farcs7,1\farcs0\right]$. While valid \gaap\ fluxes across all bands can be obtained with the 
lower $r_{\rm min}$ value for most sources, data with poorer seeing require the larger aperture (this point is further 
discussed in Sect.~\ref{sec:nirgaap}).

\section{NIR observations and reduction}
\label{sec:nirobservations}

One of the unique features of  \kids\  is the complementary five-band  \nir\ survey  
\viking. The utility of this overlap lies particularly in estimation of photometric redshifts and redshift distribution
calibration, resulting in higher-quality measurements across a larger redshift
baseline, as well as in classification studies generally \citep[see e.g.][]{nakoneczny/etal:2021}. 
With the extension of \kids\ to include the \kidz\ observations in \drfive, we also
required new \nir\ observations over these fields. In this section we detail the existing and recently acquired \nir\
observations in the \kidskidz\ fields, their reduction, and their quality.

\subsection{\vista\ \pawprint\ observations}\label{sec:vista}

The Visible and Infrared Survey Telescope for Astronomy (\vista) is a 4m ESO Telescope, is also located at ESO's Cerro Paranal observatory (albeit on a separate peak, roughly 1500m from the VST), and is serviced only by the 
Visible and Infrared Camera (\vircam). \vircam\ consists of 16 individual HgCdTe detectors, each with a 
$0.2\times 0.2$ square degree angular size, but which jointly span a $1\times1.2$ \sqdeg\  field of view. A single exposure of
the sky therefore contains (considerable) gaps between the detectors, in
what is referred to as the `\pawprint' pattern, and standard \vircam\ observations combine six dithers designed to fill in these gaps
\citep[see][]{dalton/etal:2006}.  Additionally, observations of a single \pawprint\ consist of a number of small
jitters, taken in quick succession, which allow the \pawprints\ to have reliably estimated backgrounds and to sample over detector defects. These exposures are stacked into a single `stacked \pawprint', and six
of these are combined to form a single contiguous $\sim 1.5$ \sqdeg\ `tile'.  The task of reducing the raw
images, particularly producing stacked \pawprints, was carried out by the Cambridge Astronomy Survey Unit
\citep[CASU;][]{gonzalez-fernandez/etal:2018,lewis/etal:2010}.

\vista\ observations used in \drfive\ span both the \kids\ survey area (with observations from \viking,
Sect.~\ref{sec:viking}), and the \kidz\ fields (with a combination of new, dedicated \viking-like observations, and
reconstructed \viking-like observations made from pre-existing deep \vista\ observations). The relevant observing programmes are listed in Table~\ref{tab:vikingkidz}. 

\begin{table} \centering
    \caption{Run numbers of all \vista\ observations taken in the \kids\ and \kidz\ fields.}
    \label{tab:vikingkidz}
    \begin{tabular}{c|ccc}
      Survey & Run Number & Start Date & $N_{\rm paw}$ \\
      \hline 
      \viking &  179.A-2004  & 2009-11-13 & 40\,741 \\
      \hline 
              &  179.A-2006 & 2009-11-05 & 168 \\
              &  179.A-2005 & 2010-02-18 &  24 \\
      \kidz   &  298.A-5015 & 2016-12-05 &  97 \\
              & 0100.A-0613 & 2017-09-17 & 204 \\
              & 0102.A-0047 & 2018-10-02 & 252 \\
      \hline
    \end{tabular}
  \tablefoot{
    Run numbers include observations taken as part of guaranteed time observations programmes and under
    director's discretionary time.}
\end{table}

\subsubsection{\viking\ observations within \kids}\label{sec:viking} 

\begin{figure*}
  \centering
  \includegraphics[width=2\columnwidth]{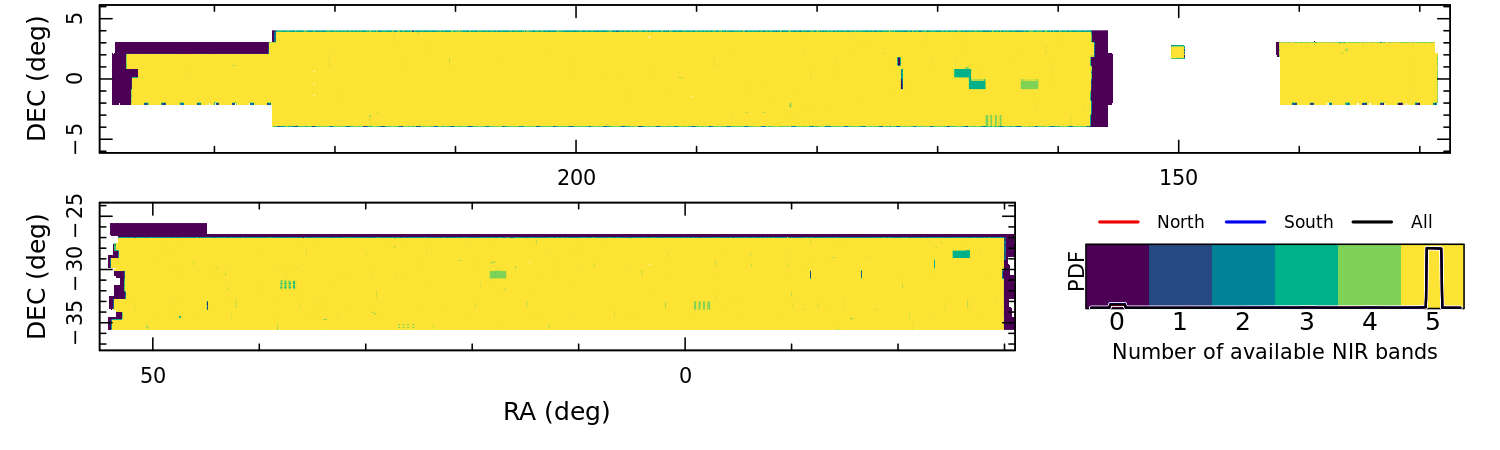}
  \caption{Coverage of \vista\ \viking\ data in the \kids\ fields, demonstrating that nearly the entire \kids\ \drfive\ footprint is covered by complete $\ztok$-band \viking\ observations.  }\label{fig:nircoverage}
\end{figure*}

As mentioned previously, the \viking\ imaging survey conducted on \vista\ was designed in combination with \kids,  to produce well-matched optical and \nir\ data.  As such,
the surveys share an almost identical footprint on-sky: Fig.~\ref{fig:nircoverage} shows the coverage of \viking\
observations within the northern and southern patches of \kids. We note the excellent overlap between the two surveys,
whereby only relatively small areas at the extreme edges of the northern and southern patches are without complete
($\ztok$) information.  

\begin{figure}
    \centering
    \includegraphics[width=\columnwidth]{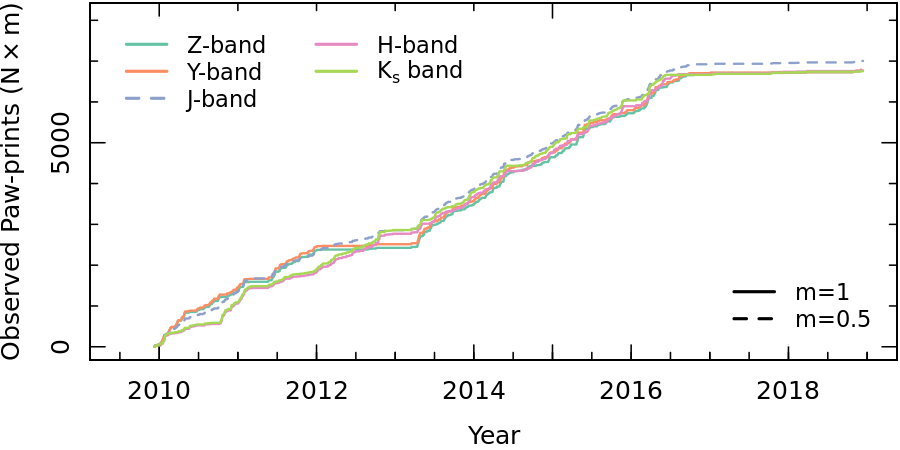}
    \caption{Progression of the NIR observations by the \viking\ survey. Note that the \jband\ line has been multiplied by a factor of $0.5$, to account 
    for the fact that it is observed with twice the frequency of the other bands. The observing strategy, combining observations of $ZYJ$-bands and $JHK_{\rm s}$-bands, 
    can be seen in the correlated increase in observed paws in the various filters.}
    \label{fig:obsprogressnir}
\end{figure}

Initial \viking\ observations were taken between 13 November 2009 and 24 August 2016, 
defining the final footprint.  Subsequently, some 
observations were repeated (primarily of data with instrumental problems), with the final observation dating from
16 February 2018. The data rate of
observations is shown in Fig.~\ref{fig:obsprogressnir}.  All \viking\ observations can be found in the
ESO archive under programme ID 179.A-2004, and a summary of the observational requirements is
provided in Table \ref{tab:viking}. 
A detailed description of the survey design and observing strategy is given in \citet{edge/etal:2013} 
and \citet{venemans/etal:2015}. 

\begin{table}
\centering
\caption{Requirements and settings for \viking\ observations with \vircam\ on \vista. }\label{tab:viking}
\begin{tabular}{c|ccccc}
Filter & $t_{\rm exp}$ & $N_{\rm exp}$ & $N_{\rm jitter}$ & $\langle N_{\rm dither}\rangle$ & $\langle t_{\rm tot} \rangle$\\
       & (s) & & & (per pix) & (s) \\
\hline
$Z $        & 60 & 1 & 4 & 2 & 480 \\
$Y $        & 25 & 2 & 4 & 2 & 400 \\
$J $        & 25 & 2 & 2 & 4 & 400 \\
$H $        & 10 & 5 & 3 & 2 & 300 \\
$K_{\rm s}$ & 10 & 6 & 4 & 2 & 480 \\
\hline
\end{tabular}
  \tablefoot{
      The number of dithers per pixel ($N_{\rm dither}$), and therefore also the total exposure time per pixel ($t_{\rm
      tot}$), varies over the contiguous $1.2$ \sqdeg tile. As such, in this table we show the $N_{\rm dither}$ and
      $t_{\rm tot}$ values for `typical' pixels, and emphasise this point by annotating these columns with `average'
      brackets ($\langle \dots \rangle$). 
     }
\end{table}

\subsubsection{\kidz\ \vista\ observing programme}\label{sec:kidzvista}

In order to produce \nir\ observations in the \kidz\ fields that are consistent with those existing in the \kids\
fields, we undertook a campaign to obtain \viking-like data over the \kidz\ fields from 5 December 2016 to 2 October 2018 
(see Table~\ref{tab:vikingkidz}), with the same
observing constraints and settings as the \viking\ survey itself (i.e.  Table~\ref{tab:viking}). 
Figure \ref{fig:tilesky_KIDZ} shows the distribution of these dedicated \vista\ observations in each of the \kidz\
fields, as green boxes. The areas of the \kidz\ fields that are not covered by these dedicated observations, but which
nonetheless contain both optical and \nir\ data (shown by the dark grey), consist of \viking-like data that was
reconstructed from pre-existing deep \vista\ observations in these fields (Sect. \ref{sec:kidznir}).

\subsubsection{Constructed \viking-like data in \kidz\ deep fields}\label{sec:kidznir}

For a number of the \kidz\ fields, there were pre-existing extremely deep \vircam\ observations obtained by the \ultravista\ \citep{mccracken/etal:2012} and \video\
\citep{jarvis/etal:2013} surveys.  Rather than re-observing these fields, we constructed \viking-quality data by selecting observations such that the total exposure time per pixel was at least as deep as in
\viking, and which contain \pawprints\ of similar seeing. The observation settings for \ultravista\ and \video\ (and how
many of such observations were selected to reach at least \viking\ depth, $\langle N_{\rm dither}\rangle$) are given in Table \ref{tab:deepvista}. We note that there are no $Z$-band 
observations within \ultravista; here new $Z$-band imaging was obtained with \vista\ using \viking\ observation parameters (included in Table \ref{tab:vikingkidz}).  
\begin{table}
\centering
\caption{Requirements and settings for deep \vircam\ with \ultravista\ and \video\ on \vista. }\label{tab:deepvista}
\begin{tabular}{c|ccccc}
Filter & $t_{\rm exp}$ & $N_{\rm exp}$ & $N_{\rm jitter}$ & $\langle N_{\rm dither}\rangle$ & $\langle t_{\rm tot} \rangle$\\
       & (s) & & & (per pix) & (s) \\
\hline
\multicolumn{5}{c}{\video} \\
  $Z $        & 50 & 1 & 10 & 1 & 500 \\
  $Y $        & 30 & 2 & 8  & 1 & 480 \\
  $J $        & 30 & 2 & 8  & 1 & 480 \\
  $H $        & 10 & 6 & 7  & 1 & 420 \\
  $K_{\rm s}$ & 10 & 6 & 7  & 2 & 840 \\
\multicolumn{5}{c}{\ultravista} \\
  $Y $        & 60 & 2 & 30 & 1 & 3600 \\
  $J $        & 30 & 4 & 30 & 1 & 3600 \\
  $H $        & 10 & 6 & 30 & 1 & 1800 \\
  $K_{\rm s}$ & 10 & 6 & 30 & 1 & 1800 \\
\hline
\end{tabular}
  \tablefoot{
      Observation settings for each paw-print ($t_{\rm exp}$, $N_{\rm exp}$, $N_{\rm jitter}$) were used to compute the
      average number of dithers per-pixel on-sky ($\langle N_{\rm dither}\rangle$) that are needed to reach at least
      \viking\ depth. We then randomly sampled $\langle N_{\rm dither}\rangle$ dithers per pixel on-sky to construct our
      \viking-like data in these deep fields. Subsequent further degradation of the photometry (in catalogue space) is
      needed to reach equivalent \viking\ depths.
    }
\end{table}
Given the very different observing strategies, it proved impossible to match the depth of the \viking\ data exactly, and we therefore further degraded the photometry in these bands to the 
expected
depth of \viking\ (a process described in \citealt{hildebrandt/etal:2017} as `magnitude adaption'). This step is clearly particularly relevant in \ultravista, where typical depths are equivalent to 
the deepest parts of \viking\ (i.e. where dither overlaps create $N_{\rm dither}=6$).

\subsection{KiDS-VIKING pipeline reduction}

The \nir\ data in \kidskidz\ were reduced with the same reduction pipeline used in previous \kids\ data releases:
\kvpipe\ \citep{wright/etal:2018}, based on the processing pipeline developed in
\citet{driver/etal:2016} for \gama. 
We opted to start from the reduced CASU \pawprints, because of the complex observing strategy and dither pattern of
\vista\ observations: within one tile there is a wide range of total exposure times (between
$1\times$ and $6\times$ the individual exposure times, $t_{\rm exp}$) and observational properties (up to $96$ different
PSFs). This variability motivated us to utilise individual stacked chips, extracted from the CASU \pawprints, as the
basis for our reduction and forced photometry, rather than co-added tiles. 

The processing pipeline first corrects the images for atmospheric extinction ($\tau$) given the observation
airmass ($\sec \chi$), removes the exposure time ($t$, in seconds) from the image units, and converts the images 
from various Vega zero-points ($Z_{\rm v}$) to a standard AB zero-point of $30$ \citep[using Vega to AB corrections, 
$X_{\rm AB}$, from][]{gonzalez-fernandez/etal:2018}. These various corrections and transformations are performed using a
single multiplicative recalibration factor $\mathcal{F}$ per \vista\ detector that is applied to all pixels in the
detector image $I$:
\begin{equation}
    I_{\rm new} = I_{\rm orig} \times \mathcal{F}. 
\end{equation}
The factor $\mathcal{F}$ is constructed as 
\begin{equation}
\begin{split}
    \log_{10}(\mathcal{F}) = -0.4[&Z_{\rm v} - 2.5 \log_{10}(1/t)  \\
    &-\tau\left(\sec\chi-1\right)+X_{\rm AB}-30]. 
\end{split}
\end{equation}
The processing pipeline also performs a re-orientation of the individual \pawprint\ detector stacks using \swarp\ 
\citep{swarp} and a background subtraction (again using \swarp). The \swarp\ background subtraction is computed 
using a $256\times256$ pixel mesh, and a $3\times3$ filter for the bicubic spline. This additional background
subtraction was considered optimal by \cite{driver/etal:2016}, who demonstrated that these settings allowed for maximal
removal of backgrounds with minimal impact on the photometry of extended sources. 

\begin{figure}
  \centering
  \includegraphics[width=\columnwidth]{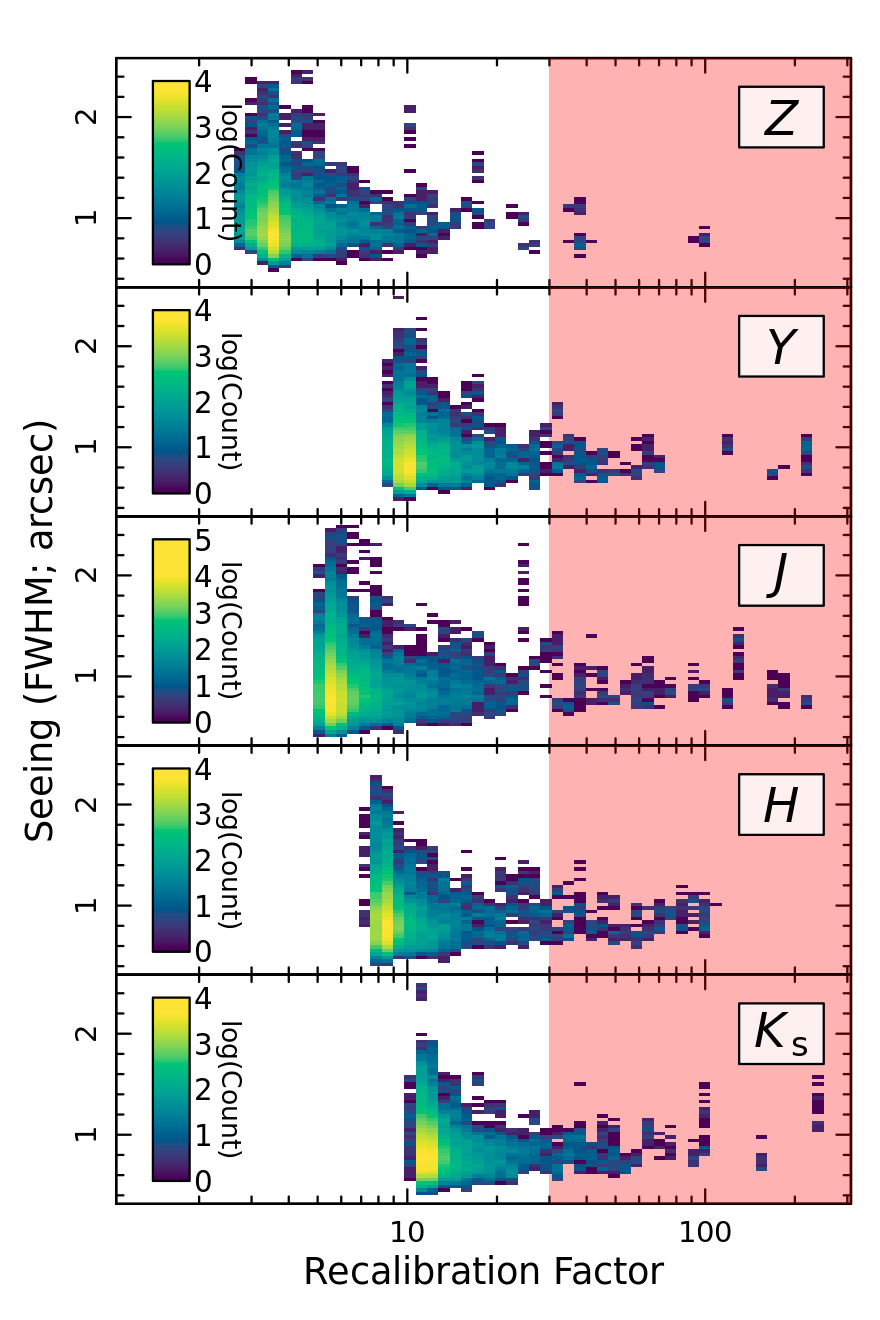}
  \caption{Distribution of recalibration factors $\cal F$, derived by \kvpipe\ using the parameters provided by \casu, for
  \viking\ and \viking-like observations in the \kidskidz\ fields. Following \protect\cite{driver/etal:2016} and
  \protect\cite{wright/etal:2019}, we applied a rejection of detectors with recalibration factors $\mathcal{F}\ge30$. There are additional
  indirect selections, however, that occur for fields with large PSF sizes (see Sect. \protect\ref{sec:strongselection}).}\label{fig:nirzpmultiplier}
\end{figure}

The distribution of the recalibration factor $\mathcal{F}$ per band is shown in Fig. \ref{fig:nirzpmultiplier}.
Following \citet{driver/etal:2016,wright/etal:2018} we applied a blanket quality-control selection of $\mathcal{F}\leq
30.0$ to remove detectors with strong persistence or other artefacts. 

\subsection{NIR quality metrics}\label{sec:vistaquality}

\begin{figure*}\centering
  \includegraphics[width=\columnwidth]{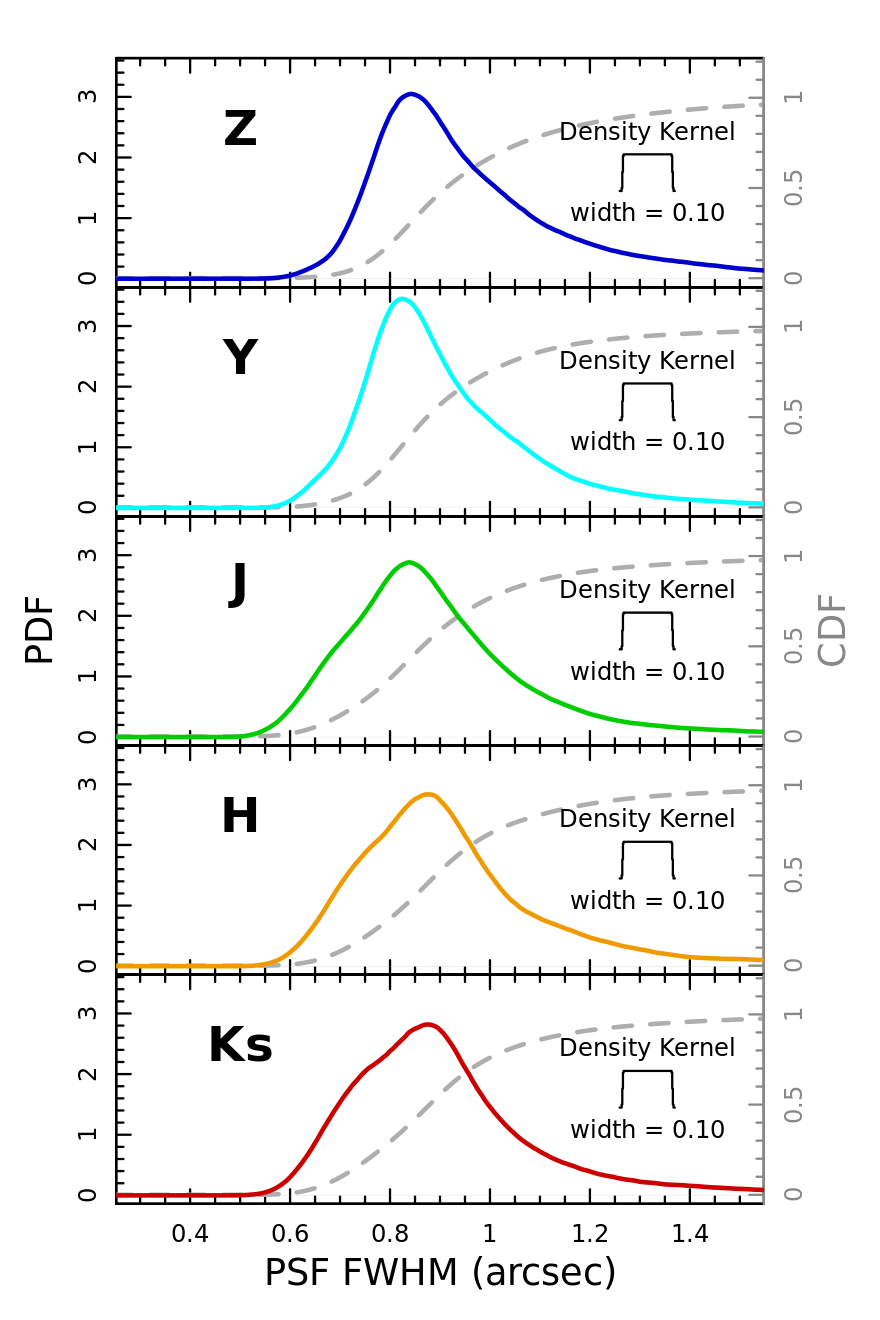}
  \includegraphics[width=\columnwidth]{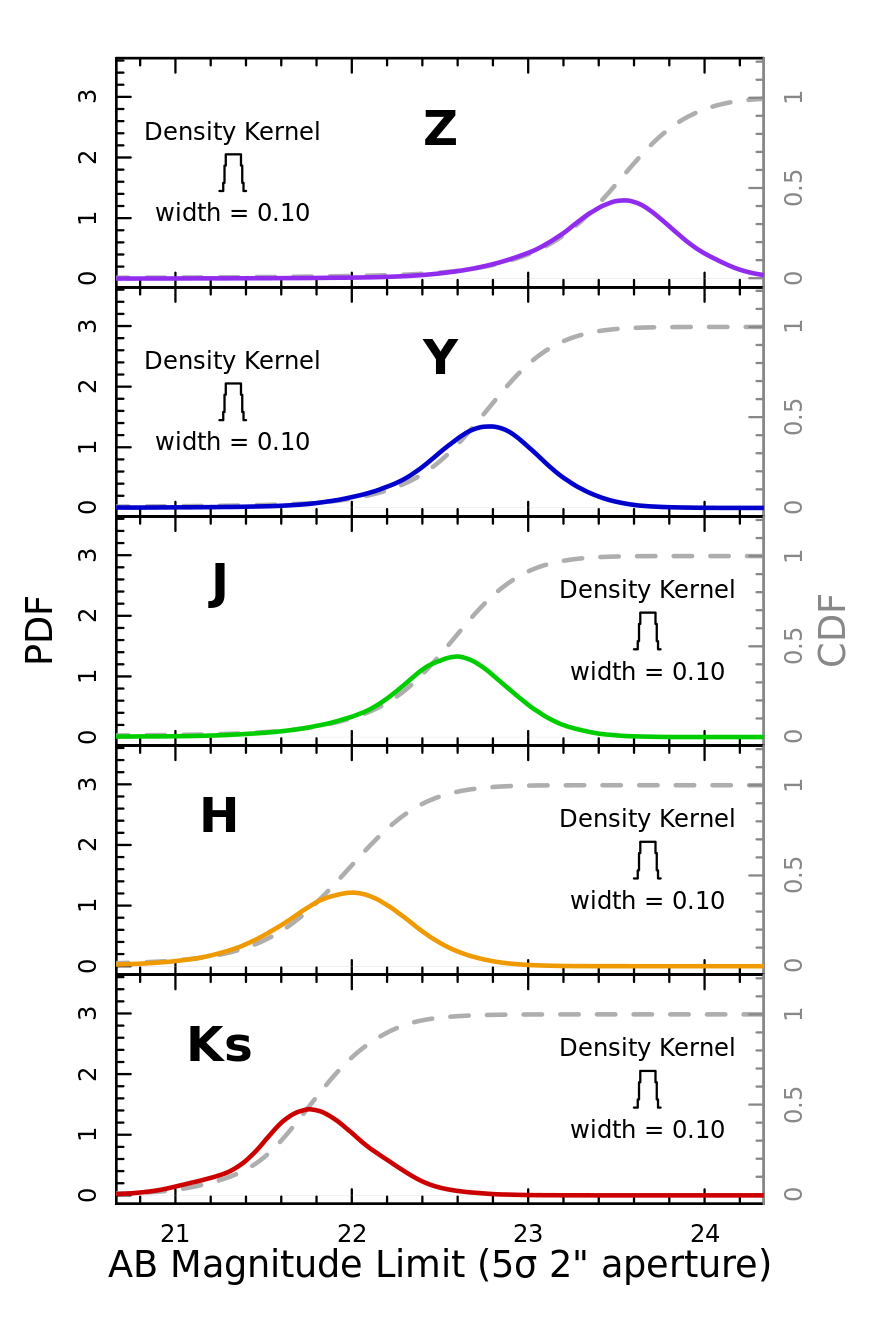}
  \caption{Primary observational properties of pointings in \kidskidz\ for observations in the five \nir\ bands. Each
  row shows the distribution of PSF sizes (as measured on each \vista\ chip; left) and limiting magnitude 
  (as determined by the magnitude of a $5\sigma$ source in a $2\arcsec$ circular aperture; right) determined with a KDE
  using the annotated kernel. The corresponding cumulative distribution functions for each panel are shown as grey
  lines.
  }\label{fig:nirpsfsizes}
\end{figure*}

We quantify the quality of the NIR reduction using the same metrics as for the optical portion of the survey: PSF FWHM 
and depth (as determined in blank apertures scattered randomly over the detectors, Appendix~\ref{sec:lambdar}). Figure
\ref{fig:nirpsfsizes} presents the summary of PSF sizes and limiting magnitudes in each of the five \nir\ bands. 

The PSF sizes in the \nir\ bands are all exceptionally consistent. This is due (at least in part) to the rapid variability
of the \nir\ sky: PSFs vary on a second-to-second timescale, as do backgrounds. This necessitates that the `typical' PSF
of any one observation approaches the mean of the seeing conditions (i.e. following the central limit theorem). However, 
the distributions also all show considerable tails to high seeing; again, a demonstration of the difficulties in
observing the \nir\ sky from the ground. This tail of poor seeing observations has some important consequences for our
analysis and sample selection further down the line in our processing pipeline (Sects. \ref{sec:nirgaap} and 
\ref{sec:strongselection}). 

Limiting magnitudes in the \nir\ filters were determined on individual detectors. However, as there are possibly many
individual detectors per source, we need a reasonable method for combining the estimated magnitude limits into a final
representative distribution of depths (i.e. that reflects the approximate depth that we achieve per source, in the correct 
proportion). We achieved this by measuring the background variance of the individual chips at a
number of locations within the chip. Each of these background estimates is tagged with the central RA and Dec used for the
estimate, and estimates (from a single band) that are within $\Delta\{{\rm RA,Dec}\}\leq0.02\deg$ of each other are
pooled into a single estimate, assuming that fluxes are combined using a simple average of their
individual measurements (which is not strictly the case; see Sect. \ref{sec:nirgaap}), and that the noise in overlapping chips is uncorrelated. This means that the 
variance of the mean stack of chips at each location on sky is simply 
\begin{equation}
  \sigma^2_{\rm final} = \frac{\sum^N_i \sigma_i^2 }{N^2},
\end{equation}  
where $N$ is the number of detectors overlapping with each other at this point on-sky, and $\sigma_i^2$ is the variance
of the $i^{\rm th}$ detector. From this calculation, we found that the median limiting magnitudes in the $\ztok$-bands are 
$\left\{23.49,22.74,22.55,21.96,21.77\right\}$. The distribution of these combined limiting magnitude estimates on-sky 
is given in Appendix~\ref{sec:vistaonskymetrics}. 

\section{\kidz\ spectroscopic compilation}\label{sec:specz}

\begin{figure*}
  \centering
  \includegraphics[width=2\columnwidth]{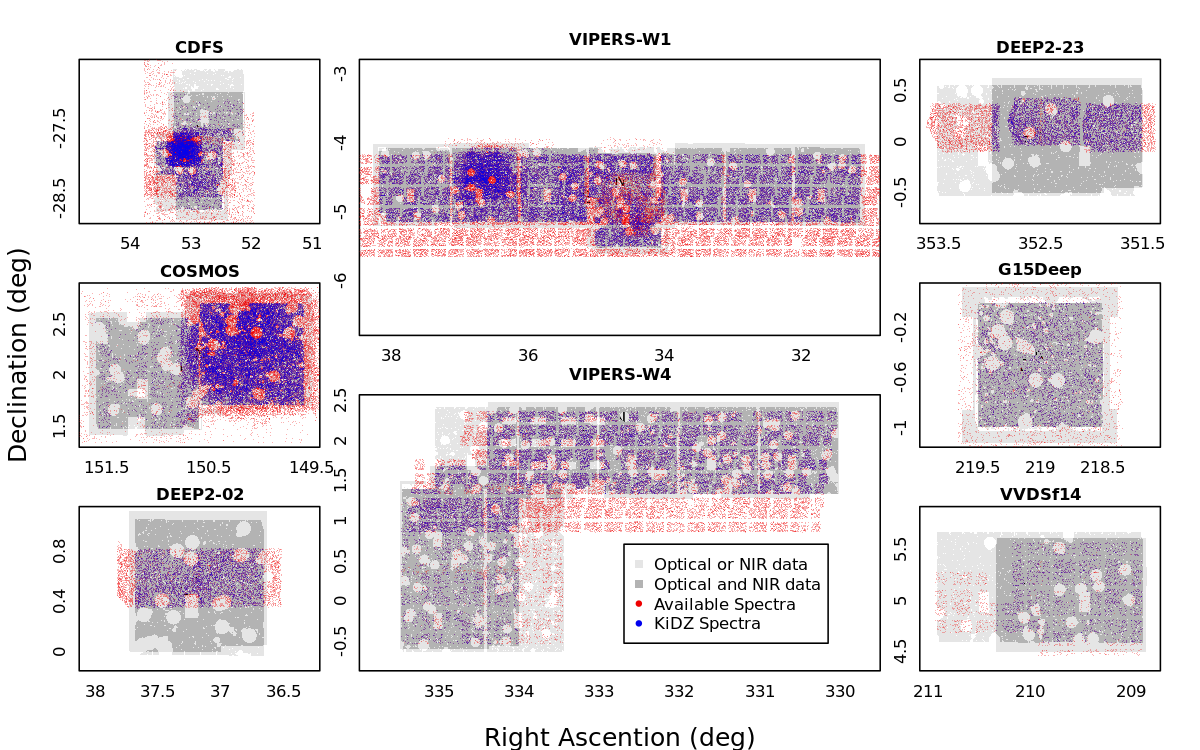}
  \caption{Distribution of \kidz\ spectroscopic redshift estimates on-sky. The figure shows the distribution of all 
  available spectroscopic redshift estimates from the full spectroscopic compilation (red) and those that are matched to unmasked 
  \kidz\ sources (blue). The available footprint
  of the \kidz\ imaging is shown in grey scale beneath the points, demonstrating where we have imaging but no available spectra
  (and vice versa).}\label{fig:kidzonsky}
\end{figure*}

As discussed in Sects. \ref{sec:kidzacquisition} and \ref{sec:kidzvista}, we undertook a targeted campaign to observe eight
spectroscopic calibration fields, which make up the `\kidz' fields. 
The associated spectroscopic redshift data used in \drfive\ is a compilation of public and proprietary measurements from a range of
surveys, which we refer to here simply as `the full compilation'. The list of spectroscopic surveys that we draw from to create 
the full compilation is provided in Table \ref{tab:spectra}. 

The compilation includes two main components. The first are redshifts from wide-angle spectroscopic surveys that 
mainly cover lower redshifts and brighter magnitudes, and which overlap at least partly with the \kids\ main survey 
area. Surveys that contribute to this portion of the dataset include \sdss\ \citep{york/etal:2000}, 
\gama\ \citep{driver/etal:2022}, \tdflens\ \citep{blake/etal:2016}, and \wigglez\ \citep{parkinson/etal:2012}. 
The second component 
comes from several deep (often pencil-beam) spectroscopic surveys that intersect the \kidz\ fields, including \vvds\ \citep{ilbert/etal:2006}, 
\vipers\ \citep{scodeggio/etal:2018}, \deeptwo\ \citep{newman/etal:2013}, \ctrtwo\ \citep{stanford/etal:2021}, and \zcosmos\ \citep{lilly/etal:2009}. 

A number of these  datasets have sources in common  (due, in part, to some of the samples being compilations themselves: 
e.g. G10\_COSMOS, \zcosmos, and \goods). Furthermore, some datasets include duplicate and/or multiple redshift estimates, with some appearing more 
than twice in their respective datasets. To remove duplicated spectroscopic redshift estimates, we adopted a strategy designed to 
retain only the highest-quality redshift estimates per source. 
Before performing this duplicate removal, however, we first homogenised the redshift quality flags among the different surveys 
(which are defined using various inconsistent criteria). This procedure resulted in a single redshift quality flag, based on the 
GAMA $\mathtt{NQ}$ nomenclature, that indicates the confidence in a redshift estimate's accuracy as  
\begin{equation}
\mathtt{NQ}=
\begin{cases}
4 & {\rm certain}\, (\,\geq 98\% \,{\rm confidence})\\ 
3 & {\rm high}\, (95{\rm -}98\% \,{\rm confidence})\\ 
2 & {\rm moderate\,confidence} \\
1 & {\rm low\,confidence} \\ 
\end{cases}.
\end{equation}
We note that in what follows we only use objects with $\mathtt{NQ}\geq3$.

This spectroscopic compilation is nominally the same as used previously in \citet{vandenbusch/etal:2022}, although the 
overlap with photometric data has increased considerably due to our dedicated \kidz\ imaging.
\citet{vandenbusch/etal:2022} describe the homogenisation process used in the construction of the compilation, and the results 
of the merger in the spectroscopic fields available during \drfour. This process is unchanged here, but is applied to the 
larger \kidz\ sample that is now available.  
Here we summarise the selection and quality flag assignment per input survey.

\begin{itemize}
  \item Arizona CDFS Environment Survey \citep[ACES;][]{cooper/etal:2012}: galaxies were selected with $\mathtt{Z\_QUALITY} \geq 3$ 
and $\mathtt{zErr}/z<0.01$. We assigned $\mathtt{NQ} = \mathtt{Z\_QUALITY}$;

\item Deep Extragalactic Visible Legacy Survey \citep[DEVILS;][]{davies/etal:2018}: galaxies were selected with spectroscopic redshifts (i.e. 
$\mathtt{zBestType}=\mathtt{spec}$), and with 
flags $\mathtt{starFlag}=0$, $\mathtt{mask}=0$, and $\mathtt{artefactFlag}=0$. We assigned 
$\mathtt{NQ}=4$ if $\mathtt{zBestSource}=\mathtt{DEVILS}$ and $\mathtt{NQ}=3$ otherwise;
 
\item Complete Calibration of the Colour-Redshift Relation (C3R2) survey: galaxies were selected from the combination of four C3R2 public datasets: 
DR1 \citep{masters/etal:2017}, DR2 \citep{masters/etal:2019}, DR3 
\citep{stanford/etal:2021}, and KMOS \citep{euclid20}. We required $\mathtt{QFLAG} \geq 3$ 
for galaxies in this sample, and assigned $\mathtt{NQ}=\mathtt{QFLAG}$;

\item Deep Imaging Multi-Object Spectrograph survey \citep[DEIMOS;][]{hasinger/etal:2018}: galaxies were selected with quality flag $\mathtt{Q}=2$. We assigned 
$\mathtt{NQ}=4$ for $\mathtt{Q_f} \in \left[ 4,14 \right]$, and $\mathtt{NQ}=3$ otherwise;

\item DEEP2 \citep{newman/etal:2013}: as in the previous KiDS papers 
\citep{hildebrandt/etal:2017,hildebrandt/etal:2020,hildebrandt/etal:2021} 
we selected galaxies from two equatorial fields (0226 \& 2330), with $\mathtt{Z\_QUALITY} \geq 3$ and  
$\mathtt{zErr}/z<0.01$. We assigned $\mathtt{NQ}=\mathtt{Z\_QUALITY}$;

\item Fiber Multi-Object Spectrograph COSMOS survey \citep[FMOS-COSMOS;][]{silverman/etal:2015}: galaxies were selected with quality flag 
$\mathtt{q\_z} \geq 2$. We assigned $\mathtt{NQ}=4$ if $\mathtt{q\_z} =4$ and 
$\mathtt{NQ}=3$ otherwise;

\item GAMA \citep{driver/etal:2022}: galaxies from the $4^{\rm th}$ Data Release were selected with 
redshift quality $\mathtt{NQ}\geq 3$ and with $z>0.002$ to avoid stellar contamination. 
We propagated $\mathtt{NQ}$ from GAMA;

\item GAMA-G15Deep \citep{kafle/etal:2018,driver/etal:2022}: galaxies were selected with input redshift 
quality $\mathtt{Z\_QUAL}\geq 3$ and with redshifts $z>0.001$ to avoid stellar contamination. We assigned 
$\mathtt{NQ} = \mathtt{Z\_QUAL}$;

\item G10-COSMOS \citep{davies/etal:2015}: galaxies were selected with $\mathtt{Z\_BEST}$ as the redshift 
value, and otherwise  following the documentation for selecting galaxy redshifts: 
$\mathtt{Z\_BEST}>0.0001$, $\mathtt{Z\_USE}<3$, and $\mathtt{STAR\_GALAXY\_CLASS}=0$. 
All sources are assigned $\mathtt{NQ}=3.5$. 

\item Great Observatories Origins Deep Survey (GOODS CDFS): galaxies were selected from the public ESO compilation of spectroscopy in the CDFS 
field\footnote{Available from \url{https://www.eso.org/sci/activities/garching/projects/goods/MasterSpectroscopy.html}} 
\citep{popesso/etal:2009,balestra/etal:2010}. Following the recommendations in the dataset 
description\footnote{\url{https://www.eso.org/sci/activities/garching/projects/goods/MASTERCAT_v3.0.dat}} we selected `secure' 
redshifts (assigning $\mathtt{NQ}=4$ to them) and `likely' redshifts ($\mathtt{NQ}=3$);

\item Hectospec COSMOS survey \citep[hCOSMOS;][]{damjanov/etal:2018}: all galaxies from the published dataset were selected, and assigned them redshift quality $\mathtt{NQ}=4$;

\item The Large Early Galaxy Astrophysics Census \citep[LEGA-C;][]{vanderwel/etal:2016}: galaxies were selected with $\mathtt{f\_use}=1$. We assigned $\mathtt{NQ}=4$ to all sources;

\item Australian Dark Energy Survey \citep[OzDES;][]{lidman/etal:2020}: galaxies were selected in two patches partly overlapping with the \kidz\ around CDFS and 
the VVDS 2h field, with required quality $\mathtt{qop} \in \lbrace3,4\rbrace$. An additional selection of $z>0.002$ was made 
to exclude stellar contaminants. We assigned $\mathtt{NQ}=\mathtt{qop}$;

\item SDSS \citep{sdss_dr14:2018}: galaxies from the $14^{\rm th}$ Data Release 
were selected with $\mathtt{zWarning}=0$ and $\mathtt{zErr}>0$. We furthermore required that $\mathtt{zErr}<0.001$, 
$\mathtt{zErr}/z<0.01$, and $z>0.001$. We assigned $\mathtt{NQ}=4$ to all such selected galaxies;

\item VANDELS survey \citep{garilli/etal:2021}: galaxies were selected with $(\mathtt{zflg}\!\mod 10) \in \lbrace 2, 3, 4\rbrace$. We 
assigned $\mathtt{NQ}=4$ if $(\mathtt{zflg}\!\mod 10) \in \lbrace 3, 4 \rbrace$, and $\mathtt{NQ}=3$ otherwise. The reassignment 
of the quality flags here was motivated by the reportedly high redshift confidence of objects with flag values of two and three.

\item VIPERS \citep{scodeggio/etal:2018}: galaxies were selected with $2\leq\mathtt{zflg}<10$ or $22\leq\mathtt{zflg}<30$. 
We assigned $\mathtt{NQ}=4$ if $3\leq\mathtt{zflg}<5$ or $23\leq\mathtt{zflg}<25$, and $\mathtt{NQ}=3$ otherwise;

\item VIMOS Ultra Deep Survey \citep[VUDS;][]{lefevre/etal:2015}: galaxies were selected with flag $\mathtt{zflags}$ ending with $\lbrace 3, 4, 9\rbrace$ 
(reliability $\geq 80\%$) and assigned $\mathtt{NQ}=4$ if $3\leq \mathtt{zflags} <5$ or 
$13 \leq \mathtt{zflags} < 25$, and $\mathtt{NQ}=3$ otherwise;

\item VVDS \citep{lefevre/etal:2005,lefevre/etal:2013}: galaxies were selected from the combined WIDE, DEEP, and UDEEP sub-samples, 
with $\mathtt{ZFLAGS} \in \lbrace3,4,23,24\rbrace$. We assigned $\mathtt{NQ}=4$ to all sources. 

\item zCOSMOS: galaxies were selected from a compilation of public \citep{trump/etal:2009,comparat/etal:2015,lilly/etal:2009} and proprietary\footnote{Sources from this sample that are proprietary 
are tagged with an additional flag, and are only provided to members of the community with allowed access (or after the 
public release of these data).} spectra in the COSMOS field, kindly 
provided to us by Mara Salvato, updated as of 1 September 2017. That dataset includes some of the surveys already included 
in our compilation, but also provides redshifts from various other campaigns. We used the provided quality flag and 
selected galaxies with $3 \leq \mathtt{Q\_f} \leq 5$, or $13 \leq \mathtt{Q\_f} \leq 15$, or 
$23 \leq \mathtt{Q\_f} \leq 25$, or $\mathtt{Q\_f} \in \lbrace 6, 10 \rbrace$. We rejected sources with low-confidence redshift estimate (e.g. from grism spectroscopy), and limit the galaxies to $\mathtt{z\_spec} > 0.002$ to avoid stellar contamination. We 
assigned redshift quality as $\mathtt{NQ} = \min((\mathtt{Q_f}\!\mod 10),4)$;

\end{itemize}

When combining the above spectroscopic samples, we removed both internal (i.e. within the same input catalogue) and
external (i.e. in different input catalogues) duplicates. In the latter case, we assigned the most reliable measurement
per source based on a specific `hierarchy'. Namely, we joined the catalogues by cross-matching objects within $1\arcsec$
radius and apply the following order of preference:
\begin{itemize}
\item GAMA takes precedence over others, followed by SDSS; then:
\item COSMOS field: G10-COSMOS > DEIMOS > hCOSMOS > VVDS > Lega-C > FMOS > VUDS > C3R2 > DEVILS > zCOSMOS;
\item CDFS field: ACES > VANDELS > VVDS > VUDS > GOODS CDFS > DEVILS > OzDES;
\item VIPERS\_W1 field: VIPERS > VVDS > C3R2 > DEVILS > OzDES.
\end{itemize}

This hierarchy was followed regardless of the relative quality flags between different surveys. 
For objects with multiple spectroscopic measurements within a particular survey, we either selected the redshift 
with the highest quality flag or, if various entries for the same source have the same quality flag and the 
reported redshifts differ by no more than 0.005, we used the average of the provided redshift estimates. 
If the reported redshifts have the same quality flag but differ by more than 0.005, we excluded the source 
from the compilation.

The final spectroscopic sample consists of $635\,099$ spectroscopic redshift estimates taken from $22$ samples.
Table~\ref{tab:spectra} lists the statistics for this full compilation, prior to matching with sources 
detected in our \kidz\ imaging. The table presents the number of spectroscopic redshift estimates, redshift range, mean
redshift, and redshift scatter (computed using the NMAD) for each sample. Figure \ref{fig:kidzonsky} shows the
distribution of spectroscopic redshift estimates from the full compilation (blue and red) in the context of the
individual \kidz\ fields on-sky. The spectroscopic data can be seen to extend beyond the spatial extent of the \kidz\
optical and NIR data (grey scale), to maximise the cross-match between available spectroscopic redshift estimates and
extracted \kidz\ photometric sources. 

Finally, the spectroscopic redshift estimates were cross-matched to the sources detected in the \kidz\ \atlas-reduced
images using a simple sky match at $1\arcsec$ radial tolerance\footnote{In practice, the matching of the spectroscopic
compilation to the \kidz\ sources happens after the computation of multi-band information for the \kidz\ fields,
including masks, discussed in Sect. \ref{sec:multiband}. Nonetheless we include this step here for simplicity.}. The
matched sources are shown in Fig.~\ref{fig:kidzonsky} as blue points. The final sample of spectroscopic redshift
estimates after matching to the \kidz\ photometric data contains $126\,085$ sources; the statistics for these redshift
estimates are also provided in Table~\ref{tab:spectra}.

\begin{figure*}
  \centering
  \includegraphics[width=2\columnwidth]{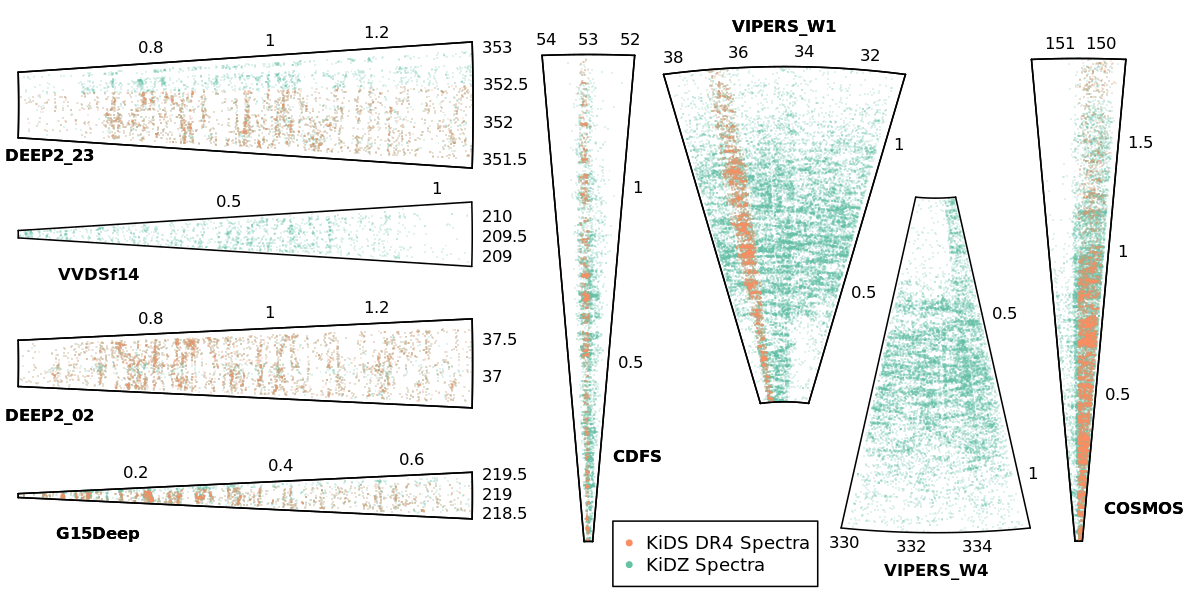}
  \caption{Distribution of \kidz\ spectroscopic redshift estimates in RA-$z$ space. Sources in orange indicate estimates that were available for
  previous \kids\ analyses, and sources in cyan show new estimates added here.}\label{fig:kidzraz}
\end{figure*}

Figure \ref{fig:kidzraz} shows the cross-matched \kidz\ spectroscopic redshift estimates in the RA-$z$ plane,
demonstrating the relative depth of the various fields, and large-scale structures contained within them. For
comparison, the spectroscopic redshift estimates that were available for previous \kids\ analyses are shown in orange.   

\begin{table*}
 \caption{Statistics for the \kidz\ spectroscopic sample.}\label{tab:spectra}
 \resizebox{\textwidth}{!}{
  \begin{tabular}{c|rccc|rcccc|l}
    & \multicolumn{4}{c|}{Full Compilation} & \multicolumn{5}{c|}{Matched to \kidz\ sources} &  \\
    Survey & ${N_{\rm spec}}$ & $[z_{\rm min},z_{\rm max}]$ & $\langle z\rangle$ & ${\rm NMAD}(z)$ & ${N_{\rm spec}}$ & $[z_{\rm min},z_{\rm max}]$ & $\langle z\rangle$ & ${\rm NMAD}(z)$ & $\langle r_{\rm auto}\rangle$ & Key reference \\
    \hline
          VIPERS  &  76\,519 & $[0.008,2.148]$  &  0.712 &  0.181 & 43\,382 & $ [0.045,2.148] $ &   0.713 &  0.177 &  22.6 &                                     \citet{scodeggio/etal:2018} \\ 
      G10\_COSMOS  &  23\,545 & $[0.007,4.596]$  &  0.582 &  0.322 & 20\,214 & $ [0.007,4.596] $ &   0.583 &  0.324 &  22.4 &                                        \citet{davies/etal:2015} \\ 
          DEVILS  &  18\,213 & $[0.004,4.786]$  &  0.581 &  0.333 & 15\,103 & $ [0.005,4.786] $ &   0.569 &  0.333 &  21.4 &                                        \citet{davies/etal:2018} \\ 
            VVDS  &  15\,059 & $[0.003,4.540]$  &  0.609 &  0.295 & 12\,598 & $ [0.010,4.116] $ &   0.609 &  0.295 &  22.6 &                             \citet{lefevre/etal:2013} \\ 
           DEEP2  &  15\,888 & $[0.009,2.161]$  &  0.957 &  0.227 & 11\,889 & $ [0.009,2.161] $ &   0.962 &  0.233 &  23.7 &                                        \citet{newman/etal:2013} \\ 
            ACES  &   4\,276 & $[0.037,2.838]$  &  0.592 &  0.250 &  4\,157 & $ [0.037,2.838] $ &   0.591 &  0.250 &  22.4 &                                        \citet{cooper/etal:2012} \\ 
            C3R2  &   4\,348 & $[0.030,4.503]$  &  1.014 &  0.517 &  3\,132 & $ [0.030,4.175] $ &   0.967 &  0.538 &  24.0 &         \citet{stanford/etal:2021} \\ 
         zCOSMOS  &   3\,496 & $[0.006,4.445]$  &  1.396 &  1.155 &  2\,585 & $ [0.006,4.445] $ &   1.490 &  1.110 &  23.6 &                  \textit{priv. comm. (M. Salvato)} \\ 
          DEIMOS  &   4\,197 & $[0.009,6.604]$  &  1.272 &  0.421 &  2\,457 & $ [0.009,4.654] $ &   1.003 &  0.381 &  23.5 &                                      \citet{hasinger/etal:2018} \\ 
           OZDES  &  12\,511 & $[0.002,4.541]$  &  0.574 &  0.352 &  2\,288 & $ [0.008,4.541] $ &   0.697 &  0.372 &  21.2 &                                        \citet{lidman/etal:2020} \\ 
        G15\_Deep  &   1\,840 & $[0.006,1.193]$  &  0.357 &  0.163 &  1\,840 & $ [0.006,1.193] $ &   0.357 &  0.163 &  21.2 &                                \citet{kafle/etal:2018} \\ 
            GAMA  & 241\,130 & $[0.003,3.870]$  &  0.234 &  0.115 &  1\,755 & $ [0.006,0.729] $ &   0.217 &  0.103 &  19.0 &   
                                \citet{driver/etal:2022}\\ 
           GOODS  &   2\,517 & $[0.002,6.277]$  &  1.382 &  0.765 &  1\,703 & $ [0.002,4.029] $ &   1.187 &  0.644 &  23.8 &                             \textit{ESO compilation} \\ 
         hCOSMOS  &   1\,081 & $[0.012,1.265]$  &  0.302 &  0.132 &     876 & $ [0.012,1.265] $ &   0.303 &  0.132 &  20.1 &                                      \citet{damjanov/etal:2018} \\ 
            SDSS  & 109\,459 & $[0.002,1.002]$  &  0.369 &  0.251 &     525 & $ [0.004,0.817] $ &   0.355 &  0.285 &  19.0 &                                \citet{sdss_dr14:2018}\\ 
         VANDELS  &   1\,701 & $[0.309,5.784]$  &  3.086 &  0.765 &     458 & $ [0.309,4.537] $ &   2.346 &  0.875 &  24.5 &                                       \citet{garilli/etal:2021} \\ 
     FMOS\_COSMOS  &    \,383 & $[0.969,2.486]$  &  1.568 &  0.083 &     325 & $ [1.313,2.486] $ &   1.568 &  0.080 &  24.1 &                                     \citet{silverman/etal:2015} \\ 
         WIGGLEZ  &  55\,076 & $[0.002,6.159]$  &  0.598 &  0.204 &     308 & $ [0.006,1.776] $ &   0.586 &  0.190 &  21.6 &                                  \citet{parkinson/etal:2012} \\ 
          LEGA\_C  &    \,352 & $[0.346,1.085]$  &  0.798 &  0.155 &     302 & $ [0.519,1.050] $ &   0.806 &  0.153 &  22.9 &                                     \citet{vanderwel/etal:2016} \\ 
            VUDS  &    \,331 & $[0.098,6.091]$  &  2.219 &  1.488 &     186 & $ [0.098,4.161] $ &   2.003 &  1.294 &  24.7 &                                       \citet{lefevre/etal:2015} \\ 
         2dFLens  &  43\,177 & $[0.004,4.688]$  &  0.361 &  0.249 &       2 & $ [0.273,0.336] $ &   0.304 &  0.046 &  18.4 &  
                                        \citet{blake/etal:2016}\\ 
    \hline
  \end{tabular}
  }
 \tablefoot{Sources and statistics of the spectroscopic redshift estimates contained within the full spectroscopic
  compilation used in \kidz\ (`Full Compilation') and after matching to sources extracted from \kidz\ imaging (including
  masking, `Matched to \kidz\ sources'). The numbers in the ${N_{\rm spec}}$ column refer to samples with duplicates
  removed (both internally and between overlapping surveys). Statistics for each sample shown are the minimum and
  maximum redshift in the sample ($z_{\rm min}$, $z_{\rm max}$), the mean redshift of the sample ($\langle z\rangle$),
  and the NMAD scatter of the redshifts in the sample (${\rm NMAD}(z)$). For each survey we provide a single key
  reference; additional relevant papers are provided in the main text.}
\end{table*}

\section{Optical and \nir\ }
\label{sec:multiband}

At this stage in the reduction process we have calibrated optical imaging and masks in the $\utoi$-bands from
\theliraw\ and \astrowise, source catalogues extracted with \sourceextractor\ from \theliraw\ imaging, 
forced \gaap\ photometry in the $\utoi$-bands from \astrowise, and calibrated $\ztok$-band\ detectors from
\kvpipe. These data products are prepared for all pointings in \kids\ and \kidz. The remaining tasks are therefore the
measurement of \gaap\ photometry for all sources in the $\ztok$-bands (on individual \viking\ chips), combination of per-chip 
flux estimates for individual sources (per band), correction of fluxes for Galactic extinction, estimation of 
photometric redshifts, and construction of final ten-band\ masks per tile. Each of these tasks is performed within our
new post-processing pipeline \photopipe.

\subsection{The \photopipe\ repository} \label{sec:photopipe}

The \photopipe\ repository was constructed from scripts used for \kids\ \drfour, and is publicly available on
GitHub\footnote{\url{https://github.com/KiDS-WL/PhotoPipe}}. The pipeline serves one primary function: to ensure that
reduction is performed consistently across a wide array of complex steps and processes, even under significant changes
to underlying datasets and/or methodology choices. 

With regard to the scripts used for \drfour, significant changes implemented in \photopipe\ include updated selection
of source apertures used for photometry (Sect. \ref{sec:nirgaap}), new masking of pathological photometric failures
(Sect. \ref{sec:strongselection}), and updated calibration for \theli\ `auto' magnitudes.

\subsection{\nir\ \gaap\ photometry} \label{sec:nirgaap}

The \viking\ \nir\ data are treated differently than the optical data due to the inherently different dither patterns of
\kids\ and \viking. As discussed in Sect.~\ref{sec:vst}, \kids\ was designed with small dithers that ensure similar PSF
properties are combined at each point in the focal plane, thereby improving the stability of the final co-added images
and resulting in stacks that mostly have smoothly varying PSF. In contrast, as discussed in
Sect.~\ref{sec:nirobservations}, \viking\ tiles require dithering by a fair fraction of the field of view, resulting in
large discontinuous variations in the PSF across the mosaicked tile. Rather than modelling these complex 
PSF patterns directly (required for the Gaussianisation process performed by \gaap), we opted to extract the 
NIR photometry from the individual chips of each exposure, and perform an optimal averaging of the flux measurements at
the catalogue level; that is, we performed our mosaic construction in catalogue space. 

GAaP yields a flux measurement for each object and each exposure on which it appears, typically resulting in two observations for each object in the $ZYHK_{\rm s}$-bands, and four observations in the $J$-band. These fluxes are
then combined using an approximately\footnote{Insofar as the optimal estimator would require knowledge of the true
measurement variance of the sources, whereas the flux error reported by \gaap\ is only a partial reflection of the true
variance, as it does not account for all sources of flux uncertainty.} optimal inverse variance weighting, based on the
flux error reported by \gaap. 

One complication in this process, already described in \citet{kuijken/etal:2019}, is the fact that \gaap\ will fail to
produce a flux estimate in band $X$ if the aperture size, which is set by the $r$-band seeing, is smaller than the 
$X$-band seeing. Furthermore, short of complete failure, the \gaap\ flux measurements also become
extremely noisy when the aperture size is comparable to the PSF size residual, as flux information is being determined
by few data pixels. Such circumstances are infrequent but nonetheless bothersome: inspection of the PSF distributions in
Figs.~\ref{fig:psfsizes} and \ref{fig:nirpsfsizes} demonstrate that the \nir\ PSFs consistently exceed the median \rband\ 
seeing by more than $0\farcs5$ for roughly $10\%$ of the observations (closer to $15\%$ in the \zband). To remedy this
behaviour the pipeline runs \gaap\ twice, implementing two different choices of minimum aperture sizes: \zeropseven\ and \onepzero. 
This forces the smallest sources to have larger apertures in the \onepzero\ case, in an effort to suppress the bias and
failure rate (the \nir\ PSF sizes are only more than $1\arcsec$ larger than the median \rband\ seeing in less than $1\%$
of observations). Once all flux measurements are available in all bands, the choice of which aperture to use for flux
measurement is taken on an object-by-object basis, using the following criterion: 
\begin{equation}
  \begin{cases} 
    \left[\min(R) < 1/\max(R)\right] \,\|\, \left[\max(R)<0\right]) & {\rm use}\,r_{\rm min}=1\farcs0 \\
    {\rm otherwise} & {\rm use}\,r_{\rm min}=0\farcs7 \\
  \end{cases},
\end{equation}
where $R$ is the vector of flux error ratios for the two apertures in each band:
$R=\{\sigma_X^{0.7}/\sigma_X^{1.0}\}\,\forall\,X\in\{ugri_1i_2ZYJHK_{\rm s}\}$ (a negative $R$ identifies cases where 
all 0.7" apertures fail to yield a flux.) This approach maximises the number of
objects with high-S/N \gaap\ flux measurements in all bands, and yields the best possible photo-$z$ from the combined
\kidsviking\ data.

\begin{figure*}
  \centering
  \includegraphics[width=2\columnwidth]{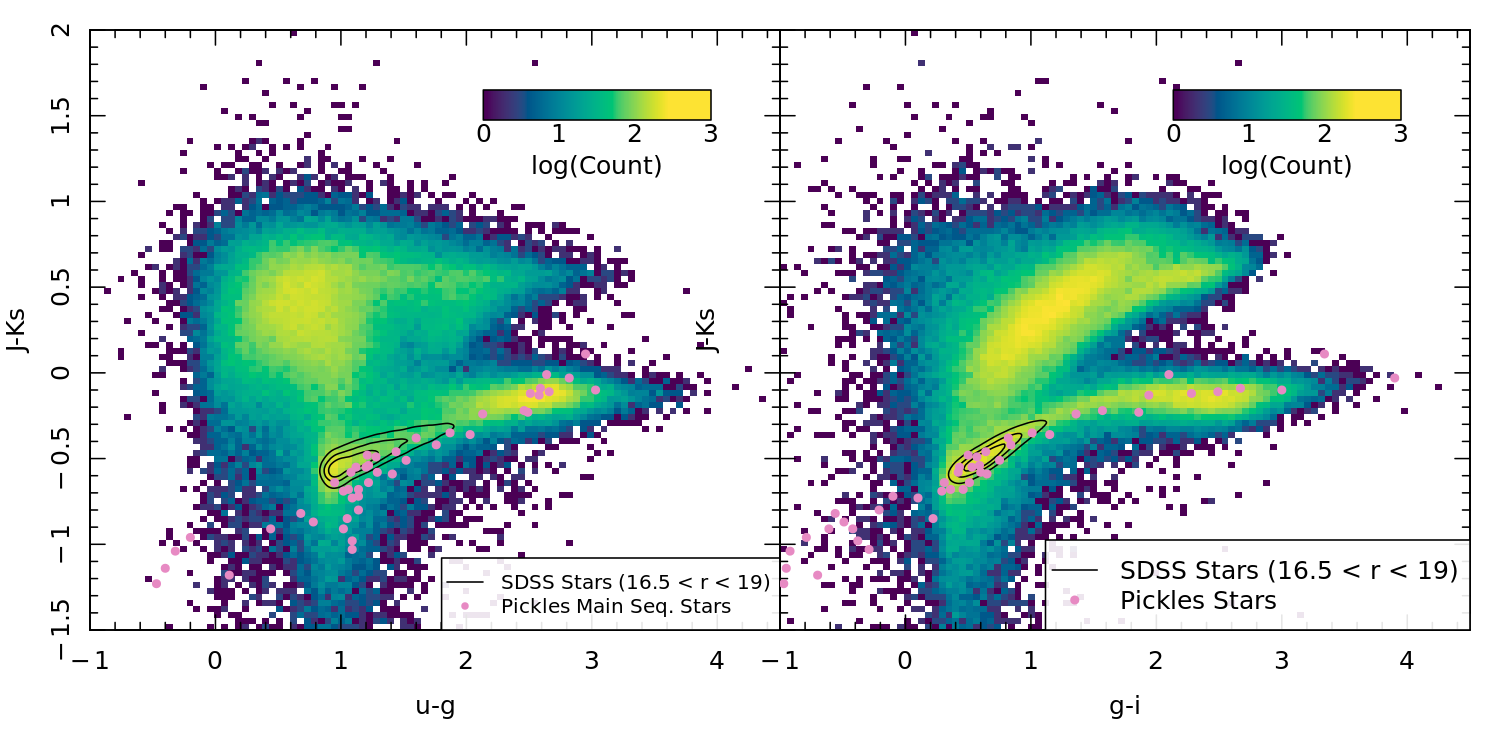}
  \caption{Colour-colour diagrams for \kids\ \drfive\ sources. The stellar locus can be identified by the distinct
  clouds of data that are coincident with both SDSS stars (black contours) and Pickles stellar templates (pink dots).
  }\label{fig:legacycolourcolour}
\end{figure*}

To demonstrate the quality of the ten-band\ photometric catalogues and the cross-survey calibration, in 
Fig.~\ref{fig:legacycolourcolour} we present an optical and \nir\ colour-colour diagram for \kids\ \drfive\ sources. The
figure shows the $u-g$ and $g-i$ optical colours versus the $J-K_{\rm s}$ colour from \viking. The figure shows the standard
stellar locus and galaxy colour distributions, with the sources that match to known SDSS stars highlighted with red
contours. We also overlay the colours of all Pickles main sequence stellar templates \citep{pickles:1998}, as observed through the relevant
optical and \nir\ filters. We note the excellent agreement between the Pickles templates and the observed stellar
locus: the only potential discrepancy is a small offset ($\Delta u\leq 0.05$) in the \uband, where (uncertain) modelling 
of metallicity in the stellar population can cause significant bias. Nonetheless, the \uband\ discrepancy is consistent with our
systematic uncertainty of $\Delta u_{\rm sys}=0.05$ (Sect. \ref{sec:vstcalibration}).      

\subsection{Photometric redshift estimation} \label{sec:photoz}

Photometric redshifts were estimated in a very similar way as in \kids\ \drfour, with the primary difference being the 
incorporation of the second pass \iband\ information. We ran the Bayesian photo-$z$ \citep[\bpz;][]{benitez:2000} code on the 
ten-band\ \gaap\ photometry using the Bayesian prior
presented in \citet{raichoor/etal:2014}, a maximum redshift of $z=7$, and a redshift stepping of $\Delta z=0.01$. The
two \iband\ measurements are treated independently, relying on BPZ to optimally use the information based on the
two \gaap\ magnitude errors. Providing
the two fluxes independently in this way is mathematically equivalent to providing \bpz\ with a single variance-weighted 
average of the two \iband\ flux measurements. 

\begin{figure*}
  \centering
  \includegraphics[width=2\columnwidth]{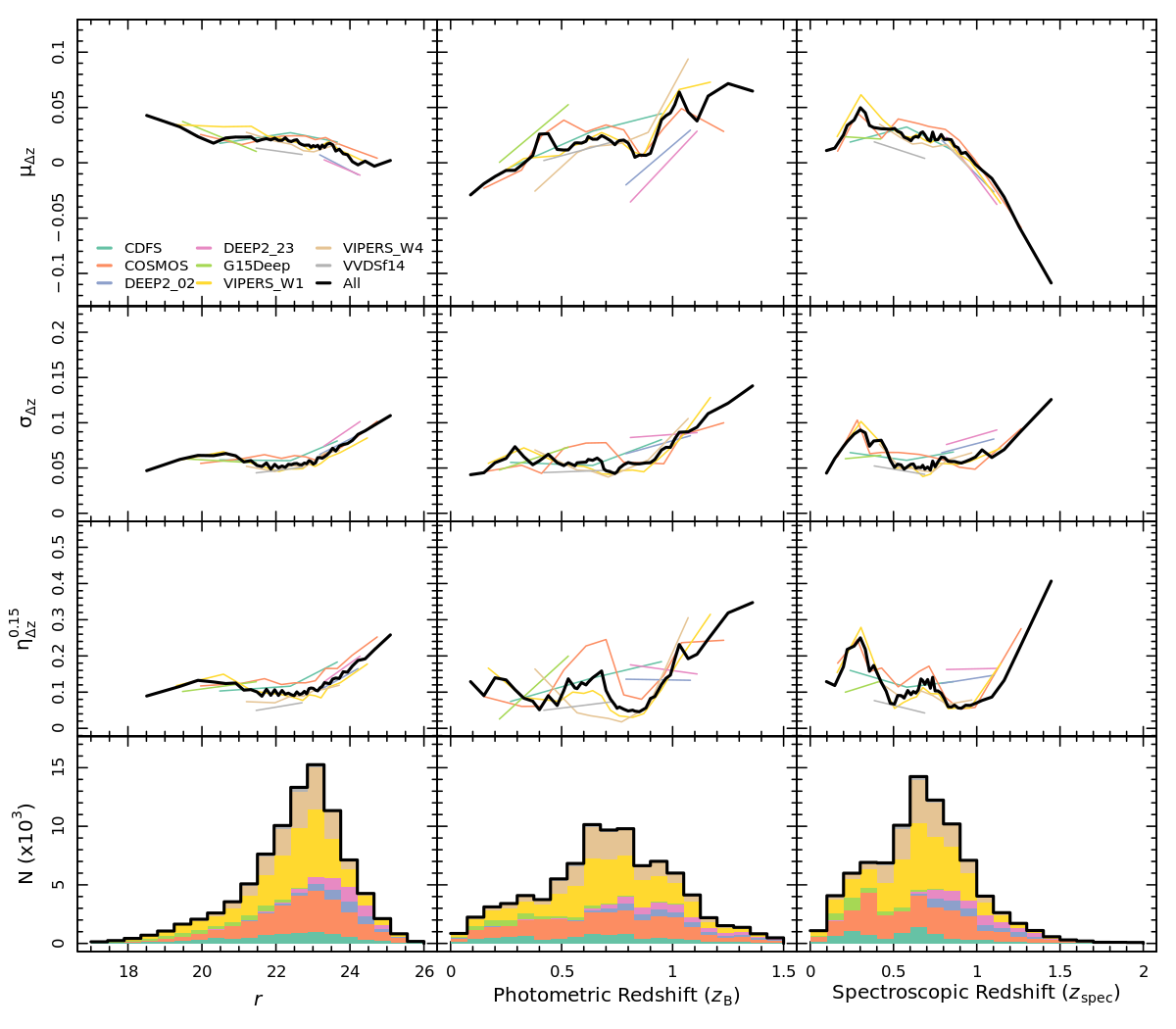}
  \caption{Quality metrics for \photoz\ point-estimates produced by \bpz\ in \photopipe. 
  Each row shows one quality metric, computed from the distribution of $\Delta z =
  \left(\zb-\zspec\right)/\left(1+\zspec\right)$: the running
  median (`bias', $\mu$), the running normalised median-absolute-deviation (`scatter', $\sigma$), and the fraction of
  sources with $|\Delta z|\,>0.15$ (`outlier rate', $\eta^{0.15}$). The columns show these statistics computed as a
  function of \rband\ magnitude ({\em left}), \photoz\ point-estimate ($\zb$, {\em centre}), and spectroscopic redshift
  ($\zspec$, {\em right}).}\label{fig:photoz}
\end{figure*}

The resulting \photoz\ point-estimates were compared to the available \kidz\ \specz\ (Sect.~\ref{sec:specz}), and summary statistics for
the various fields and full compilation are presented in Fig.~\ref{fig:photoz}. The figure contains both summary statistics 
computed for the entire compilation of \specz\ present in all of the \kidz\ 
fields and for statistics computed in the individual \kidz\ fields. In all cases the samples are binned into quasi-equal-N bins, which contain at most $2000$ sources per bin. In the event that a field has fewer than $4000$ spectra, the sample is split into two equal-N bins. This has the effect of producing short straight lines for some fields in the 
figures. 

We quantify the \photoz\ quality via the statistics of 
$\Delta z=(z_{\mathrm{B}}-z_{\mathrm{spec}})/(1+z_{\mathrm{spec}})$. 
In particular, we report in Fig.~\ref{fig:photoz} the median, NMAD scatter, and rate of outliers (objects with $|\Delta z|\,>0.15$) as a function of \rband\ magnitude,
spectroscopic redshift, and photometric redshift. The well-known dependence of \photoz\ quality on photometric S/N,
here parametrised by the $r$-band magnitude, is clearly visible for $r>22$. For $r<22$ the photo-$z$ scatter and outlier
rate are roughly constant, at $\sigma_{\Delta z}=0.05$ and $\eta^{0.15}_{\Delta z}=0.1,$ respectively, indicating that other sources of 
error limit the precision of the photo-$z$ at bright magnitudes. Of further note is the increase in photo-$z$ bias seen a magnitudes 
brighter than $r=20$; here the photometric pipeline, and photo-$z$ priors, used for the \kids-\drfive\ sample are not optimal, having been 
chosen to optimise performance at fainter magnitudes (see e.g. Figure 10 of \citealt{wright/etal:2019}). 
 
The dependence of \photoz\ quality on spectroscopic and photometric redshift is more complicated, showing several
features that we attribute to the non-trivial interplay of the specific filter set, depth in the different bands, the
spectral energy distribution templates, and the real galaxy population. Only at high redshift and/or photo-$z$ can the behaviour be easily attributed to
S/N effects, visible by the consistent degradation of quality in all \kidz\ fields. 

We find \photoz\ biases at the few percent level over all magnitudes, but which exacerbate at the extremes in photo-$z$ ($\mu_{\Delta z} > 0.05$) and 
redshift ($\mu_{\Delta z} < -0.05$). These biases are typically accompanied by an increase in photo-$z$ scatter, shown in the second row of the 
figure. Beyond spectroscopic and photometric redshifts (i.e. at $z_{\rm B}\geq1$ and $z_{\rm spec}\geq1$), the scatter in the photo-$z$ can be 
seen to deteriorate for all available surveys due to the increased photometric noise of these distant objects. Of particular note, though, is the 
increased in scatter and outlier rate observed in the COSMOS sample at $z_{\rm B}\approx0.6$ and $z_{\rm spec}\approx0.7$. Indeed, in this region 
the COSMOS sample of sources display roughly twice the number of outliers ($\eta^{0.15}_{\Delta z}\approx 0.25$) as other samples 
($\eta^{0.15}_{\Delta z}\lesssim 0.13$). This may simply be indicative of different selection effects in the COSMOS field spectroscopic sample, 
compared to the other fields (as the COSMOS sample is made up of many individual spectroscopic samples with a range of targeting criteria). However,
the increase in scatter is coincident with an overdensity in the large-scale-structure, visible as a spike in the COSMOS number counts in the 
bottom row (and visually in the redshift-axis diagrams shown in Fig.~\ref{fig:kidzraz}). This increase in scatter may therefore be caused by the 
overdensity, which causes sources at this redshift to over-sample a particular subset of colour redshift degeneracies (specifically those 
degenerate with red galaxy spectra at redshift $z\approx0.7$). 

\begin{figure*}
  \centering
  \includegraphics[width=2\columnwidth]{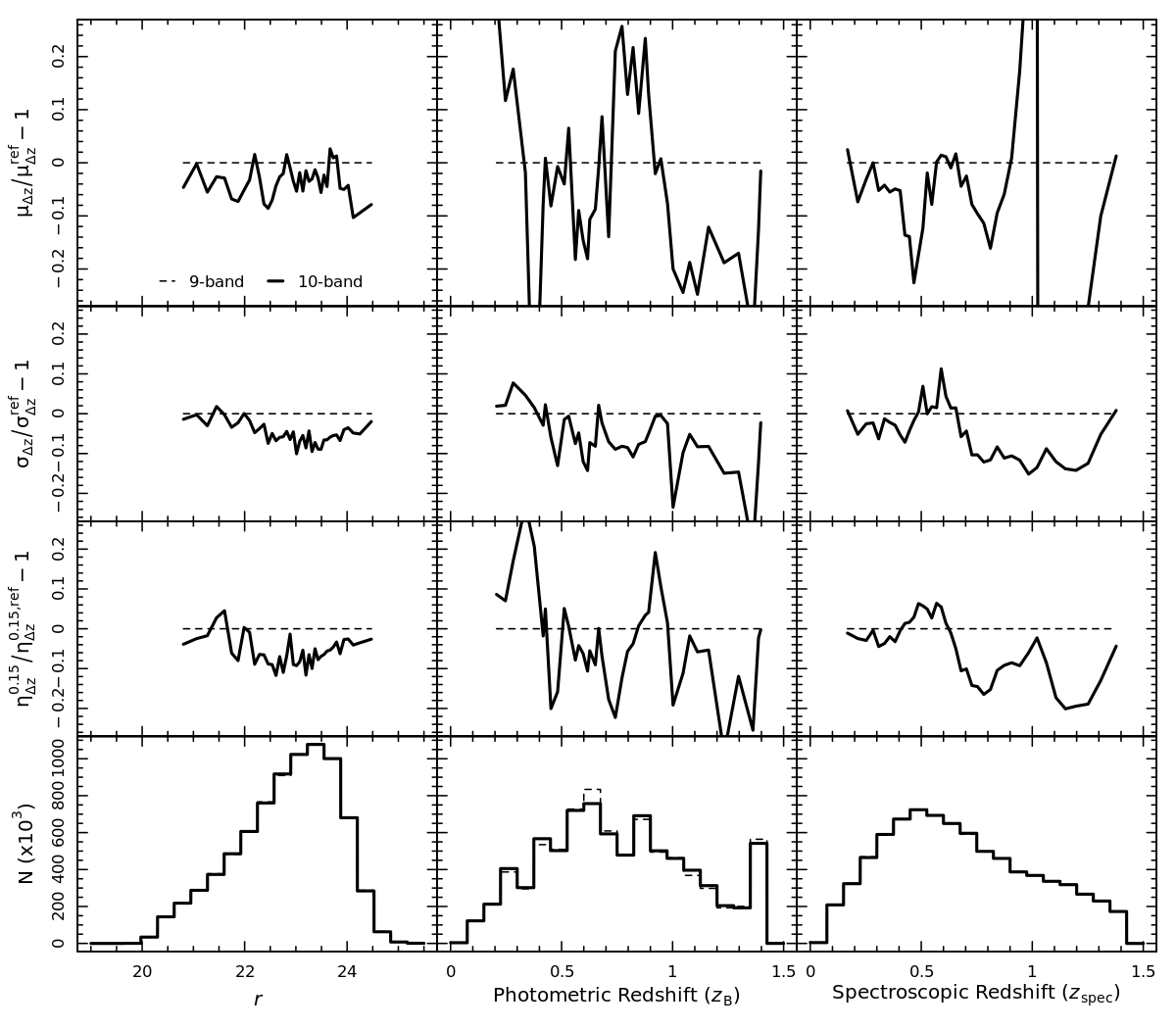}
  \caption{Differences between \photoz\ point-estimates produced by \bpz\ in \photopipe\ computed without the second 
  pass $i$-band (`$9$-band') and with the second pass $i$-band (`$10$-band'), using a full simulated \kids\ wide-field sample 
  described in \protect \cite{vandenbusch/etal:2020}. Metrics here are not directly comparable to Fig.~\ref{fig:photoz} due to this 
  sample being simulated, having a different redshift baseline, and representing a full wide-field sample (unmatched to spectroscopy). 
  Therefore, only relative differences between the nine- and ten-band cases are relevant. 
  Metrics are computed as in Fig.~\ref{fig:photoz}, except metrics here are shown as percentage difference with respect to the nine-band case. 
  }\label{fig:photoz_i2}
\end{figure*}

Regarding the inclusion of the additional $i$-band information in the computation of the photo-$z$, we find that the summary statistics presented here 
change only slightly when moving between nine and ten bands. We find, however, that this is primarily due to the strong selection effects present in the 
sample that we use to construct this data-side photo-$z$ statistics. As a demonstration of the expected performance improvements in the photo-$z$ for the wide-field photometric, we can utilise simulations. Using the simulated KiDS galaxy sample presented in 
\cite{vandenbusch/etal:2020}, we computed our BPZ photo-$z$ with and without an additional $i$-band photometric realisation. These results are presented 
in Fig.~\ref{fig:photoz_i2}. We find a consistent $5-10\%$ reduction in 
photometric scatter and outlier rate at magnitudes fainter than $r=22$, where the additional noise realisation provides useful information. More 
significantly, however, the improvement seen in the photo-$z$ as a function of true redshift: we see a consistent $10-20\%$ reduction in scatter 
at $z>0.7$, coupled with $5-20\%$ reduction in outlier rate in the same redshift range. This is understandable given the propagation of the major 
spectral features as a function of redshift, mainly the $4000\AA$ and Balmer breaks: these features enter the $i$-band at approximately $z=0.7$, and 
beyond this redshift the additional photometric information allows us to better localise the break (and its absence, for $z>1.2$). 

We stress here that these biases are not reflective of the bias imprinted on 
typical cosmological measurements from weak lensing with \kids. The template-based \photoz\ shown here are typically only
used to define subsamples of galaxies with redshift distributions that are largely localised and distinct along the line of sight. 
The redshift distributions themselves are determined
via empirical methods making direct use of the \specz\ compilation, which here is only employed for validation. Furthermore, 
the underlying magnitude, colour, and morphology distributions of the sources that Fig.~\ref{fig:photoz} is
based on are inherently different from typical weak lensing source samples, due to selection effects and weighting. 
As a result, the comparison of \photoz\ and \specz\ shown in
Fig.~\ref{fig:photoz} does not directly translate into the \photoz\ quality of galaxy samples used for cosmological
measurements, but simply illustrates the raw performance on the \specz\ without accounting for these differences.

Finally, we note that there are ongoing efforts to produce additional \photoz\ estimates within the \kids\ collaboration, 
aimed at improving \photoz\ performance via sample selection \citep[see e.g.][]{vakili/etal:2020}, 
using machine learning of fluxes \citep[see e.g.][]{bilicki/etal:2021}, and using machine learning 
directly on images \citep[see e.g.][]{lirui/etal:2022}. These efforts, however, will not produce 
\photoz\ estimates that are included in the formal \drfive\ ESO release. Nonetheless, such additional photo-$z$ estimates 
will undoubtedly be of value to the community as they are designed to produce higher accuracy and precision for (in particular) 
samples brighter than $r=20$.

\subsection{Ten-band\ mask construction} \label{sec:strongselection}

\begin{table*}
  \caption{\kidskidz\ MASK bits, their names, and additional information.}\label{tab:maskbits}
  \begin{tabular}{cr|ccc}
    Bit & (value) & Band & Name & Additional Info \\
    \hline
     0 & (1)&  \theliraw\ \rband & starhalo (cons.) & $10.5<r<11.5$ of UCAC4 and GSC1 stellar catalogue \\
     1 & (2)&  \theliraw\ \rband & star \& starhalo & $r<10.5\,(14.0)$ of UCAC4 and GSC1 for stars (starhalos) \\
     2 & (4)&  \theliraw\ \rband & manual mask & \\
     3 & (8)&  \theliraw\ \rband & void \& asteroid & contains weight==0 (saturation, chip gap, etc) selection \\
     4 & (16)&  \astrowise\ \uband\ & auto masks & star \& starhalo \& weight \& manual \\
     5 & (32)&  \astrowise\ \gband\ & auto masks & star \& starhalo \& weight \& manual \\
     6 & (64)&  \astrowise\ \rband\ & auto masks & star \& starhalo \& weight \& manual \\
     7 & (128)&  \astrowise\ \ione\  & auto masks & star \& starhalo \& weight \& manual \\
     8 & (256)&  \astrowise\ \itwo\  & auto masks & star \& starhalo \& weight \& manual \\
     9 & (512)&  VISTA \zband & footprint mask & contains strong-selection mask  \\
     10& (1024)&  VISTA \yband & footprint mask & contains strong-selection mask  \\
     11& (2048)&  VISTA \jband & footprint mask & contains strong-selection mask  \\
     12& (4096)&  VISTA \hband & footprint mask & contains strong-selection mask  \\
     13& (8192)&  VISTA \ksband & footprint mask & contains strong-selection mask  \\
     14& (16384)&  \theliraw\ \rband & WCS tiling cuts &  \\
     15& (32768)&  Reserved, unused & & \\
    \hline
  \end{tabular}
\end{table*}

We adopt a bit-masking scheme for the FITS masks that encodes the availability and quality of data at a given sky
position. The 16 bits corresponding to different image defects are described in Table \ref{tab:maskbits}. Compared to
previous KiDS data releases, we reorder the bits to facilitate the masking of a fiducial sample for typical weak lensing
applications by selecting objects with
\begin{equation}\label{eq:fiducial_mask}
    \mathrm{MASK}\,\le 1.
\end{equation}
For compatibility with previous data releases and codes, the fiducial mask as a bitwise-logical statement is\begin{equation}\label{eq:bitwise_mask}
    \mathrm{MASK}\, \&\, 32\,766
\end{equation}
(corresponding to $0111\ 1111\ 1111\ 1110_2$). 
The first four bits encode objects and defects that are present in the \theliraw\ \rband\ detection image. The zeroth bit
($\mathrm{MASK} =1$) corresponds to faint halos created by reflected starlight that typically do not significantly bias
flux and shape measurements, and is therefore excluded in the fiducial masking scheme given in  Eq.~\ref{eq:fiducial_mask}. The first bit
($\mathrm{MASK} =2$) corresponds to brighter halos from reflections and the direct bright starlight. The second bit
($\mathrm{MASK} =4$) encodes areas that are masked out manually due to defects that are not captured by any of the
automatic algorithms, and also rejects area that is affected by resolved dwarf galaxies or globular clusters, which
confuse our automatic star-galaxy separation and subsequent PSF measurement algorithms due to the abnormally high
stellar density. The third bit ($\mathrm{MASK}=8$) encodes areas that are removed from consideration by the \theliraw\ masking 
of regions with abnormally low number densities (the `void mask'), sources automatically flagged as being asteroids, 
and areas of \theliraw\ imaging with zero weight (caused by, for example, chip gaps and detector saturation). 
The following ten bits ($\mathrm{MASK}=16$ to $\mathrm{MASK}=8\,192$) correspond to defects in the ten
$\utok$-bands that the \gaap\ photometry is extracted from. The final, $14^{\rm th}$ bit that is used
($\mathrm{MASK}=16\,384$) corresponds to area that is outside the limits of each pointing, chosen to knit together the
overlapping \vst\ tiles without source duplication. These are referred to as the `WCS' cuts, as they are defined using
constant limits of RA and Dec in the World Coordinate System (WCS) of the individual tiles. 

One significant change in the masks from previous \kids\ releases is the removal of redundancy between the \rband\ mask 
created for the \theliraw\ detection images (bits $0-3$) and the $\utoi$-band masks created for the \astrowise\ images that
are used for the optical \gaap\ photometry (bits $4-8$). Previous releases ignored the faint stellar reflection halos from
the \astrowise\ \rband\ masks (bit $5$), instead relying on the ability of the \theliraw\ \rband\ mask to capture those
effects in bits $1$ and $0$ ($\mathrm{MASK}=\{1,2\}$). This choice was made because the default \astrowise\ masking was
slightly too conservative, correlating more with the \theli\ conservative mask ($\mathrm{MASK}=1$) than with the
standard \theliraw\ mask ($\mathrm{MASK}=2$). This effect was largely resolved with the update of the \pulcinella\ mask
parameters (Sect.~\ref{sec:pulcinella}), but nonetheless persists at low levels. In DR5, we apply all \astrowise\ mask
bits in our fiducial mask. If we were to remove the \astrowise\ \rband\ masks from the fiducial
masking, as in DR4, we would free up $16.0$ \sqdeg\ of data over the full survey area ($\sim 1.6\%$).  The final correlations
between the various bits of the \mask\ are provided in Appendix~\ref{sec:maskcorrelations}. 

Another addition to the masks for \drfive\ is the inclusion of a `strong-selection' mask, which is designed
to remove areas of the survey where there have been pathological photometric measurement failures within \gaap. These
failures are related to the difference in the PSF sizes between the detection \rband\ images and the other bands, as
discussed in Sect.~\ref{sec:nirgaap}. As discussed, this effect led initially to the requirement of measuring fluxes in
apertures with two minimum radii (see Sect. \ref{sec:gaap}), but in some cases even the use of the larger aperture is
not sufficient to stop \gaap\ from being unable to measure a flux. As these failures are related to the size of the
object aperture, when such failures occur, they preferentially affect the smallest objects first. This can imprint a
redshift-dependent bias on the source distribution that is localised in particular areas on-sky. Such a selection bias
is problematic for (primarily) photometric clustering studies, but also for cosmic shear (albeit at higher order) as 
it exacerbates variable depth \citep[see e.g.][]{heydenreich/etal:2020}. 

\begin{figure}
  \centering
  \includegraphics[width=\columnwidth]{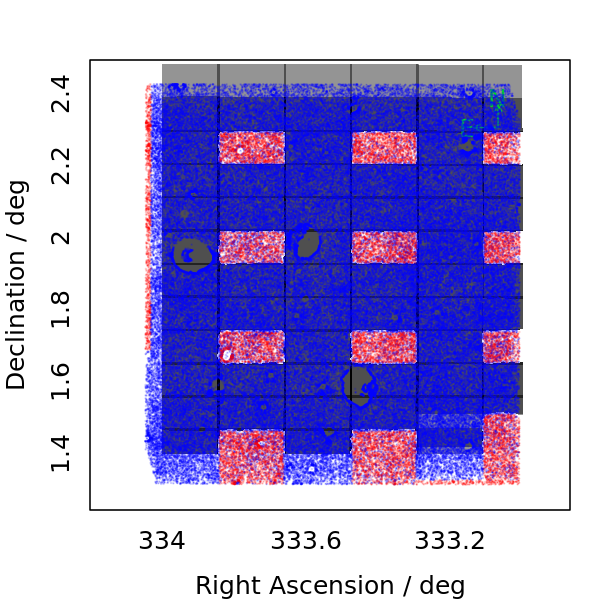}
  \caption{Demonstration of the strong-selection masking in the \kidskidz\ catalogues for \kidz\ pointing
  {\tt KIDZ_333p0_1p9} in the \yband. All sources detected by \sourceextractor\ are shown coloured by the
  number of missing photometric bands as determined by \gaap\ (0: blue, 1: red). The mask is determined by the fraction
  of sources per unit area on sky that are missing one or more bands, as
  determined using a ratio of KDEs constructed on a $1\arcmin\times1\arcmin$ grid with a $1\arcmin$ Gaussian kernel. 
  The strong-selection mask removes regions with missing-source fractions greater than $20\%$. The effect of this mask 
  can be seen in the background, which shows the sum image of this field's \yband\ data after masking: areas that show the 
  strong selection effect (i.e. red dots) have a zero value in the sum image (i.e. white), and are therefore masked.  
  }\label{fig:strongselection}
\end{figure}

We therefore opted to remove areas of the survey where such pathological failures on-sky are detected. We quantify this
failure using a 2D kernel-density estimate (KDE) measurement of the source density on-sky before and after selection due
to \gaap\ photometric failures. If the ratio of these two KDEs is less than $0.8$ (i.e. more than $20\%$ of the 
sources, per square arcminute, have been removed due to photometric failures), we flag these regions and remove them
from the survey. This masking can be seen in Fig. \ref{fig:strongselection}, where we show the effect of this masking
for one of the \kidz\ tiles in \vipers. The figure shows the distribution of all sources in the tile, without any
preselection. The sources are coloured by whether or not they have a photometric failure in the \yband: blue have
successful photometric measurements, and red have failed measurements. In the background of the image we show the
\yband\ sum-image for this field, after the application of the WCS cuts (bit $14$). This shows the regular tile pattern
formed by the dithering of many paw-prints (Sect. \ref{sec:nirobservations}). It is apparent from the photometric
failures that there are two paw-prints that have particularly poor seeing, which causes sources covered only by these paws 
to have failed photometric measurements. These areas of the sky have been identified by the strong
selection algorithm, and these areas have been masked (leading to the sum-image being zero, white, in these areas). We note
that the sources have not all failed in these regions; however, the pathological nature of the selection means that
we have nonetheless removed these areas from consideration. It should also be stressed that this strong-selection masking 
does not correlate with cosmic large-scale structure, as it is driven by observing conditions. As such, it is 
unlikely to introduce any bias in cosmological measurements from weak lensing.
\subsection{Mosaic mask construction} \label{sec:mosaics}\begin{table*}
  \caption{Mosaic areas computed from the $6\arcsec$ mosaic masks.}\label{tab:areas}
  \begin{tabular}{c|lrrrrr}
    Selection Name    & Selection Bits                  & Selection & North Area & South Area & Total Area \\
                      &                                 &   Sum     &   (\sqdeg) & (\sqdeg)   & (\sqdeg)  \\
    \hline
    WCS                       & 14                                  & 16\,384       &  681.098 & 649.883 & 1\,330.981 \\
    WCS+\astrowise            & 4,5,6,7,8,14                        & 16\,880       &  608.668 & 588.015 & 1\,196.682 \\
    WCS+\nir                  & 9,10,11,12,13,14                    & 32\,256       &  645.583 & 608.975 & 1\,254.558 \\
    WCS+AW+\nir               & 4,5,6,7,8,9,10,11,12,13,14          & 32\,752       &  577.319 & 550.902 & 1\,128.220 \\
    WCS+\theliraw             & 1,2,3,14                            & 16\,398       &  583.059 & 562.348 & 1\,145.407 \\
    WCS+\theliraw\ Cons.      & 0,1,2,3,14                          & 16\,399       &  524.231 & 508.495 & 1\,032.726 \\
    WCS+AW+\theliraw          & 1,2,3,4,5,6,7,8,14                  & 16\,894       &  544.850 & 529.992 & 1\,074.842 \\
    WCS+\theliraw+\nir        & 1,2,3,9,10,11,12,13,14              & 32\,270       &  553.478 & 527.155 & 1\,080.633 \\
    {\bf WCS+THELI+AW+NIR}    & 1,2,3,4,5,6,7,8,9,10,11,12,13,14    & {\bf 32\,766} &  {\bf 517.342} & {\bf 496.689} & {\bf 1\,014.031} \\
    WCS+TH.Cons.+AW+NIR       & 0,1,2,3,4,5,6,7,8,9,10,11,12,13,14  & 32\,767       &  468.845 & 452.350 &    921.195 \\
    WCS+AW $u$         & 4,14                                       & 16\,400       &  672.796 & 639.317 & 1\,312.113 \\
    WCS+AW $g$         & 5,14                                       & 16\,416       &  647.942 & 622.772 & 1\,270.714 \\
    WCS+AW $r$         & 6,14                                       & 16\,448       &  626.232 & 607.595 & 1\,233.827 \\
    WCS+AW $i_1$       & 7,14                                       & 16\,512       &  636.101 & 610.085 & 1\,246.186 \\
    WCS+AW $i_2$       & 8,14                                       & 16\,640       &  644.359 & 619.996 & 1\,264.355 \\
    WCS+NIR $Z$        & 9,14                                       & 16\,896       &  649.522 & 612.493 & 1\,262.016 \\
    WCS+NIR $Y$        & 10,14                                      & 17\,408       &  648.805 & 612.570 & 1\,261.375 \\
    WCS+NIR $J$        & 11,14                                      & 18\,432       &  650.089 & 612.589 & 1\,262.678 \\
    WCS+NIR $H$        & 12,14                                      & 20\,480       &  647.457 & 610.845 & 1\,258.302 \\
    WCS+NIR $K_{\rm s}$& 13,14                                      & 24\,576       &  647.737 & 610.384 & 1\,258.120 \\
    \hline
  \end{tabular}
    \tablefoot{
        The fiducial \kids\ \drfive\ and \kidslegacy\ MASK selection is given in bold.
    }
\end{table*}

Finally, we constructed mosaic masks for science use and calculation of the fiducial survey area for \kids.  Mosaic masks
are constructed using \swarp\, combining all individual ten-band\ masks using a `minimum' combination. Masks are
constructed on a predefined WCS, using an Aitoff projection and a $6\arcsec$ pixel resolution. 

Using these mosaic masks, we are able to calculate the area of the \kids\ \drfive\ data when selecting specific mask
bits. Table \ref{tab:areas} presents a compilation of the available survey areas when masking specific bits, and
combinations of bits. We note a few particular examples. First, the area of the survey when considering all available
tiles after removing overlaps (`WCS', bit $14$) is $1331.0$ \sqdeg.  This area is then further reduced when considering
masking of artefacts, stars, stellar reflections, and missing chips in the \theli\ \rband\ imaging (`WCS+\theli', bits
$\{1$-$4,14\}$) to $1145.4$ \sqdeg. Considering only the area available to all of the $\utoi$-bands
(`WCS+\theli+\astrowise', bits $\{1$-$8,14\}$), the survey area reduces to $1074.8$ \sqdeg. After removal of
area that does not have overlap with the \vista\ $\ztok$-bands, the survey area is further reduced to the `fiducial'
survey area of $1014.0$ \sqdeg (`WCS+THELI+AW+NIR', bits $\{1$-$14\}$).

\section{Shape measurement and legacy sample construction}\label{sec:shapemeasurement}

In this section we detail the construction of the lensing sample that will be used for the fiducial \kids\ \drfive\
cosmological analyses. This `\kidslegacy'\  sample is a subset of the full \drfive\ sample, determined (primarily) by the availability of reliable shape measurements. We first describe the new version of our fiducial shape
measurement code \lensfit\ \citep[][Sect. \ref{sec:lensfit}]{miller/etal:2013}, followed by the
 selection of the \kidslegacy\ sample (Sect. \ref{sec:sampleselection}). 

\subsection{\drfive\ \lensfit}\label{sec:lensfit} 

Fiducial shape measurement in \kids\ \drfive\ was performed with \lensfit\ $v321$. Demonstration of the properties of
this \lensfit\ version are provided in \cite{li/etal:2022}, where the accuracy of the measured shapes were tested with
simulations. Regarding the use of \lensfit\ for selecting the \drfive\ lensing sample, there are two important
changes with respect to \drfour\ (which was produced using \lensfit\ version $v309$): the values of the \fitclass\ flags
have been modified, and the calibration procedure for the lensing weight has been modified. 

\subsubsection{\fitclass}\label{sec:fitclass}

\begin{table}
  \caption{\fitclass\ definitions and statistics output by \lensfit\ $v321$.   
  }\label{tab:fitclass}
  \begin{tabular}{c|cccc}
      Value   & Flag    & Source   & $f_{\rm MASK}$ & $f_{\rm avail}$ \\
              & Meaning & Excl.? &  ($\%$) & ($\%$)  \\
    \hline
    $-10$ & Multiple Objects      & Yes &  2.91 &  3.85 \\
     $-9$ & Object too-large      & No  &  0.41 &  0.55 \\
     $-8$ & {\it deprecated}      & No  &  0.00 &  -    \\
     $-7$ & Centroid mismatch     & Yes &  0.06 &  0.05 \\
     $-6$ & {\it deprecated}      & No  &  3.85 &  -    \\
     $-5$ & {\it deprecated}      & -   &  -    &  -    \\
     $-4$ & Poor Model fit        & Yes &  0.32 &  0.41 \\
     $-3$ & Unable to fit         & Yes &  0.57 &  0.66 \\
     $-2$ & {\it deprecated}      & -   &  -    &  -    \\
     $-1$ & Insufficient Data     & Yes & 20.69 &  -    \\  
     $ 0$ & Well Modelled Galaxy  & No  & 64.43 & 85.42 \\
     $ 1$ & Star or Point-Source  & Yes &  4.59 &  6.17 \\
     $ 2$ & Star or Point-Source  & Yes &  2.16 &  2.90 \\
    \hline
      \multicolumn{3}{r}{\bf Total Removed:} & 31.31 & 14.03 \\
    \hline
  \end{tabular}
  \tablefoot{
      The percentage of the full $\mask\leq1$ dataset that is assigned to each of the \fitclass\ values is shown
      ($f_{\rm MASK}$). The percentage of course that are actually present in the sample prior to the \fitclass\
      selection (see Sect. \ref{sec:sampleselection} and Table \ref{tab:samplecuts}) is also shown ($f_{\rm avail}$).
      Classes with $-$ were never assigned to a source within the relevant sample.
    }
\end{table}

The new definitions of the
\fitclass\ flags in \lensfit\ $v321$ are presented in Table \ref{tab:fitclass}. These new
flags require the selection to be updated, and the choice of which flags to reject is also presented in the table. 
A total of $31.31\%$ of the available $\mask\leq1$ sample is flagged by the various \fitclass\ selections, the most
severe of which is the `insufficient data' flag, which is assigned to $20.69\%$ of the $\mask\leq1$ sources. However, 
the majority of these sources are not modelled by \lensfit\ because they are too bright (see Sect. \ref{sec:sampleselection}). 
In practice, the \fitclass\ selection removes $14.03\%$ of sources that would otherwise exist in the sample; the majority 
of which are either stars ($9.07\%$) or have multiple sources detected within a single segmentation region (and so are 
deemed to be blended, $3.85\%$).

\subsubsection{Weight recalibration}\label{sec:weightrecal} 

Galaxy shapes are able to be expressed as ellipticities $\beps$, quantified using moments 
of the light distribution or using direct model 
fits to galaxy images, and take the form of complex quantities combining the source axis ratio ($q$) and the major axis 
orientation angle ($\theta$):   
\begin{equation}
{\beps} \equiv  \eps_1+{\rm i}\eps_2 = \, \frac{1-q}{1+q}\exp{\left(2{\rm i}\theta\right)}.
\end{equation}
The magnitude of any source's ellipticity is therefore simply 
\begin{equation} 
    |\eps|\,=\sqrt{\left(\eps_1\right)^2+\left(\eps_2\right)^2}. 
\end{equation}
Ellipticities are measured by \lensfit\ using a model-fitting approach that is described at length 
in \cite{miller/etal:2013}. Source ellipticities are reported at their posterior-informed maximum likelihood 
\citep[see section 3.5 of][]{miller/etal:2013}, and the 2D measurement variance of the likelihood in ellipticity  
is provided as $\nu_\eps\equiv\sigma^2_\eps$.   
Faint sources have inherently noisier ellipticity estimates than their brighter counterparts, 
and this is reflected in them having a broader 
likelihood surface and a larger variance $\nu_\eps$. To translate this 
uncertainty in shape measurement to subsequent scientific analyses, each source $i$ is therefore tagged with 
a shape measurement weight $w_i$, which is related to the inverse of the 2D variance estimate. Individual weight 
estimates are computed as
\begin{equation}\label{eqn:shapeweight}
    w_{i} = \left[ \frac{\nu_{\eps,i}\,\eps^2_{\rm max}}{
     2\eps^2_{\rm max}-4\nu_{\eps,i}}+\nu_{\eps,{\rm pop}}\right]^{-1},
\end{equation}
where $|\eps|_{\rm max}$ is the maximum allowed ellipticity (determined by the intrinsic thickness of edge-on disk 
galaxies), and $\nu_{\eps,{\rm pop}}$ is the population variance in 
ellipticity of all galaxies in the sample. 

Unfortunately, shapes measured by the \lensfit\ algorithm are not free of systematic biases, and therefore require recalibration. 
Of particular concern is the systematic imprint of the PSF shape on the estimated properties of galaxies, as such systematic 
contamination of estimated ellipticities and weights would ultimately introduce bias into shear correlation functions (which are 
a primary probe of cosmology).  
As such, our recalibration procedure (described in \citealt{li/etal:2022}, and with minor modifications explained below) is designed to remove 
residual imprints of the PSF on the data (although the resulting shape estimates nonetheless require a correction for 
multiplicative shear bias; see Sect.~\ref{sec:mbias}). 

Although the \kids\ PSF in the \rband\ is relatively round, any anisotropy can lead to inferred galaxy shapes 
that are correlated with the orientation of the PSF. This `PSF leakage' affects both the ellipticities per source, 
and the associated weights. PSF leakage 
in measured source weights refers to the tendency of sources to be preferentially up-weighted when their on-sky orientation 
happens to be aligned with the PSF. PSF leakage in the measured source shapes refers to the tendency 
of detected sources to be preferentially aligned with the PSF, as they have higher effective surface 
brightnesses in this regime (and thus are less likely to have been missed during source detection). 

In \kidslegacy, we corrected for PSF leakage in the weights of sources by measuring the 
linear relationship between the variance of the shape measurement per source $\nu_\eps$  
and the scalar projection $S$ of the PSF ellipticity $\beps_{\rm PSF}$ in the direction of 
the galaxy ellipticity $\beps$ per source: 
\begin{equation}\label{eqn:proj}
S_\eps=\frac{{\rm Real}\left({\beps_{\rm PSF}\,\beps^\ast}\right)}{|\eps|}. 
\end{equation}
The relationship between $\nu_\eps$ and $S_\eps$ is highly dependent both on the resolution of the source
and on the S/N of the \lensfit\ model.
Here resolution is defined as the ratio of the squared PSF radius (standard deviation) $r^2_{{\rm PSF}}$ to the quadrature sum of the PSF radius  
and the circularised galaxy effective radius $r_{\rm ab}=r_{\rm e}\sqrt{q}$ (where $r_{\rm e}$ is the effective radius) per source: 
\begin{align}\label{eqn:resolution}
    \mathcal{R}\equiv&\,\frac{r_{{\rm PSF}}^2}{r_{{\rm PSF}}^2+r_{{\rm ab}}^2}.
\end{align}
We therefore opted to fit for a linear relationship between $\nu_{\eps}$ and $S_{\eps}$ in bins of $\mathcal{R}$ and S/N, which 
we constructed to each contain an equal number of sources. 
We fitted our linear regression (between all galaxies $i$ that reside within a bin) as
\begin{equation}\label{eqn:varfit}
    \nu_{\eps,i} = \alpha_{\rm S}\, S_{\eps,i} + \langle S_\eps\rangle + \mathcal{N}(0,\sigma_{\rm S}), 
\end{equation}
where $\langle S_\eps\rangle$ is the additive component of the model, and the sources are expected to be normally distributed 
about the regression given a standard deviation $\sigma_{\rm S}$; that is, assuming homoskedasticity of the residuals in the 
fitting process. 
This regression provides an estimate of the PSF leakage into the shape measurement variance for sources in this bin of 
$\mathcal{R}$ and S/N, parameterised by the multiplicative coefficient $\alpha_{\rm S}$. 
Panel (a) of Fig. \ref{fig:recal} shows the value of $\alpha_{\rm S}$ in our bins of 
$\mathcal{R}$ and S/N, demonstrating the non-negligible systematic PSF leakage into the weights over this plane.

One development that we have made with respect to the method presented in \citet{li/etal:2022} regards these linear fits. 
Inspection of the data within each bin demonstrates that the assumption of homoskedasticity is poor: measurement variance is 
significantly larger for sources with small intrinsic ellipticities, and therefore the residuals in the fit balloon in the 
middle of the fitting region. Additionally, the variance is a truncated variable, which can lead to further heteroskedasticity 
in the regression residuals. To combat the influence of these two effects, we implemented a conservative clipping of 
data with exceptionally high variance, and a correction for heteroskedasticity in the form of iterative regression with 
residual weighting. The result of these processes (compared to simple linear regression) is presented in Panels (b) and (c) of 
Fig.~\ref{fig:recal}. 

Once the value of $\alpha_{\rm S}$ is computed per bin, the original measurement variances (for sources in that bin) are 
corrected by subtracting the inferred leakage amplitude per source $i$: 
\begin{equation}
    \nu^\prime_{\eps,i} = \nu_{\eps,i}-\alpha_{\rm S}\,S_{\eps,i}.
\end{equation}
The distribution of $\alpha_{\rm S}$ values measured after this correction (i.e. re-fitting Eq.~\ref{eqn:varfit} 
with $\nu_{\eps,i}\rightarrow\nu^\prime_{\eps,i}$) are shown in Panel (d) of Fig.~\ref{fig:recal}. 
The corrected \lensfit\ measurement variances $\nu^\prime_{\eps,i}$ are then used to define a corrected shape measurement 
weight per source $w^\prime_i$ using a modified version of Eq.~\ref{eqn:shapeweight}:
\begin{equation}
    w^\prime_{i} = \left[ \frac{\nu^\prime_{\eps,i}\,\eps^2_{\rm max}}{
     2\eps^2_{\rm max}-4\nu^\prime_{\eps,i}}+\nu_{\eps,{\rm pop}}\right]^{-1}.
\end{equation}

\subsubsection{Ellipticity correction}
The final step in our recalibration process is the correction of source ellipticities, designed to remove any 
residual PSF leakage into the 
distribution of measured shapes. This process is similarly performed in bins of resolution and S/N, but also 
in bins of \photoz\ (which are used for cosmic shear tomography). As such, this process formally does not contribute 
to the sample statistics calculated here, as the choice of tomographic bins for \kidslegacy\ is decided during 
the cosmological analysis. Nonetheless, we document the process here for posterity. Furthermore, the ellipticity 
correction does not fully encapsulate biases inherent to the shape measurement process, as multiplicative shear biases 
remain and must be corrected for (see Sect.~\ref{sec:mbias}). 

The first step in our shape recalibration process is to perform a 
weighted linear fit (using our recalibrated shape weights) between the distribution of source ellipticities $\beps=(\eps_1,\eps_2)$ 
and measured PSF ellipticities $\beps_{\rm PSF}=(\eps_{{\rm PSF},1},\eps_{{\rm PSF},2})$, per source $i$, assuming the model
\begin{align}
    \beps_i =&\, \bm{\alpha}\,\beps_{{\rm PSF},i}+\bm{c}+\mathcal{N}(0,\bm{\sigma}_{\eps}),
\end{align}
where $\bm{c}=(c_1,c_2)$ are the additive components of the regression models for each ellipticity component, 
and $\bm{\sigma}_{\eps}=(\sigma_{\eps,1},\sigma_{\eps,2})$ are the standard deviations of the 
residuals in the fits to each ellipticity component, which again are assumed to be Gaussian. 
These fits provide us with an estimate of the residual PSF leakage 
in the individual ellipticity components, $\bm{\alpha}=(\alpha_1,\alpha_2)$, and an associated fit uncertainty 
$\bm{\sigma}_{\alpha}=(\sigma_{\alpha,1},\sigma_{\alpha,2})$. 

\begin{figure*}
  \centering
  \includegraphics[width=2\columnwidth]{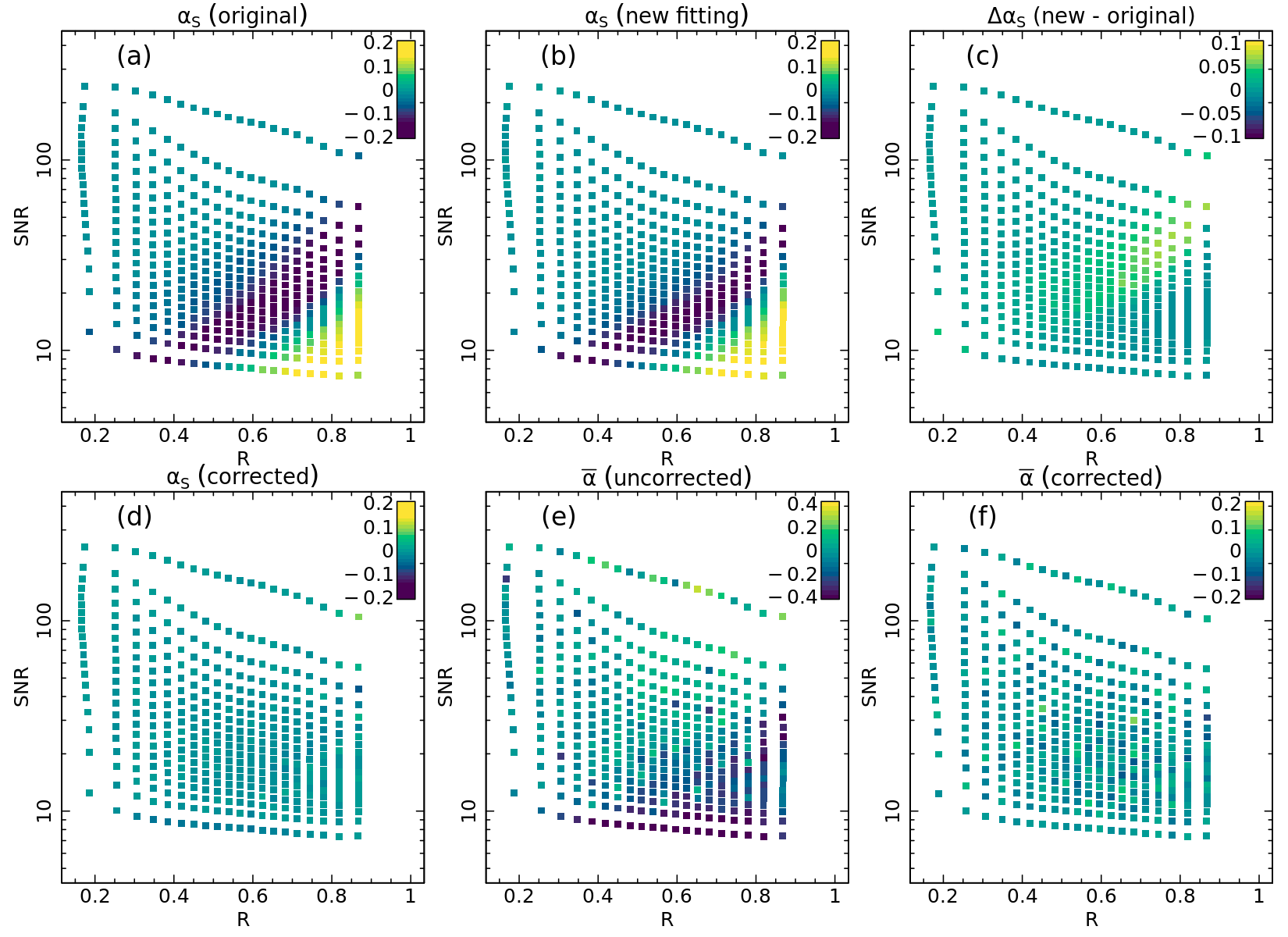}
  \caption{Sources in the resolution-\snr\ plane that is used for recalibration of \lensfit\ weights and shape
  estimates. {\em Top row:} Estimates of the PSF leakage into the shape measurement variance, estimated with simple 
  linear regression ({\em left}) and with clipped, iterative, residual-weighted linear regression ({\em centre}). {\em Right}: Difference between the leakage estimates.
  {\em Bottom row:} PSF leakage estimates after correction of shape variances and weights ({\em left}). 
  The estimates of PSF leakage into the source shape distributions is shown before ({\em centre}) and after ({\em right}) correction 
  of estimated ellipticities. 
  }\label{fig:recal}
\end{figure*}

The distribution of $\bm{\alpha}$ prior to any correction, expressed as the arithmetic mean of the two components 
$\bar{\alpha}=(\alpha_1+\alpha_2)/2$, is shown in Panel (e) of Fig. \ref{fig:recal}. 
In most bins we observe modest biases: amplitudes are typically $\bar{\alpha}\lesssim0.2$, and vary 
smoothly over the plane of resolution and S/N. 
However, in low S/N\ cases the biases drop sharply to negative values, resulting in significant 
within-bin variation. As a result, the direct per-bin correction approach introduced in 
Sect.~\ref{sec:weightrecal} does not satisfactorily correct for all biases within each bin. 
Instead, we invoked a hybrid approach that first removes a continuous polynomial bias over the 
$\mathcal{R}$ and S/N\ plane, followed by a direct bin-wise correction to account for 
remaining (incoherent) biases. 

To correct for the overall trend, we first estimated the amplitude of PSF contamination by fitting 
a polynomial to the estimated distribution of $\bm{\alpha}$ (per component) and resolution vs S/N:
\begin{equation}
\bm{\alpha} = \bm{c}_0 + \bm{c}_1\,\hbox{\snr}^{-2} + \bm{c}_2\,\hbox{\snr}^{-3} + 
    \bm{c}_3\,\mathcal{R} + \bm{c}_4\,\mathcal{R}\,\hbox{\snr}^{-2},
\end{equation}
where the coefficients $\bm{c}_x=(c_{x,1},c_{x,2})$ are fit for using weighted least squares, per component and with inverse variance weights 
$(\bm{\sigma}^{-2}_{\alpha})$.  
The result of this polynomial fit is a contiguously estimated $\hat{\bm{\alpha}}$ per source that is used to remove 
the overall PSF leakage into the source ellipticities: 
\begin{equation}
\hat{\beps} = \beps - \hat{\bm{\alpha}}\, \beps_{\rm PSF}.
\end{equation}
In practice $\hat{\beps}$ is an intermediate product because, as mentioned previously, the ellipticities after this 
correction still contain residual biases due to higher-order bin-wise fluctuations in 
the distribution of $\bm{\alpha}$. These residual fluctuations are removed on a per-bin basis, using a direct correction  
for the residual bias in the ellipticities per bin, 
\begin{equation}
\hat{\beps} = {\bm{\alpha}}_{\rm r}\,\beps_{\rm PSF}+\bm{c}+\mathcal{N}(0,\bm{\sigma}_{\hat{\beps}}),
\end{equation} 
resulting in the final ellipticity measurements used for science: 
\begin{equation}
\beps^{\prime} = \hat{\beps} - \bm{\alpha}_{\rm r}\, \beps_{i,{\rm PSF}}.
\end{equation}
The distribution of measured biases after both corrections is also shown in Panel (f) of Fig.~\ref{fig:recal}. 

After this recalibration procedure, sources with non-physical ellipticities (i.e. $|\epsilon_i|\,>1$) are removed from the 
sample; however, this number is typically very small 
\citep[][found four sources out of more than $26$ million were removed by this selection in their mock analyses]{li/etal:2022}. 

Finally, we include the on-sky distribution of shape measurement statistics 
in Appendix~\ref{sec:lensing_properties}. 
These diagnostics are useful for understanding the variability of source shapes, 
PSF sizes, and possible systematic effects present in the lensing catalogue.

\subsubsection{Multiplicative bias}\label{sec:mbias}
The ellipticity estimates from \lensfit\ suffer from small percent-level multiplicative biases. These biases originate from different physical effects such as source detection, blending, noise, model imperfections, and so on, and strongly depend on galaxy properties such as size, S/N, shape of the light profile, and so on.
Ultimately, an estimate of the total multiplicative bias requires the choice of a source sample and the careful simulation of the shape measurement process on that sample. Such simulations are presented in \citet{li/etal:2022} for the \kids-1000 source sample and the tomographic bins used in the \kids-1000 cosmic shear analysis \citep{asgari/etal:2021}. We expect similar multiplicative biases for the \kids-\drfive\ source sample presented here and refer the reader to \citet{li/etal:2022} for orientation. However, the actual estimation of these biases will differ slightly due to the differences between \drfour\ and \drfive\ and possible differences in the upcoming \kidslegacy\ analyses compared to \kids-1000. Hence, we do not present actual numbers for the multiplicative bias here but defer this to the forthcoming science papers.

\subsection{\kidslegacy\ sample selection}\label{sec:sampleselection} 

\begin{table*}\centering
  \caption{Statistics of the additional selections used for \kidslegacy\ sample definition.}\label{tab:samplecuts}
  \begin{tabular}{c|lrrrrrr}
    Selection Name & Formula & $N^{\rm r}_{\rm s}$   & $f^{\rm r}_{\rm tot}$    & $N^{\rm r}_{\rm s,run}$  &  $f^{\rm r}_{\rm run}$ &  $N^{\rm k}_{\rm s,run}$ & $f^{\rm k}_{\rm run}$    \\
    \hline
    MASK        & Eq. \ref{eqn:fidmask}      &             -  &    -        &             -  &    -         & $100\,744\,685$ & $ 100.00 \% $ \\
    FLAG\_GAAP  & Eq. \ref{eqn:gaapmask}     & $     85\,192$ & $  0.08 \% $& $     85\,192$ & $  0.08 \% $ & $100\,659\,493$ & $  99.92 \% $ \\
    ASTEROIDS   & Eq. \ref{eqn:astmask}      & $    951\,580$ & $  0.94 \% $& $    951\,411$ & $  0.95 \% $ & $ 99\,708\,082$ & $  98.97 \% $ \\
    UNMEASURED  & Eq. \ref{eqn:psfmask}      & $20\,844\,833$ & $ 20.69 \% $& $20\,196\,904$ & $ 20.26 \% $ & $ 79\,511\,178$ & $  78.92 \% $ \\
    BLENDING    & Eq. \ref{eqn:blendmask}    & $26\,486\,769$ & $ 26.29 \% $& $ 5\,560\,307$ & $  6.99 \% $ & $ 73\,950\,871$ & $  73.40 \% $ \\
    FITCLASS    & Eq. \ref{eqn:fitmask}      & $31\,545\,314$ & $ 31.31 \% $& $10\,376\,131$ & $ 14.03 \% $ & $ 63\,574\,740$ & $  63.10 \% $ \\
    BINARY      & Eq. \ref{eqn:binary}       & $    143\,790$ & $  0.14 \% $& $     14\,221$ & $  0.02 \% $ & $ 63\,560\,519$ & $  63.09 \% $ \\
    MAG\_AUTO   & Eq. \ref{eqn:magmask}      & $ 2\,658\,703$ & $  2.64 \% $& $    751\,064$ & $  1.18 \% $ & $ 62\,809\,455$ & $  62.35 \% $ \\
    RESOLUTION  & Eq. \ref{eqn:resolutioncut}& $43\,904\,820$ & $ 43.58 \% $& $12\,658\,667$ & $ 20.15 \% $ & $ 50\,150\,788$ & $  49.78 \% $ \\
    WEIGHT      & Eq. \ref{eqn:weight}       &              - &          -  & $ 6\,945\,632$ & $ 13.85 \% $ & $ 43\,205\,363$ & $  42.89 \% $ \\
    \hline
    Full Legacy Sample     & - & - & - &                         - & - & $ 43\,205\,363$ & $  42.89 \% $ \\
    Astrometric masking    & - & - & - & $ 1\,796\,184$ & $  4.16 \% $ & $ 41\,409\,179$ & $  41.10 \% $ \\
    $0.1<z_{\rm B}\leq2.0$ & - & - & - & $    458\,572$ & $  1.11 \% $ & $ 40\,950\,607$ & $  40.65 \% $ \\
    $0.1<z_{\rm B}\leq1.2$ & - & - & - & $ 5\,383\,505$ & $ 13.00 \% $ & $ 36\,025\,674$ & $  35.76 \% $ \\
    \hline
  \end{tabular}
  \tablefoot{ 
  Superscript indices specify whether a number refers to sources that are removed from the sample (`r') or kept in the
  sample (`k'). 
  $N^{\rm r}_{\rm s}$ is the total number of sources (from the full ${\rm MASK}\leq 1$ sample) that are flagged for removal by
  the selection $s$. $f^{\rm r}_{\rm tot}$ is $N^{\rm r}_{\rm s}$ expressed as a fraction of the full ${\rm MASK}\leq 1$ sample
  $(N^{\rm k}_{\rm MASK})$. $N^{\rm r}_{\rm s,run}$ is the number of sources from the currently un-flagged population
  that are newly flagged after applying selection $s$ (assuming flags are applied in row-order). $f^{\rm r}_{\rm run}$
  is $N^{\rm r}_{\rm s,run}$ expressed as a fraction of the currently un-flagged population. $N^{\rm k}_{\rm s,run}$ is
  the number of sources remaining un-flagged after the flagging of selection $s$. $f^{\rm k}_{\rm run}$ is $N^{\rm
  k}_{\rm s,run}$ expressed as a fraction of the full ${\rm MASK}\leq 1$ sample.}
\end{table*}

\begin{figure}\centering
  \includegraphics[width=\columnwidth]{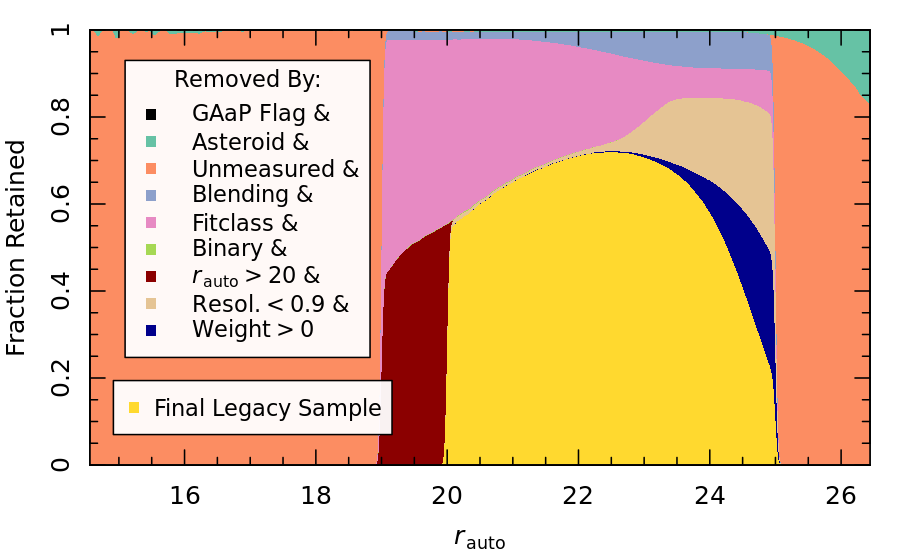}
  \caption{Selections applied to the \kids\ \drfive\ ${\rm MASK}\leq 1$ sample, and how they modify the available number
  of sources as a function of \rband\ magnitude, relative to the total number counts. The number of sources that are removed 
  by each selection is given in Table \ref{tab:samplecuts}.}\label{fig:selections}
\end{figure}

The final \kidslegacy\ sample is defined as the lensing sources that satisfy a number of quality
criteria. These criteria are summarised here, in Table \ref{tab:samplecuts}, and in Fig. \ref{fig:selections}. The table
presents details of the selections in terms of the number of sources removed (and remaining) in the \kidslegacy\ sample
after each successive selection. Similarly, Fig. \ref{fig:selections} presents the fraction of sources that are retained
in the sample as a function of \rband\ magnitude after each selection. 

The initial sample was constructed by applying the fiducial mask selection (Sect. \ref{sec:strongselection}):
\begin{equation}\label{eqn:fidmask}
  {\rm MASK}\leq 1.
\end{equation}
Next, we removed all sources that do not have successful \gaap\ measurements in all ten photometric
bands $x\in{\utok}$: 
\begin{equation}\label{eqn:gaapmask}
  N_{\rm flag} = \sum_{\hbox{\tt x}} {\tt Flag\_GAAP\_x}==0. \\
\end{equation}
This selection removes less than $0.1\%$ of the total ${\rm MASK}\leq 1$ sample. 

We then removed asteroids and other transients present in the detection imaging, using a
combination of extreme colour selections \citep{kuijken/etal:2015}: 
\begin{equation}\label{eqn:astmask}
  (g-r \leq 1.5) \;\hbox{or}\; (i_1-r \leq 1.5) \;\hbox{or}\; (i_2-r \leq 1.5).
\end{equation}
The asteroid mask removes $0.94\%$ of sources from the ${\rm MASK}\leq 1$ sample,
consisting of both bright objects and a portion of the faintest sources in the sample (due to high photometric noise). 

As the \kidslegacy\ sample is primarily intended for lensing science, we next selected on lensing-related quantities. First,
we removed sources for which \lensfit\ does not produce a shape estimate. Such sources retain a placeholder-value of exactly 
zero for the PSF moments (among other parameters), and are therefore able to be removed by selecting only galaxies with  
\begin{equation}\label{eqn:psfmask}
  ({\tt PSF\_Q11} \neq 0.0) \;\hbox{{\rm and}}\; ({\tt PSF\_Q22} \neq 0.0).
\end{equation}
This causes a significant reduction of the sample, removing over $20$ million
sources. Figure \ref{fig:selections} demonstrates, however, 
that this reduction is caused by a blanket cut of sources in \rband\ magnitude that are brighter than $r_{\rm auto}=19$
or fainter than $r_{\rm auto}=25$. These limits are internally imposed by \lensfit. Within this magnitude window, very few 
sources are removed by this selection. 

The next lensing-related selection removes sources that are considered to be blended. This selection is performed using
the \lensfit\ {\tt contamination\_radius} parameter, 
\begin{equation}\label{eqn:blendmask}
  {\tt contamination\_radius}>4.25\,{\rm pix},
\end{equation}
and removes sources that are contaminated by extracted neighbours. This selection removed $26.3\%$ of the ${\rm
MASK}\leq 1$ sample; however, most of these sources are already removed by previous selections (particularly the `unmeasured' 
selection). As such, this blending selection only removes a further $6.99\%$ of the available sample. 

As described in Sect. \ref{sec:lensfit}, the \lensfit\ code provides a range of flags related to the quality of
the fitted model, called \fitclass. \fitclass\ selection is particularly useful for the rejection of point-like sources
from the sample, which might otherwise contaminate our lensing measurements. For the lensing sample, we applied the
fiducial \fitclass\ selection: 
\begin{equation}\label{eqn:fitmask}
  \hbox\fitclass\notin\{-10,-7,-4,-3,-1,1,2\},
\end{equation}
where the meanings of the various flags are given in Table \ref{tab:fitclass}. This \fitclass\ selection removes more
than $31\%$ of the ${\rm MASK}\leq 1$ sample from consideration; however, more than half of these sources are already removed
by previous flags. The remaining $14\%$ of sources that are removed are preferentially bright, as can be seen from Fig. 
\ref{fig:selections}. This is consistent with the majority of these sources being stellar contaminants; a conclusion that 
is supported by the \fitclass\ selection producing \rband\ number counts that more consistent with a simple power-law. 

\begin{figure*}
  \centering
  \includegraphics[width=2\columnwidth]{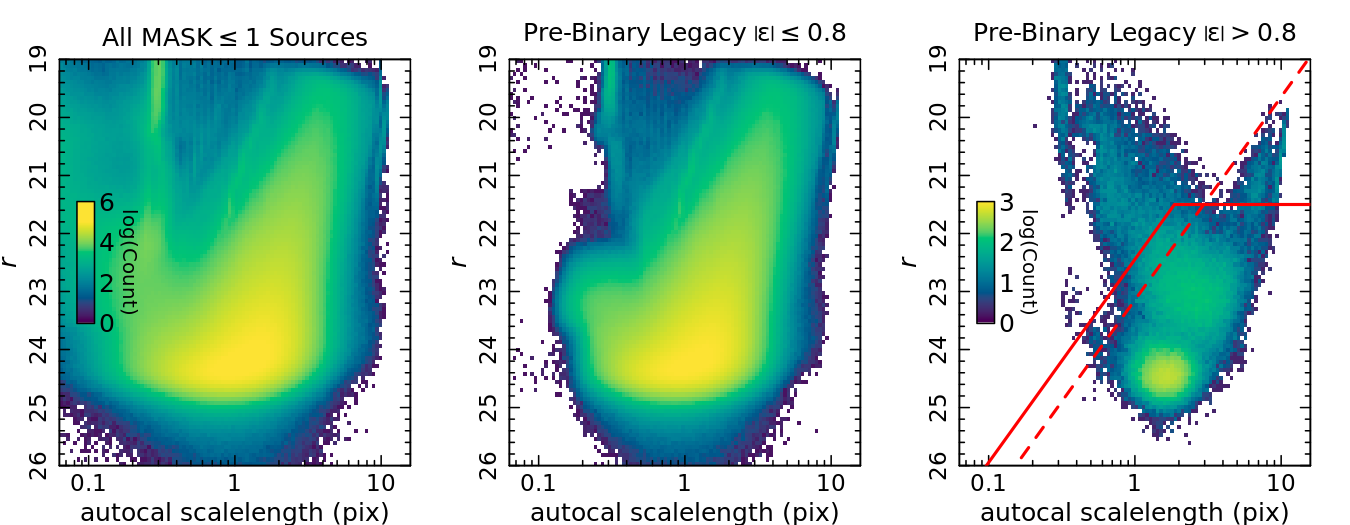}
  \caption{Selection of binary stars in construction of the \kidslegacy\ lensing sample. 
  {\it Left}: Distribution of all \drfive\ sources in size and apparent magnitude. 
  {\it Centre}: Distribution of all sources remaining prior to the rejection of binary stars, which have ellipticities 
  $|\epsilon|\,\leq 0.8$. The colour scale in the centre panel is the same as in the left panel. 
  {\it Right}: As in the centre panel, but for sources with ellipticities $|\epsilon|\,> 0.8$. This is the space in which the 
  binary rejection is performed. The binary rejection criteria used in previous \kids\ releases is shown as the dashed red line. 
  The new binary rejection criteria for sources in \drfive\ is shown as the solid red line (sources above the line are discarded).
  }\label{fig:binarycut}
\end{figure*}

Next we applied a further selection to remove unresolved binary stars from the sample. This binary star selection has also
been updated from that used in \drfour. Figure \ref{fig:binarycut} shows the distribution of objects from the
${\rm MASK}\leq 1$ sample in \rband\ magnitude and (a parameter akin to) angular extent on-sky. The centre and right panels of
the figure show the sample prior to binary rejection, split by ellipticity. The highly elliptical sample ($|\epsilon|\,>0.8$) is
used for the identification of unresolved binaries, using a cut in the size-brightness diagram, quantified using the \rband\ scale-length measured by \lensfit\ ($R_s={\tt autocal\_scalelength\_pixels}$), and the \rband\ magnitude measured by \gaap. 
In \drfour, the binary selection was made using a simple linear cut in this plane (the dashed red line). For \drfive, we revised the cut
to now perform a slightly modified linear rejection, coupled with a brightness cut: 
\begin{align}\label{eqn:binary}
  &\log_{10}(2R_s)\geq (24.9-r_{\rm GAaP})/3.5\\
  \begin{split}
  &\hbox{and}\\
  &r_{\rm GAaP}>21.5.
  \end{split}
\end{align}
This selection removes only $0.14\%$ of
sources in total, and only $0.02\%$ of sources that would otherwise be currently in the lensing sample. Nonetheless,
visual inspection of the $r-i$ versus $g-r$ colour-colour-space demonstrates that these selected sources are well localised
on the stellar locus. 

Next we applied a blanket magnitude in the detection band: 
\begin{equation}\label{eqn:magmask}
  {\tt MAG\_AUTO} > 20.0. 
\end{equation}
This selection causes a $2.64\%$ reduction in the total source counts, and a $1.18\%$ reduction in the available sample
for lensing. The origin of this selection lies in the previous \lensfit\ version, which did not model the shapes for sources
brighter than $r_{\rm auto}=20$. The fraction of sources lost to this selection is relatively small ($1.18\%$), and 
therefore we opted to keep this selection to be consistent with previous releases. 

Next we applied a resolution selection that is designed to improve weight and shape recalibration (Sect.
\ref{sec:weightrecal}):  
\begin{equation}\label{eqn:resolutioncut}
 \mathcal{R}<0.9,
\end{equation}
where $\mathcal{R}$ is defined in Eq. \ref{eqn:resolution}. This selection is a significant one, and removes 
$43.58\%$ of the total ${\rm MASK}\leq 1$ sample. Of sources
otherwise in the lensing sample, this selection removes $20.15\%$ of the sources. However, these sources are (by
definition) the least resolved sources in \kids, which means that they also have the lowest shear-responses and thus the
lowest lensing weight. The sources are also preferentially faint, as can be seen from Fig. \ref{fig:selections}. 
Therefore, while this selection seems pathological (causing a $>20\%$ reduction in source number), 
\cite{li/etal:2022} found (for \drfour) that the cut only reduced the effective number density of 
lensing sources (i.e. including lensing weights) by $\sim 2\%$. In \kidslegacy\ the selection has a slightly larger
effect, reducing the effective number density by $4.6\%$ (see Appendix~\ref{sec:weighted_selection},
Table~\ref{tab:samplecuts_wtd}).

Finally, we applied a selection on whether sources have a non-zero shape measurement weight: 
\begin{equation}\label{eqn:weight}
  {\tt AlphaRecalC\_weight}>0. 
\end{equation}
This selection is highly correlated with the resolution selection, but can only be applied {\rm after} the
resolution selection has been applied (because the resolution selection is invoked in the weight recalibration, Sect.
\ref{sec:weightrecal}). Similarly, because of the weight recalibration, the weight selection can only be applied to the
subset of the data that also satisfy the resolution selection, and so it is not meaningful to quote the fraction of the
${\rm MASK}\leq 1$ dataset that satisfy the weight requirement. Furthermore, as can be seen in Fig. \ref{fig:selections},
these are primarily faint sources. Of the currently available dataset, the weight selection removes the $13.85\%$ of the
sources. We note, though, that the removal of these sources has no influence on the effective number density of lensing
sources, because they carry zero lensing weight (see Appendix \ref{sec:weighted_selection}).  

These selections combine to produce a \kidslegacy\ sample that consists of $43\,205\,363$ sources drawn from $1014.0$ \sqdeg\ of sky. 
Using the 
effective number density formulation of \citet{heymans/etal:2012}, the sample has an effective number density for lensing 
of $n_{\rm eff}=8.92$ arcmin$^{-2}$. 
We then perform one final selection to produce the fiducial \kidslegacy\ sample, based on the additional masking of
areas impacted by systematic astrometric failures identified during the cosmological analysis of Legacy (see {Wright et al.
submitted for detail)}. The masking of systematic astrometric residuals causes a further down-selection of $4.16\%$ of
the survey sample via on-sky masks, leading to a final \kidslegacy\ sample that consists of $41\,409\,179$ sources drawn from $967.4$
\sqdeg.  
Limiting this sample to the \photoz\ range expected for the \kidslegacy\ cosmological 
analyses ($0.1<z_{\rm B}\leq 2.0$) reduces the sample size by $1.1\%$ (to $40\,950\,607$), corresponding to an effective number density 
$n_{\rm eff}=8.94$ arcmin$^{-2}$. 
This is the fiducial effective number density for the \kidslegacy\ sample, and is what 
we show for \kids\ in Fig. \ref{fig:surveycomp}. 
Finally, for direct comparison with previous \kids\ releases, the effective number 
density of sources in the \photoz\ range $0.1<z_{\rm B}\leq 1.2$ is $n_{\rm eff}=8.00$ arcmin$^{-2}$. This represents a
modest increase from the measured effective number density in \kids\ \drfour, which was $n_{\rm eff}=7.66$ arcmin$^{-2}$
\citep{hildebrandt/etal:2021} in the same \photoz\ window. We attribute this increase in effective number primarily to
the application of the strong-selection mask (Sect. \ref{sec:strongselection}), which masks areas of the survey with
systematically reduced galaxy number densities. 
 
\section{Data release and catalogues}\label{sec:datarelease}

The public release of \kids\ data has been made through the ESO archive. As with previous releases, the data products 
made available via the archive include optical imaging (include science, weight, flag, and sum images), single-band 
detection catalogues, and multi-band catalogues (including \nir\ information from \viking). However, changes to the data that 
are available within the survey (specifically regarding the two \iband\ passes) require changes to the format of the data 
products that are stored at ESO. Columns that are contained within the multi-band catalogues are presented in 
Appendix~\ref{sec:eso_catalogues} (Table~\ref{tab:eso_colourcats}), as well as specific columns that have changed names and/or 
meanings (Table~\ref{tab:eso_changes}).

\section{Summary}\label{sec:summary}

In this paper we present the fifth and final data release of \kids. The data release presents optical and NIR photometry
over $1370$ square degrees of sky: $1347$ square degrees of wide-field imaging designed for use in weak lensing studies,
and $27$ square degrees covering deep spectroscopic survey fields (with $4$ deg$^2$ overlap). The release improves upon
previous \kids\ data releases in more ways than simply the $34\%$ increase in survey area: multi-epoch photometry in the
\iband\ allows for improved \photoz\ and temporal science; improved calibration samples from the larger spectroscopic
survey overlap allow for a better quantification of survey performance and systematic effects; the improved calibration
of photometry and astrometry produces higher-quality imaging and derived data products, improved masking reduces
unnecessary loss of data (and the remaining data are of higher quality); and the updated shape measurement (and
calibration) leads to reductions in systematic biases in downstream data products. 

The data forming this release include images (science, weight, flag, sum, and mask frames), independent catalogues
(single-band source extractions per tile), multi-band catalogues (forced photometry and \photoz\ catalogues per tile),
and mosaic catalogues (the \kidslegacy\  catalogue for weak-lensing science and the \kidz\  catalogue for calibration
efforts).  These data are made publicly available via the ESO archive and the \kids\ collaboration web pages. 

The fiducial \kidslegacy\ sample consists of $41\,409\,179$ sources drawn from $967.4$ \sqdeg\ of sky, with an effective number
density of $n_{\rm eff}=8.87$ arcmin$^{-2}$. After limiting the sample to the expected tomographic limits for
\kidslegacy\ cosmological analyses ($0.1<z_{\rm B}\leq 2.0$), the effective number density decreases slightly to $n_{\rm
eff}=8.94$ arcmin$^{-2}$.  The \kidz\ sample consists of $126\,085$ sources extracted from \kids-depth imaging, with
spectroscopic and photometric redshift estimates. This dataset represents a significant increase in the calibration
sample available for future cosmological analyses with \kids. 

This paper is intended as a one-stop reference for current and future members of the community wishing to utilise \kids\
data for their science. It is the sincere hope of the authors that this release will cement the legacy value of this
unique dataset for years to come. 
 
\begin{acknowledgements}
We thank the anonymous referee for their comments, which have undoubtedly improved the quality of the manuscript. 
AHW, HHi, AD, CM, RR, and JLvdB are supported by an European Research Council Consolidator Grant (No. 770935). 
AHW \& HHi acknowledge funding from the German Science Foundation DFG, via the Collaborative 
Research Center SFB1491 ``Cosmic Interacting Matters - From Source to Signal".
AHW is supported by the Deutsches Zentrum für Luft- und Raumfahrt (DLR), made possible by the Bundesministerium für Wirtschaft und Klimaschutz.
HHi is also supported by a Heisenberg grant of the Deutsche Forschungsgemeinschaft (Hi 1495/5-1). 
KK acknowledges support from the Royal Society and Imperial College. 
MR acknowledges financial support from the INAF mini-grant 2022 "GALCLOCK". 
MBi, PJ, \& GK are supported by the Polish National Science Center through grant no. 2020/38/E/ST9/00395. MBi is also supported by the Polish National Science Center through grant no. 2018/30/E/ST9/00698, 2018/31/G/ST9/03388 and 2020/39/B/ST9/03494, and by the Polish Ministry of Science and Higher Education through grant DIR/WK/2018/12.
CH acknowledges support from the European Research Council under grant number 647112, and the UK Science and Technology Facilities Council (STFC) under grant ST/V000594/1. 
CH, BS, ZY, \& MY acknowledge support from the Max Planck Society and the Alexander von Humboldt Foundation in the framework of the Max Planck-Humboldt Research Award endowed by the Federal Ministry of Education and Research. 
HHo acknowledges support from Vici grant 639.043.512, financed by the Netherlands Organisation for Scientific Research (NWO).
SSL is supported by NOVA, the Netherlands Research School for Astronomy.
LM acknowledges support from the UK STFC under grant ST/N000919/1. 
MBr acknowledges the INAF PRIN-SKA 2017 program 1.05.01.88.04 and the funding from MIUR Premiale 2016: MITIC.
PB acknowledges support from the German Academic Scholarship Foundation.
GC acknowledges the support from the grant ASI n.2018-23-HH.0.
JTAdJ is supported by the NWO through grant 621.016.402. 
BG acknowledges the support of the Royal Society through an Enhancement Award (RGF/EA/181006) and the Royal Society of Edinburgh for support through the Saltire Early Career Fellowship (ref. number 1914).
CG thanks the support from INAF theory Grant 2022: Illuminating Dark Matter using Weak Lensing by Cluster Satellites.
JHD acknowledges support from an STFC Ernest Rutherford Fellowship (project reference ST/S004858/1). 
BJ acknowledges support by STFC Consolidated Grant ST/V000780/1.
LL has been supported by the Deutsche Forschungsgemeinschaft through the project SCHN 342/15-1 and DFG SCHN 342/13. 
LM acknowledges support from the grants PRIN-MIUR 2017 WSCC32 and ASI n.2018-23-HH.0.
SJN is supported by the Polish National Science Center through grant UMO-2018/31/N/ST9/03975.
LP acknowledges support from the DLR grant 50QE2002.
CS acknowledges support from the Agencia Nacional de Investigaci\'on y Desarrollo (ANID) through FONDECYT grant no.\ 11191125 and BASAL project FB210003.
HYS acknowledges the support from CMS-CSST-2021-A01 and CMS-CSST-2021-B01, NSFC of China under grant 11973070, and Key Research Program of Frontier Sciences, CAS, Grant No. ZDBS-LY-7013.
TT acknowledges funding from the Swiss National Science Foundation under the Ambizione project PZ00P2_193352.
GVK acknowledges financial support from the Netherlands Research School for Astronomy (NOVA) and Target. Target is supported by Samenwerkingsverband Noord Nederland, European fund for regional development, Dutch Ministry of economic affairs, Pieken in de Delta, Provinces of Groningen and Drenthe.
JY acknowledges the support of the National Science Foundation of China (12203084), the China Postdoctoral Science Foundation (2021T140451), and the Shanghai Post-doctoral Excellence Program (2021419).
YHZ acknowledges support from the UK STFC. 
Based on data obtained from the ESO Science Archive Facility with DOI: 
\url{https://doi.org/10.18727/archive/37}, and
\url{https://doi.eso.org/10.18727/archive/59} and on data products produced by the KiDS consortium. 
The KiDS production team acknowledges support from: 
  Deutsche Forschungsgemeinschaft, ERC, NOVA and NWO-M grants; 
  Target; the University of Padova, and the University Federico II (Naples).
This work was performed in part at Aspen Center 
for Physics, which is supported by National Science Foundation grant PHY-1607611.
\\
\textit{Author Contributions:} 
All authors contributed to the development and writing of this paper. The
authorship list is given in three groups: the lead authors (AHW, KK, HHi, MR, and MB), 
followed by two alphabetical groups. The first alphabetical
group includes those who are key contributors to both the scientific analysis
and the data products of this manuscript and release. The second group covers those who 
have either made a significant contribution to the preparation of data products or to the 
scientific analyses of \kids\ since its inception.
\end{acknowledgements}

\bibpunct{(}{)}{;}{a}{}{,}
\bibliographystyle{aa}
\newcommand{\noopsort}[1]{}

\begin{appendix} %First  appendix
\section{Magnitude limit estimates with \lambdar}\label{sec:lambdar}
\begin{figure*}[htpb]
  \centering
  \includegraphics[width=1.9\columnwidth]{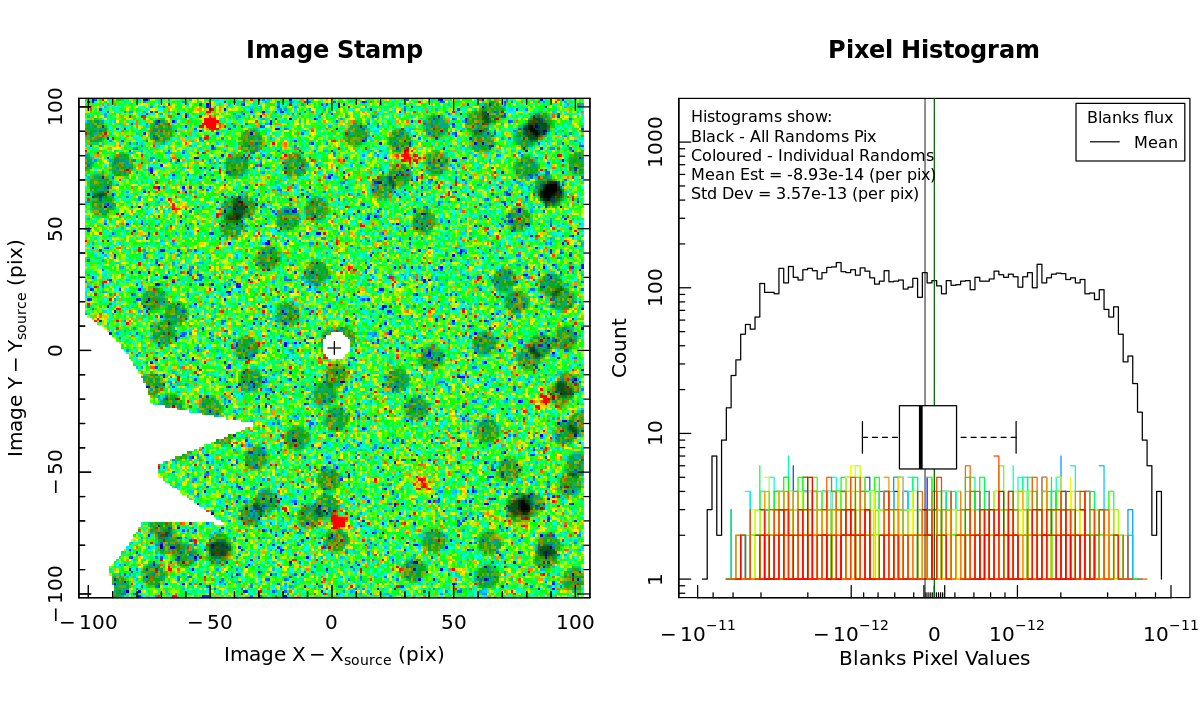}
  \caption{Correlated background estimated using the \lambdar\ randoms routine. {\em Left:} Local region of the image under analysis, 
  with the $100$ realisations of the random aperture shown as translucent black circles. The image has been masked using the fiducial mask, 
  seen as the missing triangular region (a diffraction spike) at the bottom of the image. {\em Right:} Distribution of pixel values 
  measured within the random apertures, for all apertures (black) and the individual apertures (colours). There are some apertures that are 
  coincident with sources, resulting in large positive fluxes. These are limited in the computation of the random noise estimate through the use 
  of median statistics. The distribution of aperture fluxes from the randoms is shown by the box-and-whisker plot, with the outlier fluxes shown as 
  stars. This figure is a direct output of the \lambdar\ code. }\label{fig:blanks}
\end{figure*}
\begin{figure*}[htpb]
  \centering
  \includegraphics[width=1.9\columnwidth]{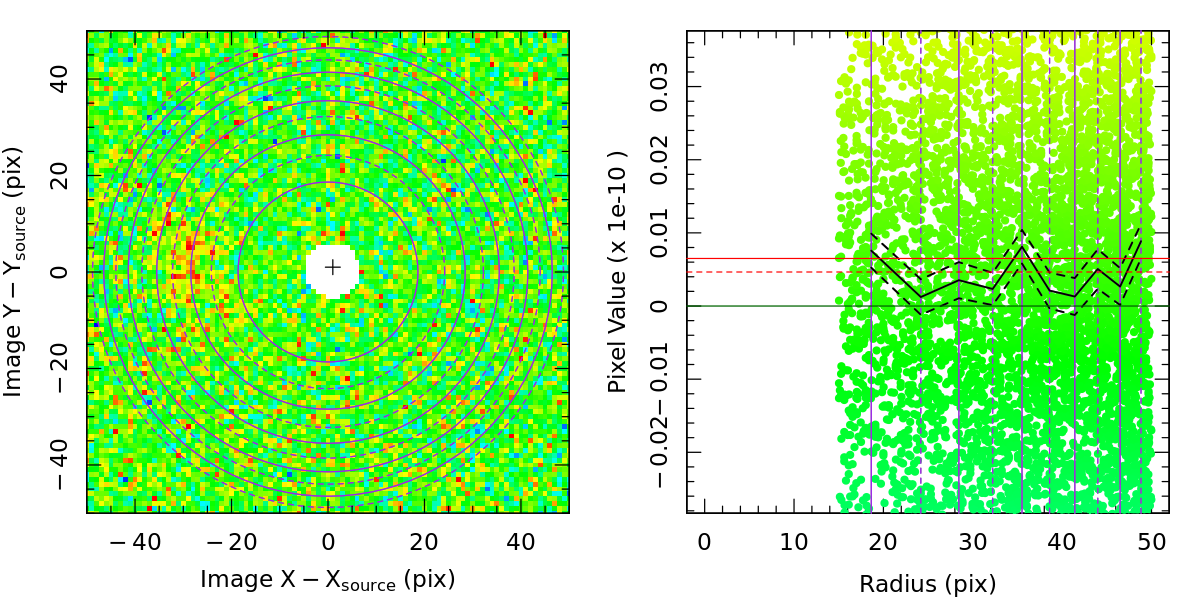}
  \caption{Uncorrelated background estimated using the \lambdar\ sky-estimate routine. {\em Left:} Annuli used to estimate 
  the background estimates, shown against the input image (colour mapping for pixels in the left panel is given by the point colouring 
  in the right panel). 
  {\em Right:} Distribution of pixel values (points) as a function of 
  radius from the random location chosen. The black lines show the median sky value per annulus (solid) and the uncertainty on the 
  median (dashed). The final mean (median) sky estimate is shown as the solid (dashed) red line. This figure is a direct output of  
  the \lambdar\ code. }\label{fig:skyest}
\end{figure*}
Background estimates in \kidskidz\ were made using \lambdar~\citep{wright/etal:2016}. \lambdar\ provides 
a convenient tool for estimating the correlated and uncorrelated noise present in the various sets of imaging that 
are present in \kids-\drfive, as it can be ran on images of arbitrary pixel scale. This allows us 
to simply specify the on-sky locations that we wish to sample with our sky-estimates, and allow the code to report back 
the noise estimates at each location on-sky for these positions. For the grid of on-sky sampling positions we chose a simple 
rectangular grid that sampled each of the $\sim 1$ deg$^2$ pointings with a spacing of $0.02$ degrees in both RA and declination. 
We estimate the background level of imaging data using the two methods built into the \lambdar\ code: randoms estimation and 
sky estimation. 

The randoms estimate of noise is the fiducial one that we use for quantifying our imaging depth in \drfive, as it includes the 
influence of correlated noise. We summarise the computation of the randoms noise estimation process here, and direct the interested reader to 
Sect. 3.6 of \citet{wright/etal:2016} for a complete description of the randoms measurement process. 

The randoms noise estimate is made by measuring the flux contained within the chosen aperture 
(i.e. circular with a $2\arcsec$ diameter), shifted to 
multiple random location within the local area of the image under analysis. An example of this is shown in Fig.~\ref{fig:blanks}. 
After randomly shifting the location of the source aperture, the flux measured in the aperture at this new location is measured and 
recorded. After many realisations, the distribution of random fluxes therefore encodes the expected flux contained within an 
aperture of this specific size for a source randomly placed in this area of imaging. This measurement is performed on the image 
after application of the fiducial mask, as can be seen in Fig.~\ref{fig:blanks} where a diffraction spike has been masked at the 
bottom of the image. However, this estimate does not include source-masking: this means that the distribution of measured random 
fluxes also includes contamination from detectable sources. As such, the expectation and scatter of the random flux distribution is 
computed with median statistics, to suppress bias caused by outliers in the tails of the flux distribution (shown as stars in the 
right panel of Fig.~\ref{fig:blanks}).

The second estimate of the background noise is made without considering the on-sky correlation present in the background. We leverage this 
sky estimation method as a robustness check for our fiducial noise estimation method, and to estimate the implied degree of on-sky correlation 
present in the \kids\ imaging. We briefly describe the sky estimation process here, and direct the interested reader to Sect. 3.5 of 
\citet{wright/etal:2016} for a complete description of the measurement process. 

The sky-estimate routine in \lambdar\ assumes that the pixel-noise can be computed in annular radii around the input source, 
without consideration of the relative correlation between adjacent pixels. The sky estimate is made on the masked image (but, again, without source 
masking in our application). Figure~\ref{fig:skyest} demonstrates the estimation of the uncorrelated sky, for the same location shown in Fig.~\ref{fig:blanks}. 
The sky estimate is made on a smaller scale than in the randoms case, and is similarly constructed to be robust to contamination: estimates of the mean/median/RMS sky are 
made in annuli around the requested location, and annuli with anomalously high/low values are discarded. In the case of Fig.~\ref{fig:skyest}, all bins were acceptable. 
The resulting mean/median sky estimates, and the associated RMS scatter, are provided for this location. The values of the sky at the requested locations are therefore 
tabulated in this method without consideration of pixel-to-pixel noise correlation. 

Previous data releases in \kids\ have leveraged background estimates similar to the uncorrelated method used by \lambdar. This means that the background estimates (and therefore reported magnitude limits) in previous \kids\ releases should be similar to those computed using \lambdar's uncorrelated estimate. The magnitude limits estimated using this method for \drfive\ are $\{24.26\pm0.10,25.15\pm0.12,25.07\pm0.14,23.66\pm0.25,23.73\pm0.3\}$ for the $\{u,g,r,i_1,i_2\}$-bands, respectively, in very good agreement with the estimates presented in previous releases. 

\begin{figure*}[h]
  \centering
  \includegraphics[width=2.0\columnwidth]{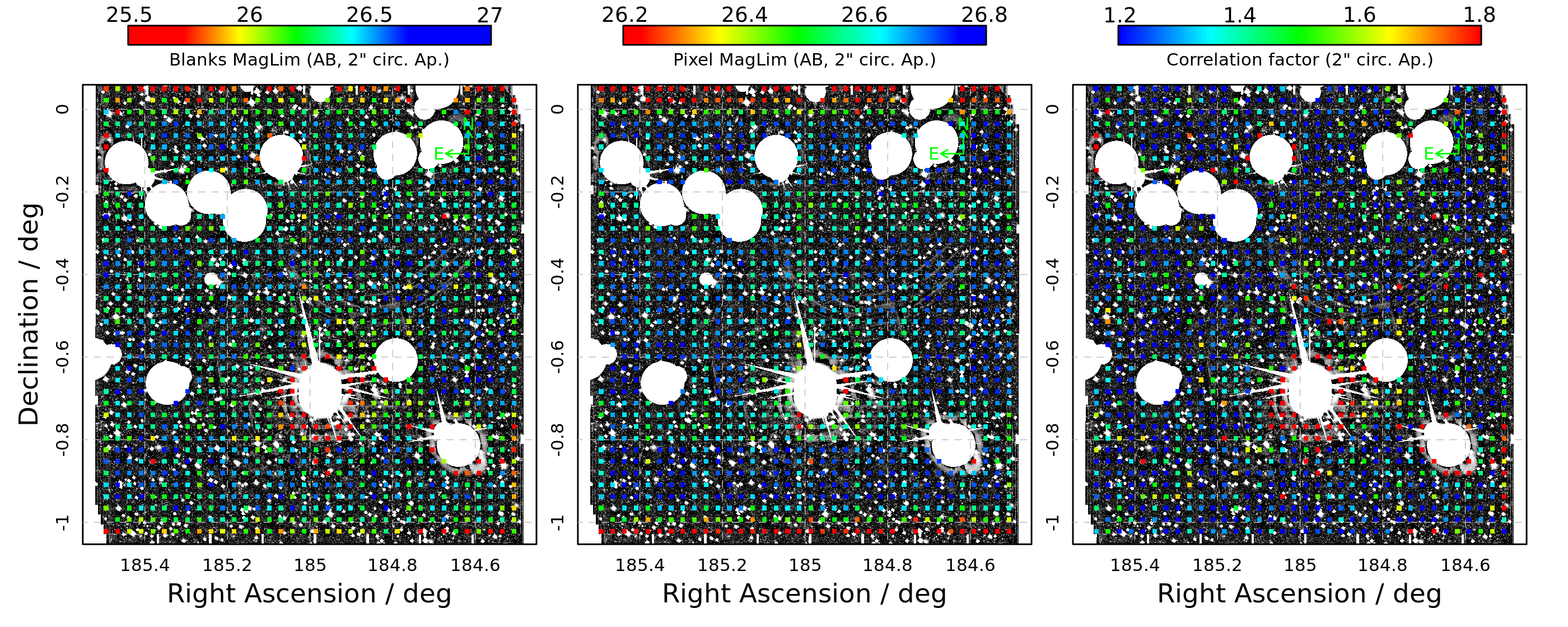}
  \caption{Demonstration of the background estimates produced by \lambdar\ for pointing {\tt KIDS\_185p0\_m0p5} in the
  \rband. Each panel shows a grey-scale image of the pointing, smoothed with a $2\arcsec$ Gaussian kernel and after
  masking. Overlaid on this image are various estimates of the image noise level, shown with rectangles covering the
  extent of the area used for the estimate. 
  {\em Left:} Magnitude limits estimated with blanks. {\em Centre:} Magnitude limits estimated with the pixel RMS. {\em
  Right:} Correlation factor, estimated as the ratio between the aperture noise RMSs estimated with blanks and
  pixels. This factor clearly recovers the regions of the image with increased noise correlation caused by residual
  bright artefacts. }\label{fig:skycorr}
\end{figure*}

An example of the estimation of these two background levels for a single pointing of \kids\ is shown in Fig.~\ref{fig:skycorr}. The figure shows the 
estimate of the correlated sky noise with randoms, the estimate of the uncorrelated noise, and ratio between these two noise estimates. This 
ratio is an estimate of the degree of correlation present in the imaging at that location on sky. The distribution of uncorrelated sky values can be 
seen to closely trace the dither pattern of an observation in \kids: the lowest noise values occur where pixels were
observed by all five exposures, and 
the highest noise values occur in the upper and lower extremes of the image where only one exposure is available per pixel. The correlated noise estimate 
shows much less of this structure, suggesting that the expected $\sqrt{N_{\rm exp}}$ reduction in noise is being hampered by other effects. The cause of this 
is suggested by the third panel, which shows the noise correlation factor: 
\begin{equation}
    C_{\rm n} = \sigma_{\rm rand} / \sigma_{\rm uncorr}. 
\end{equation}
The correlation factor can be seen to increase in regions of the image where there is considerable background contamination from, for example, residual reflections 
from bright stars. 

\FloatBarrier

\section{\vst\ on-sky quality metrics}\label{sec:vstonskymetrics}

\begin{figure*}
  \centering
  \includegraphics[width=1.8\columnwidth]{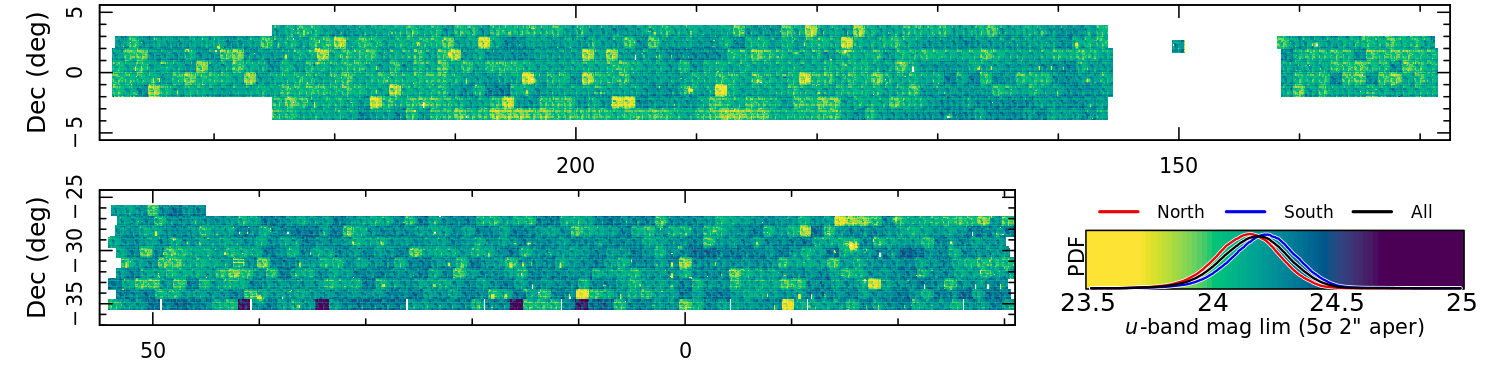}
  \includegraphics[width=1.8\columnwidth]{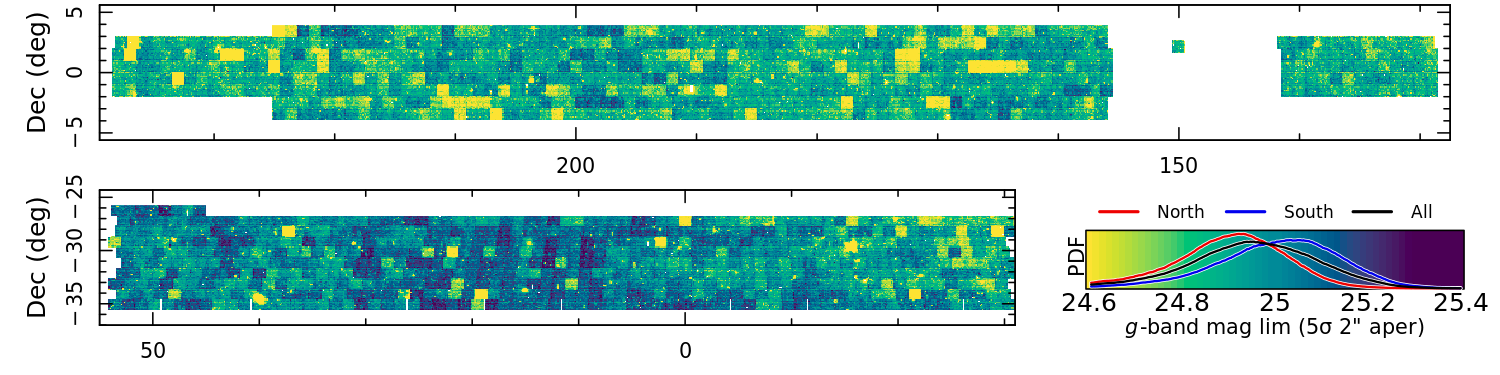}
  \includegraphics[width=1.8\columnwidth]{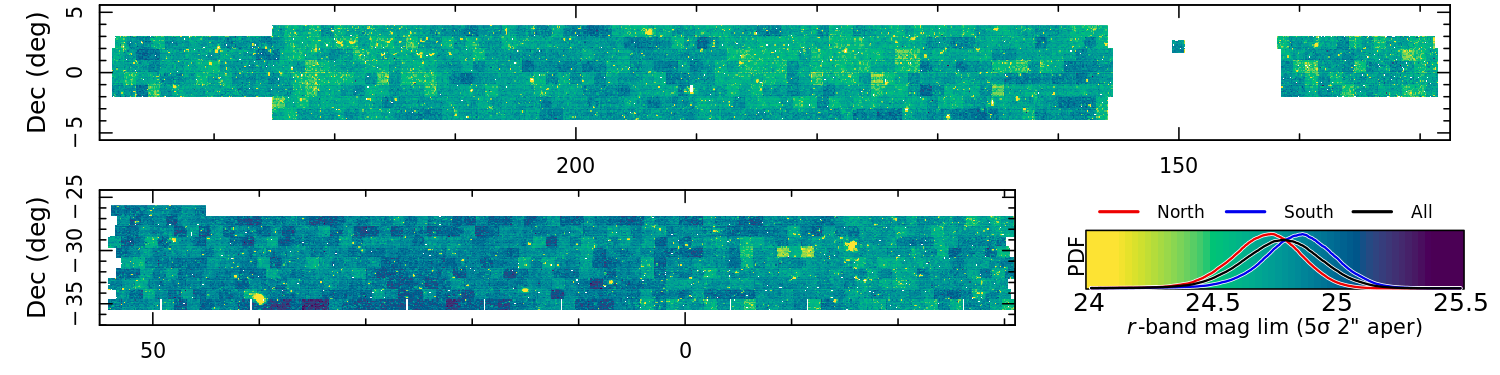}
  \includegraphics[width=1.8\columnwidth]{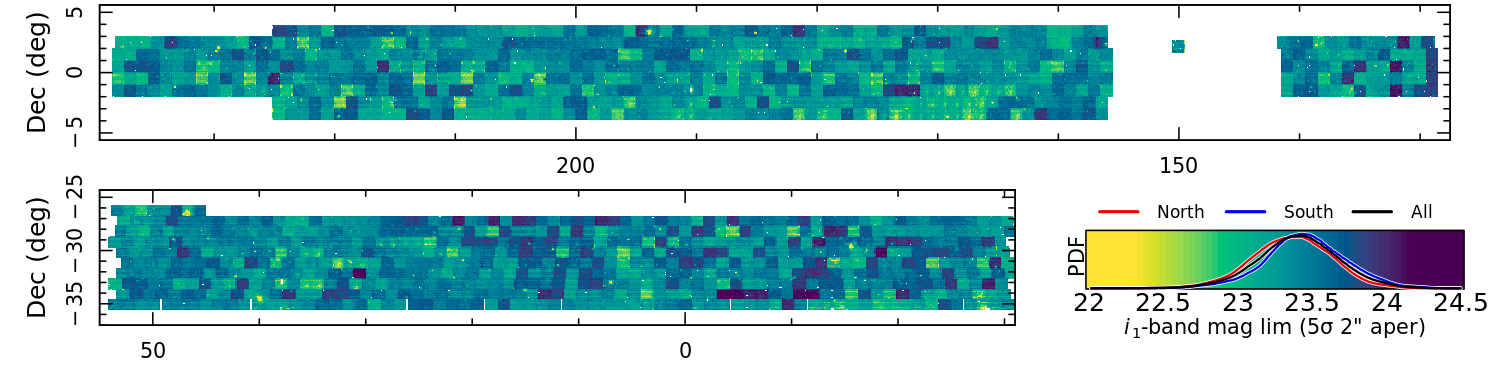}
  \includegraphics[width=1.8\columnwidth]{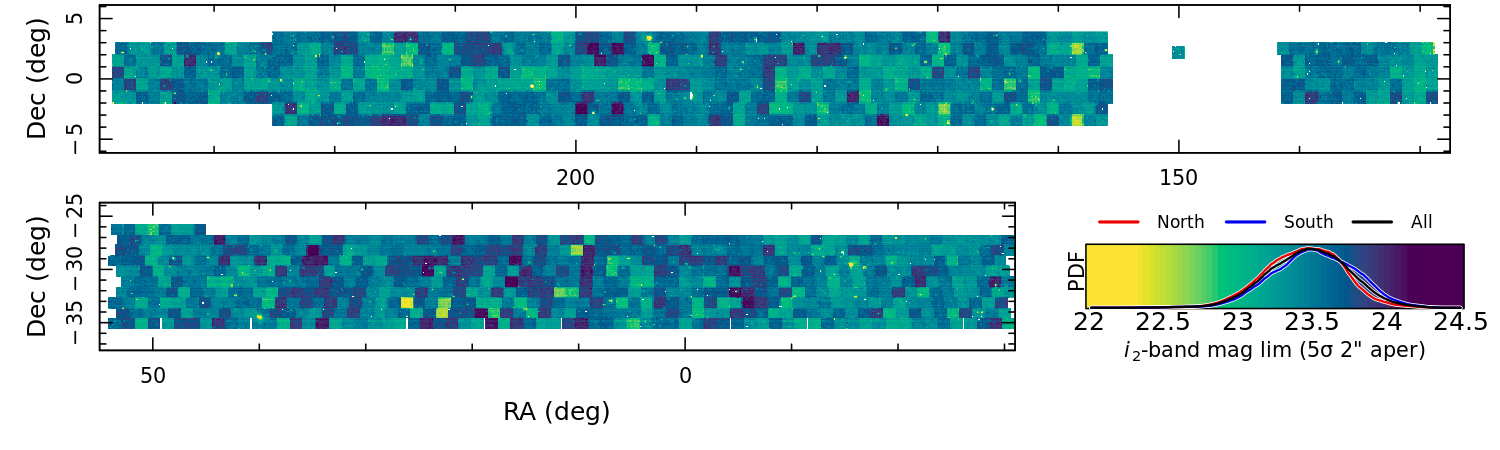}
  \caption{Distribution of limiting magnitudes in the optical bands for the \kids\ fields. The background levels measured using 
  $2\arcsec$ circular apertures, as described in Appendix~\ref{sec:lambdar}. To construct the on-sky mosaic, estimates per-pointing are 
  combined using the same WCS cuts as are applied to the data (Sect.~\ref{sec:mosaics}).}\label{fig:vstbanddepth}
\end{figure*}

\begin{figure*}
  \centering
  \includegraphics[width=1.8\columnwidth]{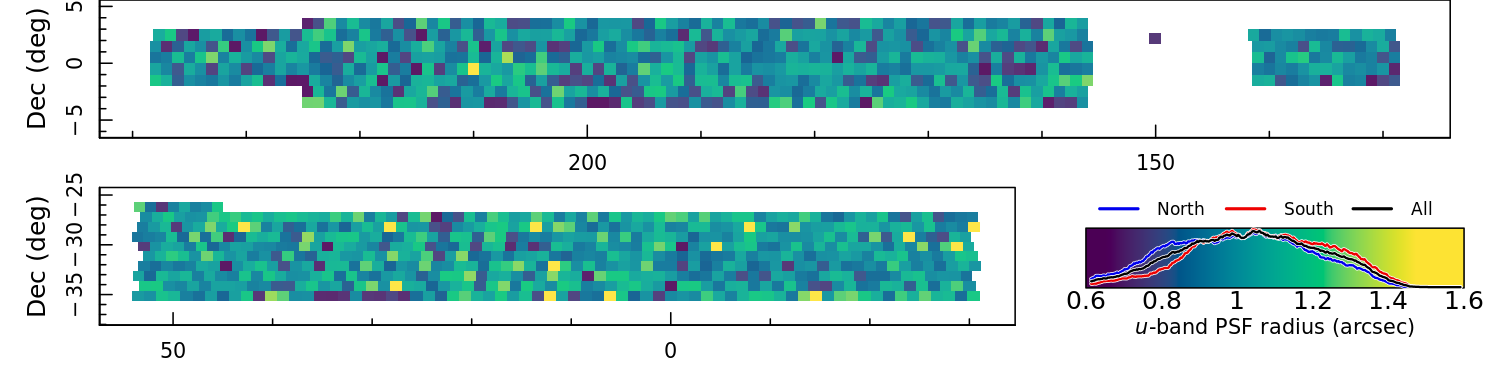}
  \includegraphics[width=1.8\columnwidth]{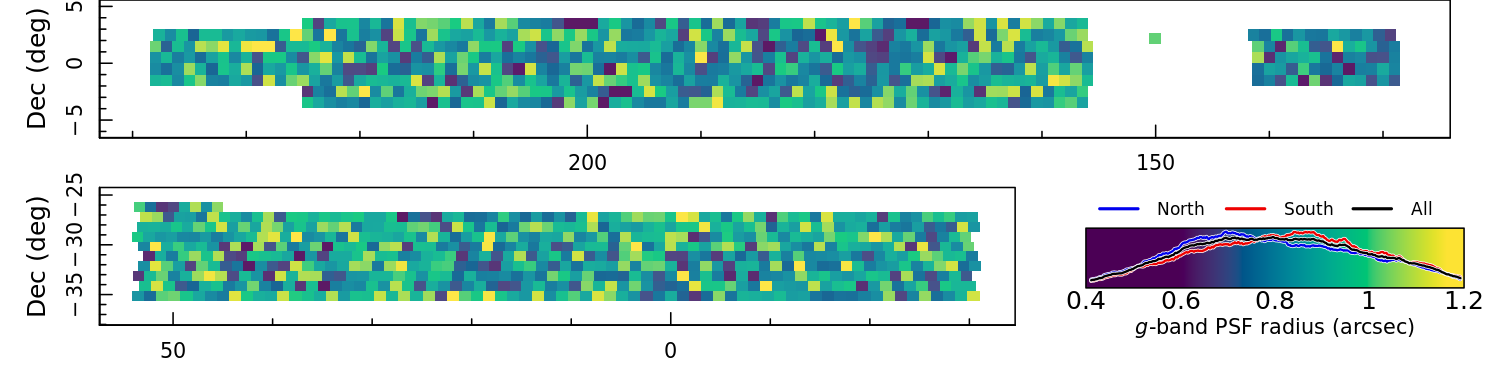}
  \includegraphics[width=1.8\columnwidth]{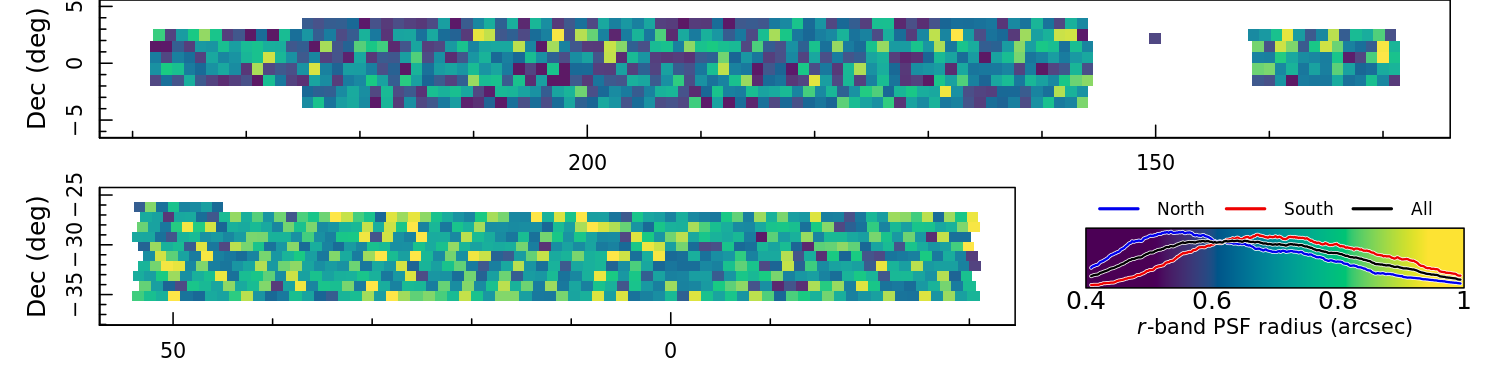}
  \includegraphics[width=1.8\columnwidth]{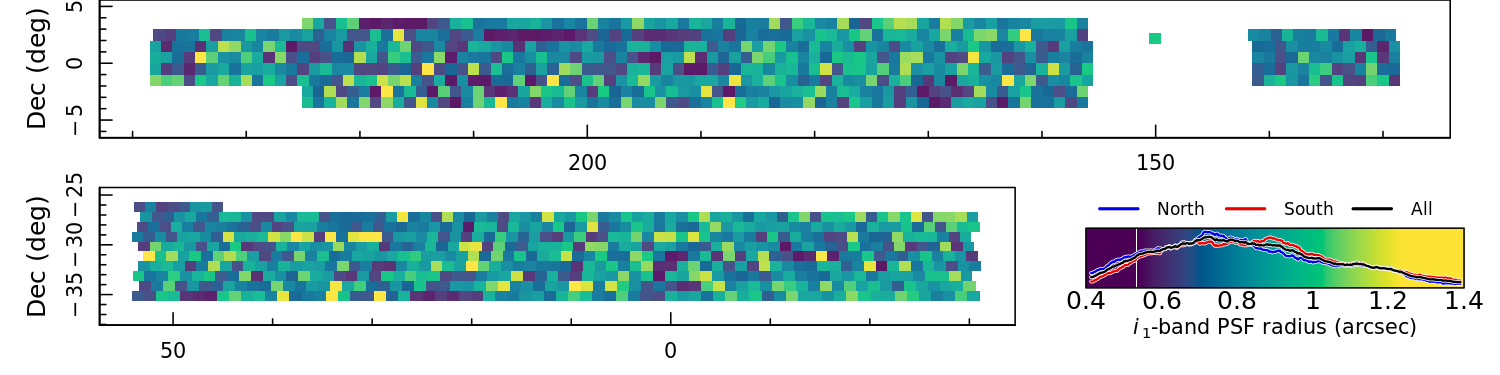}
  \includegraphics[width=1.8\columnwidth]{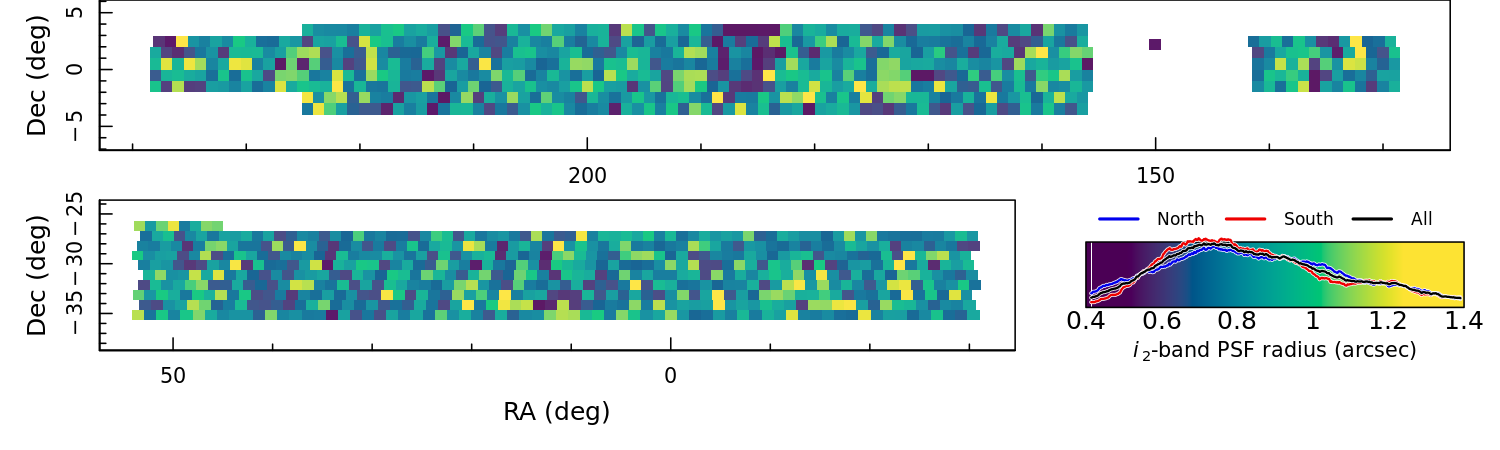}
  \caption{Distribution of PSF FWHM sizes (in arcsec) as reported by \astrowise\ for each pointing within \kids. 
  }\label{fig:vstbandpsf}
\end{figure*}

For many scientific applications, an understanding of the systematic differences in 
imaging quality on-sky is relevant. For imaging in \kids\ the primary parameters of relevance are 
the magnitude limit (estimated using the method described in Appendix~\ref{sec:lambdar}), and the 
PSF size in each band. As such, in Fig.~\ref{fig:vstbanddepth} we show the distribution of magnitude 
limits estimated in each band on-sky, and in Fig.~\ref{fig:vstbandpsf} we show the distribution of 
average PSF sizes (per pointing, as reported by \astrowise). 
Each figure shows the parameters per filter, with the two \iband\ passes shown separately.   

\FloatBarrier

\section{Astrometric residuals compared to \gaia}
\label{sec:astrometry_residuals}

\begin{figure}
    \centering
    \includegraphics[width=\columnwidth]{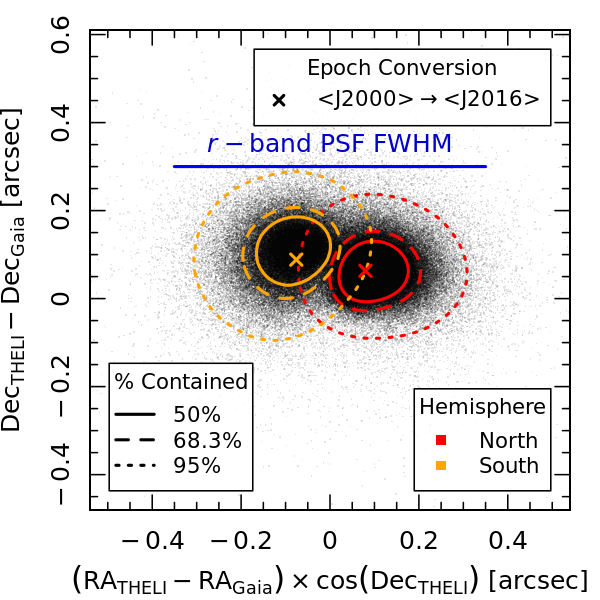}
    \caption{Astrometric residuals between \kids\ and \gaia\ when comparing to J2016 epoch stellar positions. The systematic 
    difference is caused by the systematic shift in the positions of stars between 2000 and 2016, caused by the motion of the 
    Solar System through the Milky Way.}
    \label{fig:astrometrygaia}
\end{figure}

As mentioned in Sect. \ref{sec:astrometry}, the \kids\ data are calibrated to \gaia\ stars at the J2000 epoch. 
In practice, this means that the positions of stars used to calibrate \kids\ have been shifted (using their observed positions 
and estimated proper motions) back to where they {would have been} in the year 2000. Crucially, though, this does not just have 
a random effect per star: as the Solar System moves through the Milky Way, an additional coherent shift is imprinted on the stellar 
distribution. The purpose of calibrating to J2000 epoch stellar positions is to remove this coherent drift. 

However, \kids\ observations were not taken in the year 2000. Rather, the typical \kids\ observation was taken around 2015--2016. As 
a result, the observed positions of stars in \kids\ imaging are actually much more similar to the observed positions of \gaia\ stars 
at epoch J2016. 

The fact that \kids\ is calibrated to J2000 epoch stars, but that \kids\ images were taken around epoch J2016 (similar to \gaia) 
leads to an apparent systematic bias when performing a direct sky match between the \kids\ \drfive\ catalogues and \gaia\ stars 
at their epoch J2016 positions: there appears to be a systematic bias in the two hemispheres, at the level of 
$|\Delta {\rm RA}|\,\sim0\farcs1$, which has different sign in the 
northern and the southern hemispheres. This can be seen in Fig. \ref{fig:astrometrygaia}. Of particular note, though, is the 
low scatter in each of the systematically offset clouds. This is demonstrative that, modulo the difference in overall coordinates 
between epoch J2000 and epoch J2016 positions, the relative position of stars in \kids\ and \gaia\ are very similar 
(because the observations were actually taken at similar epochs). 

\begin{figure}
    \centering
    \includegraphics[width=\columnwidth]{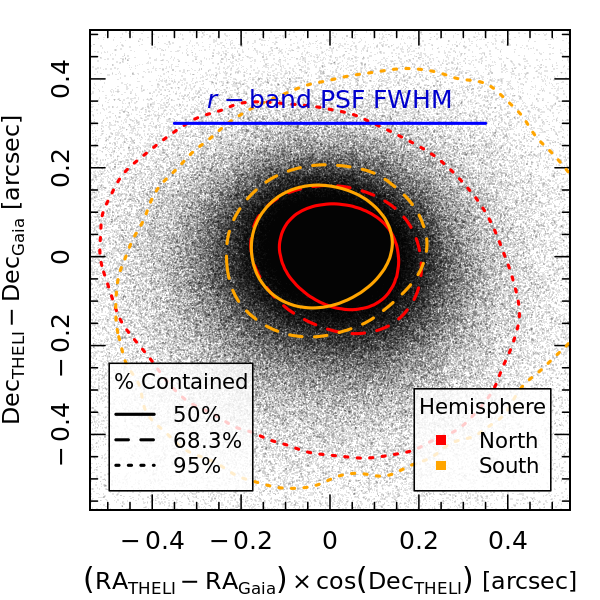}
    \caption{Astrometric residuals between \kids\ and \gaia\ when comparing to J2000 epoch stellar positions. The systematic 
    difference is no longer present; however, a considerable increase in scatter is visible. This is due to the \kids\ stars 
    being relatively positioned according to the J2016 epoch (determined by when the images were taken).}
    \label{fig:astrometrygaiaj2000}
\end{figure}

Figure \ref{fig:astrometrygaiaj2000} shows the same as Fig. \ref{fig:astrometrygaia}, but now using J2000 epoch positions. 
When performing the comparison of per-star astrometry between the \kids\ data and the J2000 epoch \gaia\ stars, one expects that 
the systematic offsets would disappear, which it does. However, we also see a considerable increase in scatter: this is due to the 
proper motions of stars introducing noise, as they shuffle positions between epoch J2000 and epoch J2016. 

\FloatBarrier

\section{List of reprocessed \uband\ co-adds}
\label{sec:reprocess_list}

As discussed in Sect. \ref{sec:co-add}, \uband\ data were flagged for reprocessing with a modified astrometric 
distortion polynomial order, to reduce pathological errors in co-add construction. The full list of reprocessed 
\uband\ co-adds is given in Table \ref{tab:reprocesseduband}. 

\begin{table*}
    \caption{Co-adds for which the \uband\ was reprocessed with linear polynomial order for the astrometric distortion definition.}\label{tab:reprocesseduband}
    \centering
    \begin{tabular}{c|c|c|c|c|c}
KiDS_0.0_-27.2 & KiDS_0.0_-28.2 & KiDS_0.0_-30.2 & KiDS_0.0_-31.2 & KiDS_1.1_-27.2 & KiDS_1.2_-30.2 \\
KiDS_10.3_-29.2 & KiDS_10.4_-30.2 & KiDS_10.6_-32.1 & KiDS_11.3_-28.2 & KiDS_11.4_-29.2 & KiDS_11.5_-30.2 \\
KiDS_11.9_-33.1 & KiDS_12.0_-34.1 & KiDS_12.4_-28.2 & KiDS_12.5_-29.2 & KiDS_12.8_-31.2 & KiDS_13.2_-34.1 \\
KiDS_13.5_-27.2 & KiDS_13.5_-28.2 & KiDS_13.7_-29.2 & KiDS_14.1_-32.1 & KiDS_14.4_-34.1 & KiDS_14.6_-35.1 \\
KiDS_14.7_-28.2 & KiDS_14.8_-29.2 & KiDS_15.0_-30.2 & KiDS_15.1_-31.2 & KiDS_15.3_-32.1 & KiDS_15.4_-33.1 \\
KiDS_15.7_-27.2 & KiDS_15.8_-28.2 & KiDS_15.8_-35.1 & KiDS_158.4_3.5 & KiDS_16.3_-31.2 & KiDS_16.5_-32.1 \\
KiDS_16.8_-27.2 & KiDS_16.8_-34.1 & KiDS_16.9_-28.2 & KiDS_161.0_0.5 & KiDS_163.0_-1.5 & KiDS_164.0_-0.5 \\
KiDS_164.0_1.5 & KiDS_165.5_3.5 & KiDS_167.5_-3.5 & KiDS_167.5_2.5 & KiDS_168.0_-1.5 & KiDS_168.5_3.5 \\
KiDS_17.0_-35.1 & KiDS_17.1_-29.2 & KiDS_17.3_-30.2 & KiDS_17.5_-31.2 & KiDS_17.6_-32.1 & KiDS_17.8_-33.1 \\
KiDS_170.0_-1.5 & KiDS_171.0_0.5 & KiDS_171.5_-3.5 & KiDS_173.0_-0.5 & KiDS_173.5_3.5 & KiDS_175.0_-0.5 \\
KiDS_175.0_1.5 & KiDS_175.5_2.5 & KiDS_177.0_0.5 & KiDS_177.5_2.5 & KiDS_177.5_3.5 & KiDS_178.0_1.5 \\
KiDS_179.0_-1.5 & KiDS_179.5_3.5 & KiDS_18.0_-34.1 & KiDS_18.1_-28.2 & KiDS_18.2_-35.1 & KiDS_18.4_-30.2 \\
KiDS_18.6_-31.2 & KiDS_18.8_-32.1 & KiDS_181.0_0.5 & KiDS_181.0_1.5 & KiDS_181.5_-2.5 & KiDS_183.0_-1.5 \\
KiDS_185.0_-0.5 & KiDS_186.0_1.5 & KiDS_186.5_2.5 & KiDS_187.0_0.5 & KiDS_19.2_-28.2 & KiDS_19.2_-34.1 \\
KiDS_19.4_-29.2 & KiDS_19.5_-35.1 & KiDS_190.0_-1.5 & KiDS_192.5_3.5 & KiDS_193.0_-0.5 & KiDS_193.0_0.5 \\
KiDS_193.0_1.5 & KiDS_194.5_3.5 & KiDS_196.0_0.5 & KiDS_2.4_-32.1 & KiDS_20.0_-32.1 & KiDS_20.2_-27.2 \\
KiDS_20.2_-33.1 & KiDS_20.3_-28.2 & KiDS_20.4_-34.1 & KiDS_20.5_-29.2 & KiDS_201.6_-3.5 & KiDS_204.0_-0.5 \\
KiDS_207.6_2.5 & KiDS_21.0_-31.2 & KiDS_21.2_-32.1 & KiDS_21.3_-27.2 & KiDS_21.4_-28.2 & KiDS_21.4_-33.1 \\
KiDS_21.6_-29.2 & KiDS_21.6_-34.1 & KiDS_21.9_-30.2 & KiDS_219.6_2.5 & KiDS_22.1_-31.2 & KiDS_22.4_-32.1 \\
KiDS_22.6_-33.1 & KiDS_22.8_-29.2 & KiDS_22.8_-34.1 & KiDS_23.0_-30.2 & KiDS_23.6_-27.2 & KiDS_23.7_-28.2 \\
KiDS_23.8_-33.1 & KiDS_23.9_-29.2 & KiDS_24.5_-31.2 & KiDS_24.7_-32.1 & KiDS_24.8_-28.2 & KiDS_25.0_-33.1 \\
KiDS_25.1_-29.2 & KiDS_25.2_-34.1 & KiDS_25.3_-30.2 & KiDS_25.5_-35.1 & KiDS_25.6_-31.2 & KiDS_25.8_-27.2 \\
KiDS_25.9_-32.1 & KiDS_26.0_-28.2 & KiDS_26.1_-33.1 & KiDS_26.2_-29.2 & KiDS_26.4_-34.1 & KiDS_26.8_-31.2 \\
KiDS_27.1_-28.2 & KiDS_27.1_-32.1 & KiDS_27.3_-33.1 & KiDS_27.6_-30.2 & KiDS_27.6_-34.1 & KiDS_28.0_-27.2 \\
KiDS_28.0_-31.2 & KiDS_28.2_-32.1 & KiDS_28.5_-29.2 & KiDS_28.8_-34.1 & KiDS_29.1_-31.2 & KiDS_29.2_-27.2 \\
KiDS_29.2_-35.1 & KiDS_29.3_-28.2 & KiDS_29.4_-32.1 & KiDS_29.6_-29.2 & KiDS_3.4_-28.2 & KiDS_3.5_-32.1 \\
KiDS_3.6_-33.1 & KiDS_3.6_-34.1 & KiDS_30.0_-34.1 & KiDS_30.3_-31.2 & KiDS_30.5_-28.2 & KiDS_30.6_-32.1 \\
KiDS_30.8_-29.2 & KiDS_31.1_-30.2 & KiDS_31.4_-27.2 & KiDS_31.6_-28.2 & KiDS_31.6_-35.1 & KiDS_31.8_-32.1 \\
KiDS_31.9_-29.2 & KiDS_32.1_-33.1 & KiDS_32.2_-30.2 & KiDS_32.5_-27.2 & KiDS_32.8_-35.1 & KiDS_32.9_-32.1 \\
KiDS_33.0_-29.2 & KiDS_33.4_-30.2 & KiDS_33.8_-31.2 & KiDS_33.9_-28.2 & KiDS_335.5_-31.2 & KiDS_336.5_-32.1 \\
KiDS_34.2_-29.2 & KiDS_34.8_-27.2 & KiDS_346.3_-29.2 & KiDS_346.5_-27.2 & KiDS_347.6_-28.2 & KiDS_348.7_-28.2 \\
KiDS_349.2_-34.1 & KiDS_349.6_-30.2 & KiDS_35.0_-31.2 & KiDS_35.3_-35.1 & KiDS_35.6_-33.1 & KiDS_35.9_-27.2 \\
KiDS_350.6_-32.1 & KiDS_352.1_-28.2 & KiDS_355.3_-32.1 & KiDS_355.4_-30.2 & KiDS_355.5_-27.2 & KiDS_356.4_-33.1 \\
KiDS_356.4_-35.1 & KiDS_356.5_-31.2 & KiDS_356.6_-27.2 & KiDS_356.6_-29.2 & KiDS_357.6_-33.1 & KiDS_358.8_-35.1 \\
KiDS_358.9_-28.2 & KiDS_36.0_-34.1 & KiDS_36.1_-28.2 & KiDS_36.1_-31.2 & KiDS_36.5_-29.2 & KiDS_36.5_-32.1 \\
KiDS_36.5_-35.1 & KiDS_37.2_-34.1 & KiDS_37.6_-29.2 & KiDS_37.6_-32.1 & KiDS_37.7_-35.1 & KiDS_38.0_-30.2 \\
KiDS_38.0_-33.1 & KiDS_38.1_-27.2 & KiDS_38.4_-28.2 & KiDS_38.4_-31.2 & KiDS_38.4_-34.1 & KiDS_38.9_-35.1 \\
KiDS_39.2_-33.1 & KiDS_39.3_-27.2 & KiDS_39.6_-31.2 & KiDS_39.9_-29.2 & KiDS_4.5_-28.2 & KiDS_4.6_-29.2 \\
KiDS_4.7_-31.2 & KiDS_4.7_-32.1 & KiDS_4.9_-35.1 & KiDS_40.0_-32.1 & KiDS_40.1_-35.1 & KiDS_40.3_-30.2 \\
KiDS_40.4_-33.1 & KiDS_40.8_-31.2 & KiDS_40.8_-34.1 & KiDS_41.0_-29.2 & KiDS_41.5_-27.2 & KiDS_41.8_-28.2 \\
KiDS_41.9_-31.2 & KiDS_42.2_-29.2 & KiDS_42.4_-32.1 & KiDS_42.6_-27.2 & KiDS_42.6_-35.1 & KiDS_42.8_-33.1 \\
KiDS_43.1_-31.2 & KiDS_43.3_-29.2 & KiDS_43.5_-32.1 & KiDS_43.7_-27.2 & KiDS_43.8_-35.1 & KiDS_44.0_-28.2 \\
KiDS_44.4_-29.2 & KiDS_44.7_-32.1 & KiDS_44.9_-27.2 & KiDS_45.0_-35.1 & KiDS_45.1_-28.2 & KiDS_45.1_-33.1 \\
KiDS_45.4_-31.2 & KiDS_45.6_-26.2 & KiDS_45.6_-29.2 & KiDS_46.0_-27.2 & KiDS_46.3_-28.2 & KiDS_47.1_-32.1 \\
KiDS_47.5_-33.1 & KiDS_47.8_-29.2 & KiDS_48.0_-34.1 & KiDS_48.2_-32.1 & KiDS_48.9_-26.2 & KiDS_48.9_-31.2 \\
KiDS_49.2_-34.1 & KiDS_49.3_-27.2 & KiDS_49.9_-33.1 & KiDS_49.9_-35.1 & KiDS_5.6_-27.2 & KiDS_5.7_-29.2 \\
KiDS_5.8_-31.2 & KiDS_50.8_-28.2 & KiDS_51.1_-35.1 & KiDS_51.3_-31.2 & KiDS_52.2_-26.2 & KiDS_52.3_-33.1 \\
KiDS_52.9_-32.1 & KiDS_53.0_-28.2 & KiDS_53.3_-26.2 & KiDS_53.5_-29.2 & KiDS_6.1_-35.1 & KiDS_6.8_-28.2 \\
KiDS_6.8_-29.2 & KiDS_7.3_-35.1 & KiDS_7.9_-27.2 & KiDS_7.9_-28.2 & KiDS_8.0_-29.2 & KiDS_8.1_-30.2 \\
KiDS_8.2_-32.1 & KiDS_8.4_-34.1 & KiDS_8.5_-35.1 & KiDS_9.0_-27.2 & KiDS_9.0_-28.2 & KiDS_9.1_-29.2 \\
KiDS_9.3_-31.2 & KiDS_9.4_-32.1 & KiDS_9.5_-33.1 & KiDZ_209.4_5.1 & KiDZ_31.6_-4.6 & KiDZ_32.6_-4.6 \\
KiDZ_34.6_-4.6 & & & & & \\
    \end{tabular}
\end{table*} 

\FloatBarrier

\section{\vista\ on-sky quality metrics}\label{sec:vistaonskymetrics}

\begin{figure*}
  \centering
  \includegraphics[width=1.8\columnwidth]{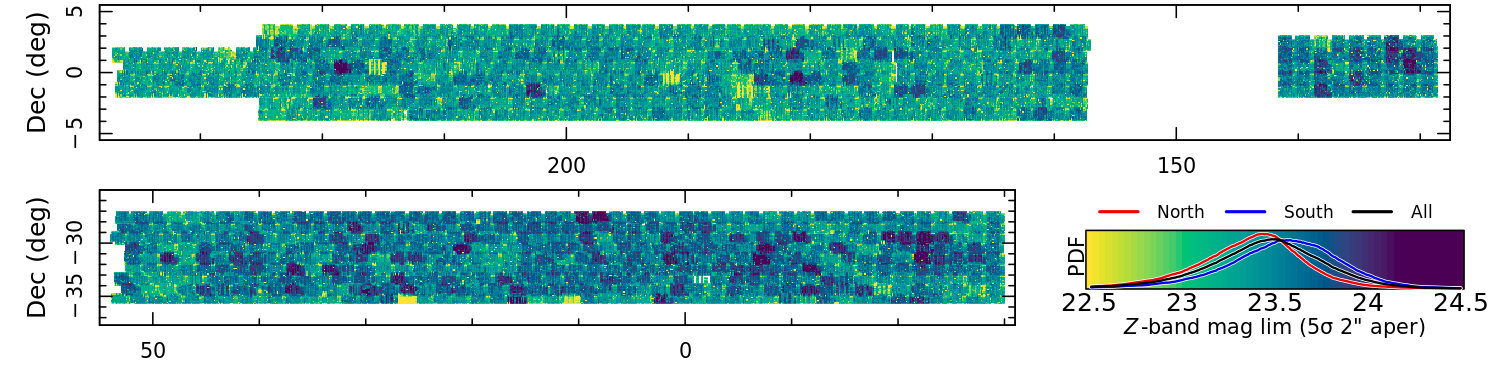}
  \includegraphics[width=1.8\columnwidth]{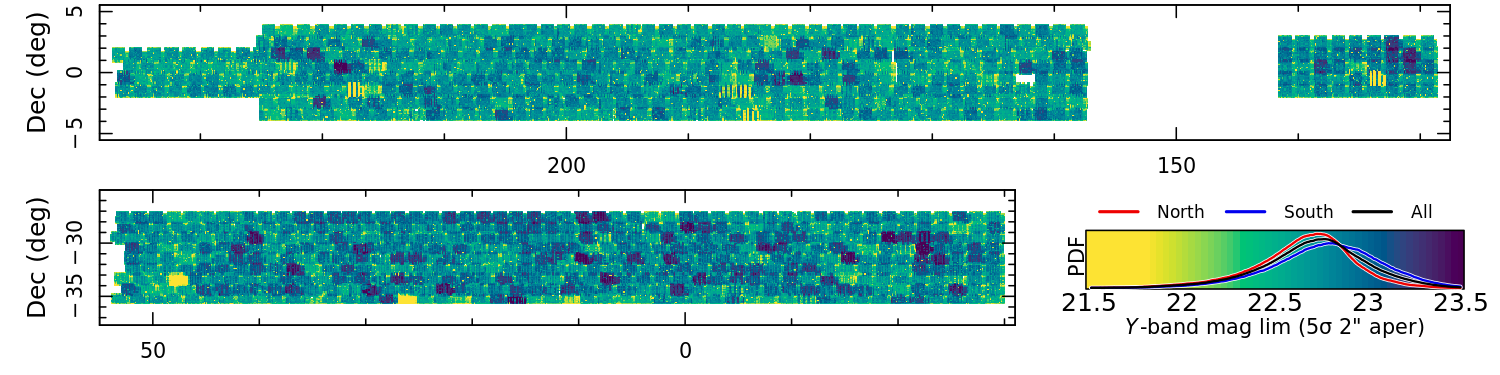}
  \includegraphics[width=1.8\columnwidth]{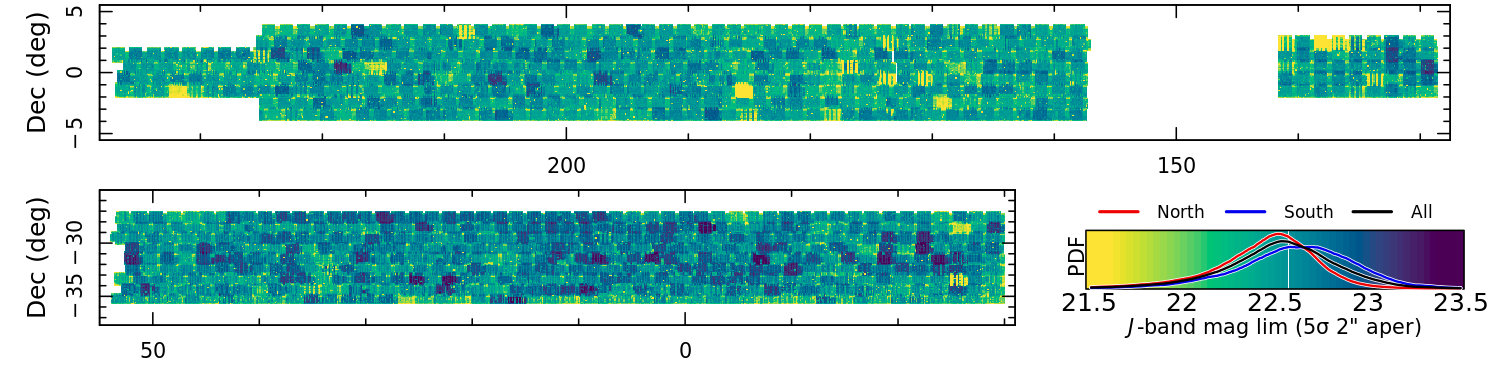}
  \includegraphics[width=1.8\columnwidth]{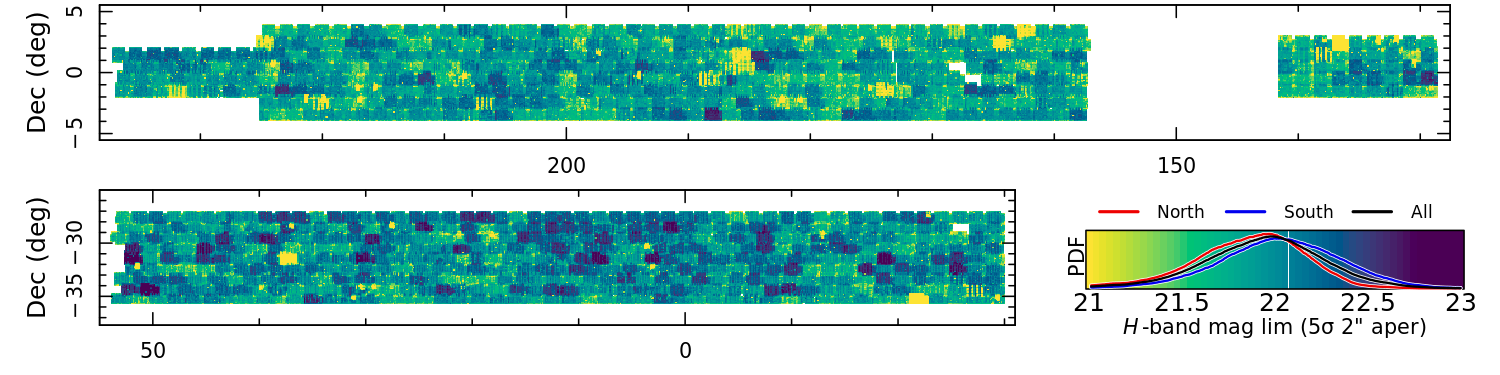}
  \includegraphics[width=1.8\columnwidth]{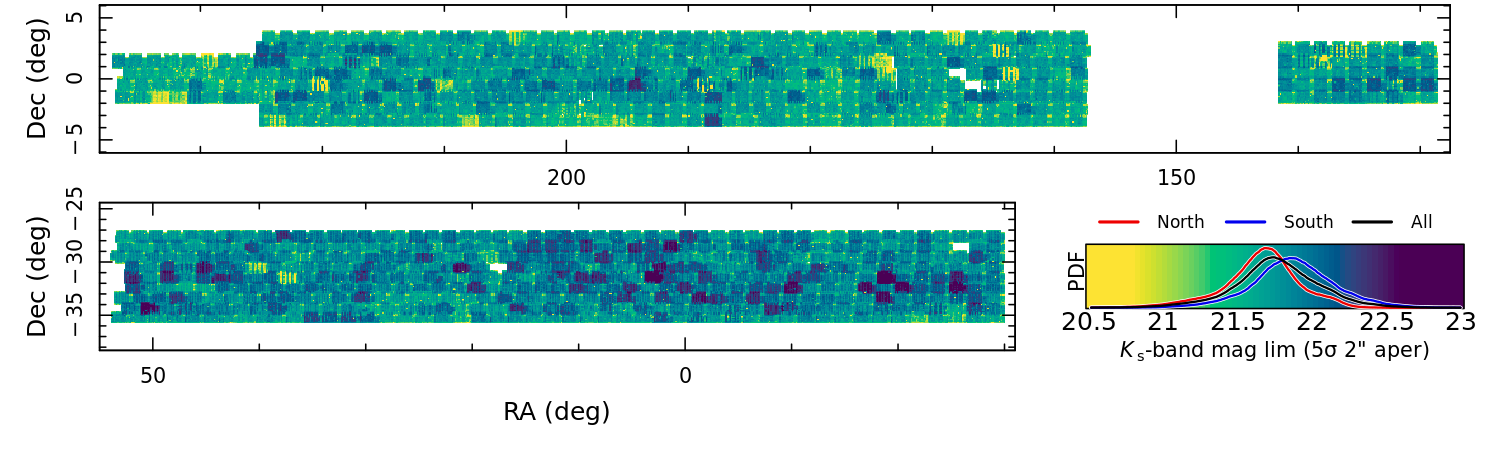}
  \caption{Distribution of effective \nir\ magnitude limits in the \kids\ fields, 
  computed using \lambdar\ with $2\arcsec$ circular apertures.}\label{fig:vistabanddepth}
\end{figure*}

As with Appendix~\ref{sec:vstonskymetrics}, the distribution of magnitude limits and 
PSF sizes on-sky in the NIR bands is a relevant statistics for understanding systematic 
variations of photometric quality over the \kids-\drfive\ survey. Figure~\ref{fig:vistabanddepth} shows the 
distribution of depths in each of the five NIR filters, computed using the method of Appendix~\ref{sec:lambdar} and 
combined as described in Sect.~\ref{sec:vistaquality}. The dither pattern of the \vista\ observations is visible, 
particularly in cases where sequences of observations are particularly shallow (leading to stripes and rectangles in the 
image). 

\begin{figure*}
  \centering
  \includegraphics[width=1.8\columnwidth]{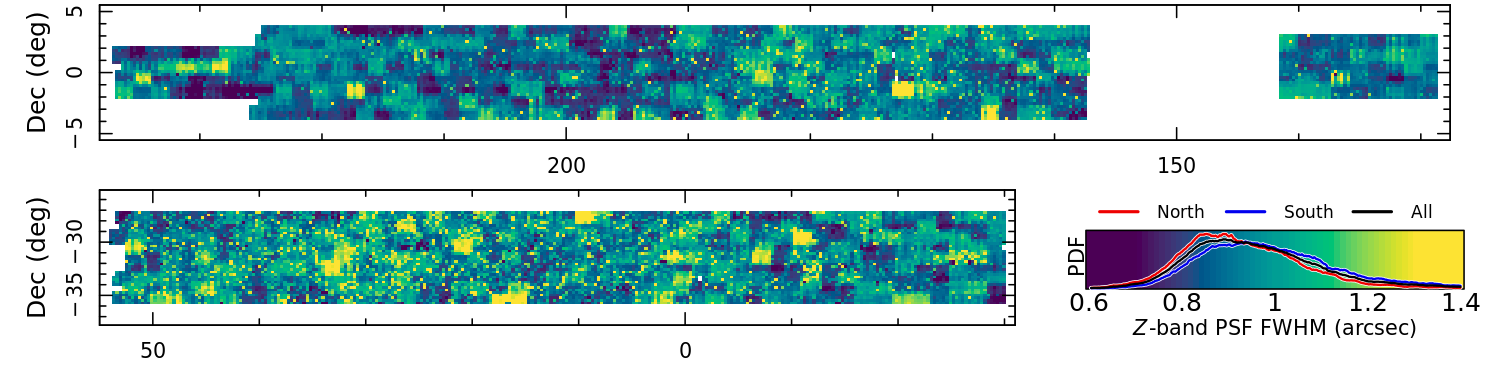}
  \includegraphics[width=1.8\columnwidth]{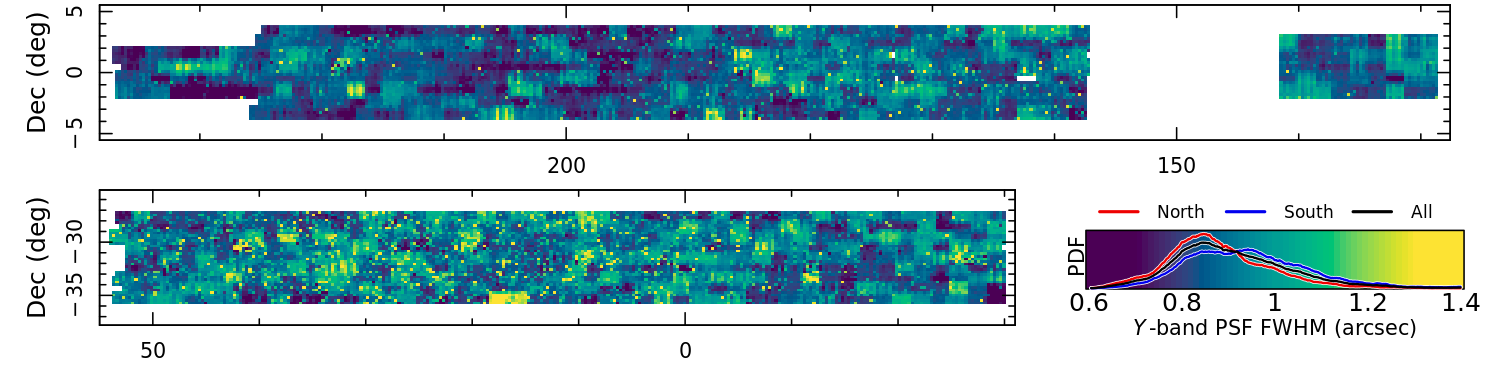}
  \includegraphics[width=1.8\columnwidth]{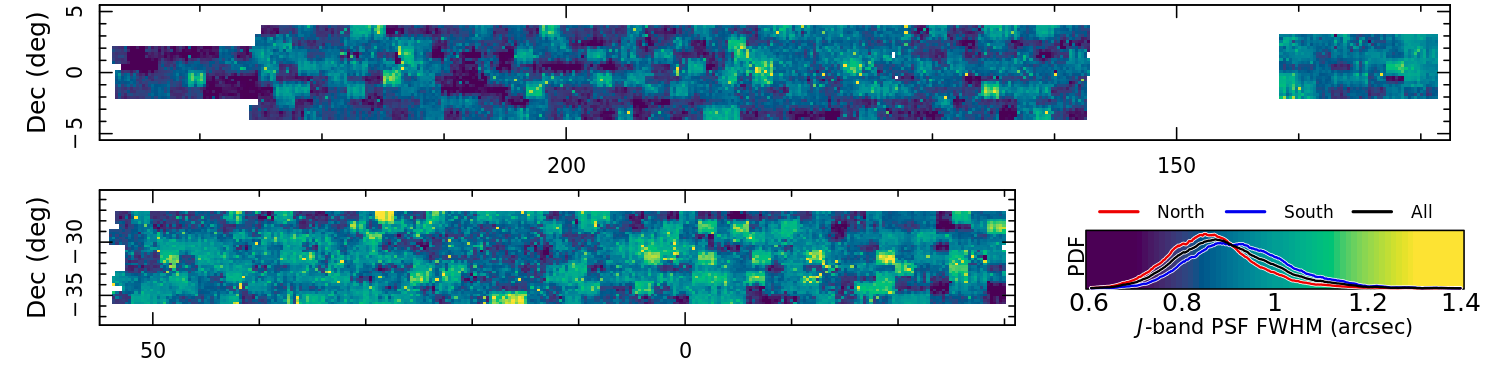}
  \includegraphics[width=1.8\columnwidth]{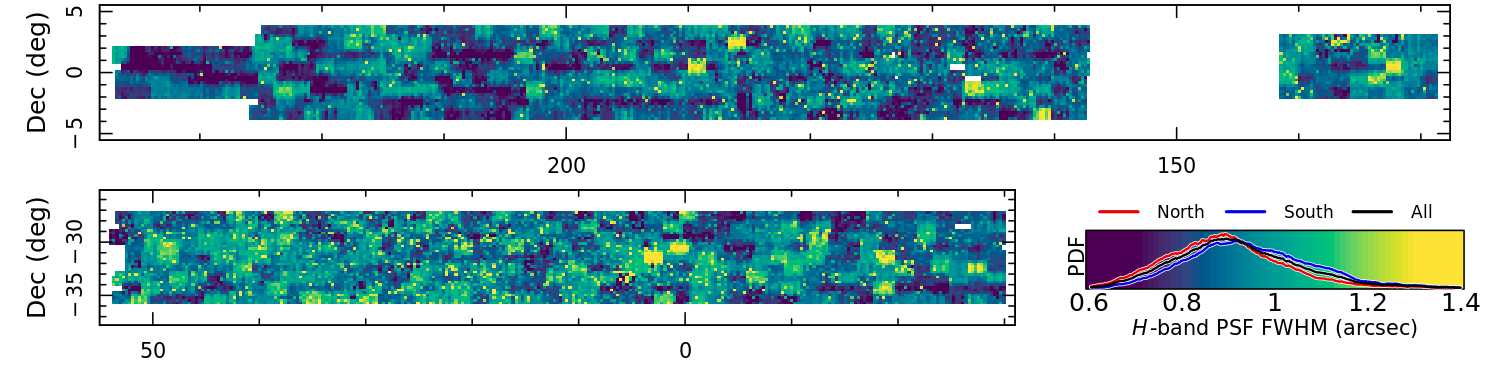}
  \includegraphics[width=1.8\columnwidth]{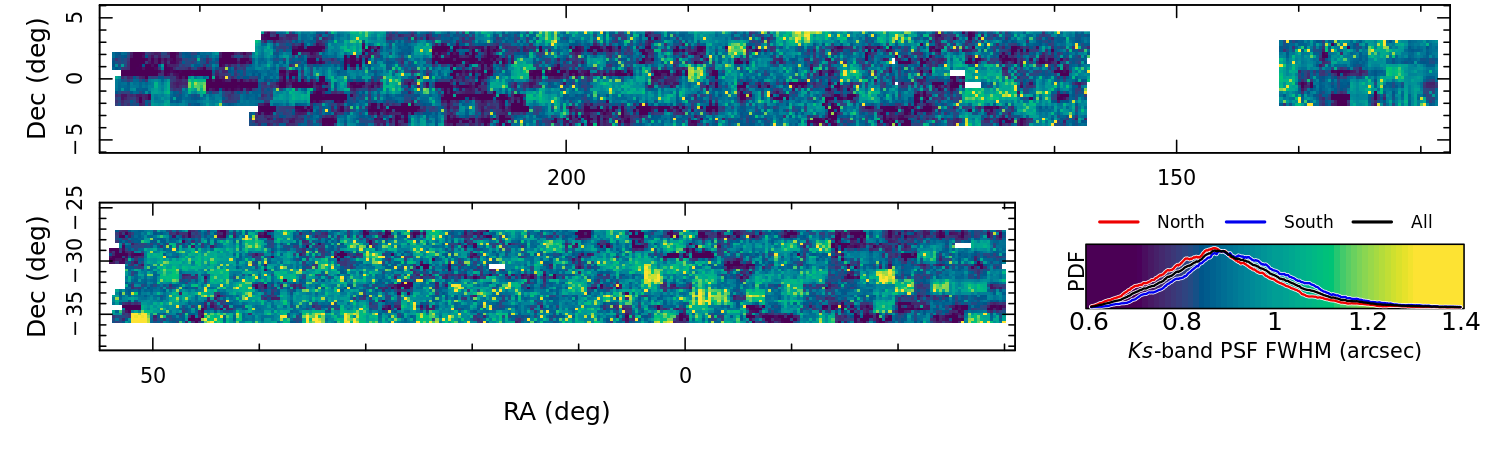}
  \caption{Distribution of mean PSF FWHM sizes (in arcsec) as reported by ESO for each \pawprint\ observation of the
  \kids\ fields. Binning is made is steps of $\{\delta {\rm RA},\delta{\rm Dec}\} = \{0.25/\cos{(\rm Dec}),0.25\}$ deg, as the spacing between 
  adjacent chips in the \pawprint\ is $\sim0.2$ deg. 
  }\label{fig:vistabandpsf}
\end{figure*}

In Fig.~\ref{fig:vistabandpsf} we also show the distribution of \vista\ PSF sizes on-sky, as reported by ESO for each \pawprint. 
PSF sizes have been combined on-sky using the same procedure as for the magnitude limits. Again, the paw-print pattern of the 
\vista\ telescope is visible. We note, however, that the distribution here is not necessarily representative of the information 
contained in the photometry, as the measurement of photometry by \gaap\ is PSF-size dependent. In cases where the PSF is particularly 
poor for an observation, that observation cannot be used in the final flux calculation (as \gaap\ failed to produce a flux measurement in 
that case).  

\FloatBarrier

\section{Masking correlations}\label{sec:maskcorrelations} 

The areas masked by each of our various mask criteria are not unique. An area marked for flagging because of stellar reflections 
in the \rband, for example, are likely highly correlated with areas marked for similar reflections in the \gband, because the same 
stars cause the artefacts in both images. Furthermore, the \kids\ observing strategy exacerbates this effect because each optical filter tiles the 
sky with essentially the same dither pattern. As such, it is worth noting the degree of correlation between areas masked (per pointing) by each 
of the bits in our mask. 

\begin{figure*}
    \centering
    \includegraphics[width=0.8\textwidth]{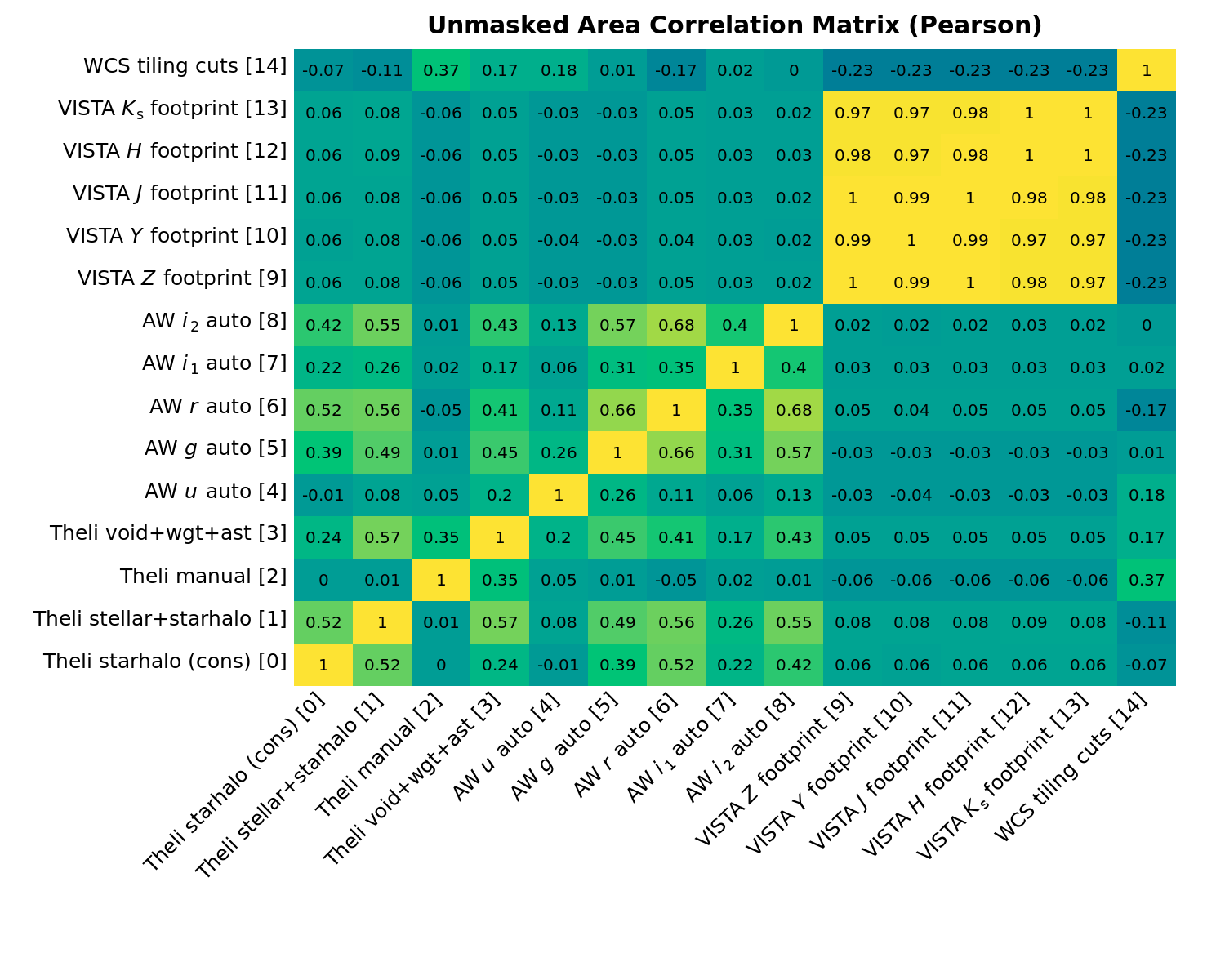}
    \caption{Correlations between masked area per pointing, for each of the individual mask bits. Correlations are computed between the area 
    available per pointing after the application of the relevant mask bits, for the full \kids-\drfive\ survey.}
    \label{fig:mask_correlations}
\end{figure*}

Figure \ref{fig:mask_correlations} shows the correlation matrix of masked area per pointing, for individual masking bits. 
Correlations are computed between the total area masked per pointing under each condition. As such, the correlations are indirect 
(i.e. they do not necessarily mean the same pixels are masked in each case), but nonetheless are indicative of the approximate survey area masked 
in each of the cases. 

The figure shows that the footprint mask in each of the NIR bands are highly correlated, as \viking\ naturally observed a consistent footprint in each filter. 
Slight deviations from unity in the correlations of the NIR footprint masks come from the additional masking that we perform in the NIR (due to 
the PSF size, for example; see Sect.~\ref{sec:strongselection}). Otherwise, there is a surprising lack of correlation in the masks, indicating 
that information contained in each of the masks is somewhat independent. This is particularly interesting in the case of the \rband, where masks 
constructed within \astrowise\ and \theliraw\ are maximally correlated at the level of $\sim 50\%$, despite being constructed from the same raw 
images. This motivates our fiducial choice of masks: individual masks from \astrowise\ and \theliraw\ flag somewhat different artefacts, and so 
the combination of these masks is the most appropriate (and conservative) choice. 

\FloatBarrier

\section{Shape measurement properties}\label{sec:lensing_properties}

\begin{figure*}
 \centering
  \includegraphics[width=1.8\columnwidth]{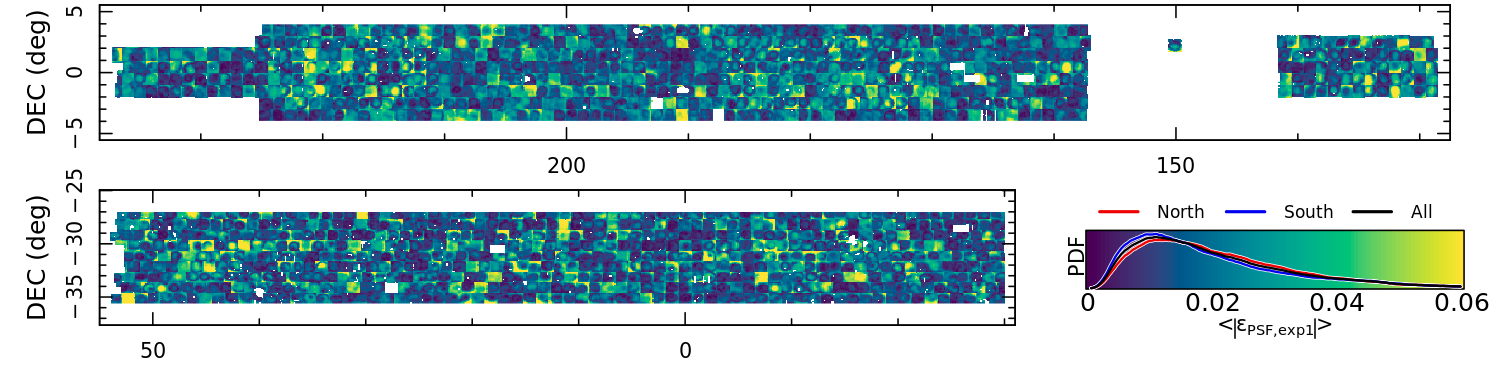}
  \includegraphics[width=1.8\columnwidth]{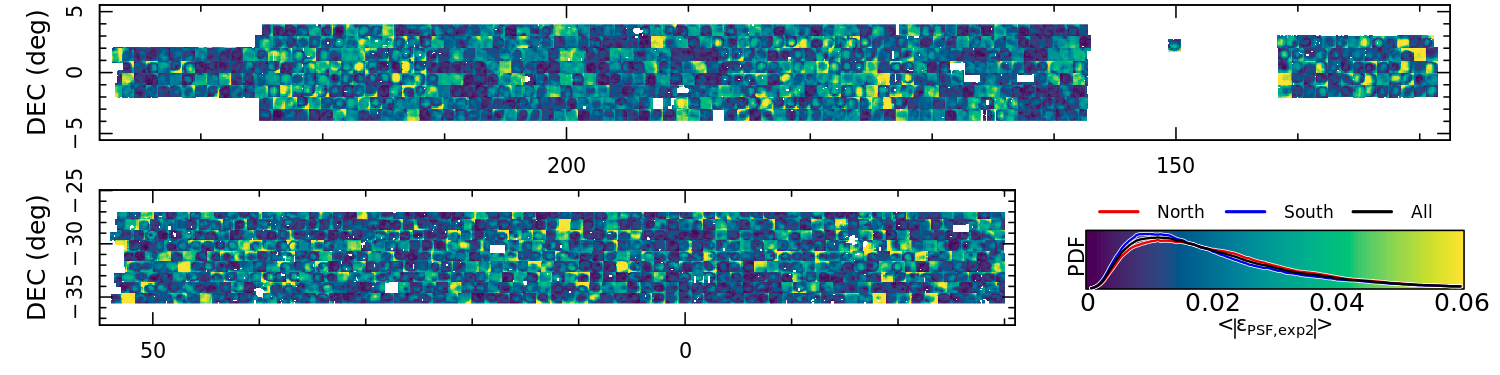}
  \includegraphics[width=1.8\columnwidth]{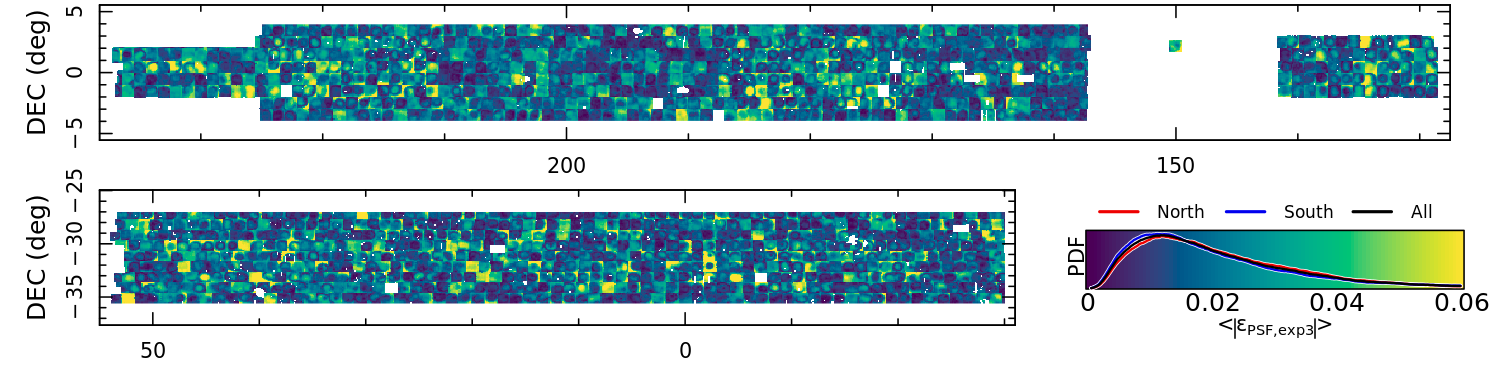}
  \includegraphics[width=1.8\columnwidth]{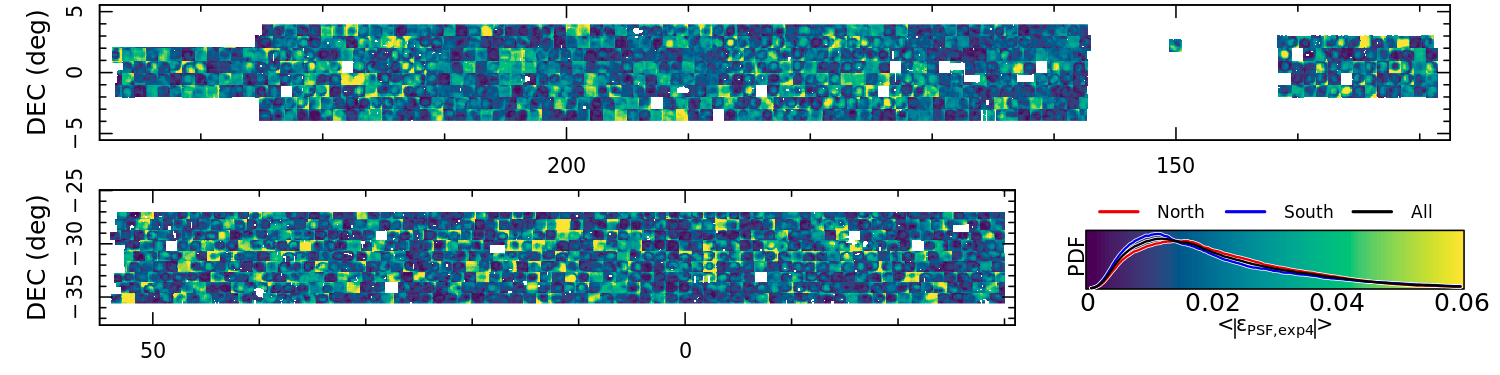}
  \includegraphics[width=1.8\columnwidth]{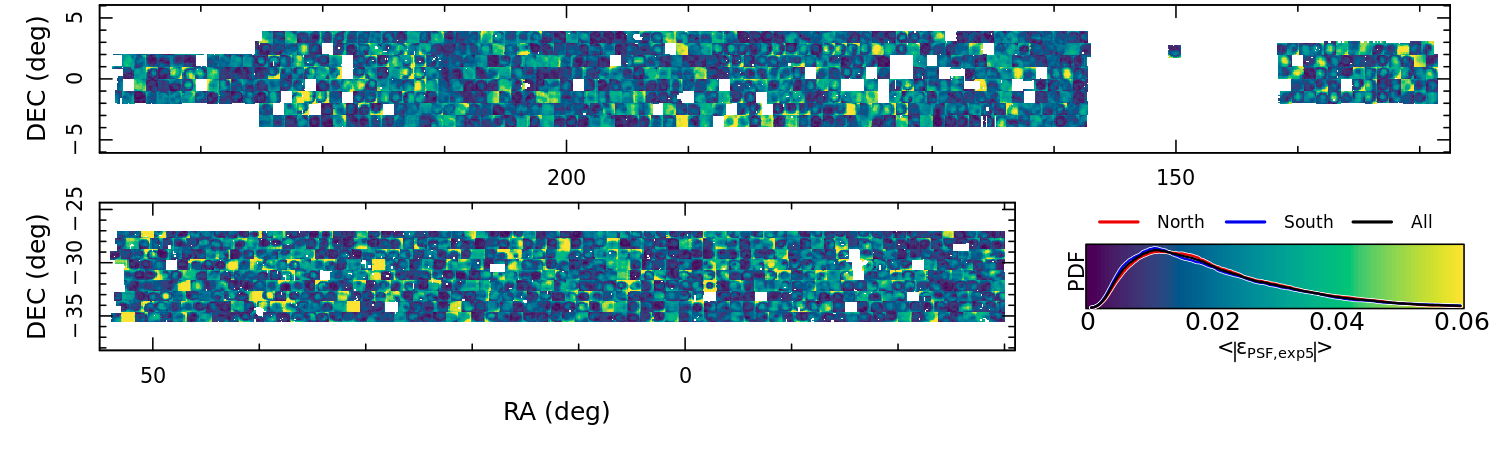}
  \caption{Distribution of PSF ellipticities ($|\epsilon_{\rm PSF}|$) measured in each of the five \rband\ exposures 
  used for shape measurement by \lensfit. The panels show the relative homogeneity of the PSF measured across all exposures, and the amount of variability between exposures of the same pointing. }\label{fig:psf_emod}
\end{figure*}

\begin{figure*}
 \centering
  \includegraphics[width=1.8\columnwidth]{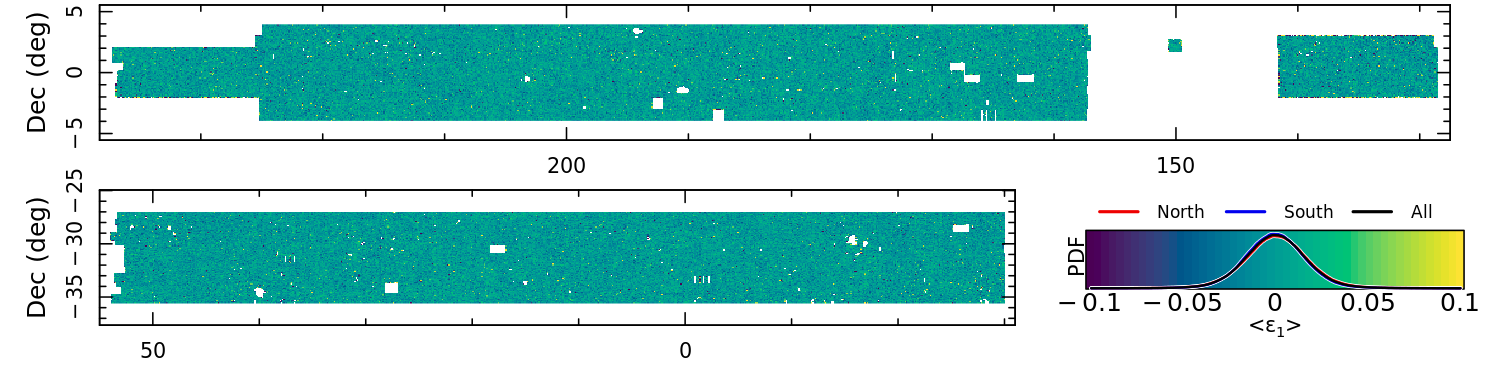}
  \includegraphics[width=1.8\columnwidth]{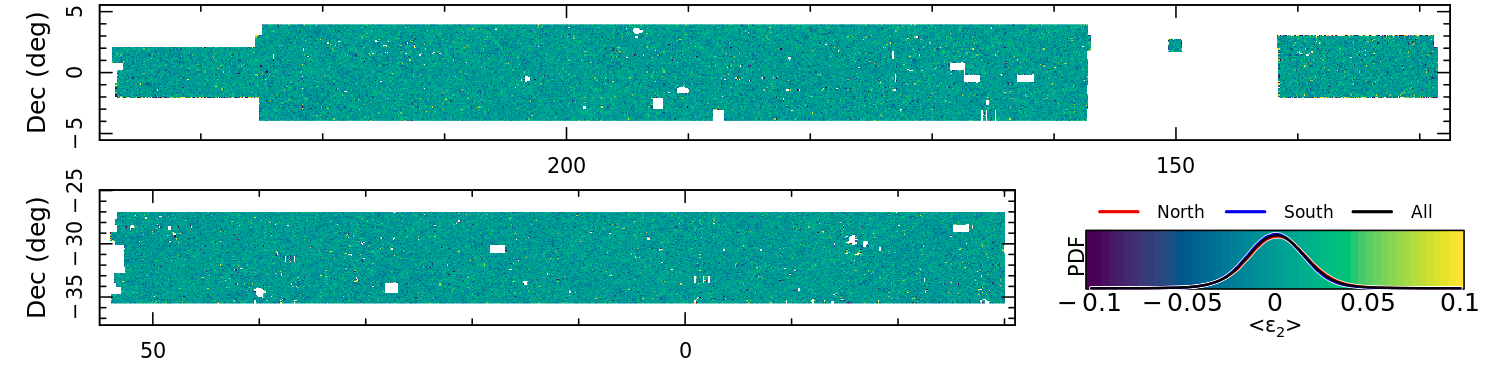}
  \includegraphics[width=1.8\columnwidth]{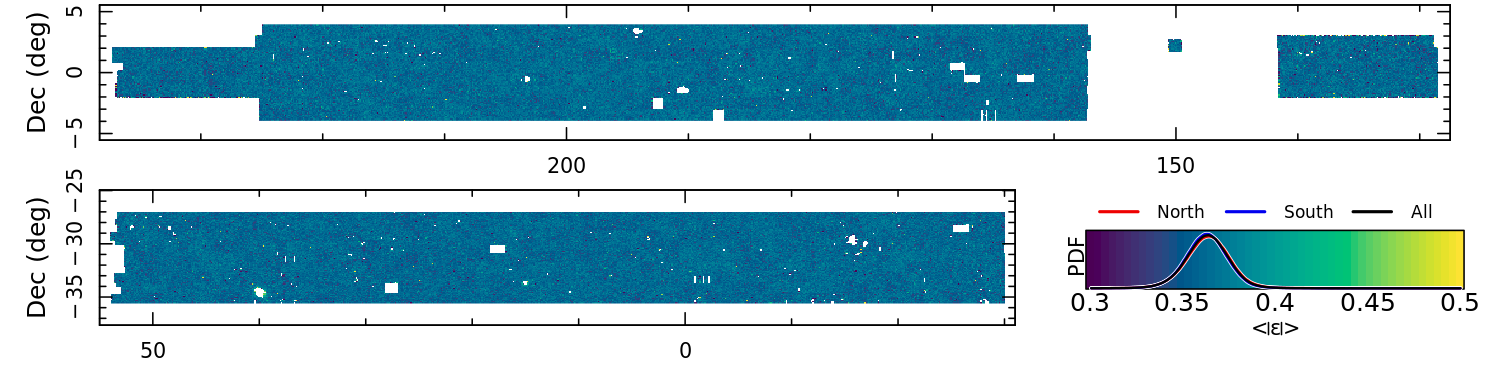}
  \includegraphics[width=1.8\columnwidth]{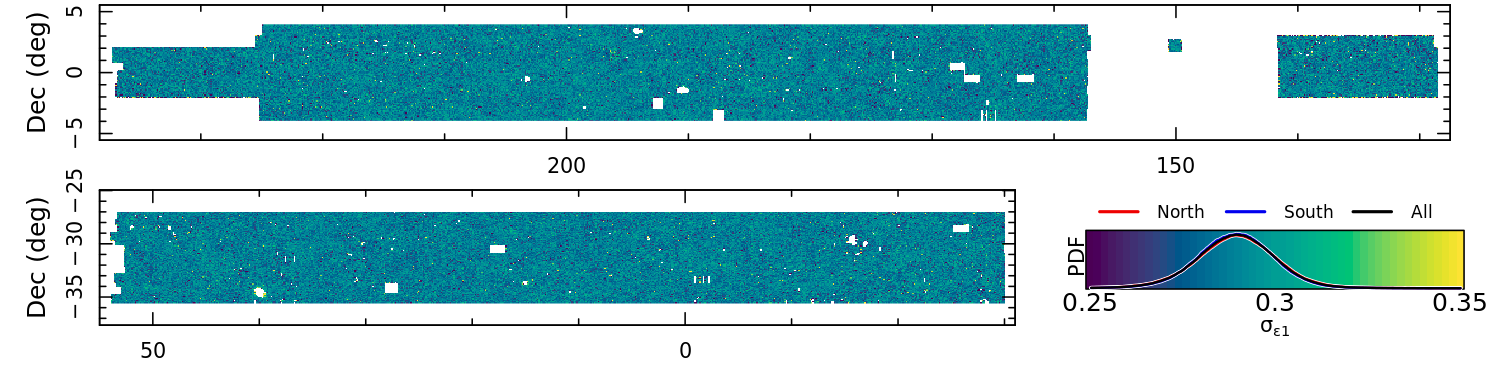}
  \includegraphics[width=1.8\columnwidth]{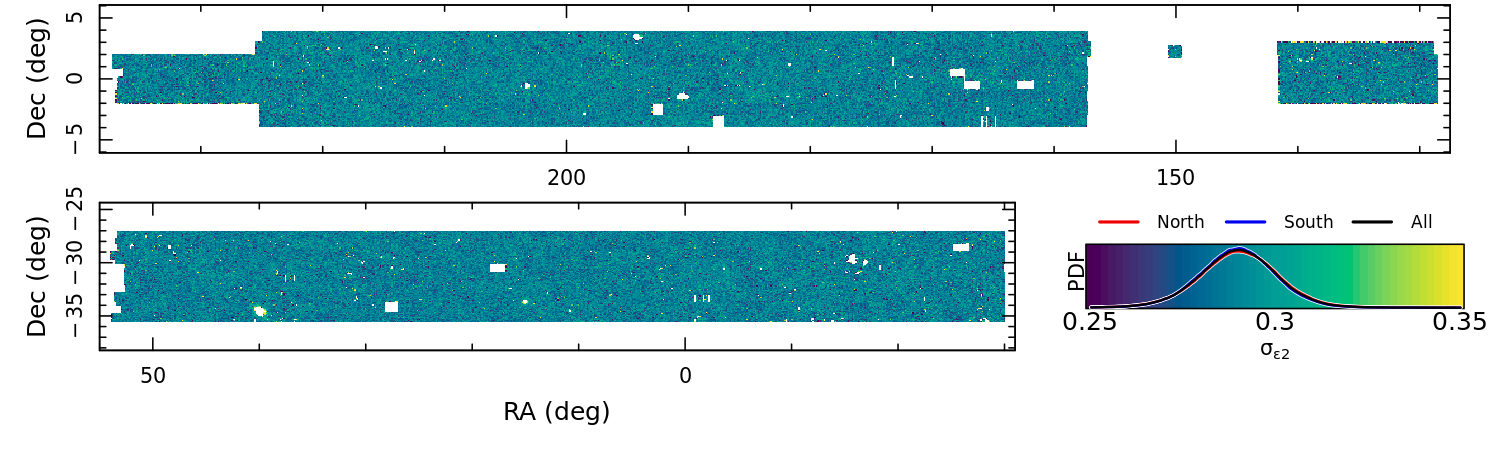}
  \caption{Distribution of $\epsilon_1$, $\epsilon_2$, $|\epsilon|$, 
  $\sigma_{\epsilon,1}$, and $\sigma_{\epsilon,2}$ measured by \lensfit\ after recalibration. All estimates are unweighted. }\label{fig:emod}
\end{figure*}

\begin{figure*}
 \centering
  \includegraphics[width=1.8\columnwidth]{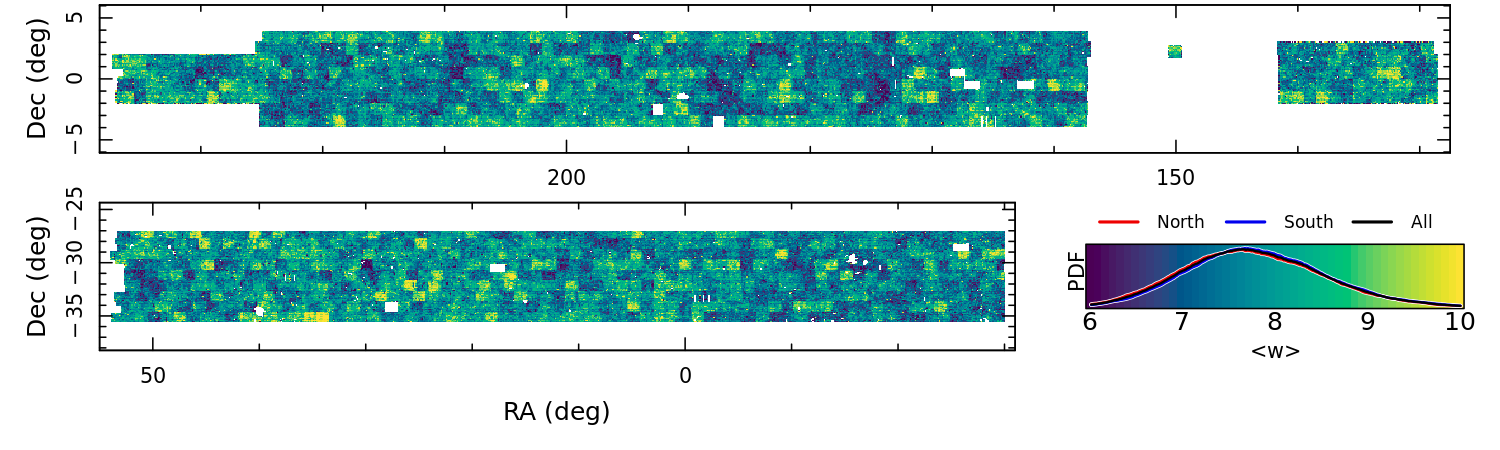}
   \caption{Distribution of recalibrated shape-measurement weights on-sky. }\label{fig:weight}
\end{figure*}

The distribution of shape measurement properties on-sky is a useful diagnostic of 
the level of variability in the lensing sample, after recalibration, which can 
influence modelling and hint at possible systematic effects. Figure \ref{fig:psf_emod} 
shows the on-sky distribution of PSF ellipticity in each of the five exposures (per 
pointing) that contribute to our shape measurement. Figure \ref{fig:emod} shows the 
distribution of measured shapes, and their uncertainty. Finally, Fig. \ref{fig:weight} 
shows the on-sky sum of the recalibrated shape-measurement weight.

\FloatBarrier

\section{Sample selection with weighting} 
\label{sec:weighted_selection}

Here we provide the details of the sample selections used in \kidslegacy, as presented in Table \ref{tab:samplecuts}, 
but computed using weighted values rather than direct source counts, and expressed in terms of the effective number 
density of sources $\eta=A^{-1}(\sum w)^2/(\sum w^2)$, computed using the uncalibrated lensing weight $w$ and the survey area in square-arcminutes $A$. 
This provides an indication of the most significant selections, when considering their influence on the 
weighted source sample that is used for computation of, for example, shear correlation functions. 
The weighted selections are shown in Table \ref{tab:samplecuts_wtd}.

\begin{table*}\centering
  \caption{Statistics of the \kidslegacy\ sample definition, specified in terms of 
           the effective number density. }\label{tab:samplecuts_wtd}
  \begin{tabular}{c|lrrrrrr}
    Selection Name & Formula & $\eta^{\rm r}_{\rm s}$   & $f^{\rm r}_{\rm tot}$    & $\eta^{\rm r}_{\rm s,run}$  & $f^{\rm r}_{\rm run}$ &  $\eta^{\rm k}_{\rm s,run}$ & $f^{\rm k}_{\rm run}$    \\
    \hline
    MASK        & Eq. \ref{eqn:fidmask}   &  -  &    -      &  -  &    -       & 10.94 & $ 100.00 \% $ \\
    FLAG\_GAAP  & Eq. \ref{eqn:gaapmask}  & 0.01 & $ 0.08 \% $& 0.01 & $ 0.08 \% $ & 10.93 & $ 99.92 \% $ \\
    ASTEROIDS   & Eq. \ref{eqn:astmask}   & 0.03 & $ 0.26 \% $& 0.03 & $ 0.26 \% $ & 10.90 & $ 99.66 \% $ \\
    UNMEASURED  & Eq. \ref{eqn:psfmask}   &  -  &    -      &  -  &    -       & 10.90 & $ 99.66 \% $ \\
    BLENDING    & Eq. \ref{eqn:blendmask} & 0.53 & $ 4.82 \% $& 0.52 & $ 4.75 \% $ & 10.39 & $ 95.00 \% $ \\
    FITCLASS    & Eq. \ref{eqn:fitmask}   & 0.97 & $ 8.88 \% $& 0.95 & $ 9.09 \% $ & 9.45 & $ 86.36 \% $ \\
    BINARY      & Eq. \ref{eqn:binary}    & 0.03 & $ 0.30 \% $& 0.00 & $ 0.04 \% $ & 9.44 & $ 86.33 \% $ \\
    MAG\_AUTO   & Eq. \ref{eqn:magmask}   & 0.37 & $ 3.35 \% $& 0.21 & $ 2.19 \% $ & 9.26 & $ 84.60 \% $ \\
    RESOLUTION  & Eq. \ref{eqn:resolution}& 0.82 & $ 7.47 \% $& 0.43 & $ 4.60 \% $ & 8.92 & $ 81.54 \% $ \\
    WEIGHT      & Eq. \ref{eqn:weight}    &  -  &    -      &  -  &    -       & 8.92 & $ 81.53 \% $ \\
    \hline
    Full Legacy Sample     & - & - & - & - & - & 8.92 & $ 81.53 \% $ \\
    Astrometric masking    & - & - & - & - & - & 8.91 & $ 81.48 \% $ \\
    $0.1<z_{\rm B}\leq2.0$ & - & - & - & - & - & 8.81 & $ 80.55 \% $ \\
    $0.1<z_{\rm B}\leq1.2$ & - & - & - & - & - & 7.99 & $ 73.07 \% $ \\
  \end{tabular}
  \tablefoot{ 
   Effective number density is defined as $\eta=A^{-1}(\sum w)^2/(\sum w^2)$, computed using the uncalibrated shape
   measurement weight, $w,$ and the survey area in square-arcmin, $A$.
  Superscript indices specify whether a number refers to sources that are removed from the sample (`r') or kept in the
  sample (`k'). 
  $\eta^{\rm r}_{s}$ is the effective number density of sources (from the full ${\rm MASK}\leq 1$ sample) that are flagged for removal by
  the selection $s$. $f^{\rm r}_{\rm tot}$ is $\eta^{\rm r}_{s}$ expressed as a fraction of the full ${\rm MASK}\leq 1$ sample
  $(\eta^{\rm k}_{\rm MASK})$. $\eta^{\rm r}_{s,{\rm run}}$ is the effective number density of sources from the currently un-flagged population
  that are newly flagged after applying selection $s$ (assuming flags are applied in row-order). $f^{\rm r}_{\rm run}$
  is $\eta^{\rm r}_{s,{\rm run}}$ expressed as a fraction of the currently un-flagged population. $\eta^{\rm k}_{s,{\rm run}}$ is
  the effective number density of sources remaining un-flagged after the flagging of selection $s$. $f^{\rm k}_{\rm run}$
  is $\eta^{\rm k}_{s,{\rm run}}$ expressed as a fraction of the full ${\rm MASK}\leq 1$ sample.}
\end{table*} 
The table shows that the most significant selections, in terms of lensing weight, are the blending, {\tt fitclass}, 
magnitude, and resolution selections. These remove, respectively: $5\%$, $9\%$, $2\%$, and $5\%$ of the available 
effective number density. 

\begin{figure}
  \centering
  \includegraphics[width=\columnwidth]{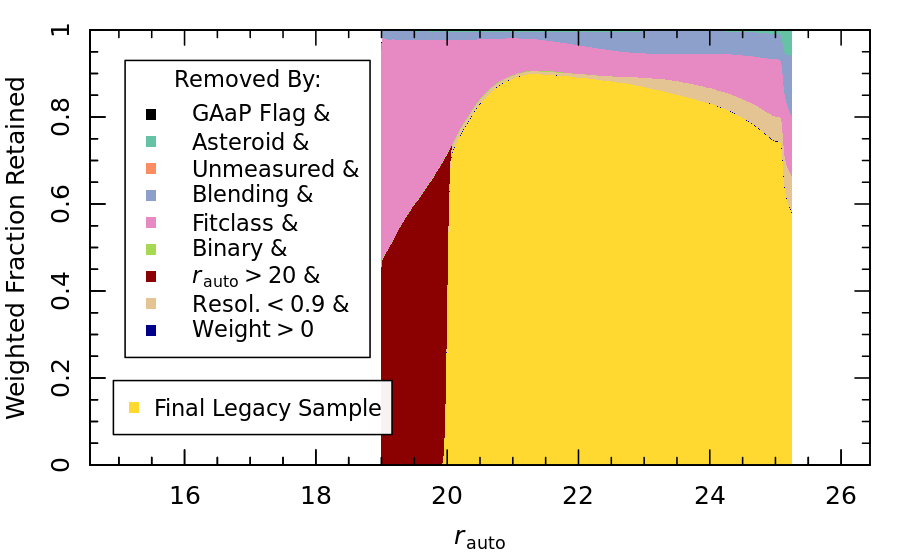}
  \caption{As in Figure \protect \ref{fig:selections}, but now computed using sources weighted by their uncalibrated shape measurement
  weight. The distribution is truncated to the limits of analysed by \lensfit\ ($19\leq r\leq 25.5$), as the 
  weights are not defined outside this region.}\label{fig:weighted_sel}
\end{figure}

Finally, we reproduce the fractional selection of sources as a function of \rband\ magnitude shown in Fig.
\ref{fig:selections}, now computed as a weighted fraction in Fig. \ref{fig:weighted_sel}. The distribution is truncated 
to the limits that are analysed by \lensfit\ (i.e. $19\leq r\leq 25.5$).

\FloatBarrier

\section{ESO data products}\label{sec:eso_catalogues}

The catalogues and images that are available within the ESO database have changed between \drfour\ and \drfive, 
as the format of the data has changed. In particular, the addition of the second-pass \iband\ necessitates a 
modification to the catalogue (and metadata) formats. The description of the columns contained within the 
ESO multi-band catalogues is provided in Table~\ref{tab:eso_colourcats}, the columns contained within the single-band 
catalogues are provided in Table \ref{tab:eso_singlecats}. Columns that have changed meaning between 
\drfour\ and \drfive\ are specifically highlighted in Table~\ref{tab:eso_changes}. Details of the image headers 
that are released are provided in the Table \ref{tab:eso_headers}. 

%\clearpage
\onecolumn

\begin{center}
\begin{longtable}{llll}
\caption{\label{tab:eso_colourcats} Columns provided in the ten-band catalogue.}\\
\hline\hline
Label & Unit & Format & Description \\
\hline
\endfirsthead

\multicolumn{4}{c}{\tablename\ \thetable{} continued.}\\
\hline\hline
Label & Unit & Format & Description \\
\hline
\endhead

\hline
\multicolumn{4}{r}{{Continued on next page}}\\
\endfoot

\hline
\endlastfoot
\\
\multicolumn{4}{c}{Identifiers per source and Pointing on-sky:}\\
\\
{\tt ID }&   & 30A & ESO ID  \\
{\tt KIDS_TILE }&   & 16A & Name of the pointing in {\sc Astro-WISE} convention \\
{\tt THELI_NAME }&   & 16A & Name of the pointing in {\sc theli} convention \\
{\tt SeqNr }&   & 1J & Running object number within the catalogue         \\
{\tt SLID }&   & 1J & {\sc Astro-WISE} Source list ID           \\
{\tt SID }&   & 1J & {\sc Astro-WISE} Source ID within the source list        \\
\\
\multicolumn{4}{c}{Parameters derived from the {\sc theli} $r$-band detection image:}\\
\\
{\tt FLUX_AUTO }& Jy  & 1E & \rband\ flux             \\
{\tt FLUXERR_AUTO }& Jy  & 1E & Error on  {\tt FLUX_AUTO}            \\
{\tt MAG_AUTO }& mag  & 1E & \rband\ magnitude            \\
{\tt MAGERR_AUTO }& mag  & 1E & Error on  {\tt MAG_AUTO}           \\
{\tt KRON_RADIUS }& pixel& 1E & Scaling radius of the ellipse for magnitude measurements\\
{\tt BackGr }& Jy  & 1E & Background counts at centroid position         \\
{\tt Level }& Jy  & 1E & Detection threshold above background         \\
{\tt MU_THRESHOLD }& mag $\cdot$ arcsec$^{-2}$& 1E &Detection threshold above background\\
{\tt MaxVal }& Jy  & 1E & Peak flux above background          \\
{\tt MU_MAX }& mag $\cdot$ arcsec$^{-2}$  & 1E & Peak surface brightness above background\\
{\tt ISOAREA_WORLD }& deg$^2$  & 1E & Isophotal area above analysis threshold         \\
{\tt Xpos }& pixel  & 1E & Centroid x position in the {\sc theli} image       \\
{\tt Ypos }& pixel  & 1E & Centroid y position in the {\sc theli} image        \\
{\tt RAJ2000 }& deg  & 1D & Centroid sky position right ascension (J2000)        \\
{\tt DECJ2000 }& deg  & 1D & Centroid sky position declination (J2000)          \\
{\tt A_WORLD }& deg  & 1E & Profile RMS along major axis          \\
{\tt B_WORLD }& deg  & 1E & Profile RMS along minor axis          \\
{\tt THETA_J2000 }& deg  & 1E & Position angle (west of north)          \\
{\tt THETA_WORLD }& deg  & 1E & Position angle (Counterclockwise from world x-axis)   \\
{\tt ERRA_WORLD }& deg  & 1E & World RMS position error along major axis        \\
{\tt ERRB_WORLD }& deg  & 1E & World RMS position error along minor axis        \\
{\tt ERRTHETA_J2000 }& deg  & 1E & Error on  {\tt THETA_J2000}            \\
{\tt ERRTHETA_WORLD }& deg  & 1E & Error on  {\tt THETA_WORLD}            \\
{\tt FWHM_IMAGE }& pixel  & 1E & FWHM assuming a Gaussian object profile         \\
{\tt FWHM_WORLD }& deg  & 1E & FWHM assuming a Gaussian object profile         \\
{\tt Flag }&   & 1I & \sourceextractor\ extraction flags            \\
{\tt FLUX_RADIUS }& pixel  & 1E & Half-light radius             \\
{\tt CLASS_STAR }&   & 1E & Star-galaxy classifier             \\
{\tt MAG_ISO }& mag  & 1E & \rband\ isophotal Magnitude      \\
{\tt MAGERR_ISO }& mag  & 1E & Error on  {\tt MAG_ISO}        \\
{\tt FLUX_ISO }& Jy  & 1E & \rband\ isophotal Flux           \\
{\tt FLUXERR_ISO }& Jy  & 1E & Error on  {\tt FLUX_ISO}            \\
{\tt MAG_ISOCOR }& mag  & 1E & \rband\ corrected Isophotal Magnitude      \\
{\tt MAGERR_ISOCOR }& mag  & 1E & Error on  {\tt MAG_ISOCOR}        \\
{\tt FLUX_ISOCOR }& Jy  & 1E & \rband\ corrected Isophotal Flux       \\
{\tt FLUXERR_ISOCOR }& Jy  & 1E & Error on  {\tt FLUX_ISOCOR }        \\
{\tt NIMAFLAGS_ISO }&   & 1I & Number of flagged pixels over the isophotal profile    \\
{\tt IMAFLAGS_ISO }&   & 1I & FLAG-image flags ORed over the isophotal profile      \\
{\tt XMIN_IMAGE }& pixel  & 1I & Minimum x-coordinate among detected pixels     \\
{\tt YMIN_IMAGE }& pixel  & 1I & Minimum y-coordinate among detected pixels     \\
{\tt XMAX_IMAGE }& pixel  & 1I & Maximum x-coordinate among detected pixels      \\
{\tt YMAX_IMAGE }& pixel  & 1I & Maximum y-coordinate among detected pixels      \\
{\tt X_WORLD }& deg  & 1D & Barycentre position along world x axis         \\
{\tt Y_WORLD }& deg  & 1D & Barycentre position along world y axis         \\
{\tt X2_WORLD }& deg$^2$  & 1E & Variance of position along  {\tt X_WORLD}  (alpha)      \\
{\tt Y2_WORLD }& deg$^2$  & 1E & Variance of position along  {\tt Y_WORLD}  (delta)      \\
{\tt XY_WORLD }& deg$^2$  & 1E & Covariance of position  {\tt X_WORLD,Y_WORLD}       \\
{\tt ERRX2_WORLD }& deg$^2$  & 1E & Error on  {\tt X2_WORLD }           \\
{\tt ERRY2_WORLD }& deg$^2$  & 1E & Error on  {\tt Y2_WORLD }           \\
{\tt ERRXY_WORLD }& deg$^2$  & 1E & Error on  {\tt XY_WORLD }           \\
{\tt CXX_WORLD }& deg$^{-2}$  & 1E & \sourceextractor\ Cxx object ellipse parameter         \\
{\tt CYY_WORLD }& deg$^{-2}$  & 1E & \sourceextractor\ Cyy object ellipse parameter         \\
{\tt CXY_WORLD }& deg$^{-2}$  & 1E & \sourceextractor\ Cxy object ellipse parameter         \\
{\tt ERRCXX_WORLD }& deg$^{-2}$  & 1E & Error on  {\tt CXX_WORLD}           \\
{\tt ERRCYY_WORLD }& deg$^{-2}$  & 1E & Error on  {\tt CYY_WORLD}           \\
{\tt ERRCXY_WORLD }& deg$^{-2}$  & 1E & Error on  {\tt CXY_WORLD}           \\
{\tt A_IMAGE }& pixel  & 1D & Profile RMS along x-axis          \\
{\tt B_IMAGE }& pixel  & 1D & Profile RMS along y-axis          \\
{\tt ERRA_IMAGE }& pixel  & 1E & Error on  {\tt A_IMAGE}            \\
{\tt ERRB_IMAGE }& pixel  & 1E & Error on  {\tt B_IMAGE}            \\
{\tt S_ELLIPTICITY }&   & 1E & \sourceextractor\ ellipticity  {\tt (1-B_IMAGE/A_IMAGE)}           \\
{\tt S_ELONGATION }&   & 1E & \sourceextractor\ elongation  {\tt (A_IMAGE/B_IMAGE) }          \\
{\tt MAG_APER_4 }& mag  & 1E & \rband\ magnitude within a circular aperture of 4 pixels \\
{\tt MAGERR_APER_4 }& mag  & 1E & Error on  {\tt MAG_APER_4 }          \\
{\tt FLUX_APER_4 }& Jy  & 1E & \rband\ flux within a circular aperture of 4 pixels  \\
{\tt FLUXERR_APER_4 }& Jy  & 1E & Error on  {\tt FLUX_APER_4 }        \\
\multicolumn{4}{l}{\ldots}\\
\multicolumn{4}{l}{Similarly for radii 6, 8, 10, 14, 20, 30, 40, 60 pixels, up to}\\
\multicolumn{4}{l}{\ldots}\\
{\tt MAG_APER_100 }& mag  & 1E & \rband\ magnitude within a circular aperture of 100 pixels \\
{\tt MAGERR_APER_100 }& mag  & 1E & Error on  {\tt MAG_APER_100 }          \\
{\tt FLUX_APER_100 }& Jy  & 1E & \rband\ flux within a circular aperture of 100 pixels  \\
{\tt FLUXERR_APER_100 }& Jy  & 1E & Error on  {\tt FLUX_APER_100 }         \\
{\tt ISO0 }& pixel$^2$  & 1I & Isophotal area at level zero         \\
{\tt ISO1 }& pixel$^2$  & 1I & Isophotal area at level one       \\
{\tt ISO2 }& pixel$^2$  & 1I & Isophotal area at level two        \\
{\tt ISO3 }& pixel$^2$  & 1I & Isophotal area at level three     \\
{\tt ISO4 }& pixel$^2$  & 1I & Isophotal area at level four      \\
{\tt ISO5 }& pixel$^2$  & 1I & Isophotal area at level five       \\
{\tt ISO6 }& pixel$^2$  & 1I & Isophotal area at level six     \\
{\tt ISO7 }& pixel$^2$  & 1I & Isophotal area at level seven   \\
{\tt ALPHA_J2000 }& deg  & 1D & \sourceextractor\ centroid right ascension (J2000)\\
{\tt DELTA_J2000 }& deg  & 1D & \sourceextractor\ centroid declination (J2000) \\
{\tt SG2DPHOT }&   & 1I & 2DPhot StarGalaxy classifier (1 for high confidence star)    \\
{\tt HTM }&   & 1J & Hierarchical Triangular Mesh (level 25)          \\
{\tt FIELD_POS }&   & 1I & Reference number to field parameters          \\
\\
\multicolumn{4}{c}{List-driven \gaap\ photometry on the {\sc Astro-WISE} co-added KiDS images and the pawprint VIKING images:}\\
\\
{\tt Agaper_0p7 }& arcsec  & 1E & Major axis of \gaap\ aperture  {\tt MIN_APER}  0\farcs7       \\
{\tt Bgaper_0p7 }& arcsec  & 1E & Minor axis of \gaap\ aperture  {\tt MIN_APER}  0\farcs7       \\
{\tt Agaper_1p0 }& arcsec  & 1E & Major axis of \gaap\ aperture  {\tt MIN_APER}  1\farcs0       \\
{\tt Bgaper_1p0 }& arcsec  & 1E & Minor axis of \gaap\ aperture  {\tt MIN_APER}  1\farcs0       \\
{\tt PAgaap }& deg  & 1E & Position angle of major axis of \gaap\ aperture (north of west)\\
\\
\multicolumn{4}{l}{$\qquad\qquad$and then for each band  $x \in \{u,g,r,i_1,i_2,Z,Y,J,H,K_{\rm s}\}$:\tablefootmark{a}}\\
\\
{\tt FLUX_GAAP_0p7_x }& Jy  & 1E & \gaap\ flux in band  $x$  with  {\tt MIN_APER} =0\farcs7         \\
{\tt FLUXERR_GAAP_0p7_x }& Jy  & 1E & Error on  {\tt FLUX_GAAP_0p7_x }    \\
{\tt MAG_GAAP_0p7_x }& mag  & 1E &  $x$-band \gaap\ magnitude with  {\tt MIN_APER} =0\farcs7        \\
{\tt MAGERR_GAAP_0p7_x }& mag  & 1E & Error on  {\tt MAG_GAAP_0p7_x} \\
{\tt FLAG_GAAP_0p7_x }&   & 1J & \gaap\ Flag for  $x$-band photometry with  {\tt MIN_APER} =0\farcs7 \\
{\tt FLUX_GAAP_1p0_x }& Jy  & 1E & \gaap\ flux in band  x  with  {\tt MIN_APER} =1\farcs0        \\
{\tt FLUXERR_GAAP_1p0_x }& Jy  & 1E & Error on  FLUX_GAAP_1p0_x    \\
{\tt MAG_GAAP_1p0_x }& mag  & 1E &  $x$-band \gaap\ magnitude with  {\tt MIN_APER} =1\farcs0     \\
{\tt MAGERR_GAAP_1p0_x }& mag  & 1E & Error on  MAG_GAAP_1p0_x \\
{\tt FLAG_GAAP_1p0_x }&   & 1J & \gaap\ Flag for  $x$-band photometry with  {\tt MIN_APER} =1\farcs0  \\
\\
\multicolumn{4}{c}{Optimal-aperture \gaap\ ten-band\ photometry including interstellar extinction corrections}\\
\\
{\tt Agaper }& arcsec  & 1E & Major axis of \gaap\ aperture for optimal  {\tt MIN_APER}  \\
{\tt Bgaper }& arcsec  & 1E & Minor axis of \gaap\ aperture for optimal  {\tt MIN_APER}  \\
\\
\multicolumn{4}{c}{and then for each band $x \in \{u,g,r,i_1,i_2,Z,Y,J,H,K_{\rm s}\}$:\tablefootmark{a}}\\
\\
{\tt EXTINCTION_x }& mag  & 1E & Galactic extinction in band $x$         \\
{\tt MAG_GAAP_x }& mag  & 1E &  Corrected $x$-band \gaap\ magnitude with optimal  {\tt MIN_APER}\\
{\tt MAGERR_GAAP_x }& mag  & 1E & Error on  {\tt MAG_GAAP_x} \\
{\tt FLUX_GAAP_x }& Jy  & 1E &  $x$-band \gaap\ flux with optimal  {\tt MIN_APER}           \\
{\tt FLUXERR_GAAP_x }& Jy  & 1E & Error on  {\tt FLUX_GAAP_x} \\
{\tt FLAG_GAAP_x }&   & 1I & \gaap\ Flag for  $x$-band photometry with optimal  {\tt MIN_APER} \\
{\tt MAG_LIM_x  }& mag   & 1E &  $x$-band limiting magnitude for optimal  {\tt MIN_APER} \\
\\
\multicolumn{4}{c}{Ten-band\ photometric redshifts (BPZ):}\\
\\
{\tt Z_B}&   & 1D & Ten-band\ BPZ redshift estimate (posterior mode)     \\
{\tt Z_B_MIN }&   & 1D & Lower bound of the 68\% confidence interval of  {\tt Z_B}      \\
{\tt Z_B_MAX }&   & 1D & Upper bound of the 68\% confidence interval of  {\tt Z_B}      \\
{\tt T_B }&   & 1D & Spectral type corresponding to  {\tt Z_B}          \\
{\tt ODDS }&   & 1D & Empirical ODDS of  {\tt Z_B}            \\
{\tt Z_ML }&   & 1D & Ten-band\ BPZ maximum likelihood redshift          \\
{\tt T_ML }&   & 1D & Spectral type corresponding to  {\tt Z_ML} \tablefootmark{b}         \\
{\tt CHI_SQUARED_BPZ }&   & 1D & chi squared value associated with  {\tt Z_B}         \\
{\tt M_0 }& mag  & 1D & Reference magnitude for BPZ prior          \\
{\tt BPZ_FILT }&   & 1J & filters with good photometry (BPZ)          \\
{\tt NBPZ_FILT }&   & 1J & number of filters with good photometry (BPZ)        \\
{\tt BPZ_NONDETFILT }&   & 1J & filters with faint photometry (BPZ)          \\
{\tt NBPZ_NONDETFILT }&   & 1J & number of filters with faint photometry (BPZ)        \\
{\tt BPZ_FLAGFILT }&   & 1J & flagged filters (BPZ)            \\
{\tt NBPZ_FLAGFILT }&   & 1J & number of flagged filters (BPZ) \\
{\tt SG_FLAG }&   & 1E & Star/Gal Classifier\\
{\tt MASK }&   & 1J & Ten-band\ mask information \tablefootmark{c}           \\
\\
\multicolumn{4}{c}{PSF estimates (\emph{lens}fit)}\\
\\
{\tt PSF_e1 }&   & 1E & mean ellipticity of PSF, component one            \\
{\tt PSF_e2 }&   & 1E & mean ellipticity of PSF, component two            \\
{\tt PSF_Strehl_ratio }&   & 1E & Pseudo-Strehl ratio of PSF             \\
{\tt PSF_Q11 }&   & 1E & model PSF moment Q11             \\
{\tt PSF_Q22 }&   & 1E & model PSF moment Q22             \\
{\tt PSF_Q21 }&   & 1E & model PSF moment Q21             \\
\\
\multicolumn{4}{c}{Stellar Mass estimates (LePhare)\tablefootmark{d}}\\
\\
\multicolumn{4}{c}{for each band $x \in \{u,g,r,i_1,i_2,Z,Y,J,H,K_{\rm s}\}$:}\\
\\
{\tt MAGABS_GAAP_x}&   & 1D & rest-frame x-band magnitude  \\
{\tt KCOR_x}&   & 1D & x-band k-correction  \\
\\
{\tt LUM_GAAP_r_bestfit}&   & 1D & r-band luminosity  \\
{\tt mstar_bestfit}&   & 1D & stellar mass  \\
{\tt mstar_med}&   & 1D & posterior median stellar mass  \\
{\tt mstar_lower}&   & 1D & posterior $16^{\rm th}$ percentile stellar mass  \\
{\tt mstar_upper}&   & 1D & posterior $84^{\rm th}$ percentile stellar mass  \\
{\tt sfr_bestfit}&   & 1D & star-formation rate  \\
{\tt sfr_med}&   & 1D & posterior median star-formation rate  \\
{\tt sfr_lower}&   & 1D & posterior $16^{\rm th}$ percentile star-formation  \\
{\tt sfr_upper}&   & 1D & posterior $84^{\rm th}$ percentile star-formation \\
\\
\multicolumn{4}{c}{Astrometric flagging parameters}\\
{\tt max_dRADec_AW}& deg  & 1D & Max separation between THELI and AW coadd  \\
{\tt max_dRADec_THELI}& deg  & 1D & Max internal separation exposure centroid and stack  \\
{\tt ast_flag}&   & 1B & Fiducial astrometric accuracy selection \\
\\
\multicolumn{4}{c}{Dereddened colours based on optimal-aperture \gaap\ photometry}\\
\\
{\tt COLOUR_GAAP_u_g }& mag  & 1E & $u-g$ colour index  (dereddened)          \\
{\tt COLOUR_GAAP_g_r }& mag  & 1E & $g-r$ colour index  (dereddened)          \\
{\tt COLOUR_GAAP_r_i1 }& mag  & 1E & $r-i_1$ colour index  (dereddened)        \\
{\tt COLOUR_GAAP_r_i2 }& mag  & 1E & $r-i_2$ colour index  (dereddened) \\
{\tt COLOUR_GAAP_i1_Z }& mag  & 1E & $i_1-Z$ colour index   (dereddened)  \\
{\tt COLOUR_GAAP_i2_Z }& mag  & 1E & $i_2-Z$ colour index   (dereddened) \\
{\tt COLOUR_GAAP_Z_Y }& mag  & 1E & $Z-Y$ colour index   (dereddened)         \\
{\tt COLOUR_GAAP_Y_J }& mag  & 1E & $Y-J$ colour index   (dereddened)         \\
{\tt COLOUR_GAAP_J_H }& mag  & 1E & $J-H$ colour index  (dereddened)          \\
{\tt COLOUR_GAAP_H_Ks }& mag  & 1E & $H-K_{\rm s}$ colour index   (dereddened)         \\
{\tt DIFF_GAAP_i1_i2 }& mag  & 1E & $i_1-i_2$ $i$-band magnitude difference         \\

\hline
\end{longtable}
\tablefoot{
\tablefoottext{a}{See Sect.~\ref{sec:mosaics} for the definitions of the flux and magnitude zero-points.}
\tablefoottext{b}{Definition of the spectral types: 1=CWW-Ell, 2=CWW-Sbc, 3=CWW-Scd, 4=CWW-Im, 5=KIN-SB3, 6=KIN-SB2}
\tablefoottext{c}{For the meaning of the mask see Table~\ref{tab:maskbits}}
\tablefoottext{d}{Stellar masses are computed in the same manner as described in \citet{wright/etal:2019}}
}

\end{center}

% \clearpage
%\twocolumn

%\clearpage
%\onecolumn
\begin{center}
\begin{longtable}{llll}
\caption{\label{tab:eso_singlecats} Columns provided in the single-band source lists.}\\
\hline\hline
Label & Format & Unit & Description \\
\hline
\endfirsthead

\multicolumn{4}{c}{\tablename\ \thetable{} continued.}\\
\hline\hline
Label & Format & Unit & Description \\
\hline
\endhead

\hline
\multicolumn{4}{r}{{Continued on next page}}\\
\endfoot

\hline
\endlastfoot

2DPHOT & J & Source & classification \citep[see sect.~4.5.1 in][]{dejong/etal:2015}\\
X\_IMAGE & E & pixel & Object position along x      \\
Y\_IMAGE & E & pixel & Object position along y      \\
NUMBER & J &  & Running object number        \\
CLASS\_STAR & E &  & {\sc SExtractor} S/G classifier        \\
FLAGS & J &  & Extraction flags         \\
IMAFLAGS\_ISO & J &  & FLAG-image flags summed over the iso. profile\\
NIMAFLAG\_ISO & J &  & Number of flagged pixels entering IMAFLAGS\_ISO\\
FLUX\_RADIUS & E & pixel & Radius containing half of the flux \\ KRON\_RADIUS & E & pixel & Kron apertures in units of A or B  \\
FWHM\_IMAGE & E & pixel & FWHM assuming a Gaussian core     \\
ISOAREA\_IMAGE & J & pixel$^2$ & Isophotal area above Analysis threshold     \\
ELLIPTICITY & E &  & 1 - B\_IMAGE/A\_IMAGE        \\
THETA\_IMAGE & E & deg & Position angle (CCW/x)       \\
MAG\_AUTO & E & mag & Kron-like elliptical aperture magnitude      \\
MAGERR\_AUTO & E & mag & RMS error for AUTO magnitude     \\
ALPHA\_J2000 & D & deg & Right ascension of barycentre (J2000)     \\
DELTA\_J2000 & D & deg & Declination of barycentre (J2000)      \\
FLUX\_APER\_2 & E & count & Flux within circular aperture of diameter 2 pixels  \\
... & ... & ... & ...         \\
FLUX\_APER\_200 & E & count & Flux within circular aperture of diameter 200 pixels   \\
FLUXERR\_APER\_2 & E & count & RMS error for flux within aperture of diameter 2 pixels  \\
... & ... & ... & ...         \\
FLUXERR\_APER\_200 & E & count & RMS error for flux within aperture of diameter 200 pixels  \\
MAG\_ISO & E & mag & Isophotal magnitude        \\
MAGERR\_ISO & E & mag & RMS error for isophotal magnitude     \\
MAG\_ISOCOR & E & mag & Corrected isophotal magnitude    (deprecated)   \\
MAGERR\_ISOCOR & E & mag & RMS error for corrected isophotal magnitude    \\
MAG\_BEST & E & mag & Best of MAG\_AUTO and MAG\_ISOCOR     \\
MAGERR\_BEST & E & mag & RMS error for MAG\_BEST      \\
BACKGROUND & E & count & Background at centroid position      \\
THRESHOLD & E & count & Detection threshold above background      \\
MU\_THRESHOLD & E & arcsec$^{-2}$ & Detection threshold above background      \\
FLUX\_MAX & E & count & Peak flux above background      \\
MU\_MAX & E & arcsec$^{-2}$ & Peak surface brightness above background     \\
ISOAREA\_WORLD & E & deg$^2$ & Isophotal area above Analysis threshold     \\
XMIN\_IMAGE & J & pixel & Minimum x-coordinate among detected pixels     \\
YMIN\_IMAGE & J & pixel & Minimum y-coordinate among detected pixels     \\
XMAX\_IMAGE & J & pixel & Maximum x-coordinate among detected pixels     \\
YMAX\_IMAGE & J & pixel & Maximum y-coordinate among detected pixels     \\
X\_WORLD & D & deg & Barycentre position along world x axis    \\
Y\_WORLD & D & deg & Barycentre position along world y axis    \\
XWIN\_IMAGE & E & pixel & Windowed position estimate along x     \\
YWIN\_IMAGE & E & pixel & Windowed position estimate along y     \\
X2\_IMAGE & D & pixel$^2$ & Variance along x       \\
Y2\_IMAGE & D & pixel$^2$ & Variance along y       \\
XY\_IMAGE & D & pixel$^2$ & Covariance between x and y     \\
X2\_WORLD & E & deg$^2$ & Variance along X-WORLD (alpha)      \\
Y2\_WORLD & E & deg$^2$ & Variance along Y-WORLD (delta)      \\
XY\_WORLD & E & deg$^2$ & Covariance between X-WORLD and Y-WORLD     \\
CXX\_IMAGE & E & pixel$^{-2}$ & Cxx object ellipse parameter      \\
CYY\_IMAGE & E & pixel$^{-2}$ & Cyy object ellipse parameter      \\
CXY\_IMAGE & E & pixel$^{-2}$ & Cxy object ellipse parameter      \\
CXX\_WORLD & E & deg$^{-2}$ & Cxx object ellipse parameter (WORLD units)    \\
CYY\_WORLD & E & deg$^{-2}$ & Cyy object ellipse parameter (WORLD units)    \\
CXY\_WORLD & E & deg$^{-2}$ & Cxy object ellipse parameter (WORLD units)    \\
A\_IMAGE & D & pixel & Profile RMS along major axis     \\
B\_IMAGE & D & pixel & Profile RMS along minor axis     \\
A\_WORLD & E & deg & Profile RMS along major axis (WORLD units)   \\
B\_WORLD & E & deg & Profile RMS along minor axis (WORLD units)   \\
THETA\_WORLD & E & deg & Position angle (CCW/world-x)       \\
THETA\_J2000 & E & deg & Position angle (east of north) (J2000)    \\
ELONGATION & E & deg & A\_IMAGE/B\_IMAGE         \\
ERRX2\_IMAGE & E & pixel$^2$ & RMS error on variance of position along x     \\
ERRY2\_IMAGE & E & pixel$^2$ & RMS error on variance of position along y     \\
ERRXY\_IMAGE & E & pixel$^2$ & RMS error on covariance between x and y position  \\
ERRX2\_WORLD & E & deg$^2$ & RMS error on variance of position along X-WORLD (alpha)    \\
ERRY2\_WORLD & E & deg$^2$ & RMS error on variance of position along Y-WORLD (delta)    \\
ERRXY\_WORLD & E & deg$^2$ & RMS error on covariance between X-WORLD and Y-WORLD      \\
ERRCXX\_IMAGE & E & pixel$^{-2}$ & Cxx error ellipse parameter      \\
ERRCYY\_IMAGE & E & pixel$^{-2}$ & Cyy error ellipse parameter      \\
ERRCXY\_IMAGE & E & pixel$^{-2}$ & Cxy error ellipse parameter      \\
ERRCXX\_WORLD & E & deg$^{-2}$ & Cxx error ellipse parameter (WORLD units)    \\
ERRCYY\_WORLD & E & deg$^{-2}$ & Cyy error ellipse parameter (WORLD units)    \\
ERRCXY\_WORLD & E & deg$^{-2}$ & Cxy error ellipse parameter (WORLD units)    \\
ERRA\_IMAGE & E & pixel & RMS position error along major axis    \\
ERRB\_IMAGE & E & pixel & RMS position error along minor axis    \\
ERRA\_WORLD & E & deg & World RMS position error along major axis   \\
ERRB\_WORLD & E & deg & World RMS position error along minor axis   \\
ERRTHETA\_IMAGE & E & deg & Error ellipse position angle (CCW/x)     \\
ERRTHETA\_WORLD & E & deg & Error ellipse position angle (CCW/world-x)     \\
ERRTHETA\_J2000 & E & deg & J2000 error ellipse pos. angle (east of north)  \\
FWHM\_WORLD & E & deg & FWHM assuming a Gaussian core     \\
ISO0 & J & pixel$^2$ & Isophotal area at level zero     \\
ISO1 & J & pixel$^2$ & Isophotal area at level one   \\
ISO2 & J & pixel$^2$ & Isophotal area at level two   \\
ISO3 & J & pixel$^2$ & Isophotal area at level three     \\
ISO4 & J & pixel$^2$ & Isophotal area at level four  \\
ISO5 & J & pixel$^2$ & Isophotal area at level five  \\
ISO6 & J & pixel$^2$ & Isophotal area at level six   \\
ISO7 & J & pixel$^2$ & Isophotal area at level seven \\
SLID & K &  & \textsc{Astro-WISE} SourceList identifier       \\
SID & K &  & \textsc{Astro-WISE} source identifier     \\
HTM & K &  & Hierarchical Triangular Mesh (level 25)      \\
FLAG & K &  & Not used         \\
\hline
\end{longtable}
\end{center} 
%\clearpage
\twocolumn

\begin{table*}
    \centering
    \caption{Changes to variable names and definitions in \drfive\ compared to \drfour.}
    \begin{tabular}{ccc}
    \multicolumn{3}{c}{Variables Name Changes} \\
    Previous Name & New Name & Definition\\
    \hline
      {\tt MAG_AUTO}   & {\tt MAG_AUTO_raw}    &  Uncorrected \rband\ AUTO magnitude \\
      {\tt MAG_GAAP_i} & -- & Deleted \\
      -- & {\tt MAG_GAAP_i1} & New column\\
      -- & {\tt MAG_GAAP_i2} & New column\\
      -- & {\tt FLUX_GAAP_i1} & New column\\
      -- & (etc. for errors) & \\
         &    &   \\
    \multicolumn{3}{c}{} \\
    \end{tabular}
    \begin{tabular}{ccc}
    \multicolumn{3}{c}{} \\
    \multicolumn{3}{c}{Variables Definition Changes}\\
    Name & Old Definition & New Definition\\
    \hline
      {\tt MASK}   & Defined as in \citet{kuijken/etal:2019} & Defined as in Table \ref{tab:maskbits}    \\
    \end{tabular}
    \label{tab:eso_changes}
\end{table*} 

\begin{table*}
\caption{Main keywords in the ten-band catalogue headers}\label{tab:eso_headers}
\begin{tabular}{ll}
\hline\hline
Keyword & Description \\
\hline
{\tt RA    }  &                 Field centre (J2000) (deg) \\
{\tt DEC   }  &                Field centre (J2000) (deg) \\
{\tt PROV[1,2,3,4] }  & Originating $[u,g,r,i_1,i_2]$-band co-add file name\\
{\tt FPRA[1,2,3,4] }  &         Footprint [SE,NE,NW,SW] corner RA (deg) \\
{\tt FPDE[1,2,3,4] }  &       Footprint [SE,NE,NW,SW] corner DEC (deg) \\
{\tt CALSTARS }&   Number of stars used for \gaia\ calibration ({\tt MIN_APER}=0\farcs7) \\
{\tt DMAG_[U,G,R,I1]}  &   SLR+Gaia  $[u,g,r,i_1]$-band offset  ({\tt MIN_APER}=0\farcs7) \\
{\tt DMAG_[U,G,R,I2]}  &   SLR+Gaia  $[u,g,r,i_2]$-band offset  ({\tt MIN_APER}=0\farcs7) \\
{\tt CALSTR_1 }& Number of stars used for \gaia\ calibration ({\tt MIN_APER}=1\farcs0) \\
{\tt DMAG_[U,G,R,I1]_1} &  SLR+Gaia  $[u,g,r,i_1]$-band offset  ({\tt MIN_APER}=1\farcs0) \\
{\tt DMAG_[U,G,R,I1]_2} &  SLR+Gaia  $[u,g,r,i_2]$-band offset  ({\tt MIN_APER}=1\farcs0) \\
{\tt OB[U,G,R,I1,I2]_STRT} &  $[u,g,r,i_1,i_2]$-band Observing Block start \\
{\tt ASSON1 } & Associated ten-band mask\\
{\tt ASSON2 } & Associated {\sc theli} \rband\ detection image \\
{\tt ASSON3 } & Associated {\sc theli} \rband weight image \\
\hline
\end{tabular}
\end{table*}

\end{appendix}

\end{document}